\documentclass[11pt]{article}
\usepackage[english]{babel}
\usepackage{geometry}
\usepackage{amsmath,mathtools,amssymb}
\usepackage{color}
\usepackage{nccmath}
\usepackage{xstring,stringstrings}
\usepackage{pgf}
\usepackage{calc}
\usepackage{comment}
\usepackage{tensor}
\usepackage{bbold}
\usepackage[colorlinks,linkcolor=blue]{hyperref}
\usepackage{cite}

\DeclareMathAlphabet{\mathdutchcal}{U}{dutchcal}{m}{n}
\DeclareMathAlphabet{\mathpzc}{OT1}{pzc}{m}{it}

\numberwithin{equation}{section}

\newcommand{\sd}[1]{{\dot{#1}}}
\newcommand{\su}[1]{{\underline{#1}}}
\newcommand{\ct}[0]{\dagger}
\newcommand{\cc}[0]{*}
\newcommand{\p}[0]{|}
\newcommand{\pp}[0]{\Arrowvert}
\newcommand{\exd}[0]{\mathrm{d}}
\newcommand{\intp}[0]{\iota}
\newcommand{\lied}[0]{L}
\newcommand{\clied}[0]{\mathcal{L}}
\newcommand{\del}[0]{\partial}
\newcommand{\delb}[0]{\bar{\partial}}
\newcommand{\delt}[0]{\tilde{\del}}
\newcommand{\dsg}[0]{\delta^\text{\tiny{SG}}}
\newcommand{\Dsg}[0]{\Delta^\text{\tiny{SG}}}
\newcommand{\wR}[0]{w_R}
\newcommand{\lacR}[0]{{\alpha_R}}
\newcommand{\tr}[0]{{\textsf{T}}}
\newcommand{\cD}[0]{\mathcal{D}}
\newcommand{\cDb}[0]{\bar{\cD}}
\newcommand{\cDt}[0]{\tilde{\cD}{}}
\newcommand{\cA}[0]{\mathcal{A}}
\newcommand{\At}[0]{\tilde{A}}
\newcommand{\ca}[0]{\boldsymbol{a}}
\newcommand{\caR}[0]{\ca_R}
\newcommand{\cl}[0]{\boldsymbol{\lambda}}
\newcommand{\clb}[0]{\boldsymbol{\bar{\lambda}}{}}
\newcommand{\cG}[0]{\mathcal{G}}
\newcommand{\Md}[0]{\mathbf{M}}
\newcommand{\Mbd}[0]{\conj{\mathbf{M}}}
\newcommand{\bd}[0]{\mathbf{b}}
\newcommand{\Fd}[0]{\mathbf{F}}
\newcommand{\Fbd}[0]{\conj{\mathbf{F}}}
\newcommand{\Dd}[0]{\hat{\mathbf{D}}}
\newcommand{\cS}[0]{\mathcal{S}}
\newcommand{\cF}[0]{\mathcal{F}}
\newcommand{\cf}[0]{\boldsymbol{f}}
\newcommand{\cW}[0]{\mathcal{W}}
\newcommand{\cWb}[0]{\bar{\cW}}
\newcommand{\cR}[0]{\mathcal{R}}
\newcommand{\cL}[0]{\mathcal{L}}
\newcommand{\cK}[0]{\mathcal{K}}
\newcommand{\cKb}[0]{\bar{\mathcal{K}}{}}
\newcommand{\cO}[0]{\mathcal{O}}
\newcommand{\Db}[0]{\boldsymbol{D}}
\newcommand{\Dt}[0]{\tilde{D}{}}
\newcommand{\g}[1]{{(#1)}}
\newcommand{\sco}[3]{{c_{\g{#1}\g{#2}}}^\g{#3}}
\newcommand{\gen}[1]{\boldsymbol{T}_\g{#1}}
\newcommand{\gkf}[2]{f_{\g{#1}\g{#2}}}
\newcommand{\gkfb}[2]{\bar{f}_{\g{#1}\g{#2}}}
\newcommand{\soge}[0]{\Lambda}
\newcommand{\soae}[0]{L}
\newcommand{\soaet}[0]{\tilde{\soae}{}}
\newcommand{\uge}[0]{u}
\newcommand{\uae}[0]{y}

\newcommand{\lge}[0]{\boldsymbol{g}}
\newcommand{\lae}[0]{\boldsymbol{y}}
\newcommand{\lac}[1]{\alpha^\g{#1}}
\newcommand{\lact}[1]{\tilde{\alpha}{}^\g{#1}}
\newcommand{\com}[2]{[#1,#2]}
\newcommand{\acom}[2]{\{#1,#2\}}
\newcommand{\mcom}[2]{(#1,#2)}
\newcommand{\conj}[1]{\overline{#1}}
\newcommand{\sig}[0]{\sigma}
\newcommand{\sigb}[0]{\bar{\sig}{}}
\newcommand{\etab}[0]{\bar{\eta}}
\newcommand{\zetab}[0]{\bar{\zeta}}
\newcommand{\varphib}[0]{\bar{\varphi}{}}
\newcommand{\phib}[0]{\bar{\phi}{}}
\newcommand{\psib}[0]{\bar{\psi}{}}
\newcommand{\chib}[0]{\bar{\chi}{}}
\newcommand{\Phib}[0]{\conj{\Phi}{}}

\newcommand{\Xib}[0]{\conj{\Xi}{}}
\newcommand{\lambdab}[0]{\bar{\lambda}{}}

\newcommand{\Fb}[0]{\conj{F}{}}
\newcommand{\Tb}[0]{\conj{T}{}}
\newcommand{\Pb}[0]{\conj{P}{}}
\newcommand{\Wb}[0]{\conj{W}{}}
\newcommand{\Rb}[0]{\conj{R}{}}

\newcommand{\Hb}[0]{\conj{H}{}}
\newcommand{\Xb}[0]{\conj{X}{}}
\newcommand{\Ub}[0]{\conj{U}{}}
\newcommand{\Zb}[0]{\conj{Z}{}}
\newcommand{\Mb}[0]{\conj{M}{}}
\newcommand{\Pib}[0]{\conj{\Pi}{}}
\newcommand{\Sigmab}[0]{\conj{\Sigma}{}}

\newcommand{\Lambdat}[0]{\tilde{\Lambda}{}}
\newcommand{\fb}[0]{\bar{f}{}}
\newcommand{\sbb}[0]{\bar{s}{}}
\newcommand{\rb}[0]{\bar{r}{}}
\newcommand{\ib}[0]{{\bar{i}}}
\newcommand{\jb}[0]{{\bar{j}}}
\newcommand{\kb}[0]{{\bar{k}}}
\newcommand{\lb}[0]{{\bar{l}}}

\newcommand{\xib}[0]{\bar{\xi}{}}
\newcommand{\thetab}[0]{\bar{\theta}{}}
\newcommand{\deltat}[0]{\tilde{\delta}}
\newcommand{\Nu}[0]{\mathcal{V}}
\newcommand{\eps}[0]{\epsilon}
\newcommand{\so}[0]{\mathfrak{so}}
\newcommand{\uu}[0]{\mathfrak{u}}
\newcommand{\SO}[0]{\mathrm{SO}}
\newcommand{\U}[0]{\mathrm{U}}
\newcommand{\UK}[0]{\U_\mathrm{K}}
\newcommand{\UR}[0]{\U(1)_R}
\newcommand{\s}[0]{\:}
\newcommand{\sym}[2]{
	\pgfmathparse{#1 *0.4}
	\underset{\rotatebox{90}{\scalebox{1}[\pgfmathresult]{(}}}{#2}
}

\DeclareMathOperator{\sign}{sign}
\DeclareMathOperator{\diag}{diag}
\DeclareMathOperator{\re}{Re}
\DeclareMathOperator{\im}{Im}
\DeclareMathOperator{\sdet}{sdet}
\DeclareMathOperator{\trace}{tr}

\begin{document}

\begin{titlepage}

\vspace*{-15mm}
\vspace*{0.7cm}

\begin{center}

{\Large {\bf Notes on the derivation of the general supergravity/matter/Yang-Mills Lagrangian for $N=1$ supersymmetry in $d=4$ dimensions using superspace techniques}}\\[8mm]

Christian Hohl$^{\star}$\footnote{Email: \texttt{ch.hohl@unibas.ch}}

\end{center}

\vspace*{0.20cm}

\centerline{$^{\star}$ \it
Department of Physics, University of Basel,}
\centerline{\it
Klingelbergstr.\ 82, CH-4056 Basel, Switzerland}

\vspace*{1.2cm}

\begin{abstract}
\noindent The coupling of matter to supergravity with $N=1$ supersymmetry in $d=4$ dimensions is described in a geometric manner by K\"ahler superspace. A straightforward way to implement K\"ahler superspace is via $\U(1)$ superspace by identifying the $\U(1)$ pre-potential with the K\"ahler potential, which is a function of the matter (chiral) superfields. In this framework, the components of the supergravity multiplet are contained in the supervielbein and torsion tensor of superspace. Furthermore, interactions with the Yang-Mills (vector) multiplet are formulated by introducing a connection $1$-superform of an additional gauge structure. In these notes, the Bianchi identities in $\U(1)$ superspace are solved for a particular set of torsion constraints which lead to the minimal supergravity multiplet. Moreover, the solution of the Bianchi identities in the gauge sector is derived and K\"ahler superspace is defined. At the superfield level, the general action of the supergravity/matter/Yang-Mills system and supergravity transformations are formulated, and the equations of motion are deduced. Using projection to lowest components in superspace, the corresponding Lagrangian and supergravity transformations at the component field level are calculated, and the equations of motion of the auxiliary fields are determined. Compared to existing literature, these notes provide a self-contained and consistent step by step derivation of the general supergravity/matter/Yang-Mills Lagrangian for $N=1$ supersymmetry in $d=4$ dimensions by means of superspace techniques.
\end{abstract}
\end{titlepage}

\tableofcontents
\newpage

\section{Overview}
\begin{itemize}
\item \textbf{Organisation of the notes:}\\
The notes are organized as follows: in Section~\ref{sec:conventions} definitions and useful identities related to Pauli matrices and Weyl spinors are listed. In Section~\ref{sec:superspace} the supervielbein and superforms are defined and $\U(1)$ superspace is introduced, including the torsion tensor and the Lorentz and $\U(1)$ connection. Furthermore, the Bianchi identities in $\U(1)$ superspace are presented, which are then solved in Section~\ref{sec:bianchi_sp_solution_all}, by taking into account a particular set of torsion constraints. These constraints lead to the minimal supergravity multiplet. The Yang-Mills connection as well as the matter (chiral) superfields are introduced in Section~\ref{sec:matter_and_yang_mills}, where also the Bianchi identity with respect to the gauge group is solved, considering a set of constraints of the Yang-Mills field strength. In Section~\ref{sec:kahler_superspace} K\"ahler superspace is derived from $\U(1)$ superspace by a suitable identification of the K\"ahler potential, which is a function of the matter superfields, with the $\U(1)$ pre-potential. Moreover, supergravity transformations are defined and the general action of the supergravity/matter/Yang-Mills system at the superfield level is stated. By considering variations of the action which are compatible with the torsion and Yang-Mills field strength constraints as well as with the chirality condition of the matter superfields, the equations of motion of the superfields are derived in Section~\ref{sec:equations_of_motion}. In Section~\ref{sec:component_fields} the component fields of the supergravity, the matter and the Yang-Mills multiplet (including auxiliary fields) are defined via projection to lowest components in superspace, and the corresponding supergravity transformations are determined. In addition the general (off-shell) Lagrangian of the supergravity/matter/Yang-Mills system at the component field level is derived and the equations of motion of the auxiliary fields are calculated.
\item \textbf{Literature:}\\
The following list of references represents only a selection of the existing literature of the different topics and is far away from complete. An introduction to superspace geometry can be found in~\cite{DeWitt:1992cy,Moura:2008ra,Wess:1992cp,Buchbinder:1998qv}. Furthermore, $\U(1)$ superspace is explicitly discussed in~\cite{Binetruy:2000zx,Muller:1985vga}. The underlying mathematical framework of superspace are supermanifolds, which are considered in a mathematical context in~\cite{Groeger:2010,Fioresi:2010,Helein:2008,Goertsches:2006pna}. At the component field level, the supergravity Lagrangian, including matter and gauge multiplets, is derived in~\cite{Binetruy:2000zx,Wess:1992cp,Buchbinder:1998qv,Freedman:2012zz}. In general, the calculations in Sections~\ref{sec:bianchi_sp_solution_all}--\ref{sec:component_fields} follow the considerations in \cite{Binetruy:2000zx}.
\end{itemize}

\section{Notations and identities}
\label{sec:conventions}
\subsection{Conventions}
\begin{itemize}
\item \textbf{Minkowski metric:}\\
In all the subsequent sections the mostly plus convention for the Minkowski metric,
\begin{align}
\eta_{ab} = \eta^{ab} &= \diag(-1,+1,+1,+1)\,, \label{eq:identities_conventions_id_1}
\end{align}
is used, with the Lorentz vector indices $a,b\in\{0,1,2,3\}$. The matrix $\eta^{ab}$ is the inverse of $\eta_{ab}$. These two matrices are employed to raise and lower Lorentz vector indices.
\item \textbf{Levi-Civita tensors:}\\
The totally antisymmetric Levi-Civita tensor $\eps_{abcd}$ with four Lorentz vector indices is defined as
\begin{align}
\eps_{0123} &= +1\,, & \eps^{0123} &= -1\,, \label{eq:identities_conventions_id_2}
\end{align}
such that
\begin{align}
\eps_{abcd} &= \eta_{ae}\eta_{bf}\eta_{cg}\eta_{dh}\eps^{efgh}\,, & \eps^{abcd} &= \eta^{ae}\eta^{bf}\eta^{cg}\eta^{dh}\eps_{efgh}\,, \label{eq:identities_conventions_id_3}
\end{align}
and
\begin{align}
\eps_{abcd}\eps^{efcd} &= -2(\delta_a^e\delta_b^f - \delta_a^f\delta_b^e)\,, \label{eq:identities_conventions_id_4}\\
\eps_{abcd}\eps^{ebcd} &= -6\delta_a^e\,. \label{eq:identities_conventions_id_5}
\end{align}
Furthermore, the totally antisymmetric Levi-Civita tensor $\eps^{\alpha\beta}$ ($\eps^{\sd{\alpha}\sd{\beta}}$) with two (Lorentz) Weyl spinor indices $\alpha,\beta\in\{1,2\}$ ($\sd{\alpha},\sd{\beta}\in\{1,2\}$) is defined as follows:
\begin{align}
\eps^{12} = \eps_{21}\; (= \eps^{\sd{1}\sd{2}} = \eps_{\sd{2}\sd{1}}) = +1\,. \label{eq:identities_conventions_id_6}
\end{align}
Thus, $\eps^{\alpha\beta}$ and $\eps_{\alpha\beta}$ are the inverses of each other, namely
\begin{align}
\eps^{\alpha\gamma}\eps_{\gamma\beta} = \eps_{\beta\gamma}\eps^{\gamma\alpha} = \delta^\alpha_\beta\,, \label{eq:identities_conventions_id_7}
\end{align}
and they are used to raise and lower spinor indices. Finally, the two matrices obey the relations
\begin{align}
\eps^{\alpha\beta} &= -\eps^{\alpha\gamma}\eps^{\beta\delta}\eps_{\gamma\delta}\,, & \eps_{\alpha\beta} &= -\eps_{\alpha\gamma}\eps_{\beta\delta}\eps^{\gamma\delta}\,. \label{eq:identities_conventions_id_8}
\end{align}
Note that the identities in Eq.~\eqref{eq:identities_conventions_id_7} and \eqref{eq:identities_conventions_id_8} are also valid for $\eps^{\sd{\alpha}\sd{\beta}}$.
\item \textbf{Grassmann parity:}\\
In general, commuting and anti-commuting objects $\cO$ are referred to as $c$-type and $a$-type, respectively, with Grassmann parity $\eps(\cO)=0$ and $\eps(\cO)=1$. A general object may consist of a sum of a $c$- and an $a$-type component, consequently it has no definite commutation property. The object is called pure (of $c$- or $a$-type), if one of the components vanishes:
\begin{align}
\eps(\cO) &= \left\{
\begin{array}{l}
0 \quad\text{if}\quad \cO\text{ is }c\text{-type} \\
1 \quad\text{if}\quad \cO\text{ is }a\text{-type}
\end{array}\right. \label{eq:identities_conventions_id_9}
\end{align}
The Grassmann parity of the product of two pure objects is the sum of the two individual Grassmann parities modulo $2$:
\begin{align}
\eps(\cO_1\cO_2) &= \eps(\cO_1) + \eps(\cO_2)\quad\pmod 2\,. \label{eq:identities_conventions_id_10}
\end{align}
If the order of the two objects in the product is reversed, there is a factor $+1$ or $-1$ depending on the product of the two Grassmann parities, namely
\begin{align}
\cO_1 \cO_2 &= (-1)^{\eps(\cO_1)\eps(\cO_2)}\cO_2\cO_1 =: (-1)^{\cO_1\cO_2}\cO_2\cO_1\,. \label{eq:identities_conventions_id_11}
\end{align}
It is convenient to use the abbreviations
\begin{align}
(-1)^\cO &\equiv (-1)^{\eps(\cO)}\,, \label{eq:identities_conventions_id_12}\\
(-1)^{\cO_1\cO_2} &\equiv (-1)^{\eps(\cO_1)\eps(\cO_2)}\,. \label{eq:identities_conventions_id_13}
\end{align}
Moreover, if an object $\cO_A$ with a Lorentz index $A\in\{a,\alpha,\sd{\alpha}\}$ has the Grassmann parity $\eps(\cO_a)=0$ and $\eps(\cO_\alpha)=\eps(\cO^\sd{\alpha})=1$, it is handy to write
\begin{align}
(-1)^A &\equiv (-1)^{\eps(\cO_A)}\,, \\
(-1)^{AB} &\equiv (-1)^{\eps(\cO_A)\eps(\cO_B)}\,.
\end{align}
\item \textbf{Symmetric spinor indices:}\\
If objects $\cO_{\alpha_1...\alpha_n}$ and $\cO_{\sd{\alpha}_1...\sd{\alpha}_n}$ are symmetric in the spinor indices $\alpha_1...\alpha_n$ and $\sd{\alpha}_1...\sd{\alpha}_n$, respectively, this is often indicated by using the notation
\begin{align}
\cO_{\sym{6}{\alpha_1...\alpha_n}}\,,\quad \cO_{\sym{6}{\sd{\alpha}_1...\sd{\alpha}_n}}\,.
\end{align}
\end{itemize}

\subsection{Pauli matrices}
\begin{itemize}
\item \textbf{Definition:}\\
The Pauli matrices $\sig^a_{\beta\sd{\gamma}}$ and $\sigb^{a\sd{\beta}\gamma}$ are defined as
\begin{align}
\sig^0 = +\sigb^0 &= \begin{pmatrix}1&0\\0&1\end{pmatrix}, & \sig^1 = -\sigb^1 &= \begin{pmatrix}0&1\\1&0\end{pmatrix}, \nonumber\\\label{eq:identities_pauli_id_1}\\
\sig^2 = -\sigb^2 &= \begin{pmatrix}0&-i\\i&0\end{pmatrix}, & \sig^3 = -\sigb^3 &= \begin{pmatrix}1&0\\0&-1\end{pmatrix}. \nonumber
\end{align}
The matrices $\sig^a$ and $\sigb^a$ are Hermitian, i.e.\
\begin{align}
(\sig^c_{\alpha\sd{\beta}})^\cc &= \sig^c_{\beta\sd{\alpha}}\,, & (\sigb^{c\sd{\alpha}\beta})^\cc &= \sigb^{c\sd{\beta}\alpha}\,, \label{eq:identities_pauli_id_2}
\end{align}
and they are converted into each other by using the Levi-Civita tensor
\begin{align}
\eps^{\beta\alpha}\eps^{\sd{\beta}\sd{\alpha}}\sig^a_{\alpha\sd{\alpha}} &= \sigb^{a\sd{\beta}\beta}\,, & \eps_{\sd{\beta}\sd{\alpha}}\eps_{\beta\alpha}\sigb^{a\sd{\alpha}\alpha} &= \sig^a_{\beta\sd{\beta}}\,, \label{eq:identities_pauli_id_3}\\
\eps^{\beta\alpha}\sig^a_{\alpha\sd{\alpha}} &= \eps_{\sd{\alpha}\sd{\beta}}\sigb^{a\sd{\beta}\beta}\,, & \eps^{\sd{\beta}\sd{\alpha}}\sig^a_{\alpha\sd{\alpha}} &= \eps_{\alpha\beta}\sigb^{a\sd{\beta}\beta}\,. \label{eq:identities_pauli_id_4}
\end{align}
It is convenient to use the following notations:
\begin{align}
{(\eps\sig^a)^\beta}_\sd{\alpha} &= \eps^{\beta\alpha}\sig_{\alpha\sd{\alpha}}\,, & {(\eps\sigb^a)_\sd{\beta}}^\alpha &= \eps_{\sd{\beta}\sd{\alpha}}\sigb^{a\sd{\alpha}\alpha}\,, \label{eq:identities_pauli_id_5}\\
{(\sig^a\eps)_\alpha}^\sd{\beta} &= \sig^a_{\alpha\sd{\alpha}}\eps^{\sd{\alpha}\sd{\beta}}\,, & {(\sigb^a\eps)^\sd{\alpha}}_\beta &= \sigb^{a\sd{\alpha}\alpha}\eps_{\alpha\beta}\,. \label{eq:identities_pauli_id_6}
\end{align}
The matrices $\sig^{ab}$ and $\sigb^{ab}$ are defined as
\begin{align}
{(\sig^{ab})_\alpha}^\beta &= \frac{1}{4}{(\sig^a\sigb^b-\sig^b\sigb^a)_\alpha}^\beta\,, & {(\sigb^{ab})^\sd{\alpha}}_\sd{\beta} &= \frac{1}{4}{(\sigb^a\sig^b-\sigb^b\sig^a)^\sd{\alpha}}_\sd{\beta}\,, \label{eq:identities_pauli_id_7}
\end{align}
thus, the product of two Pauli matrices has the form
\begin{align}
{(\sig^a\sigb^b)_\alpha}^\beta &= -\eta^{ab}\delta_\alpha^\beta + 2{(\sig^{ab})_\alpha}^\beta\,, & {(\sigb^a\sig^b)^\sd{\alpha}}_\sd{\beta} &= -\eta^{ab}\delta^\sd{\alpha}_\sd{\beta} + 2{(\sigb^{ab})^\sd{\alpha}}_\sd{\beta}\,. \label{eq:identities_pauli_id_8}
\end{align}
Moreover, the definition in Eq.~\eqref{eq:identities_pauli_id_7} implies that $\sig^{ab}$ and $\sigb^{ab}$ are traceless and antisymmetric in the vector indices:
\begin{align}
\trace{\sig^{ab}} &= 0\,, & \trace{\sigb^{ab}} &= 0\,, \label{eq:identities_pauli_id_9} \\
\sig^{ba} &= -\sig^{ab}\,, & \sigb^{ba} &= -\sigb^{ab}\,. \label{eq:identities_pauli_id_10}
\end{align}
They are transformed into each other by conjugation, namely
\begin{align}
\big({(\sig^{ab})_\alpha}^\beta\big)^\cc &= -{(\sigb^{ab})^\sd{\beta}}_\sd{\alpha}\,.
\end{align}
Further identities which involve the Levi-Civita tensor from Eq.~\eqref{eq:identities_conventions_id_6} are given by
\begin{align}
{(\eps\sig^{ab}\eps)^\alpha}_\beta &= -{(\sig^{ab})_\beta}^\alpha\,, & {(\eps\sigb^{ab}\eps)_\sd{\alpha}}^\sd{\beta} &= -{(\sigb^{ab})^\sd{\beta}}_\sd{\alpha}\,, \\
(\eps\sig^{ab})^{\alpha\beta} &= (\eps\sig^{ab})^{\beta\alpha}\,, & (\eps\sigb^{ab})_{\sd{\alpha}\sd{\beta}} &= (\eps\sigb^{ab})_{\sd{\beta}\sd{\alpha}}\,, \\
(\sig^{ab}\eps)_{\alpha\beta} &= (\sig^{ab}\eps)_{\beta\alpha}\,, & (\sigb^{ab}\eps)^{\sd{\alpha}\sd{\beta}} &= (\sigb^{ab}\eps)^{\sd{\beta}\sd{\alpha}}\,,
\end{align}
where
\begin{align}
{(\eps\sig^{ab}\eps)^\alpha}_\beta &= \eps^{\alpha\gamma}{(\sig^{ab})_\gamma}^\delta\eps_{\delta\beta}\,, & {(\eps\sigb^{ab}\eps)_\sd{\alpha}}^\sd{\beta} &= \eps_{\sd{\alpha}\sd{\gamma}}{(\sigb^{ab})^\sd{\gamma}}_\sd{\delta}\eps^{\sd{\delta}\sd{\beta}}\,.
\end{align}
Using the Levi-Civita tensor from Eq.~\eqref{eq:identities_conventions_id_2}, the self-duality relations are written as
\begin{align}
\eps_{abcd}\sig^{cd} &= -2i\sig_{ab}\,, & \eps_{abcd}\sigb^{cd} &= +2i\sigb_{ab}\,, \\
\eps^{abcd}\sig_{cd} &= +2i\sig^{ab}\,, & \eps^{abcd}\sigb_{cd} &= -2i\sigb_{ab}\,.
\end{align}
\item \textbf{Identities:}\\
The following identities, which involve two Pauli matrices,
\begin{align}
\trace(\sig^a\sigb^b) &= -2\eta^{ab}\,, \\
\sig^a_{\alpha\sd{\alpha}}\sigb_a^{\sd{\beta}\beta} &= -2\delta_\alpha^\beta\delta_\sd{\alpha}^\sd{\beta}\,, \\
\sig^a_{\alpha\sd{\alpha}}\sig_{a\beta\sd{\beta}} &= -2\eps_{\alpha\beta}\eps_{\sd{\alpha}\sd{\beta}}\,, \\
\sigb^{a\sd{\alpha}\alpha}\sigb_a^{\sd{\beta}\beta} &= -2\eps^{\alpha\beta}\eps^{\sd{\alpha}\sd{\beta}}\,,
\end{align}
stem from the more general relations
\begin{align}
\begin{split}
\sig^a_{\alpha\sd{\alpha}}\sigb^{b\sd{\beta}\beta} &= -\frac{1}{2}\delta_\alpha^\beta\delta_\sd{\alpha}^\sd{\beta}\eta^{ab} + \delta_\sd{\alpha}^\sd{\beta}{(\sig^{ab})_\alpha}^\beta - \delta_\alpha^\beta{(\sigb^{ab})^\sd{\beta}}_\sd{\alpha} \\
&\quad - {({\sig^a}_c)_\alpha}^\beta{(\sigb^{bc})^\sd{\beta}}_\sd{\alpha} - {({\sig^b}_c)_\alpha}^\beta{(\sigb^{ac})^\sd{\beta}}_\sd{\alpha}\,,
\end{split} \\
\nonumber\\
\begin{split}
\sig^a_{\alpha\sd{\alpha}}\sig^b_{\beta\sd{\beta}} &= -\frac{1}{2}\eps_{\alpha\beta}\eps_{\sd{\alpha}\sd{\beta}}\eta^{ab} + \eps_{\sd{\alpha}\sd{\beta}}(\sig^{ab}\eps)_{\alpha\beta} + \eps_{\alpha\beta}(\eps\sigb^{ab})_{\sd{\alpha}\sd{\beta}} \\
&\quad + ({\sig^a}_c\eps)_{\alpha\beta}(\eps\sigb^{bc})_{\sd{\alpha}\sd{\beta}} + ({\sig^b}_c\eps)_{\alpha\beta}(\eps\sigb^{ac})_{\sd{\alpha}\sd{\beta}}\,,
\end{split} \\
\nonumber\\
\begin{split}
\sigb^{a\sd{\alpha}\alpha}\sigb^{b\sd{\beta}\beta} &= -\frac{1}{2}\eps^{\alpha\beta}\eps^{\sd{\alpha}\sd{\beta}}\eta^{ab} + \eps^{\sd{\alpha}\sd{\beta}}(\eps\sig^{ab})^{\alpha\beta} + \eps^{\alpha\beta}(\sigb^{ab}\eps)^{\sd{\alpha}\sd{\beta}} \\
&\quad + (\eps{\sig^a}_c)^{\alpha\beta}(\sigb^{bc}\eps)^{\sd{\alpha}\sd{\beta}} + (\eps{\sig^b}_c)^{\alpha\beta}(\sigb^{ac}\eps)^{\sd{\alpha}\sd{\beta}}\,.
\end{split}
\end{align}
Furthermore, from the product of three Pauli matrices, namely
\begin{align}
(\sig^a\sigb^b\sig^c)_{\alpha\sd{\beta}} &= (-\eta^{ab}\eta^{cd} + \eta^{ca}\eta^{bd} - \eta^{bc}\eta^{ad} + i\eps^{abcd})\sig_{d\alpha\sd{\beta}}\,, \\
(\sigb^a\sig^b\sigb^c)^{\sd{\alpha}\beta} &= (-\eta^{ab}\eta^{cd} + \eta^{ca}\eta^{bd} - \eta^{bc}\eta^{ad} - i\eps^{abcd})\sigb_d^{\sd{\alpha}\beta}\,,
\end{align}
the identities
\begin{align}
(\sig^{ab}\sig^c)_{\alpha\sd{\beta}} &= \frac{1}{2}(\eta^{ac}\eta^{bd} - \eta^{bc}\eta^{ad} + i\eps^{abcd})\sig_{d\alpha\sd{\beta}}\,, \\
(\sigb^{ab}\sigb^c)^{\sd{\alpha}\beta} &= \frac{1}{2}(\eta^{ac}\eta^{bd} - \eta^{bc}\eta^{ad} - i\eps^{abcd})\sigb_d^{\sd{\alpha}\beta}\,, \\
(\sig^{a}\sigb^{bc})_{\alpha\sd{\beta}} &= \frac{1}{2}(\eta^{ac}\eta^{bd} - \eta^{ab}\eta^{cd} + i\eps^{abcd})\sig_{d\alpha\sd{\beta}}\,, \\
(\sigb^a\sig^{bc})^{\sd{\alpha}\beta} &= \frac{1}{2}(\eta^{ac}\eta^{bd} - \eta^{ab}\eta^{cd} - i\eps^{abcd})\sigb_d^{\sd{\alpha}\beta}\,,
\end{align}
are derived. On the other hand, if the sum is taken over a vector index, the relations
\begin{align}
\sig_{b\beta\sd{\beta}}{(\sig^{ab})_\alpha}^\gamma &= -\delta^\gamma_\beta\sig^a_{\alpha\sd{\beta}} + \frac{1}{2}\delta^\gamma_\alpha\sig^a_{\beta\sd{\beta}}\,, \\
\sigb_b^{\sd{\beta}\beta}{(\sigb^{ab})^\sd{\alpha}}_\sd{\gamma} &= -\delta_\sd{\gamma}^\sd{\beta}\sigb^{a\sd{\alpha}\beta} + \frac{1}{2}\delta_\sd{\gamma}^\sd{\alpha}\sigb^{a\sd{\beta}\beta}\,, \\
\sig_{b\beta\sd{\beta}}{(\sigb^{ab})^\sd{\gamma}}_\sd{\alpha} &= +\delta^\sd{\gamma}_\sd{\beta}\sig^a_{\beta\sd{\alpha}} - \frac{1}{2}\delta^\sd{\gamma}_\sd{\alpha}\sig^a_{\beta\sd{\beta}}\,, \\
\sigb_b^{\sd{\beta}\beta}{(\sig^{ab})_\gamma}^\alpha &= +\delta_\gamma^\beta\sigb^{a\sd{\beta}\alpha} - \frac{1}{2}\delta_\gamma^\alpha\sigb^{a\sd{\beta}\beta}\,,
\end{align}
apply. Another relation is given by
\begin{align}
\oint_{abc}\oint_{\alpha\beta\gamma}(\eps\sig^{ab})^{\alpha\beta}\sigb^{c\sd{\gamma}\gamma} &= 0\,, & \oint_{abc}\oint_{\sd{\alpha}\sd{\beta}\sd{\gamma}}(\eps\sigb^{ab})_{\sd{\alpha}\sd{\beta}}\sig^c_{\gamma\sd{\gamma}} &= 0\,, \label{eq:notations_cyclic_identity_pauli_matrices}
\end{align}
where $\oint$ labels a cyclic sum over the indices, i.e.\ $\oint_{abc}abc=abc+bac+cab$. A selection of identities which involve four Pauli matrices is the following: The trace of the product of two matrices $\sig^{ab}$ and $\sigb^{ab}$, respectively, is given by
\begin{align}
\trace(\sig^{ab}\sig^{cd}) &= -\frac{1}{2}(\eta^{ac}\eta^{bd} - \eta^{ad}\eta^{bc} + i\eps^{abcd})\,, \\
\trace(\sigb^{ab}\sigb^{cd}) &= -\frac{1}{2}(\eta^{ac}\eta^{bd} - \eta^{ad}\eta^{bc} - i\eps^{abcd})\,.
\end{align}
Furthermore, the commutator of two such matrices reads
\begin{align}
\com{\sig^{ab}}{\sig^{cd}} &= \eta^{ac}\sig^{bd} - \eta^{ad}\sig^{bc} - \eta^{bc}\sig^{ad} + \eta^{bd}\sig^{ac}\,, \\
\com{\sigb^{ab}}{\sigb^{cd}} &= \eta^{ac}\sigb^{bd} - \eta^{ad}\sigb^{bc} - \eta^{bc}\sigb^{ad} + \eta^{bd}\sigb^{ac}\,,
\end{align}
and the anticommutator has the form
\begin{align}
{\acom{\sig^{ab}}{\sig^{cd}}_\alpha}^\beta &= \trace(\sig^{ab}\sig^{cd})\delta_\alpha^\beta\,, \\
{\acom{\sigb^{ab}}{\sigb^{cd}}^\sd{\alpha}}_\sd{\beta} &= \trace(\sigb^{ab}\sigb^{cd})\delta^\sd{\alpha}_\sd{\beta}\,.
\end{align}
If the sum is taken over the vector indices, the identities
\begin{align}
(\eps\sig^{ab})^{\alpha\beta}(\sig_{ab}\eps)_{\gamma\delta} &= -\delta^\alpha_\gamma\delta^\beta_\delta - \delta^\alpha_\delta\delta^\beta_\gamma\,, \\
(\eps\sigb^{ab})_{\sd{\alpha}\sd{\beta}}(\sigb_{ab}\eps)^{\sd{\gamma}\sd{\delta}} &= -\delta_\sd{\alpha}^\sd{\gamma}\delta_\sd{\beta}^\sd{\delta} - \delta_\sd{\alpha}^\sd{\delta}\delta_\sd{\beta}^\sd{\gamma}\,, \\
\sig^{ab}\sigb_{ab} &= 0\,,
\end{align}
are valid. Finally, the contraction of a product of four Pauli matrices with the Levi-Civita tensor from Eq.~\eqref{eq:identities_conventions_id_2} is given by
\begin{align}
-\frac{i}{4!}\eps_{abcd}{(\sig^a\sigb^b\sig^c\sigb^d)_\alpha}^\beta &= \delta_\alpha^\beta\,, & +\frac{i}{4!}\eps_{abcd}{(\sigb^a\sig^b\sigb^c\sig^d)^\sd{\alpha}}_\sd{\beta} &= \delta^\sd{\alpha}_\sd{\beta}\,.
\end{align}
\end{itemize}
\subsection{Weyl spinors}
\begin{itemize}
\item \textbf{Definition:}\\
In this section $a$-type Weyl spinors $\chi_\alpha$, $\chib^\sd{\alpha}$, with an undotted and a dotted spinor index, respectively, are considered. Under Lorentz transformations $\chi$ transforms in the representation $(\frac{1}{2},0)$, and $\chib$ in the representation $(0,\frac{1}{2})$. In particular, $\chib^\sd{\alpha}$ with an upper index transforms in the dual conjugate representation of $\chi_\alpha$ with a lower index. The spinor indices are raised and lowered with the Levi-Civita tensor from Eq.~\eqref{eq:identities_conventions_id_6}:
\begin{align}
\chi^\alpha &= \eps^{\alpha\beta}\chi_\beta\,, & \chib_\sd{\alpha} &= \eps_{\sd{\alpha}\sd{\beta}}\chib^\sd{\beta}\,. \label{eq:identities_spinors_id_1}
\end{align}
Under conjugation, a spinor with an undotted index is transformed into a spinor with a dotted index and vice versa. Moreover, if a product of two Weyl spinors is conjugated, the Grassmann parity has to be taken into account:
\begin{align}
(\chi_\alpha)^\cc &= \chib_\sd{\alpha}\,, \label{eq:identities_spinors_id_2}\\
(\chi_\alpha\psi_\beta)^\cc &= -\chib_\sd{\alpha}\psib_\sd{\beta}\,. \label{eq:identities_spinors_id_3}
\end{align}
\item \textbf{Identities:}\\
Writing the spinor indices explicitly, the following terms which contain spinors and Pauli matrices have the form
\begin{align}
\chi\psi &= \chi^\alpha\psi_\alpha\,, & \chib\psib &= \chib_\sd{\alpha}\psib^\sd{\alpha}\,, \label{eq:identities_spinors_id_4}\\
\chi\sig^a\psib &= \chi^\alpha\sig^a_{\alpha\sd{\alpha}}\psib^\sd{\alpha}\,, & \chib\sigb^a\psi &= \chib_\sd{\alpha}\sigb^{a\sd{\alpha}\alpha}\psi_\alpha\,, \label{eq:identities_spinors_id_5}\\
\chi\sig^a\sigb^b\psi &= \chi^\alpha\sig^a_{\alpha\sd{\alpha}}\sigb^{b\sd{\alpha}\beta}\psi_\beta\,, & \chib\sigb^a\sig^b\psib &= \chib_\sd{\alpha}\sigb^{a\sd{\alpha}\alpha}\sig^b_{\alpha\sd{\beta}}\psib^\sd{\beta}\,, \label{eq:identities_spinors_id_6}\\
\chi\sig^a\sigb^b\sig^c\psib &= \chi^\alpha{\sig^a_{\alpha\sd{\alpha}}}\sigb^{b\sd{\alpha}\beta}\sig^c_{\beta\sd{\gamma}}\psi^\sd{\gamma}\,, & \chib\sigb^a\sig^b\sigb^c\psi &= \chib_\sd{\alpha}\sigb^{a\sd{\alpha}\alpha}\sig^b_{\alpha\sd{\beta}}\sigb^{c\sd{\beta}\gamma}\psi_\gamma\,. \label{eq:identities_spinors_id_7}
\end{align}
For the above products the following relations are valid:
\begin{gather}
\chi\psi = \psi\chi\,, \hspace{5.6cm} \chib\psib = \psib\chib\,, \label{eq:identities_spinors_id_8}\\
\chi\sig^a\psib = -\psib\sigb^a\chi\,, \label{eq:identities_spinors_id_9}\\
\chi\sig^a\sigb^b\psi = \psi\sig^b\sigb^a\chi\,, \hspace{4cm} \chib\sigb^a\sig^b\psib = \psib\sigb^b\sig^a\chib\,, \label{eq:identities_spinors_id_10}\\
\chi\sig^a\sigb^b\sig^c\psib = -\psib\sigb^c\sig^b\sigb^a\chi\,, \label{eq:identities_spinors_id_11}
\end{gather}
thus $\chi\sig^{ab}\psi=-\psi\sig^{ab}\chi$ and $\chib\sigb^{ab}\psib=-\psib\sigb^{ab}\chib$. Furthermore, under conjugation the products transform as
\begin{gather}
(\chi\psi)^\cc = \chib\psib\,, \label{eq:identities_spinors_id_12}\\
(\chi\sig^a\psib)^\cc = \psi\sig^a\chib\,, \hspace{4cm} (\chib\sigb^a\psi)^\cc = \psib\sigb^a\chi\,, \label{eq:identities_spinors_id_13}\\
(\chi\sig^a\sigb^b\psi)^\cc = \psib\sigb^b\sig^a\chib\,, \label{eq:identities_spinors_id_14}\\
(\chi\sig^a\sigb^b\sig^c\psib)^\cc = \psi\sig^c\sigb^b\sig^a\chib\,, \hspace{2.5cm} (\chib\sigb^a\sig^b\sigb^c\psi)^\cc = \psib\sigb^c\sig^b\sigb^a\chi\,, \label{eq:identities_spinors_id_15}
\end{gather}
hence $(\chi\sig^{ab}\psi)^\cc = -\psib\sigb^{ab}\chib$. A product of four spinors $\chi_i$ and $\chib_i$ $(i=1,2,3,4)$, respectively, has the property 
\begin{align}
(\chi_1\chi_2)(\chi_3\chi_4) &= -(\chi_1\chi_3)(\chi_4\chi_2) - (\chi_1\chi_4)(\chi_2\chi_3)\,, \label{eq:identities_spinors_id_16}\\
(\chib_1\chib_2)(\chib_3\chib_4) &= -(\chib_1\chib_3)(\chib_4\chib_2) - (\chib_1\chib_4)(\chib_2\chib_3)\,, \label{eq:identities_spinors_id_17}
\end{align}
where the brackets indicate the contraction of the spinor indices. Moreover, the following products can be written in the simple form
\begin{align}
(\chi_1\sig^a\chib_2)(\chi_3\sig_a\chib_4) &= -2(\chi_1\chi_3)(\chib_2\chib_4)\,, \label{eq:identities_spinors_id_18}\\
(\chib_1\sigb^a\chi_2)(\chib_3\sigb_a\chi_4) &= -2(\chi_2\chi_4)(\chib_1\chib_3)\,, \label{eq:identities_spinors_id_19}\\
(\chi_1\sig^a\chib_2)(\chib_3\sigb_a\chi_4) &= +2(\chi_1\chi_4)(\chib_2\chib_3)\,, \label{eq:identities_spinors_id_20}
\end{align}
by contracting the vector indices of the two Pauli matrices. A similar calculation applies for the terms
\begin{align}
(\chi_1\sig^{ab}\chi_2)(\chi_3\sig_{ab}\chi_4) &= (\chi_1\chi_2)(\chi_3\chi_4) + 2(\chi_1\chi_4)(\chi_2\chi_3)\,, \label{eq:identities_spinors_id_21}\\
(\chib_1\sigb^{ab}\chib_2)(\chib_3\sigb_{ab}\chib_4) &= (\chib_1\chib_2)(\chib_3\chib_4) + 2(\chib_1\chib_4)(\chib_2\chib_3)\,, \label{eq:identities_spinors_id_22}\\
(\chi_1\sig^{ab}\chi_2)(\chib_3\sigb_{ab}\chib_4) &= 0\,. \label{eq:identities_spinors_id_23}
\end{align}
\item \textbf{Spinor vs. vector:}\\
A Lorentz vector $V_a$ transforms in the representation $(\frac{1}{2},\frac{1}{2})$ of the Lorentz group, where the index is raised and lowered with the Minkowski metric from Eq.~\eqref{eq:identities_conventions_id_1}:
\begin{align}
V^a &= \eta^{ab}V_b\,.
\end{align}
A vector index $a$ is converted into a pair of spinor indices $\alpha\sd{\alpha}$ by contracting with the Pauli matrix $\sig^a_{\alpha\sd{\alpha}}$. The vector index is recovered by using $-\frac{1}{2}\sigb^{\sd{\alpha}\alpha}$, namely:
\begin{align}
V_{\alpha\sd{\alpha}} &= \sig^a_{\alpha\sd{\alpha}}V_a\,, \label{eq:identities_spinors_id_24}\\
V_a &= -\frac{1}{2}\sigb_a^{\sd{\alpha}\alpha}V_{\alpha\sd{\alpha}}\,. \label{eq:identities_spinors_id_25}
\end{align}
Thus, in terms of spinor indices the contraction of two vector indices is written as
\begin{align}
V_a W^a &= -\frac{1}{2}V_{\alpha\sd{\alpha}}W^{\alpha\sd{\alpha}}\,, \label{eq:identities_spinors_id_26}
\end{align}
where $W^{\alpha\sd{\alpha}}=\eps^{\alpha\beta}\eps^{\sd{\alpha}\sd{\beta}}W_{\beta\sd{\beta}}$.
\end{itemize}

\section{Superspace}
\label{sec:superspace}

\subsection{Local coordinates and supervielbein}
In superspace, a local coordinate patch has the form
\begin{align}
z^M := (x^m,\theta^\mu,\thetab_\sd{\mu})\,, \label{eq:superspace_coordinates}
\end{align}
where $x^m$ are the usual, commuting ($c$-type) spacetime coordinates and $\theta^\mu$, $\thetab_\sd{\mu}$ are additional anti-commuting ($a$-type) coordinates, with $m \in \{0,1,2,3\}$ and $\mu,\sd{\mu} \in \{1,2\}$. Under conjugation the coordinates transform as follows
\begin{align}
(x^m)^\cc &= x^m\,, & (\theta^\mu)^\cc &= \eps^{\sd{\mu}\sd{\nu}} \thetab_\sd{\nu}\,, & (\thetab_\sd{\mu})^\cc &= \eps_{\mu\nu} \theta^\nu\,, \label{eq:superspace_coordinates_conjugate}
\end{align}
which means that $x^m$ is real, and $\theta^\mu$ and $\thetab_\sd{\mu}$ are the conjugate of each other. In terms of the local coordinates $z^M$, a general superfield $\Phi$ has the form
\begin{align}
\begin{split}
\Phi(z^M) &= \Phi(x^m,\theta^\mu,\thetab_\sd{\mu}) \\
&= \phi_{00} + \theta\psi_{10} + \thetab\psi_{01} + \theta\theta\phi_{20} + \thetab\thetab\phi_{02} + \thetab\sigb_a\theta{\phi_{11}}^a + \thetab\thetab\,\theta\psi_{12} + \theta\theta\,\thetab\psi_{21} + \theta\theta\,\thetab\thetab\phi_{22}\,,
\end{split} \label{eq:superspace_general_superfunction_component_fields}
\end{align}
where the component fields
\begin{align}
\phi_{ij} &\equiv \phi_{ij}(x^m)\,, & \psi_{ij} &\equiv \psi_{ij}(x^m)\,,
\end{align}
are functions of the spacetime coordinates $x^m$. Note that ${\phi_{11}}^a$ carries an index $a\in\{0,1,2,3\}$. Furthermore, the component fields $\psi_{ij}$ have the following indices:
\begin{gather}
\begin{aligned}
\psi_{10} &\equiv {\psi_{10}}^\mu\,, &\hspace{4cm} \psi_{12} &\equiv {\psi_{12}}^\mu\,,  \\
\psi_{01} &\equiv \psi_{01\sd{\mu}}\,, & \psi_{21} &\equiv \psi_{21\sd{\mu}}\,.
\end{aligned}
\end{gather}
If $\Phi$ is pure, namely even or odd, then the component fields also have definite Grassmann parity:
\begin{align}
\begin{split}
\eps(\Phi) = 0 &\quad\Leftrightarrow\quad \eps(\phi_{ij}) = 0\,,\quad \eps(\psi_{ij}) = 1\,,\\
\eps(\Phi) = 1 &\quad\Leftrightarrow\quad \eps(\phi_{ij}) = 1\,,\quad \eps(\psi_{ij}) = 0\,.
\end{split}
\end{align}
The conjugation of the product of two pure superfields $\Phi$ and $\Psi$ obeys the following rule
\begin{align}
(\Phi\Psi)^\cc &= (-1)^{\Phi\Psi} \Phi^\cc \Psi^\cc\,,
\end{align}
where the shorthand notation from Eq.~\eqref{eq:identities_conventions_id_13} is used. A double conjugation is just the identity, $(\Phi^\cc)^\cc = \Phi$. In particular, $\Phi$ and $\Psi$ can just be equal to the coordinates $z^M$. Thus, according to Eq.~\eqref{eq:superspace_coordinates_conjugate}, the superfield $\Phi$ is real, if the component fields fulfil the following identities under conjugation:
\begin{align}
\Phi^\cc = \Phi &\quad\Leftrightarrow\quad (\phi_{ij})^\cc = \phi_{ji}\,,\quad (\psi_{ij})^\cc = \psi_{ji}\,,
\end{align}
with adjusted heights of the spinor indices. In particular, if $\Phi$ is real, then the component fields $\phi_{00}$, $\phi_{11}$, $\phi_{22}$ are real as well.
\\\\
On the coordinate patch $z^M$, the canonical local basis of the tangent bundle is formed by the supervector fields
\begin{align}
\del_M = (\del_m,\del_\mu,\del^\sd{\mu}) := \frac{\del}{\del z^M} = \Big(\frac{\del}{\del x^m},\frac{\del}{\del \theta^\mu},\frac{\del}{\del \thetab_\sd{\mu}}\Big)\,. \label{eq:superspace_basis_tangent_space}
\end{align}
Since the supervector fields have definite Grassmann parity, namely $\eps(\del_M) = \eps(z^M)$, they form a pure basis. With respect to this basis, a general supervector field $\chi$ is written as $\chi = \chi^M \del_M$. The supervector fields $\del_M$ act on superfields as derivatives with respect to $z^M$, which implies
\begin{align}
\del_M z^N = \delta^N_M\,, \label{eq:superspace_coordinates_derivative}
\end{align}
or written in a more explicit form:
\begin{align}
\del_m x^n &= \delta_m^n\,, & \del_\mu \theta^\nu &= \delta_\mu^\nu\,, & \del^\sd{\mu} \thetab_\sd{\nu} &= \delta^\sd{\mu}_\sd{\nu}\,, \label{eq:superspace_coordinates_derivative_explicit}
\end{align}
where all other combinations vanish. The conjugate of $\del_M$ is specified by the identity
\begin{align}
(\del_M \Phi)^\cc &= (-1)^{M\Phi} (\del_M)^\cc \Phi^\cc\,,
\end{align}
with a pure superfield $\Phi$, which, in combination with Eq.~\eqref{eq:superspace_coordinates_conjugate} and \eqref{eq:superspace_coordinates_derivative}, leads to the expressions
\begin{align}
(\del_m)^\cc &= \del_m\,, & (\del_\mu)^\cc &= \eps_{\sd{\mu}\sd{\nu}} \del^\sd{\nu}\,, & (\del^\sd{\mu})^\cc &= \eps^{\mu\nu} \del_\nu\,. \label{eq:superspace_basis_tangent_space_conjugate}
\end{align}
Every pure basis of the tangent bundle which fulfils these conjugation properties is called standard basis. The dual local basis of the cotangent bundle is composed of the differentials
\begin{align}
\exd z^M = (\exd x^m, \exd \theta^\mu,\exd \thetab_\sd{\mu})\,, \label{eq:superspace_basis_cotangent_space}
\end{align}
with the property
\begin{align}
\del_M \exd z^N = \del_M z^N = \delta^N_M\,. \label{eq:superspace_basis_cotangent_space_derivative}
\end{align}
The differentials $\exd z^M$ form a pure basis with $\eps(\exd z^M) = \eps(z^M)$. A general supercovector field ($1$-superform) $\omega$ is then written as $\omega = \exd z^M \omega_M$, and the contraction with a vector field $\chi = \chi^N \del_N$ is given by
\begin{align}
\chi\omega = (\chi^N\del_N)(\exd z^M \omega_M) = \chi^N \delta_N^M \omega_M = \chi^M \omega_M\,. \label{eq:superspace_vector_covector_contraction}
\end{align}
The conjugation properties of the basis elements $\exd z^M$ are the ones of a standard basis, which are derived from Eq.~\eqref{eq:superspace_basis_tangent_space_conjugate} and \eqref{eq:superspace_basis_cotangent_space_derivative}:
\begin{align}
(\exd x^m)^\cc &= \exd x^m\,, & (\exd \theta^\mu)^\cc &= \eps^{\sd{\mu}\sd{\nu}} \exd \thetab_\sd{\nu}\,, & (\exd \thetab_\sd{\mu})^\cc &= \eps_{\mu\nu} \exd \theta^\nu\,. \label{eq:superspace_basis_cotangent_space_conjugate}
\end{align}
To write down conjugate quantities, it is convenient to introduce dotted superspace indices $\sd{M}$. An upper dotted index $()^\sd{M}$ is defined as follows:
\begin{align}
\begin{split}
()^\sd{M} := ()^m &\quad\text{for}\quad M=m\,,\\
()^\sd{M} := \eps^{\sd{\mu}\sd{\nu}} ()_\sd{\nu} &\quad\text{for}\quad M=\mu\,,\\
()^\sd{M} := \eps_{\mu\nu} ()^\nu &\quad\text{for}\quad M=\sd{\mu}\,.
\end{split}
\end{align}
A similar definition applies for a lower index $()_\sd{M}$:
\begin{align}
\begin{split}
()_\sd{M} := ()_m &\quad\text{for}\quad M=m\,,\\
()_\sd{M} := \eps_{\sd{\mu}\sd{\nu}} ()^\sd{\nu} &\quad\text{for}\quad M=\mu\,,\\
()_\sd{M} := \eps^{\mu\nu} ()_\nu &\quad\text{for}\quad M=\sd{\mu}\,.
\end{split}
\end{align}
Note that ${()^\sd{M}}_\sd{M} = (-1)^M {()^M}_M$ and ${()^\sd{M}}_M = (-1)^M {()^M}_\sd{M}$
With this notation, Eq.~\eqref{eq:superspace_coordinates_conjugate}, \eqref{eq:superspace_basis_tangent_space_conjugate} and \eqref{eq:superspace_basis_cotangent_space_conjugate} are efficiently written as
\begin{align}
(z^M)^\cc &= z^\sd{M}\,, & (\del_M)^\cc &= \del_\sd{M}\,, & (\exd z^M)^\cc &= \exd z^\sd{M}\,.
\end{align}

\subsubsection{Supervielbein}
The supervielbein
\begin{align}
E^A = (E^a,E^\alpha,E_\sd{\alpha})
\end{align}
is a set of $1$-superforms, which form a standard basis of the cotangent bundle. The components $E^a$ are $c$-type and $E^\alpha$, $E_\sd{\alpha}$ are $a$-type, with $a\in\{0,1,2,3\}$ and $\alpha,\sd{\alpha}\in\{1,2\}$. Furthermore, under conjugation they transform as
\begin{align}
(E^A)^\cc = E^\sd{A}\,. \label{eq:superspace_vielbein_conjugate}
\end{align}
As discussed later in Section~\ref{sec:superspace_structure_group}, the index $A$ represents a Lorentz index. In terms of the coordinate basis, the supervielbein is written as
\begin{align}
E^A = \exd z^M {E_M}^A\,,
\end{align}
with the invertible supermatrix ${E_M}^A$. The conjugate of this supermatrix is determined by Eq.~\eqref{eq:superspace_vielbein_conjugate} and the identity
\begin{align}
(E^A)^\cc = (\exd z^M {E_M}^A)^\cc = (-1)^{M(M+A)} (\exd z^M)^\cc ({E_M}^A)^\cc = (-1)^{MA} \exd z^M ({E_\sd{M}}^A)^\cc\,,
\end{align}
and is given by
\begin{align}
({E_{M}}^A)^\cc = (-1)^{\sd{M}\sd{A}} {E_\sd{M}}^\sd{A}\,. \label{eq:superspace_vielbein_components_conjugate}
\end{align}
On the tangent bundle, the inverse supervielbein
\begin{align}
E_A = (E_a,E_\alpha,E^\sd{\alpha})\,,
\end{align}
forms the dual basis of the supervielbein, i.e.\ $E_A E^B = E^B E_A = \delta_A^B$. Writing
\begin{align}
E_A = {E_A}^M \del_M\,,
\end{align}
it follows immediately that the supermatrices ${E_A}^M$ and ${E_M}^A$ are inverses of each other:
\begin{align}
{E_A}^M {E_M}^B &= \delta_A^B\,, & \quad {E_M}^A {E_A}^N &= \delta_M^N\,.
\end{align}
Using Eq.~\eqref{eq:superspace_vielbein_conjugate}, the calculation
\begin{align}
(E_\sd{A})^\cc E^B = (E_\sd{A})^\cc (E^\sd{B})^\cc = (-1)^{AB} (E_\sd{A} E^\sd{B})^\cc = \delta_A^B = E_A E^B
\end{align}
provides the conjugation property of the inverse supervielbein
\begin{align}
(E_A)^\cc = E_\sd{A}\,, \label{eq:superspace_inverse_vielbein_conjugate}
\end{align}
and
\begin{align}
({E_{A}}^M)^\cc = (-1)^{\sd{A}\sd{M}} {E_\sd{A}}^\sd{M}\,. \label{eq:superspace_inverse_vielbein_components_conjugate}
\end{align}
Writing the components of $E^A$ and $E_A$ as matrices,
\begin{align}
{E_M}^A &=
\begin{pmatrix}
{E_m}^a & {E_m}^\alpha & E_{m\sd{\alpha}} \\
{E_\mu}^a & {E_\mu}^\alpha & E_{\mu\sd{\alpha}} \\
E^{\sd{\mu}a} & E^{\sd{\mu}\alpha} & {E^\sd{\mu}}_\sd{\alpha}
\end{pmatrix},
&
{E_A}^M &=
\begin{pmatrix}
{E_a}^m & {E_a}^\mu & E_{a\sd{\mu}} \\
{E_\alpha}^m & {E_\alpha}^\mu & E_{\alpha\sd{\mu}} \\
E^{\sd{\alpha}m} & E^{\sd{\alpha}\mu} & {E^\sd{\alpha}}_\sd{\mu}
\end{pmatrix}, \label{eq:superspace_vielbein_matrix}
\end{align}
their conjugates in Eq.~\eqref{eq:superspace_vielbein_components_conjugate} and \eqref{eq:superspace_inverse_vielbein_components_conjugate} have the explicit form
\begin{align}
({E_M}^A)^\cc &=
\begin{pmatrix*}[r]
{E_m}^a & \eps^{\sd{\alpha}\sd{\beta}} E_{m\sd{\beta}} & \eps_{\alpha\beta} {E_m}^\beta \\
\eps_{\sd{\mu}\sd{\nu}} E^{\sd{\nu}a} & -\eps_{\sd{\mu}\sd{\nu}} \eps^{\sd{\alpha}\sd{\beta}} {E^\sd{\nu}}_\sd{\beta} & -\eps_{\sd{\mu}\sd{\nu}} \eps_{\alpha\beta} E^{\sd{\nu}\beta} \\
\eps^{\mu\nu} {E_\nu}^a & -\eps^{\mu\nu} \eps^{\sd{\alpha}\sd{\beta}} E_{\nu\sd{\beta}} & -\eps^{\mu\nu} \eps_{\alpha\beta} {E_\nu}^\beta
\end{pmatrix*}, \label{eq:superspace_vielbein_matrix_conjugate_1}
\\ \nonumber \\
({E_A}^M)^\cc &=
\begin{pmatrix*}[r]
{E_a}^m & \eps^{\sd{\mu}\sd{\nu}} E_{a\sd{\nu}} & \eps_{\mu\nu} {E_a}^\nu \\
\eps_{\sd{\alpha}\sd{\beta}} E^{\sd{\beta}m} & -\eps_{\sd{\alpha}\sd{\beta}} \eps^{\sd{\mu}\sd{\nu}} {E^\sd{\beta}}_\sd{\nu} & -\eps_{\sd{\alpha}\sd{\beta}} \eps_{\mu\nu} E^{\sd{\beta}\nu} \\
\eps^{\alpha\beta} {E_\beta}^m & -\eps^{\alpha\beta} \eps^{\sd{\mu}\sd{\nu}} E_{\beta\sd{\nu}} & -\eps^{\alpha\beta} \eps_{\mu\nu} {E_\beta}^\nu \\
\end{pmatrix*}. \label{eq:superspace_vielbein_matrix_conjugate_2}
\end{align}
With respect to the supervielbein and its inverse a general supercovector and supervector field are written as $\omega = E^A \omega_A$ and $\chi=\chi^A E_A$ respectively, and the contraction is given by $\chi\omega = \chi^A \omega_A$. With the supermatrix ${E_M}^A$ from Eq.~\eqref{eq:superspace_vielbein_matrix} written in the compact form
\begin{align}
{E_M}^A &= \begin{pmatrix}
{E_m}^a & {E_m}^\su{\alpha} \\
{E_\su{\mu}}^a & {E_\su{\mu}}^\su{\alpha}
\end{pmatrix}, \label{eq:superspace_vielbein_matrix_compact}
\end{align}
where $\su{\mu}$ labels $\mu$, $\sd{\mu}$ and $\su{\alpha}$ labels $\alpha$, $\sd{\alpha}$ (with adjusted height for the dotted indices), the superdeterminant (also called Berezinian) is defined as
\begin{align}
\begin{split}
\sdet({E_M}^A) &:= \det({E_m}^a - {E_m}^\su{\beta}\,{E_\su{\beta}}^\su{\nu}\,{E_\su{\nu}}^a)\det({E_\su{\mu}}^\su{\alpha})^{-1} \\
&\,\,= \det({E_m}^a)\det({E_\su{\mu}}^\su{\alpha} - {E_\su{\mu}}^b\,{E_b}^n\,{E_n}^\su{\alpha})^{-1} \,.
\end{split} \label{eq:superspace_vielbein_sdet}
\end{align}
Note that the superdeterminant is defined only if the submatrices ${E_m}^a$ and ${E_\su{\mu}}^\su{\alpha}$ are invertible, which is equivalent to that the whole supermatrix ${E_M}^A$ is invertible, and that it is usually written in the short form $E=\sdet({E_M}^A)$. The conjugate of $E$ is given by
\begin{align}
\big(\sdet({E_M}^A)\big)^\cc &= \big(\det({E_m}^a - {E_m}^\su{\beta}\,{E_\su{\beta}}^\su{\nu}\,{E_\su{\nu}}^a)\big)^\cc \big(\det({E_\su{\mu}}^\su{\alpha})^\cc\big)^{-1}\,.
\end{align}
With the identities
\begin{align}
\begin{split}
\big(\det({E_m}^a - {E_m}^\su{\beta}\,{E_\su{\beta}}^\su{\nu}\,{E_\su{\nu}}^a)\big)^\cc &= \det\big(({E_m}^a)^\cc + ({E_m}^\su{\beta})^\cc\,({E_\su{\beta}}^\su{\nu})^\cc\,({E_\su{\nu}}^a)^\cc\big) \\
&= \det({E_m}^a - {E_m}^\su{\sd{\beta}}\,{E_\su{\sd{\beta}}}^\su{\sd{\nu}}\,{E_\su{\sd{\nu}}}^a) \\
&= \det({E_m}^a - {E_m}^\su{\beta}\,{E_\su{\beta}}^\su{\nu}\,{E_\su{\nu}}^a)\,,
\end{split}
\end{align}
and
\begin{align}
\begin{split}
\det({E_\su{\mu}}^\su{\alpha})^\cc &= \det({E_\mu}^\alpha - E_{\mu\sd{\beta}}{E^\sd{\beta}}_\sd{\nu}E^{\sd{\nu}\alpha})^\cc \det({E^\sd{\mu}}_\sd{\alpha})^\cc \\
&= \det\big(({E_\mu}^\alpha)^\cc - (E_{\mu\sd{\beta}})^\cc({E^\sd{\beta}}_\sd{\nu})^\cc(E^{\sd{\nu}\alpha})^\cc\big) \det\big(({E^\sd{\mu}}_\sd{\alpha})^\cc\big) \\
&= \det(-{E_\sd{\mu}}^\sd{\alpha} + E_{\sd{\mu}\beta}{E^\beta}_\nu E^{\nu\sd{\alpha}}) \det(-{E^\mu}_\alpha) \\
&= \det\big(-\eps_{\sd{\mu}\sd{\tau}}\eps^{\sd{\alpha}\sd{\gamma}}({E^\sd{\tau}}_\sd{\gamma} + E^{\sd{\tau}\beta}{E_\beta}^\nu E_{\nu\sd{\gamma}})\big) \det(-\eps^{\mu\tau}\eps_{\alpha\gamma}{E_\tau}^\gamma) \\
&= \det({E^\sd{\tau}}_\sd{\gamma} + E^{\sd{\tau}\beta}{E_\beta}^\nu E_{\nu\sd{\gamma}}) \det({E_\tau}^\gamma) \\
&= \det({E_\su{\tau}}^\su{\gamma})\,,
\end{split}
\end{align}
follows that the superdeterminant is real:
\begin{align}
E^\cc &= E\,. \label{eq:superspace_vielbein_sdet_conjugation}
\end{align}
Furthermore, if ${E_M}^A$ is a function of a parameter $\lambda$ ($c$- or $a$-type), then the derivative of $E$ with respect to $\lambda$ can be written in the form
\begin{align}
\frac{\exd}{\exd\lambda} E = (-1)^N E \Big(\frac{\exd}{\exd\lambda} {E_N}^A\Big) {E_A}^N\,, \label{eq:superspace_vielbein_sdet_jacobis_formula_derivative}
\end{align}
which is basically Jacobi's formula in superspace. Thus, an infinitesimal change $\delta{E_M}^A$ of the supermatrix results in the following change of the superdeterminant
\begin{align}
\delta E = (-1)^N E (\delta{E_N}^A) {E_A}^N\,. \label{eq:superspace_vielbein_sdet_jacobis_formula}
\end{align}

\subsubsection{Integration}
\label{sec:integration}
Integration in superspace over the coordinates $(x^m,\theta^\mu,\thetab_\sd{\mu})$ is defined by the following properties
\begin{align}
\int\exd x^m\,\theta^\nu &:= 0\,, & \int\exd x^m\,\thetab_\sd{\nu} &:= 0\,, \nonumber\\
\int\exd\theta^\mu\,\theta^\nu &:= \delta_\mu^\nu\,, & \int\exd\theta^\mu\,\thetab_\sd{\nu} &:= 0\,, & \int\exd\theta^\mu\,f(x^n) &:= 0\,, \label{eq:superspace_integration_properties_1}\\
\int\exd\thetab_\sd{\mu}\,\theta^\nu &:= 0\,, & \int\exd\thetab_\sd{\mu}\,\thetab_\sd{\nu} &:= \delta^\sd{\mu}_\sd{\nu}\,, & \int\exd\thetab_\sd{\mu}\,f(x^n) &:= 0\,, \nonumber
\end{align}
where $f$ is a general complex function of the coordinates $x^n$. In addition, $\int\exd x^m f(x^n)$ is the integral in ordinary spacetime. Furthermore, with the definitions
\begin{align}
\exd x^m\,\theta^\nu &:= +\theta^\nu\,\exd x^m\,, & \exd x^m\,\thetab_\sd{\nu} &:= +\thetab_\sd{\nu}\,\exd x^m\,, & & \nonumber\\
& & \exd\theta^\mu\,\thetab_\sd{\nu} &:= -\thetab_\sd{\nu}\,\exd\theta^\mu\,, & \exd\theta^\mu\,f(x^n) &:= +f(x^n)\,\exd\theta^\mu\,, \label{eq:superspace_integration_properties_2}\\
\exd\thetab_\sd{\mu}\,\theta^\nu &:= -\theta^\nu\,\exd\thetab_\sd{\mu}\,, & & & \exd\thetab_\sd{\mu}\,f(x^n) &:= +f(x^n)\,\exd\thetab_\sd{\mu}\,, \nonumber
\end{align}
follows from Eq.~\eqref{eq:superspace_coordinates_derivative_explicit}, that integration and differentiation with respect to the anti-commuting coordinates $\theta^\mu$, $\thetab_\sd{\mu}$ coincide, i.e.\ $\int\exd\theta^\mu=\del_\mu$ and $\int\exd\thetab_\sd{\mu}=\del^\sd{\mu}$. To perform integration over the whole superspace, it is convenient to define the quantities
\begin{align}
\exd^2\theta &:= -\frac{1}{4}\eps_{\mu\nu}\exd\theta^\mu\exd\theta^\nu\,, & \exd^2\thetab &:= -\frac{1}{4}\eps^{\sd{\mu}\sd{\nu}}\exd\theta_\sd{\mu}\exd\thetab_\sd{\nu}\,. \label{eq:superspace_integration_identity_1}
\end{align}
such that
\begin{align}
\int \exd^2\theta\,\theta\theta &= 1\,, & \int \exd^2\thetab\,\thetab\thetab &= 1\,, \label{eq:superspace_integration_identity_2}
\end{align}
where the properties from Eq.~\eqref{eq:superspace_integration_properties_1} and Eq.~\eqref{eq:superspace_integration_properties_2} are used. With these identities, the integral of a general superfield $\Phi(x^m,\theta^\mu,\theta_\sd{\mu})$, as defined in Eq.~\eqref{eq:superspace_general_superfunction_component_fields},
over all anti-commuting coordinates is given by
\begin{align}
\int\exd^2\theta\,\exd^2\thetab\,\Phi &= \phi_{22}\,. \label{eq:superspace_integration_anticommuting_coord}
\end{align}
Hence, the integration just extracts the component field $\phi_{22}(x^n)$ of the top degree coefficient $\theta\theta\,\thetab\thetab$. This implies that the integral over derivatives with respect to anti-commuting coordinates $\theta^\mu$, $\thetab_\sd{\mu}$ always vanishes:
\begin{align}
\int\exd^2\theta\,\exd^2\thetab\,\del_\nu\Phi &= 0\,, & \int\exd^2\theta\,\exd^2\thetab\,\del^\sd{\nu}\Phi &= 0\,. \label{eq:superspace_integration_derivative_1}
\end{align}
Integration over the whole superspace, i.e.\ over all coordinates $z^M=(x^m,\theta^\mu,\theta_\sd{\mu})$, is labelled by $\int_*$. In that case, according to Eq.~\eqref{eq:superspace_integration_anticommuting_coord} the integral of $\Phi(x^m,\theta^\mu,\theta_\sd{\mu})$ reads
\begin{align}
\int_*\Phi &:= \int\exd^4 x\,\exd^2\theta\,\exd^2\thetab\,\Phi = \int\exd^4 x\,\phi_{22}\,. \label{eq:superspace_integration_all_coord}
\end{align}
Usually it is assumed that surface terms of $\phi_{22}$ vanish (e.g.\ if $\Phi$ has a compact support in the spacetime coordinates $x^m$). Together with Eq.~\eqref{eq:superspace_integration_derivative_1} this implies
\begin{align}
\int_*\del_N\Phi &= 0\,. \label{eq:superspace_integration_derivative_2}
\end{align}
The infinitesimal action of a local diffeomorphism in superspace is given by the Lie derivative $\lied_\xi$, parametrized by a real, even (infinitesimal) supervector field $\xi=\xi^M\del_M$ (cf.\ Section~\ref{sec:lie_derivative}). For the superfield $\Phi$, the infinitesimal action under the diffeomorphism has the form
\begin{align}
\delta \Phi &= \xi^M\del_M\Phi\,, \label{eq:superspace_integration_diffeo_1}
\end{align}
thus, the value of the integral over $\Phi$ changes as
\begin{align}
\int_*\delta\Phi &= \int_*\xi^M\del_M\Phi = \int_*\Big((-1)^M\del_M(\xi^M\Phi) - (-1)^M(\del_M\xi^M)\Phi\Big) = \int_*-(-1)^M(\del_M\xi^M)\Phi\,, \label{eq:superspace_integration_diffeo_2}
\end{align}
where Eq.~\eqref{eq:superspace_integration_derivative_2} is used. Integrals which are invariant under diffeomorphisms are usually constructed by using densities. The canonical real density in superspace is the superdeterminant $E$ of the components ${E_M}^A$ of the supervielbein, as defined in Eq.~\eqref{eq:superspace_vielbein_sdet}. According to Eq.~\eqref{eq:superspace_lie_derivative_vielbein} the transformation of ${E_M}^A$ under the diffeomorphism $\xi$ is given by 
\begin{align}
\delta{E_M}^A &= \xi^N(\del_N{E_M}^A) + (\del_M\xi^N){E_N}^A\,, \label{eq:superspace_integration_diffeo_3}
\end{align}
which, in combination with Eq.~\eqref{eq:superspace_vielbein_sdet_jacobis_formula_derivative} and \eqref{eq:superspace_vielbein_sdet_jacobis_formula}, implies
\begin{align}
\begin{split}
\delta E &= (-1)^M E \xi^N(\del_N{E_M}^A){E_A}^M + (-1)^M E(\del_M\xi^N){E_N}^A{E_A}^M \\
&= \xi^N\del_N E + (-1)^M E(\del_M\xi^M) \\
&= (-1)^M\del_M(\xi^M E)\,.
\end{split} \label{eq:superspace_integration_diffeo_4}
\end{align}
Thus, the volume of superspace, defined as
\begin{align}
\text{vol} &:= \int_* E\,, \label{eq:superspace_integration_volume}
\end{align}
does not change under diffeomorphisms. The general form of an integral containing the superfield $\Phi$, which is invariant under superspace diffeomorphisms, is given by $\int_*E\,\Phi$. The invariance follows directly from Eq.~\eqref{eq:superspace_integration_diffeo_1} and \eqref{eq:superspace_integration_diffeo_4}, namely
\begin{align}
\begin{split}
\delta (E\,\Phi) &= (\delta E)\Phi + E(\delta\Phi) \\
&= (-1)^M\del_M(\xi^M E)\Phi + E\xi^M\del_M\Phi \\
&= (-1)^M\del_M(\xi^M E\,\Phi)\,,
\end{split} \label{eq:superspace_integration_diffeo_5}
\end{align}
which implies
\begin{align}
\int_*\delta(E\,\Phi) = 0\,. \label{eq:superspace_integration_diffeo_6}
\end{align}

\subsection{Tensors}
\label{sec:superspace_tensors}
Let $T$ be a pure tensor ($c$- or $a$-type) of rank $(s,r)$. With respect to the local bases $\del_M$ and $\exd z^M$, the components are given by
\begin{align}
{T_{M_r...M_1}}^{M_{r+1}...M_{r+s}} := (\del_{M_r},...,\del_{M_1})T(\exd z^{M_{r+1}},...,\exd z^{M_{r+s}})\,,
\end{align}
and the Grassmann parity of the components is
\begin{align}
\eps({T_{M_r...M_1}}^{M_{r+1}...M_{r+s}}) = \eps(T) + \eps(M_1) +...+ \eps(M_{r+s})\,.
\end{align}
Tensors can be constructed from supervector or supercovector fields by means of the tensor product~$\otimes$. For example, if $\chi_i$ $(i=1,...,s)$ is a set of supervector fields and $\omega^i$ $(i=1,...,r)$ a set of supercovector fields, the $(s,0)$-tensor $\chi_1 \otimes...\otimes \chi_s$ and the $(0,r)$-tensor $\omega^1 \otimes...\otimes \omega^r$ act on a tuple of supercovector and supervector fields, respectively, as follows:
\begin{align}
\chi_s \otimes...\otimes \chi_1\,(\omega^1,...,\omega^s) &:= (-1)^{\Delta_s(\omega)} (-1)^{\Delta_s(\chi,\omega)} \chi_s \omega^s \cdot...\cdot \chi_1 \omega^1\,, \label{eq:superspace_general_action_tensor_1} \\
(\chi_r,...,\chi_1)\,\omega^1 \otimes...\otimes \omega^r &:= (-1)^{\Delta_r(\chi)} (-1)^{\Delta_r(\omega,\chi)} \chi_1 \omega^1 \cdot...\cdot \chi_r \omega^r\,, \label{eq:superspace_general_action_tensor_2}
\end{align}
with
\begin{align}
\Delta_s(\omega,\chi) &:= \sum_{\underset{t<u}{t,u=1}}^s \omega^t \chi_u\,,\\
\Delta_s(\omega) &:= \Delta_s(\omega,\omega)\,,
\end{align}
and the contractions $\chi_i\omega^i$ are evaluated according to Eq.~\eqref{eq:superspace_vector_covector_contraction}. Note that tensors of rank $(1,0)$ are just supervector fields and tensors of rank $(0,1)$ are supercovector fields ($1$-superforms). Moreover, Eq.~\eqref{eq:superspace_general_action_tensor_1} and \eqref{eq:superspace_general_action_tensor_2} imply
\begin{align}
(\del_{M_r},...,\del_{M_1})\,\exd z^{N_1} \otimes...\otimes \exd z^{N_r} &= \delta_{M_1}^{N_1} \cdot...\cdot \delta_{M_r}^{N_r}\,,\\
\del_{N_1} \otimes...\otimes \del_{N_s}\,(\exd z^{M_s},...,\exd z^{M_1}) &= \delta_{N_1}^{M_1} \cdot...\cdot \delta_{N_s}^{M_s}\,,
\end{align}
thus a general tensor $T$ of rank $(s,r)$ can be written in terms of its components in the following way:
\begin{align}
T = \exd z^{N_1} \otimes...\otimes \exd z^{N_r} {T_{N_r...N_1}}^{N_{r+1}...N_{r+s}} \del_{N_{r+s}} \otimes...\otimes \del_{N_{r+1}}\,. \label{eq:superspace_tensor_coordinate_basis}
\end{align}
With the following definition for conjugation of the $(s,0)$ and $(0,r)$ tensors from Eq.~\eqref{eq:superspace_general_action_tensor_1} and \eqref{eq:superspace_general_action_tensor_2},
\begin{align}
\big( \chi_s \otimes...\otimes \chi_1 \big)^\cc &:= (-1)^{\Delta_s(\chi)} {\chi_s}^\cc \otimes...\otimes {\chi_1}^\cc\,, \\
\big( \omega^1 \otimes...\otimes \omega^r \big)^\cc &:= (-1)^{\Delta_r(\omega)} {\omega^1}^\cc \otimes...\otimes {\omega^r}^\cc \,,
\end{align}
the conjugate of the general $(s,r)$ tensor $T$ form Eq.~\eqref{eq:superspace_tensor_coordinate_basis} has the form
\begin{align}
\begin{split}
T^\cc &= \big( \exd z^{N_1} \otimes...\otimes \exd z^{N_r} {T_{N_r...N_1}}^{N_{r+1}...N_{r+s}} \del_{N_{r+s}} \otimes...\otimes \del_{N_{r+1}} \big)^\cc\\
&= (-1)^{\Delta_{r+s}(N)} (-1)^{(T+N_1+...+N_{r+s})(N_1+...+N_{r+s})}\\
&\qquad {\exd z^{N_1}}^\cc \otimes...\otimes {\exd z^{N_r}}^\cc \big({T_{N_r...N_1}}^{N_{r+1}...N_{r+s}}\big)^\cc \del_{N_{r+s}}^\cc \otimes...\otimes \del_{N_{r+1}}^\cc\\
&= (-1)^{\Delta_{r+s}(N) + T(N_1+...+N_{r+s})}\\
&\qquad \exd z^{N_1} \otimes...\otimes \exd z^{N_r} \big({T_{\sd{N}_r...\sd{N}_1}}^{\sd{N}_{r+1}...\sd{N}_{r+s}}\big)^\cc \del_{N_{r+s}} \otimes...\otimes \del_{N_{r+1}}\,.
\end{split}
\end{align}
Thus, the components of $T^\cc$ are given by
\begin{align}
{T^\cc_{N_r...N_1}}^{N_{r+1}...N_{r+s}} = (-1)^{\Delta_{r+s}(N) + T(N_1+...+N_{r+s})} \big({T_{\sd{N}_r...\sd{N}_1}}^{\sd{N}_{r+1}...\sd{N}_{r+s}}\big)^\cc\,.
\end{align}
The tensor $T$ is called real, if $T=T^\cc$, or in terms of the components:
\begin{align}
{T_{N_r...N_1}}^{N_{r+1}...N_{r+s}} = {T^\cc_{N_r...N_1}}^{N_{r+1}...N_{r+s}}\,.
\end{align}
Since in the calculations above only the standard basis properties of $\exd z^M$ and $\del_M$ are used, all the equations can be written in terms of $E^A$ and $E_A$ as well, without changing their form. For example, with respect to that basis the tensor components are given by
\begin{align}
{T_{A_r...A_1}}^{A_{r+1}...A_{r+s}} := (E_{A_r},...,E_{A_1})T(E^{A_{r+1}},...,E^{A_{r+s}})\,,
\end{align}
and the tensor is written as
\begin{align}
T = E^{A_1} \otimes...\otimes E^{A_r} {T_{A_r...A_1}}^{A_{r+1}...A_{r+s}} E_{A_{r+s}} \otimes...\otimes E_{A_{r+1}}\,.
\end{align}
Furthermore, the components of the conjugate tensor are given by
\begin{align}
{T^\cc_{A_r...A_1}}^{A_{r+1}...A_{r+s}} = (-1)^{\Delta_{r+s}(A) + T(A_1+...+A_{r+s})} \big({T_{\sd{A}_r...\sd{A}_1}}^{\sd{A}_{r+1}...\sd{A}_{r+s}}\big)^\cc\,. \label{eq:superspace_tensor_components_conjuagte}
\end{align}
The tensor components in the two different bases are related as follows:
\begin{align}
\begin{split}
{T_{A_r...A_1}}^{B_1...B_s} &= (-1)^{\Delta_r{(A)}+\Delta_r{(M,A)}} (-1)^{\Delta_s{(B)}+\Delta_r{(N,B)}} {E_{A_1}}^{M_1}...\,{E_{A_r}}^{M_r} {T_{M_r...M_1}}^{N_1...N_s} {E_{N_s}}^{B_s}...\,{E_{N_1}}^{B_1}\,,
\end{split} \\
\begin{split}
{T_{M_r...M_1}}^{N_1...N_s} &= (-1)^{\Delta_r{(M)}+\Delta_r{(A,M)}} (-1)^{\Delta_s{(N)}+\Delta_r{(B,N)}} {E_{M_1}}^{A_1}...\,{E_{M_r}}^{A_r} {T_{A_r...A_1}}^{B_1...B_s} {E_{B_s}}^{N_s}...\,{E_{B_1}}^{N_1}\,.
\end{split}
\end{align}

\subsubsection{Superforms}
A special type of $(0,p)$-tensors are $p$-superforms. They are constructed from $1$-superforms, which are just supercovector fields, by means of the wedge product. The wedge product of $p$ pure $1$-superforms $\omega^1,...,\omega^p$ is defined by
\begin{align}
\omega^1\wedge...\wedge\omega^p := \sum_{\sigma\in S_p} \sign(\sigma)\, (-1)^{\sigma(\omega)} \omega^{\sigma(1)}\otimes...\otimes\omega^{\sigma(p)}\,,
\end{align}
with the permutations $\sigma\in S_p$, and where $\sigma(\omega)$ is the number of sign changes$\pmod 2$ that occur when $\omega^{\sigma(1)}\otimes...\otimes\omega^{\sigma(p)}$ is brought back to the form $\omega^1\otimes...\otimes\omega^p$, by taking into account the Grassmann parity of the $\omega^i$. For example, the wedge product of two and three basis elements $\exd z^M$ is respectively given by
\begin{align}
\exd z^{M_1} \wedge \exd z^{M_2} &= \exd z^{M_1} \otimes \exd z^{M_2} - (-1)^{M_2 M_1} \exd z^{M_2} \otimes \exd z^{M_1}\,, \\
\nonumber\\
\begin{split}
\exd z^{M_1} \wedge \exd z^{M_2} \wedge \exd z^{M_3} &= \exd z^{M_1} \otimes \exd z^{M_2} \otimes \exd z^{M_3} \\
&\quad-(-1)^{M_3 M_2} \exd z^{M_1} \otimes \exd z^{M_3} \otimes \exd z^{M_2} \\
&\quad-(-1)^{M_2 M_1} \exd z^{M_2} \otimes \exd z^{M_1} \otimes \exd z^{M_3} \\
&\quad+(-1)^{(M_3+M_2)M_1} \exd z^{M_2} \otimes \exd z^{M_3} \otimes \exd z^{M_1} \\
&\quad+(-1)^{M_3(M_2+M_1)} \exd z^{M_3} \otimes \exd z^{M_1} \otimes \exd z^{M_2} \\
&\quad-(-1)^{M_3(M_2+M_1)+M_2 M_1} \exd z^{M_3} \otimes \exd z^{M_2} \otimes \exd z^{M_1}\,.
\end{split}
\end{align}
A general $p$-superform $\omega$ is written as
\begin{align}
\omega = \frac{1}{p!} \exd z^{M_1} \wedge ... \wedge \exd z^{M_p} \omega_{M_p...M_1}, \label{eq:superspace_general_superform}
\end{align}
where the components have the symmetry
\begin{align}
\omega_{M_p...M_i M_{i+1}...M_1} = -(-1)^{M_{i+1} M_i} \omega_{M_p...M_{i+1} M_i...M_1}\,,
\end{align}
for two adjacent indices $M_i$ and $M_{i+1}$. For example, the components of a $2$-superform have the symmetry
\begin{align}
\omega_{M_1 M_2} = -(-1)^{M_2 M_1} \omega_{M_2 M_1}\,. \label{eq:superspace_2form_components_symmetry}
\end{align}
As for general $(0,p)$-tensors, the components of $\omega$ are calculated by
\begin{align}
\omega_{M_p...M_1} = (\del_{M_p},...,\del_{M_1})\,\omega\,,
\end{align}
and the components of the conjugated $p$-superform $\omega^\cc$ read
\begin{align}
\omega^\cc_{N_p...N_1} = (-1)^{\Delta_p(N) + \omega(N_1+...+N_p)} (\omega_{\sd{N}_p...\sd{N}_1})^\cc\,.
\end{align}
Let $\omega$ be a $p$-superform and $\tau$ a $q$-superform, both pure. Using Eq.~\eqref{eq:superspace_general_superform} to express superforms, the following symmetry property of the wedge product is derived:
\begin{align}
\begin{split}
\omega \wedge \tau &= \Big( \frac{1}{p!} \exd z^{M_1} \wedge...\wedge \exd z^{M_p} \omega_{M_p...M_1} \Big) \wedge \Big( \frac{1}{q!} \exd z^{N_1} \wedge...\wedge \exd z^{N_q} \tau_{N_q...N_1} \Big) \\
&= \frac{1}{p!q!} (-1)^{\tau(\omega+M_1+...+M_p)} \exd z^{M_1} \wedge...\wedge \exd z^{M_p} \wedge \exd z^{N_1} \wedge...\wedge \exd z^{N_q} \tau_{N_q...N_1} \omega_{M_p...M_1} \\
&= (-1)^{pq+\omega\tau} \tau \wedge \omega\,.
\end{split}
\end{align}

\subsubsection{Exterior derivative}
The exterior derivative $\exd$ is an even derivative of degree $1$ which maps $p$-superforms onto $(p+1)$-superforms. It has the following properties
\begin{align}
\exd(\omega c) &= (\exd\omega)c\,,\quad \exd(\omega + \omega') = \exd\omega + \exd\omega'\,, \label{eq:superspace_exterior_derivative_property_1} \\
\exd(\omega \wedge \tau) &= \omega \wedge \exd\tau + (-1)^q \exd\omega \wedge \tau\,,\label{eq:superspace_exterior_derivative_property_2} \\
\exd \circ \exd &= 0\,,
\label{eq:superspace_exterior_derivative_property_3}
\end{align}
where $\omega$ and $\omega'$ are $p$-superforms and $\tau$ is a $q$-superform, and where $c$ is a supernumber, all pure. Note that superfields are considered as $0$-superforms. The identity in Eq.~\eqref{eq:superspace_exterior_derivative_property_2} is called Leibniz rule. The exterior derivative is uniquely defined by Eqs.~\eqref{eq:superspace_exterior_derivative_property_1}--\eqref{eq:superspace_exterior_derivative_property_3} and the statement that $\exd\Phi$ is just the ordinary differential when $\Phi$ is $0$-superform. Furthermore, $\exd$ is a real operator, i.e.\
\begin{align}
(\exd \omega)^\cc = \exd \omega^\cc\,. \label{eq:superspace_exterior_derivative_conjugate}
\end{align}
With the notation $\exd = \exd z^N \del_N$ it is straightforward to calculate the components of the $(p+1)$-superform $\exd \omega$,
\begin{align}
\begin{split}
\exd \omega &= \frac{1}{p!} \exd z^{M_1} \wedge ... \wedge \exd z^{M_p} \wedge \exd z^N \del_N \omega_{M_p...M_1}\,, \\
&= \frac{1}{(p+1)!} \exd z^{M_1} \wedge ... \wedge \exd z^{M_{p+1}} \sum_{r=1}^{p+1} \sign(\sigma_r)\, (-1)^{\sigma_r(M)} \del_{M_{\sigma_r(p+1)}} \omega_{M_{\sigma_r(p)}...M_{\sigma_r(1)}}\,,
\end{split} \label{eq:superspace_exterior_derivative_general_superform}
\end{align}
with the transposition $\sigma_r = (p+1,r)$, and where $\sigma_r(M)$ is the number of sign changes$\pmod 2$ that occur when $M_{\sigma_r(p+1)}...M_{\sigma_r(1)}$ is brought back to the form $M_{p+1}...M_1$, by taking into account the Grassmann parity of the $M_i$. For example, the exterior derivative of a $1$-superform $\omega$ and a $2$-superform $\tau$ is respectively given by
\begin{align}
d\omega &= \frac{1}{2} \exd z^{M_1} \wedge \exd z^{M_2} \big( \del_{M_2}\omega_{M_1} - (-1)^{M_2 M_1} \del_{M_1}\omega_{M_2} \big)\,, \\
\nonumber\\
\begin{split}
d\tau &= \frac{1}{6} \exd z^{M_1} \wedge \exd z^{M_2} \wedge \exd z^{M_3} \big( \del_{M_3} \tau_{M_2 M_1} -(-1)^{M_3 M_2} \del_{M_2} \tau_{M_3 M_1} -(-1)^{M_3(M_2+M_1) + M_2 M_1} \del_{M_1} \tau_{M_2 M_3} \big)\,.
\end{split}
\end{align}
Since, in contrast to $\exd(\exd z^M)=0$, the exterior derivative of the supervielbein does not vanish in general,
\begin{align}
\exd E^A = \exd z^M \wedge \exd {E_M}^A = \frac{1}{2} \exd z^M \wedge \exd z^N (\del_N {E_M}^A -(-1)^{MN} \del_M {E_N}^A)\,,
\end{align}
the formula for the components of $\exd\omega$ in Eq.~\eqref{eq:superspace_exterior_derivative_general_superform} is not valid in the basis $E^A$.
\subsubsection{Interior product}
For a pure supervector field $\xi$ the interior product $\intp_\xi$ is defined. It is a derivative of degree $-1$ and it maps $p$-superforms onto $(p-1)$-superforms, with the properties
\begin{align}
\intp_\xi (\omega \Phi) &= (-1)^{\xi \Phi}(\intp_\xi \omega)\Phi \,,\quad \intp_\xi (\omega + \omega') = \intp_\xi \omega + \intp_\xi \omega'\,, \label{eq:superspace_interior_product_property_1}\\
\intp_\xi (\omega \wedge \tau) &= \omega \wedge \intp_\xi \tau + (-1)^{q+\xi\tau} \intp_\xi \omega \wedge \tau\,, \label{eq:superspace_interior_product_property_2}
\end{align}
where $\omega$, $\omega'$ and $\tau$ are the same objects as above, and $\Phi$ is a pure superfield. Using the identity
\begin{align}
\intp_\xi \exd z^M = (-1)^{\xi M} \xi^M\,,
\end{align}
the interior product of a pure $p$-superform $\omega$ is given by
\begin{align}
\intp_\xi \omega = (-1)^{\xi(\omega+M_1+...+M_{p-1})} \frac{1}{(p-1)!} \exd z^{M_1} \wedge ... \wedge \exd z^{M_{p-1}} \xi^{M_p} \omega_{M_p...M_1}\,. \label{eq:superspace_interior_product_general_superform}
\end{align}
This is consistent with the following definition of the interior product
\begin{align}
(\chi_{p-1},...,\chi_1) \intp_\xi \omega := (-1)^{\xi(\omega+\chi_1+...+\chi_{p-1})} (\xi,\chi_{p-1},...,\chi_1) \omega\,,
\end{align}
where the $\chi_i$ are supervector fields. Note that the interior product of a $0$-form vanishes, i.e.\ $\intp_\chi\Phi=0$. The conjugate of the interior product is given by
\begin{align}
(\intp_\xi\omega)^\cc = (-1)^{\xi\omega} \intp_{\xi^\cc}\omega^\cc\,. \label{eq:superspace_interior_product_conjugate}
\end{align}
Furthermore, with the identity
\begin{align}
\intp_\xi E^A = (-1)^{\xi A} \xi^A\,,
\end{align}
Eq.~\eqref{eq:superspace_interior_product_general_superform} has the same form with respect to the basis $E^A$:
\begin{align}
\intp_\xi \omega = (-1)^{\xi(\omega+A_1+...+A_{p-1})} \frac{1}{(p-1)!} E^{A_1} \wedge ... \wedge E^{A_{p-1}} \xi^{A_p} \omega_{A_p...A_1}\,.
\end{align}
Note, for the pure supervector fields $\xi_1$ and $\xi_2$ the following identity applies:
\begin{align}
\intp_{\xi_1}\intp_{\xi_2} + (-1)^{\xi_1\xi_2}\intp_{\xi_2}\intp_{\xi_1} &= 0\,. 
\end{align}

\subsubsection{Lie derivative}
\label{sec:lie_derivative}
The Lie derivative $\lied_\xi$ along a pure supervector field $\xi$ acting on a pure superform $\omega$ can be expressed in terms of the exterior derivative $\exd$ and the interior product $\intp_\xi$, according to Cartan's formula:
\begin{align}
\lied_\xi \omega := \intp_\xi \exd \omega + \exd \intp_\xi \omega\,. \label{eq:superspace_lie_derivative_superform}
\end{align}
In particular, the Lie derivatives of a superfield $\Phi$ and of the $1$-superform $\exd z^M$ have the simple forms
\begin{align}
\lied_\xi \Phi &= \intp_\xi \exd \Phi = (-1)^{\xi \Phi} \xi^M \del_M \Phi\,, \label{eq:superspace_lie_derivative_id_1}\\
\lied_\xi \exd z^M &= \exd \intp_\xi \exd z^M = (-1)^{\xi M} \exd \xi^M\,. \label{eq:superspace_lie_derivative_id_2}
\end{align}
The conjugation property of the Lie derivative follows directly from Eq.~\eqref{eq:superspace_exterior_derivative_conjugate} and \eqref{eq:superspace_interior_product_conjugate}:
\begin{align}
(\lied_\xi \omega)^\cc = (-1)^{\xi\omega} \lied_{\xi^\cc} \omega^\cc\,. \label{eq:superspace_lie_derivative_conjugation}
\end{align}
If $\omega$ is a pure $p$-superform and $\tau$ a pure $q$-superform, then, according to Eq.~\eqref{eq:superspace_exterior_derivative_property_2} and \eqref{eq:superspace_interior_product_property_2}, the Lie derivative of their wedge product is given by
\begin{align}
\lied_\xi(\omega\wedge\tau) = \omega\wedge\lied_\xi\tau + (-1)^{\xi\tau}\lied_\xi\omega\wedge\tau\,. \label{eq:superspace_lie_derivative_wedge_product}
\end{align}
According to Eq.~\eqref{eq:superspace_lie_derivative_id_1}, \eqref{eq:superspace_lie_derivative_id_2} and \eqref{eq:superspace_lie_derivative_wedge_product} the Lie derivative of the supervielbein $E^A$ is calculated as follows:
\begin{align}
\begin{split}
\lied_\xi E^A &= \lied_\xi(\exd z^M{E_M}^A) \\
&= \exd z^M(\lied_\xi{E_M}^A) + (-1)^{\xi(M+A)}(\lied_\xi\exd z^M){E_M}^A \\
&= (-1)^{\xi A}\exd z^M\big((-1)^{\xi M}\xi^N(\del_N{E_M}^A) + (\del_M\xi^N){E_N}^A\big)\,.
\end{split} \label{eq:superspace_lie_derivative_vielbein}
\end{align}
Furthermore, the Lie derivative along $\xi$ of a pure supervector field $\chi$ is defined as
\begin{align}
\lied_\xi \chi := (-1)^{\xi\chi} \mcom{\xi}{\chi} = (-1)^{\xi\chi} \big( \xi\chi -(-1)^{\xi\chi} \chi\xi \big)\,, \label{eq:superspace_lie_derivative_supervector}
\end{align}
where $\mcom{\xi}{\chi}$ is the supercommutator, whose components are given by
\begin{align}
\mcom{\xi}{\chi}^M &= \xi^N(\del_N\chi^M) -(-1)^{\xi\chi} \chi^N(\del_N\xi^M)\,.
\end{align}
In general, a straightforward calculation shows that for two pure supervector fields $\xi_1$ and $\xi_2$ the supercommutator of the corresponding Lie derivatives is equal to the Lie derivative of the supercommutator of these supervector fields, namely
\begin{align}
\mcom{\lied_{\xi_1}}{\lied_{\xi_2}} &= \lied_{\mcom{\xi_1}{\xi_2}}\,.
\end{align}
Moreover, the supercommutator of an interior product and a Lie derivative is again an interior product:
\begin{align}
\mcom{\intp_{\xi_1}}{\lied_{\xi_2}} &= \intp_{\mcom{\xi_1}{\xi_2}}\,.
\end{align}
Note that the Lie derivative for general tensors is uniquely defined by Eq.~\eqref{eq:superspace_lie_derivative_superform} and \eqref{eq:superspace_lie_derivative_supervector}, and the requirement, that it obeys the Leibniz rule with respect to the tensor product. If the (infinitesimal) supervector field $\xi$ is real and $c$-type, its flow defines a (local) diffeomorphism of the superspace. The Lie derivative $\lied_\xi$ then provides the infinitesimal action of this diffeomorphism on tensors (see e.g.\ \cite{DeWitt:1992cy}).

\subsection{Structure group of $\U(1)$ superspace}
\label{sec:superspace_structure_group}
Based on ordinary spacetime, the structure group in supergravity contains the Lorentz group $\SO(1,3)$, which acts simultaneously on the vectorial and on the spinorial components of the supervielbein. There is the possibility to introduce beside the Lorentz group an additional group, usually called internal group, which acts only on the spinorial components, since they do not have an analogue in ordinary spacetime. In $\U(1)$ superspace the additional group is the group $\U(1)$, and the structure group is given by the direct product $\SO(1,3)\times\U(1)$. This extra $\U(1)$ is used to formulate K\"ahler superspace ($\UK(1)$ superspace), which is the appropriate framework to describe the coupling of matter to supergravity in a geometric way for $N=1$ supersymmetry. K\"ahler superspace is obtained from $\U(1)$ superspace by a suitable identification of the $\U(1)$ pre-potential and pre-gauge transformations with the K\"ahler potential and K\"ahler transformations, respectively, which are functions of the matter superfields, as discussed in Section~\ref{sec:kahler_superspace}.
\begin{itemize}
\item\textbf{Lorentz group:}\\
With respect to the Lorentz group, the vectorial components $E^a$ of the supervielbein transform in the $4$-dimensional irreducible representation $(\frac{1}{2},\frac{1}{2})$, whereas the spinorial components $E^\alpha$ and $E_\sd{\alpha}$ transform in the $2$-dimensional representations $(\frac{1}{2},0)$ and $(0,\frac{1}{2})$, respectively. Thus, in the basis $E^A = (E^a,E^\alpha,E_\sd{\alpha})$ the six basis elements $G_{de}$, with $d,e\in\{0,1,2,3\}$ and $G_{ed} = -G_{de}$, of $\so(1,3)$ have the form
\begin{align}
{(G_{de})_B}^A = \begin{pmatrix}
{(M_{de})_b}^a & & \\
& {(S_{de})_\beta}^\alpha & \\
& & -{(S^\dag_{de})^\sd{\beta}}_\sd{\alpha}
\end{pmatrix}, \label{eq:superspace_lorentz_generator}
\end{align}
which represents the reducible representation $(\frac{1}{2},\frac{1}{2})\oplus(\frac{1}{2},0)\oplus(0,\frac{1}{2})$. Note that $-S^\dag_{de}$ is the dual conjugate representation of $S_{de}$. The basis elements $G_{de}$ fulfil the commutation relations
\begin{align}
\com{G_{ab}}{G_{de}} &= - \eta_{ad}G_{be} + \eta_{ae}G_{bd} + \eta_{bd}G_{ae} - \eta_{be}G_{ad}\,,
\end{align}
which are realised by the particular set of matrices
\begin{align}
{(M_{de})_b}^a &= \eta_{db}\delta^a_e - \eta_{eb}\delta^a_d\,, \\
{(S_{de})_\beta}^\alpha &= -{(\sig_{de})_\beta}^\alpha\,, \\
{(S^\dag_{de})^\sd{\beta}}_\sd{\alpha} &= +{(\sigb_{de})^\sd{\beta}}_\sd{\alpha}\,.
\end{align}
A general element $\soae\in\so(1,3)$ in the representation of Eq.~\eqref{eq:superspace_lorentz_generator} is written as
\begin{align}
\soae = \frac{1}{2} K^{de} G_{de}\,, \label{eq:superspace_general_element_lorentz_algebra}
\end{align}
and a general element $\soge\in\SO(1,3)$ locally around the identity is given by
\begin{align}
\soge = \exp\Big(\frac{1}{2} K^{de} G_{de}\Big)\,. \label{eq:superspace_general_element_lorentz_group}
\end{align}
In order to describe gauge transformations with respect to the Lorentz group, the coefficients $K^{de}$ are real, even superfields with $K^{ed} = -K^{de}$. The action of the gauge transformation $\soge$ on the supervielbein is given by
\begin{align}
E^A \mapsto E^B {\soge_B}^A = \big( E^b {\soge_b}^a, E^\beta {\soge_\beta}^\alpha, E_\sd{\beta} {\soge^\sd{\beta}}_\sd{\alpha} \big)\,. \label{eq:superspace_vielbein_transformation_lorentz}
\end{align}
The inverse supervielbein transforms in the dual representation of the supervielbein, namely:
\begin{align}
E_A \mapsto {(\soge^{-1})_A}^B E_B = \big( {(\soge^{-1})_a}^b E_b, {(\soge^{-1})_\alpha}^\beta E_\beta, {(\soge^{-1})^\sd{\alpha}}_\sd{\beta} E^\sd{\beta} \big)\,. \label{eq:superspace_inverse_vielbein_transformation_lorentz}
\end{align}
For infinitesimal gauge transformations $\soge \approx \mathbb{1} + \soae$, Eq.~\eqref{eq:superspace_vielbein_transformation_lorentz} and \eqref{eq:superspace_inverse_vielbein_transformation_lorentz} read
\begin{align}
E^A &\mapsto E^A + E^B {\soae_B}^A\,, \\
E_A &\mapsto E_A - {\soae_A}^B E_B\,.
\end{align}
In the basis of Eq.~\eqref{eq:superspace_lorentz_generator}, the invariant metric $\rho_{BA}$ with respect to Lorentz transformations has the form
\begin{align}
\rho_{BA} = \begin{pmatrix}
\eta_{ba} & & \\
& \eps_{\beta\alpha} & \\
& & \eps^{\sd{\beta}\sd{\alpha}}
\end{pmatrix}.
\end{align}
It is straightforward to check that
\begin{align}
{\soge_B}^D {\soge_A}^C \rho_{DC} &= \rho_{BA}\,, \label{eq:superspace_metric_lorentz_transformation}
\end{align}
thus
\begin{align}
\rho^{AC}{\soae_C}^D\rho_{DB} &= -{\soae_B}^A\,. \label{eq:superspace_metric_lorentz_transformation_infinitesimal}
\end{align}
Writing $M_{de}$, $S_{de}$ and $S^\ct_{de}$, defined in Eq.~\eqref{eq:superspace_lorentz_generator}, in terms of lower indices
\begin{align}
(M_{de})_{ba} &= \eta_{ac}{(M_{de})_b}^c\,, \\
(S_{de})_{\sym{2}{\beta\alpha}} &= \eps_{\alpha\gamma} {(S_{de})_\beta}^\gamma\,, \\
(S^\dag_{de})_{\sym{2}{\sd{\beta}\sd{\alpha}}} &= \eps_{\sd{\beta}\sd{\gamma}} {(S^\dag_{de})^\sd{\gamma}}_\sd{\alpha}\,,
\end{align}
they obey the relations
\begin{align}
(S_{de})_{\sym{2}{\beta\alpha}} &= +\frac{1}{2} (\sig^{ba}\eps)_{\beta\alpha} (M_{de})_{ba}\,, \label{eq:superspace_irrep_generator_relation_1}\\
(S^\dag_{de})_{\sym{2}{\sd{\beta}\sd{\alpha}}} &= +\frac{1}{2} (\eps\sigb^{ba})_{\sd{\beta}\sd{\alpha}} (M_{de})_{ba}\,, \label{eq:superspace_irrep_generator_relation_2}\\
(M_{de})_{ba} &= -(\eps\sig_{ba})^{\beta\alpha} (S_{de})_{\sym{2}{\beta\alpha}} - (\sigb_{ba}\eps)^{\sd{\beta}\sd{\alpha}} (S^\ct_{de})_{\sym{2}{\sd{\beta}\sd{\alpha}}}\,. \label{eq:superspace_irrep_generator_relation_3}
\end{align}
and
\begin{align}
\big((M_{de})_{ba}\big)^\cc &= +(M_{de})_{ba}\,, \label{eq:superspace_irrep_generator_relation_4}\\
\big((S_{de})_{\sym{2}{\beta\alpha}}\big)^\cc &= +(S^\dag_{de})_{\sym{2}{\sd{\beta}\sd{\alpha}}}\,. \label{eq:superspace_irrep_generator_relation_5}
\end{align}
Note that $(M_{de})_{ba}$ is antisymmetric in $b,a$, whereas $(S_{de})_{\sym{2}{\beta\alpha}}$ and $(S^\dag_{de})_{\sym{2}{\sd{\beta}\sd{\alpha}}}$ are symmetric in $\beta,\alpha$ and $\sd{\beta},\sd{\alpha}$ respectively. Plugging Eqs.~\eqref{eq:superspace_irrep_generator_relation_1}--\eqref{eq:superspace_irrep_generator_relation_3} in Eq.~\eqref{eq:superspace_lorentz_generator}, the following relations for the components of $G_{de}$ are obtained:
\begin{align}
(G_{de})_{\sym{2}{\beta\alpha}} &= +\frac{1}{2} (\sig^{ba}\eps)_{\beta\alpha} (G_{de})_{ba}\,, \label{eq:superspace_generator_components_relation_1}\\
(G_{de})_{\sym{2}{\sd{\beta}\sd{\alpha}}} &= -\frac{1}{2} (\eps\sigb^{ba})_{\sd{\beta}\sd{\alpha}} (G_{de})_{ba}\,, \label{eq:superspace_generator_components_relation_2}\\
(G_{de})_{ba} &= -(\eps\sig_{ba})^{\beta\alpha} (G_{de})_{\sym{2}{\beta\alpha}} + (\sigb_{ba}\eps)^{\sd{\beta}\sd{\alpha}} (G_{de})_{\sym{2}{\sd{\beta}\sd{\alpha}}}\,, \label{eq:superspace_generator_components_relation_3}
\end{align}
where the last equation can be written more efficiently by using spinor indices
\begin{align}
(G_{de})_{\beta\sd{\beta}\s\alpha\sd{\alpha}} = +2\eps_{\sd{\beta}\sd{\alpha}} (G_{de})_{\sym{2}{\beta\alpha}} - 2\eps_{\beta\alpha} (G_{de})_{\sym{2}{\sd{\beta}\sd{\alpha}}}\,. \label{eq:superspace_generator_components_relation_4}
\end{align}
Furthermore, according to Eq.~\eqref{eq:superspace_irrep_generator_relation_4} and \eqref{eq:superspace_irrep_generator_relation_5} the conjugated components read
\begin{align}
\big((G_{de})_{ba}\big)^\cc &= +(G_{de})_{ba}\,, \label{eq:superspace_generator_components_relation_5}\\
\big((G_{de})_{\sym{2}{\beta\alpha}}\big)^\cc &= -(G_{de})_{\sym{2}{\sd{\beta}\sd{\alpha}}}\,, \label{eq:superspace_generator_components_relation_6}
\end{align}
which can be summarized as
\begin{align}
\big({(G_{de})_B}^A\big)^\cc = (-1)^B {(G_{de})_\sd{B}}^\sd{A}\,. \label{eq:superspace_generator_components_relation_7}
\end{align}
Note that Eqs.~\eqref{eq:superspace_generator_components_relation_1}--\eqref{eq:superspace_generator_components_relation_7} hold true for the components of a general element $\soae\in\so(1,3)$ as defined in Eq.~\eqref{eq:superspace_general_element_lorentz_algebra}.
\item\textbf{$\U(1)$ group:}\\
A gauge transformation $\uge$ with respect to the $\U(1)$ group is given by
\begin{align}
\uge = \exp(\uae)\,, \label{eq:superspace_general_element_u1_group}
\end{align}
where $\uae$ is an imaginary, even superfield, i.e.\ $\uae^\cc = -\uae$. The irreducible representations of $\U(1)$ are classified by weights $w\in\mathbb{Z}$. The $\U(1)$ factor of the structure group acts only on the spinorial components of the supervielbein. In particular, the weights of the supervielbein components are defined as
\begin{align}
w(E^a) &= 0\,, & w(E^\alpha) &= +1\,, & w(E_\sd{\alpha}) &= -1\,.  \label{eq:superspace_vielbein_weight_u1_group}
\end{align}
Since the inverse supervielbein transforms in the dual representation, its components have opposite weights
\begin{align}
w(E_a) &= 0\,, & w(E_\alpha) &= -1\,, & w(E^\sd{\alpha}) &= +1\,. \label{eq:superspace_inversevielbein_weight_u1_group}
\end{align}
The action of an $\U(1)$ gauge transformation $\uge$ on $E^A$ and $E_A$ is thus given by
\begin{align}
E^A &\mapsto \uge^{w(E^A)}E^A = (E^a,\uge\,E^\alpha,\uge^{-1}E_\sd{\alpha})\,, \label{eq:superspace_vielbein_transformation_u1}\\
E_A &\mapsto \uge^{w(E_A)}E_A = (E_a,\uge^{-1}E_\alpha,\uge\,E^\sd{\alpha})\,, \label{eq:superspace_inverse_vielbein_transformation_u1}
\end{align}
and for an infinitesimal $\uge\approx 1+\uae$ it reads
\begin{align}
E^A &\mapsto E^A + w(E^A)\,\uae\,E^A\,, \label{eq:superspace_vielbein_transformation_u1_infinitesimal}\\
E_A &\mapsto E_A + w(E_A)\,\uae\,E_A\,. \label{eq:superspace_inverse_vielbein_transformation_u1_infinitesimal}
\end{align}
\end{itemize}
Generally speaking, under (infinitesimal) Lorentz transformations upper an lower Lorentz indices $()^A$ and $()_A$ transform as
\begin{align}
()^A &\mapsto ()^B {\soge_B}^A\,, & ()^A &\mapsto ()^A + ()^B {\soae_B}^A\,,\\
()_A &\mapsto {(\soge^{-1})_A}^B ()_B\,, & ()_A &\mapsto ()_A - {\soae_A}^B ()_B\,,
\end{align}
which is consistent with lowering and raising the indices with the metric tensor $\rho_{BA}$ and its inverse $\rho^{BA}$, cf.\ Eq.~\eqref{eq:superspace_metric_lorentz_transformation}. On the other hand, since $\rho_{BA}$ and $\rho^{BA}$ are inert under $\U(1)$ transformations, the weight is not changed. In order that $\U(1)$ transformations are compatible with Lorentz transformations, the $\U(1)$ weights of the components $()^A$ must have the form
\begin{align}
w(()^A) = \big(w(()^a),w(()^\alpha),w(()_\sd{\alpha})\big) = \big(w_1,w_2,w_3\big)\,,
\end{align}
where $w_1,w_2,w_3\in\mathbb{Z}$. The same holds true for lower Lorentz indices. By definition, a tensor $T$ of rank $(s,r)$, as defined in Section~\ref{sec:superspace_tensors}, does not transform under the structure group. Thus, the weights of the components are given by
\begin{align}
w( {T_{A_r...A_1}}^{A_{r+1}...A_{r+s}}) &= \sum_{i=1}^r w(E_{A_i}) + \sum_{i=r+1}^{r+s} w(E^{A_i})\,.
\end{align}
Note that the superspace indices $M$ do not transform under the structure group $\SO(1,3)\times\U(1)$.
\begin{itemize}
\item\textbf{Superforms with a Lorentz index:}\\
A particular type of objects, which is often used in the subsequent sections, are $p$-superforms $\chi^A$ with a Lorentz index $A$ and $\U(1)$ weight $w(\chi^A)=\big(w(\chi^a),w(\chi^\alpha),w(\chi_\sd{\alpha})\big)$. Although, in the following discussion only the case of an upper index $A$ is considered, the statements hold true for a lower index with obvious adjustments. Written as a superform, $\chi^A$ reads
\begin{align}
\chi^A &= \frac{1}{p!}E^{C_1}\wedge...\wedge E^{C_p}{\chi_{C_p...C_1}}^A\,,
\end{align}
thus the $\U(1)$ weights of the components are given by 
\begin{align}
w\big({\chi_{C_p...C_1}}^A\big) &= w(\chi^A) + \sum_{i=1}^p w(E_{C_i})\,.
\end{align}
Under (infinitesimal) transformations of the structure group, $\chi^A$ transforms as
\begin{align}
\chi^A &\mapsto \uge^{w(\chi^A)}\chi^B {\soge_B}^A\,, & \chi^A &\mapsto \chi^A + \chi^B {\soae_B}^A + w(\chi^A)\,\uae\,\chi^A\,. \label{eq:superspace_gauge_transformation_vector_components}
\end{align}
Note, if the weights of $\chi^A$ take the values $w(\chi^A)=w(E^A)$, the object $\chi^A\otimes E_A$ represents a tensor of rank $(1,p)$. In particular, if $\chi^A$ is a $0$-superform then a supervector field $\chi^A E_A$ is represented. Furthermore, by definition the supervielbein $E^A$ is a $1$-superform with an upper Lorentz index $A$. The corresponding $(1,1)$ tensor $E^A\otimes E_A$ is usually referred to as solder form or fundamental $1$-form.
\item\textbf{Lorentz invariant superforms:}\\
An other type of objects, which are often used in the following, are $p$-superforms $\omega$ which are invariant under Lorentz transformations, but have a $\U(1)$ weight $w(\omega)$. In particular, if $\omega$ is a $0$-superform, then a superfield is represented. The (infinitesimal) transformations of $\omega$ under the $\U(1)$ group is given by
\begin{align}
\omega &\mapsto g^{w(\omega)}\omega\,, & \omega &\mapsto \omega + w(\omega)\,\uae\,\omega\,.
\end{align}
\end{itemize}

\subsubsection{Lorentz and $\U(1)$ connection}
\begin{itemize}
\item\textbf{Lorentz connection:}\\
The Lorentz connection $\Omega$ is an even $1$-superform which takes values in the Lie algebra of the Lorentz group:
\begin{align}
\Omega = \exd z^M \Omega_M = E^C \Omega_C\,, \quad\text{with}\quad \Omega_C = \frac{1}{2}\tilde{\Omega}_C{}^{de} G_{de}\,.
\end{align}
The basis elements $G_{de}$ of $\so(1,3)$ are defined in Eq.~\eqref{eq:superspace_lorentz_generator} and the $1$-superforms $\tilde{\Omega}^{de} = E^C \tilde{\Omega}_C{}^{de}$ are real. The Lorentz indices are explicitly written as
\begin{align}
{\Omega_B}^A = E^C {\Omega_{CB}}^A\,, \quad\text{with}\quad {\Omega_{CB}}^A = \frac{1}{2}\tilde{\Omega}_C{}^{de} {(G_{de})_B}^A\,,
\end{align}
which shows that ${\Omega_{CB}}^A$ inherits the block diagonal structure of ${(G_{de})_B}^A$, i.e.\
\begin{align}
{\Omega_{CB}}^A = \begin{pmatrix}
{\Omega_{Cb}}^a & & \\
& {\Omega_{C\beta}}^\alpha & \\
& & \Omega\indices{_C^{\sd{\beta}}_{\sd{\alpha}}}
\end{pmatrix}.
\end{align}
Thus, according to Eq.~\eqref{eq:superspace_metric_lorentz_transformation_infinitesimal} the Lorentz connection fulfils the identity
\begin{align}
\rho^{AC}{\Omega_C}^D\rho_{DB} &= - {\Omega_B}^A\,. \label{eq:superspace_lorentz_connection_components_index_height}
\end{align}
From Eqs.~\eqref{eq:superspace_generator_components_relation_1}--\eqref{eq:superspace_generator_components_relation_4} the following relations can be read off:
\begin{align}
\Omega_{C\sym{2}{\beta\alpha}} &= +\frac{1}{2}(\sig^{ba}\eps)_{\beta\alpha} \Omega_{Cba}\,, \label{eq:superspace_lorentz_connection_components_relation_1}\\
\Omega_{C\sym{2}{\sd{\beta}\sd{\alpha}}} &= -\frac{1}{2}(\eps\sigb^{ba})_{\sd{\beta}\sd{\alpha}} \Omega_{Cba}\,, \label{eq:superspace_lorentz_connection_components_relation_2}\\
\Omega_{Cba} &= -(\eps\sig_{ba})^{\beta\alpha} \Omega_{C\sym{2}{\beta\alpha}} + (\sigb_{ba}\eps)^{\sd{\beta}\sd{\alpha}} \Omega_{C\sym{2}{\sd{\beta}\sd{\alpha}}}\,, \label{eq:superspace_lorentz_connection_components_relation_3}\\
\Omega_{C\s\beta\sd{\beta}\s\alpha\sd{\alpha}} &= +2\eps_{\sd{\beta}\sd{\alpha}} \Omega_{C\sym{2}{\beta\alpha}} - 2\eps_{\beta\alpha} \Omega_{C\sym{2}{\sd{\beta}\sd{\alpha}}}\,. \label{eq:superspace_lorentz_connection_components_relation_4}
\end{align}
Furthermore, from Eqs.~\eqref{eq:superspace_generator_components_relation_5}--\eqref{eq:superspace_generator_components_relation_7} follows
\begin{align}
\big(\Omega_{Cba}\big)^\cc &= +\Omega_{\sd{C}ba}\,, \label{eq:superspace_lorentz_connection_components_relation_5}\\
\big(\Omega_{C\sym{2}{\beta\alpha}}\big)^\cc &= -\Omega_{\sd{C}\sym{2}{\sd{\beta}\sd{\alpha}}}\,, \label{eq:superspace_lorentz_connection_components_relation_6}
\end{align}
and
\begin{align}
({\Omega_{B}}^A)^\cc = (-1)^B {\Omega_{\sd{B}}}^\sd{A}\,,\qquad ({\Omega_{CB}}^A)^\cc = (-1)^B {\Omega_{\sd{C}\sd{B}}}^\sd{A}\,, \label{eq:superspace_lorentz_connection_components_relation_7}
\end{align}
where $(\tilde{\Omega}^{de})^\cc = \tilde{\Omega}^{de}$ is used. The Lorentz connection is inert under $\U(1)$ transformations, i.e.\ $w({\Omega_B}^A)=0$. Under Lorentz transformations $\soge$, as defined in Eq.~\eqref{eq:superspace_general_element_lorentz_group}, ${\Omega_B}^A$ transforms as
\begin{align}
{\Omega_B}^A \mapsto {(\soge^{-1})_B}^D \big({\Omega_D}^C {\soge_C}^A - \exd {\soge_D}^C\big)\,. \label{eq:superspace_lorentz_connection_transformation_lorentz}
\end{align}
For infinitesimal transformations $\soge\approx\mathbb{1}+\soae$ this reads
\begin{align}
{\Omega_B}^A \mapsto {\Omega_B}^A + {\com{\Omega}{\soae}_B}^A - \exd {\soae_B}^A\,, \label{eq:superspace_lorentz_connection_transformation_lorentz_infinitesimal}
\end{align}
with the commutator ${\com{\Omega}{\soae}_B}^A = {\Omega_B}^C {\soae_C}^A - {\soae_B}^C {\Omega_C}^A$.
\item\textbf{$\U(1)$ connection:}\\
The $\U(1)$ connection $A$ is an even $1$-superform which takes values in the Lie algebra $\uu(1)$. Thus, the components are written as
\begin{align}
A = \exd z^M A_M = E^A A_A\,,
\end{align}
and $A$ is imaginary:
\begin{align}
A^\cc = -A\,,\qquad (A_A)^\cc = -A_\sd{A}\,. \label{eq:superspace_u1_connection_components_relation_1}
\end{align}
The $\U(1)$ connection is a singlet with respect to $\SO(1,3)$. Under $\U(1)$ transformations, as defined in Eq.~\eqref{eq:superspace_general_element_u1_group}, $A$ transforms as
\begin{align}
A \mapsto \uge^{-1} A \uge - \uge^{-1}\exd \uge = A - \uge^{-1}\exd \uge \,, \label{eq:superspace_u1_connection_transformation_u1}
\end{align}
which takes the following form
\begin{align}
A \mapsto  A - \exd \uae \,, \label{eq:superspace_u1_connection_transformation_u1_infinitesimal}
\end{align}
for infinitesimal transformations $\uge\approx 1+\uae$.
\end{itemize}

\subsubsection{Covariant derivative}
\begin{itemize}
\item\textbf{Superforms with a Lorentz index:}\\
Consider a $p$-superform $\chi^A$ with an upper Lorentz index and $\U(1)$ weight $w(\chi^A)$. The covariant derivative $\cD$, with respect to the Lorentz and the $\U(1)$ connection, maps $\chi^A$ onto a $(q+1)$-superform with the same Lorentz index and $\U(1)$ weight. Motivated by the transformation properties of $\chi^A$ under gauge transformations in Eq.~\eqref{eq:superspace_gauge_transformation_vector_components}, the covariant derivative is defined as follows
\begin{align}
\cD \chi^A &= \exd\chi^A + \chi^C\wedge{\Omega_C}^A + w(\chi^A)\chi^A\wedge A\,. \label{eq:superspace_covariant_derivative_vector_component}
\end{align}
By taking into account Eq.~\eqref{eq:superspace_gauge_transformation_vector_components} and \eqref{eq:superspace_lorentz_connection_transformation_lorentz}, it is straightforward to check that the components $\cD \chi^A$ transforms the same way as $\chi^A$ under gauge transformations, namely
\begin{align}
\cD \chi^A \mapsto \uge^{w(\chi^A)}\cD \chi^B {\soge_B}^A\,. \label{eq:superspace_covariant_derivative_vector_component_transf}
\end{align}
Using the identity
\begin{align}
\exd\chi^A &= \frac{1}{p!}E^{C_1}\wedge...\wedge E^{C_p}\wedge\exd{\chi_{C_p...C_1}}^A + \frac{1}{p!}\sum_{i=1}^p (-1)^{p-i} E^{C_1}\wedge...\wedge \exd E^{C_i}...\wedge E^{C_p}{\chi_{C_p...C_1}}^A\,, \label{eq:superspace_covariant_derivative_vector_id_1}
\end{align}
the covariant derivative $\cD\chi^A$ can also be written in the form
\begin{align}
\begin{split}
\cD\chi^A &= \frac{1}{p!}E^{C_1}\wedge...\wedge E^{C_p}\wedge\cD{\chi_{C_p...C_1}}^A + \frac{1}{p!}\sum_{i=1}^p (-1)^{p-i} E^{C_1}\wedge...\wedge \cD E^{C_i}...\wedge E^{C_p}{\chi_{C_p...C_1}}^A\\
&= \frac{1}{p!}E^{C_1}\wedge...\wedge E^{C_p}\wedge\cD{\chi_{C_p...C_1}}^A + \frac{1}{(p-1)!}E^{C_1}\wedge...\wedge E^{C_{p-1}}\wedge\cD E^{C_p}{\chi_{C_p...C_1}}^A\,, \label{eq:superspace_covariant_derivative_vector_id_2}
\end{split}
\end{align}
with
\begin{align}
\cD{\chi_{C_p...C_1}}^A &= \exd{\chi_{C_p...C_1}}^A - \sum_{i=1}^p {\Omega_{C_i}}^{B_i}{\chi_{C_p...B_i...C_1}}^A + {\chi_{C_p...C_1}}^B{\Omega_B}^A + w({\chi_{C_p...C_1}}^A){\chi_{C_p...C_1}}^A A\,.
\end{align}
Note, a change of the height of an index is consistent with the identity in Eq.~\eqref{eq:superspace_lorentz_connection_components_index_height} for the Lorentz connection. For example, according to Eq.~\eqref{eq:superspace_covariant_derivative_vector_component} the covariant derivative of $\chi_A = \rho_{AB}\chi^B$ is given by
\begin{align}
\cD\chi_A = \exd\chi_A - \chi_C\wedge{\Omega_A}^C + w(\chi_A)\chi_A\wedge A\,,
\end{align}
where $w(\chi_A)=w(\chi^A)$.
\item\textbf{Lorentz invariant superforms:}\\
For a Lorentz invariant $p$-superform $\omega$ with $\U(1)$ weight $w(\omega)$ the covariant derivative maps $\omega$ onto a $(p+1)$-superform with the same weight and is given by
\begin{align}
\cD\omega &= \exd\omega + w(\omega)\omega\wedge A\,.
\end{align}
It is straightforward to check that $\cD\omega$ transforms the same way under the $\U(1)$ group as $\omega$, namely
\begin{align}
\cD\omega \mapsto \uge\,\cD\omega\,.
\end{align}
\end{itemize}
Like the exterior derivative, the covariant derivative is an even derivative of degree $1$. As such, it is linear and obeys the Leibniz rule (see Eq.~\eqref{eq:superspace_exterior_derivative_property_1} and \eqref{eq:superspace_exterior_derivative_property_2}), but in general $\cD\cD \neq 0$ as discussed in Section~\ref{sec:torsion_lorentz_curvature_u1_fieldstrength}. For example, for the $p$-superform $\chi^A$ and the $q$-superform $\xi^B$, both with an upper Lorentz index, the covariant derivative of their wedge product $\chi^A\wedge\xi^B$ is given by
\begin{align}
\begin{split}
\cD(\chi^A\wedge\xi^B) &= \exd(\chi^A\wedge\xi^B) + \chi^C\wedge\xi^B\wedge{\Omega_C}^A + \chi^A\wedge\xi^C\wedge{\Omega_C}^B \\
&\quad\; + \big(w(\chi^A)+w(\xi^B)\big) \chi^A\wedge\xi^B\wedge A \\
&= \chi^A\wedge\big(\exd\xi^B + \xi^C\wedge{\Omega_C}^B + w(\xi^B)\xi^B\wedge A\big) \\
&\quad\; + (-1)^q \big(\exd\chi^A + \chi^C\wedge{\Omega_C}^A + w(\chi^A) \chi^A\wedge A\big)\wedge\xi^B \\
&= \chi^A\wedge\cD\xi^B + (-1)^q \cD\chi^A\wedge\xi^B\,.
\end{split}
\end{align}
Furthermore, for the Lorentz invariant $p$-superform $\omega$ and $q$-superform $\tau$ with $\U(1)$ weights $w(\omega)$ and $w(\tau)$, respectively, a similar calculation shows
\begin{align}
\cD(\omega\wedge\xi^B) &= \cD\omega\wedge\xi^B + (-1)^q \omega\wedge\cD\xi^B\,, \\
\cD(\omega\wedge\tau) &= \cD\omega\wedge\tau + (-1)^q \omega\wedge\cD\tau\,.
\end{align}

\subsubsection{Torsion tensor, Lorentz curvature and $\U(1)$ field strength}
\label{sec:torsion_lorentz_curvature_u1_fieldstrength}
The structure equations with respect to the structure group $\SO(1,3)\times\U(1)$ are given by
\begin{align}
T^A &:= \cD E^A = \exd E^A + E^B \wedge {\Omega_B}^A + w(E^A) E^A \wedge A\,, \label{eq:superspace_structure_equation_torsion}\\
{R_B}^A &:= \exd {\Omega_B}^A + {\Omega_B}^C \wedge {\Omega_C}^A\,, \label{eq:superspace_structure_equation_lorentz_curvature}\\
F &:= \exd A\,, \label{eq:superspace_structure_equation_u1_fieldstrength}
\end{align}
where $T^A$ is the torsion tensor, ${R_B}^A$ the Lorentz curvature tensor and $F$ the $\U(1)$ field strength. Written as $2$-superforms, they have the form
\begin{align}
T^A &= \frac{1}{2} E^B \wedge E^C {T_{CB}}^A\,, \label{eq:superspace_torsion_definition}\\
{R_B}^A &= \frac{1}{2} E^C \wedge E^D {R_{DCB}}^A\,, \label{eq:superspace_lorentz_curvature_definition}\\
F &= \frac{1}{2} E^C \wedge E^D F_{DC}\,. \label{eq:superspace_u1_fieldstrength_definition}
\end{align}
\begin{itemize}
\item \textbf{Torsion tensor}\\
According to Eq.~\eqref{eq:superspace_2form_components_symmetry}, the components of the torsion tensor have the following symmetry property
\begin{align}
{T_{BC}}^A = -(-1)^{CB} {T_{CB}}^A\,.
\end{align}
From the definition of the torsion tensor in Eq.~\eqref{eq:superspace_structure_equation_torsion} follows that the $\U(1)$ weight of $T^A$ and ${T_{CB}}^A$ is given by
\begin{align}
w(T^A) &= w(E^A)\,, & w({T_{CB}}^A) = w(E_C) + w(E_B) + w(E^A)\,.
\end{align}
Furthermore, under conjugation $T^A$ transforms as
\begin{align}
\begin{split}
(T^A)^\cc &= (\cD E^A)^\cc \\
&= \exd E^\sd{A} + (-1)^B E^\sd{B} \wedge {\Omega_\sd{B}}^\sd{A} - w(E^A) E^\sd{A} \wedge A \\
&= \exd E^\sd{A} + E^B \wedge {\Omega_B}^\sd{A} + w(E^\sd{A}) E^\sd{A} \wedge A \\
&= \cD E^\sd{A} = T^\sd{A}\,,
\end{split} \label{eq:superspace_torsion_conjugate}
\end{align}
where Eq.~\eqref{eq:superspace_lorentz_connection_components_relation_7} and \eqref{eq:superspace_u1_connection_components_relation_1} are used for the conjugation of the Lorentz and the $\U(1)$ connection, respectively. By taking into account the Grassmann parity $\eps(T^A) = \eps(E^A)$, Eq.~\eqref{eq:superspace_tensor_components_conjuagte} and \eqref{eq:superspace_torsion_conjugate} imply that the conjugation of the components is given by
\begin{align}
({T_{CB}}^A)^\cc &= (-1)^{CB+CA+BA} {T_{\sd{C}\sd{B}}}^\sd{A}\,. \label{eq:superspace_torsion_components_conjugate}
\end{align}
\item \textbf{Lorentz curvature tensor}\\
The Lorentz curvature tensor is an $\so(1,3)$-valued $2$-superform and it inherits the block diagonal structure from the Lorentz connection:
\begin{align}
{R_{DCB}}^A = \begin{pmatrix}
{R_{DCb}}^a & & \\
& {R_{DC\beta}}^\alpha & \\
& & R\indices{_{DC}^{\sd{\beta}}_{\sd{\alpha}}}
\end{pmatrix}.
\end{align}
According to Eq.~\eqref{eq:superspace_structure_equation_lorentz_curvature} the components of the Lorentz curvature have the explicit form
\begin{align}
{R_{DCB}}^A = \cD_D {\Omega_{CB}}^A -(-1)^{DC} \cD_C {\Omega_{DB}}^A + {T_{DC}}^F{\Omega_{FB}}^A + {\mcom{\Omega_D}{\Omega_C}_B}^A\,,
\end{align}
where ${\Omega_{CB}}^A$ is considered as the component of a $1$-superform in the covariant derivative, namely
\begin{align}
\cD_D{\Omega_{CB}}^A &= D_D{\Omega_{CB}}^A - {\Omega_{DC}}^F{\Omega_{FB}}^A - w(E^C)A_D{\Omega_{CB}}^A\,,
\end{align}
using the notation $\cD=E^D\cD_D$ and $\exd=E^D D_D$. Moreover, the components have the following symmetry properties:
\begin{align}
R_{CDBA} &= -(-1)^{DC} R_{DCBA}\,, \\
R_{DCAB} &= -(-1)^B R_{DCBA}\,.
\end{align}
From the identities for the components of the Lorentz connection in Eqs.~\eqref{eq:superspace_lorentz_connection_components_relation_1}--\eqref{eq:superspace_lorentz_connection_components_relation_7} follows
\begin{align}
R_{DC\sym{2}{\beta\alpha}} &= +\frac{1}{2}(\sig^{ba}\eps)_{\beta\alpha} R_{DCba}\,, \label{eq:superspace_lorentz_curvature_components_relation_1}\\
R_{DC\sym{2}{\sd{\beta}\sd{\alpha}}} &= -\frac{1}{2}(\eps\sigb^{ba})_{\sd{\beta}\sd{\alpha}} R_{DCba}\,, \label{eq:superspace_lorentz_curvature_components_relation_2}\\
R_{DCba} &= -(\eps\sig_{ba})^{\beta\alpha} R_{DC\sym{2}{\beta\alpha}} + (\sigb_{ba}\eps)^{\sd{\beta}\sd{\alpha}} R_{DC\sym{2}{\sd{\beta}\sd{\alpha}}}\,, \label{eq:superspace_lorentz_curvature_components_relation_3}\\
R_{DC\s\beta\sd{\beta}\s\alpha\sd{\alpha}} &= +2\eps_{\sd{\beta}\sd{\alpha}} R_{DC\sym{2}{\beta\alpha}} - 2\eps_{\beta\alpha} R_{DC\sym{2}{\sd{\beta}\sd{\alpha}}}\,, \label{eq:superspace_lorentz_curvature_components_relation_4}
\end{align}
and
\begin{align}
\big(R_{DCba}\big)^\cc &= +(-1)^{DC} R_{\sd{D}\sd{C}ba}\,, \label{eq:superspace_lorentz_curvature_components_relation_5}\\
\big(R_{DC\sym{2}{\beta\alpha}}\big)^\cc &= -(-1)^{DC} R_{\sd{D}\sd{C}\sym{2}{\sd{\beta}\sd{\alpha}}}\,, \label{eq:superspace_lorentz_curvature_components_relation_6}\\
({R_{DCB}}^A)^\cc &= (-1)^{DC+B} {R_{\sd{D}\sd{C}\sd{B}}}^\sd{A}\,. \label{eq:superspace_lorentz_curvature_components_relation_7}
\end{align}
Furthermore, from Eq.~\eqref{eq:superspace_lorentz_connection_transformation_lorentz} and \eqref{eq:superspace_lorentz_connection_transformation_lorentz_infinitesimal} follows that under (infinitesimal) transformations of the Lorentz group ${R_B}^A$ transforms as
\begin{align}
{R_B}^A &\mapsto {(\soge^{-1})_B}^D {R_D}^C {\soge_C}^A\,, \label{eq:superspace_lorentz_curvature_transformation}\\
{R_B}^A &\mapsto {R_B}^A + {[R,\soae]_B}^A\,, \label{eq:superspace_lorentz_curvature_transformation_infinitesimal}
\end{align}
namely in the adjoint representation of $\SO(1,3)$. The $\U(1)$ weights of ${R_B}^A$ and ${R_{DCB}}^A$ have the values
\begin{align}
w({R_B}^A) &= 0\,, & w({R_{DCB}}^A) &= w(E_D) + w(E_C)\,.
\end{align}
\item \textbf{$\U(1)$ field strength}\\
Like the torsion and the Lorentz curvature tensor, the $\U(1)$ field strength $F$ is a $2$-superform, thus
\begin{align}
F_{CD} = -(-1)^{DC} F_{DC}\,,
\end{align}
and according to Eq.~\eqref{eq:superspace_structure_equation_u1_fieldstrength} the components have the explicit form
\begin{align}
\label{eq:superspace_u1_field_strength_components}
F_{DC} = \cD_D A_C -(-1)^{DC} \cD_C A_D + {T_{DC}}^F A_F\,,
\end{align}
where the covariant derivative is given by
\begin{align}
\cD_D A_C &= D_D\cA_C - {\Omega_{DC}}^F A_F - w(E^C)A_D A_C\,.
\end{align}
Since the $\U(1)$ connection $A$ is imaginary, according to the definition in Eq.~\eqref{eq:superspace_structure_equation_u1_fieldstrength} $F$ is imaginary too:
\begin{align}
F^\cc &= -F\,, & (F_{DC})^\cc &= -(-1)^{DC} F_{\sd{D}\sd{C}}\,.
\end{align}
Eq.~\eqref{eq:superspace_u1_connection_transformation_u1} indicates, that $F$ is a singlet under the $\U(1)$ group, i.e.\ it transforms in the adjoint representation of $\U(1)$, thus
\begin{align}
w(F) &= 0\,, & w(F_{DC}) &= w(E_D) + w(E_C)\,.
\end{align}
\end{itemize}
A general result from the two structure equations Eq.~\eqref{eq:superspace_structure_equation_lorentz_curvature} and \eqref{eq:superspace_structure_equation_u1_fieldstrength} is that a double covariant derivative $\cD\cD$ is equal to the sum of the Lorentz curvature tensor and the $\U(1)$ field strength. For example, for the $p$-superform $\chi^A$ an explicit calculation shows
\begin{align}
\begin{split}
\cD\cD\chi^A &= \exd\cD\chi^A + \cD\chi^C\wedge{\Omega_C}^A + w(\chi^A)\cD\chi^A\wedge A \\
&=\exd\exd\chi^A + \exd(\chi^C\wedge{\Omega_C}^A) + w(\chi^A)\exd(\chi^A\wedge A) \\
&\quad + \exd\chi^C\wedge{\Omega_C}^A + \chi^B\wedge{\Omega_B}^C\wedge{\Omega_C}^A + w(\chi^C)\chi^C\wedge A\wedge{\Omega_C}^A \\
&\quad + w(\chi^A)\big(\exd\chi^A\wedge A + \chi^C\wedge{\Omega_C}^A\wedge A + w(\chi^A)\chi^A\wedge A\wedge A\big) \\
&= \chi^C\wedge(\exd{\Omega_C}^A + {\Omega_C}^B\wedge{\Omega_B}^A) + w(\chi^A)\chi^A\wedge\exd A \\
&= \chi^C\wedge {R_C}^A + w(\chi^A)\chi^A\wedge F \\
&= \chi^C\wedge \big({R_C}^A + w(\chi^A)F\delta^A_C\big)\,,
\end{split} \label{eq:superspace_double_covariant_derivative_vector}
\end{align}
where $w(\cD\chi^A)=w(\chi^A)$ and $A\wedge A=0$ is used. On the other hand, if $\chi^A$ is a $0$-superform, Eq.~\eqref{eq:superspace_double_covariant_derivative_vector} can also be evaluated as
\begin{align}
\begin{split}
\cD\cD \chi^A &= \cD(E^B\cD_B\chi^A) \\
&= E^B\wedge E^C \cD_C\cD_B\chi^A + T^B\cD_B\chi^A \\
&= \frac{1}{2} E^B\wedge E^C \big( \mcom{\cD_C}{\cD_B}\chi^A + {T_{CB}}^F\cD_F\chi^A \big)\,,
\end{split}
\end{align}
which leads to the following expression for the components of the double covariant derivative of $\chi^A$:
\begin{align}
\mcom{\cD_C}{\cD_B}\chi^A = -{T_{CB}}^F \cD_F\chi^A + {R_{CBF}}^A\chi^F + w(\chi^A) F_{CB}\chi^A\,, \label{eq:superspace_double_covariant_derivative_vectorfield_components}
\end{align}
where $\mcom{\cD_C}{\cD_B} = \cD_C\cD_B -(-1)^{CB}\cD_B\cD_C$ is the supercommutator. For a lower index $A$, Eq.~\eqref{eq:superspace_double_covariant_derivative_vectorfield_components} reads
\begin{align}
\mcom{\cD_C}{\cD_B}\chi_A = -{T_{CB}}^F \cD_F\chi_A - {R_{CBA}}^F\chi_F + w(\chi_A) F_{CB}\chi_A\,. \label{eq:superspace_double_covariant_derivative_1form_components}
\end{align}
In addition, the double covariant derivative of a superform $\omega$ with $\U(1)$ weight $w(\omega)$ is given by
\begin{align}
\begin{split}
\cD\cD\omega &= \exd\cD\omega + w(\Phi)\cD\omega\wedge A \\
&= \exd\exd\omega + w(\omega)\,\exd(\omega\wedge A) + w(\omega)\,\exd\omega\wedge A + w(\omega)^2\omega\wedge A\wedge A \\
&= w(\omega)\omega\wedge\exd A \\
&= w(\omega)\omega\wedge F\,.
\end{split} \label{eq:superspace_double_covariant_derivative_function}
\end{align}
If $\omega$ is a $0$-superform, the components of the double covariant derivative read
\begin{align}
\mcom{\cD_C}{\cD_B}\omega = -{T_{CB}}^F \cD_F\omega + w(\omega) F_{CB}\omega\,. \label{eq:superspace_double_covariant_derivative_function_components}
\end{align}

\subsubsection{Bianchi identities}
Taking the covariant derivative of the structure equation in Eq.~\eqref{eq:superspace_structure_equation_torsion} leads to the algebraic Bianchi identity
\begin{align}
\cD T^A = \cD\cD E^A = E^B\wedge {R_B}^A + w(E^A) E^A\wedge F\,. \label{eq:superspace_algebraic_bianchi_identity}
\end{align}
Since $T^A$ is a $2$-superform, its covariant derivative has the form
\begin{align}
\begin{split}
\cD T^A &= \frac{1}{2}E^B\wedge E^C\wedge \cD{T_{CB}}^A + E^B\wedge T^C\wedge {T_{CB}}^A \\
&= \frac{1}{2} E^B\wedge E^C\wedge E^D (\cD_D{T_{CB}}^A + {T_{DC}}^F{T_{BF}}^A)\,.
\end{split}
\end{align}
Thus, in terms of components the algebraic Bianchi identity in Eq.~\eqref{eq:superspace_algebraic_bianchi_identity} is written as:
\begin{align}
\oint_{DCB} \big( \cD_D {T_{CB}}^A + {T_{DC}}^F {T_{FB}}^A - {R_{DCB}}^A - \omega(E^A)F_{DC} \delta^A_B \big) = 0\label{eq:superspace_algebraic_bianchi_identity_components}\,,
\end{align}
where $\oint_{DCB} DCB = DCB + (-1)^{D(C+B)}\, CBD + (-1)^{(D+C)B}\, BDC$.
Furthermore, the two differential Bianchi identities are obtained by taking the covariant derivative of the structure equations in Eq.~\eqref{eq:superspace_structure_equation_lorentz_curvature} and \eqref{eq:superspace_structure_equation_u1_fieldstrength}, namely (suppressing the Lorentz indices)
\begin{align}
\begin{split}
\cD R &= \exd R + \com{R}{\Omega} \\
&= \exd\exd\Omega + \exd(\Omega\wedge\Omega) + \com{\exd\Omega}{\Omega} + \com{\Omega\wedge\Omega}{\Omega} \\
&= \Omega\wedge\exd\Omega - \exd\Omega\wedge\Omega + \com{\exd\Omega}{\Omega}\\
&=0\,,
\end{split} \label{eq:superspace_differential_bianchi_identity_1}
\end{align}
where $\com{\Omega\wedge\Omega}{\Omega}=0$ is used, and
\begin{align}
\cD F = \exd F = \exd\exd A = 0\,. \label{eq:superspace_differential_bianchi_identity_2}
\end{align}
These identities state that the covariant derivative of the Lorentz curvature and of the $\mathrm{U}(1)$ field strength vanishes. Note that $R$ and $F$ transform in the adjoint representation of $\SO(1,3)$ and $\U(1)$, respectively. In fact, Eq.~\eqref{eq:superspace_differential_bianchi_identity_1} and \eqref{eq:superspace_differential_bianchi_identity_2} are not independent of the algebraic Bianchi identity. This can be seen by taking the covariant derivative of Eq.~\eqref{eq:superspace_algebraic_bianchi_identity}:
\begin{align}
\begin{split}
0 &= \cD\cD T^A - \cD\big(E^B\wedge{R_B}^A\big) - w(E^A)\cD\big(E^A\wedge F\big)\\
&= T^B\wedge{R_B}^A + w(E^A) T^A\wedge F - E^B\wedge \cD {R_B}^A - T^B\wedge {R_B}^A \\ &\quad - w(E^A)\big( E^A\wedge \cD F + T^A\wedge F\big)\\
&= -E^B\wedge\big(\cD {R_B}^A + w(E^A) \cD F \delta_B^A \big)\,,
\end{split}
\end{align}
which implies
\begin{align}
\cD {R_B}^A + w(E^A) \cD F \delta_B^A = 0\,.
\end{align}
Since $w(E^a)=0$ and the generators of the Lorentz group in the representations $(1/2,0)$ and $(0,1/2)$ are traceless, both equations $\cD {R_B}^A=0$ and $\cD F = 0$ are fulfilled separately. Thus, if the algebraic Bianchi identity in Eq.~\eqref{eq:superspace_algebraic_bianchi_identity} is satisfied, the differential Bianchi identities in Eq.~\eqref{eq:superspace_differential_bianchi_identity_1} and \eqref{eq:superspace_differential_bianchi_identity_2} are automatically satisfied as well. In the following, the algebraic Bianchi identity is simply called Bianchi identity, unless otherwise stated.
\\\\
One of the most important implications of the Bianchi identity is, that in $\mathrm{U}(1)$ superspace the Lorentz curvature $R$ and the $\mathrm{U}(1)$ field strength $F$ are completely determined by the torsion tensor $T^A$, which is also known as Dragon's theorem (cf.~\cite{Dragon:1978nf}). This can explicitly be shown at the level of components by writing Eq.~\eqref{eq:superspace_algebraic_bianchi_identity_components} as
\begin{align}
\oint_{DCB} \big({R_{DCB}}^A + \omega(E^A)F_{DC} \delta^A_B \big) = {\Delta_{DCB}}^A\,,
\end{align}
where ${\Delta_{DCB}}^A = \oint_{DCB} (\cD_D {T_{CB}}^A + {T_{DC}}^F {T_{FB}}^A)$ just depends on the components of the torsion tensor. The components of the Lorentz curvature tensor are obtained by choosing the following tuples of indices:
\begin{align}
\begin{split}
(D,C,B,A)=(\delta,\gamma,b,a): &\quad R_{\delta\gamma ba} = \Delta_{\delta\gamma ba}\,,\\
(D,C,B,A)=(\sd{\delta},\sd{\gamma},b,a): &\quad R_{\sd{\delta}\sd{\gamma}ba} = \Delta_{\sd{\delta}\sd{\gamma}ba}\,,\\
(D,C,B,A)=(\delta,\sd{\gamma},b,a): &\quad R_{\delta\sd{\gamma}ba} = \Delta_{\delta\sd{\gamma}ba}\,,\\
(D,C,B,A)=(\delta,c,b,a): &\quad R_{\delta cba} = \frac{1}{2}\big(\Delta_{\delta cba} - \Delta_{\delta bac} + \Delta_{\delta acb}\big)\,,\\
(D,C,B,A)=(\sd{\delta},c,b,a): &\quad R_{\sd{\delta}cba} = \frac{1}{2}\big(\Delta_{\sd{\delta}cba} - \Delta_{\sd{\delta}bac} + \Delta_{\sd{\delta}acb}\big)\,,\\
(D,C,B,A)=(d,c,\beta,\alpha): &\quad R_{dc\sym{2}{\beta\alpha}} = \frac{1}{2}\big(\Delta_{dc\beta\alpha} + \Delta_{dc\alpha\beta}\big)\,,\\
(D,C,B,A)=(d,c,\sd{\beta},\sd{\alpha}): &\quad R_{dc\sym{2}{\sd{\beta}\sd{\alpha}}} = \frac{1}{2}\big(\Delta_{dc\sd{\beta}\sd{\alpha}} + \Delta_{dc\sd{\alpha}\sd{\beta}}\big)\,,
\end{split}
\label{eq:superspace_torsion_lorentz_curvature_identities}
\end{align}
where the remaining components are determined by the relations in Eqs.~\eqref{eq:superspace_lorentz_curvature_components_relation_1}--\eqref{eq:superspace_lorentz_curvature_components_relation_3}. In addition, the components of the $\mathrm{U}(1)$ field strength are calculated by taking into account the above result for the Lorentz curvature tensor:
\begin{align}
\begin{split}
(D,C,B,A)=(\delta,\gamma,\sd{\beta},\sd{\alpha}): &\quad F_{\delta\gamma} = +\frac{1}{2}{\Delta_{\delta\gamma\sd{\alpha}}}^\sd{\alpha}\,,\\
(D,C,B,A)=(\sd{\delta},\sd{\gamma},\beta,\alpha): &\quad F_{\sd{\delta}\sd{\gamma}} = -\frac{1}{2}\tensor{\Delta}{_{\sd{\delta}\sd{\gamma}}^\alpha_\alpha}\,,\\
(D,C,B,A)=(\delta,\sd{\gamma},\beta,\alpha): &\quad F_{\delta\sd{\gamma}} = -\frac{1}{3}\big(\tensor{\Delta}{_{\delta\sd{\gamma}}^\alpha_\alpha} - \frac{1}{2}{(\sig^{da})_\delta}^\alpha \Delta_{\alpha\sd{\gamma}da}\big)\,,\\
(D,C,B,A)=(\delta,c,\sd{\beta},\sd{\alpha}): &\quad F_{\delta c} = +\frac{1}{2}{\Delta_{\delta c\sd{\alpha}}}^\sd{\alpha}\,,\\
(D,C,B,A)=(\sd{\delta},c,\beta,\alpha): &\quad F_{\sd{\delta}c} = -\frac{1}{2}\tensor{\Delta}{_{\sd{\delta}c}^\alpha_\alpha}\,,\\
(D,C,B,A)=(d,c,\beta,\alpha): &\quad F_{dc} = -\frac{1}{2}\tensor{\Delta}{_{dc}^\alpha_\alpha}\,.
\end{split}
\label{eq:superspace_torsion_u1_fieldstrength_identities}
\end{align}
The identities in Eq.~\eqref{eq:superspace_torsion_lorentz_curvature_identities} and \eqref{eq:superspace_torsion_u1_fieldstrength_identities} prove that in $\mathrm{U}(1)$ superspace the Lorentz curvature tensor and the $\mathrm{U}(1)$ field strength are completely determined by the torsion tensor.\footnote{The feature that the Lorentz curvature tensor is completely determined by the torsion tensor holds true, if the structure group is just $\SO(1,3)$, without the extra $\U(1)$ factor.} This property is not present in ordinary spacetime, where the Riemann curvature and the torsion tensor are independent. In general relativity, for example, the torsion is switched off without imposing a vanishing curvature. In $\mathrm{U}(1)$ superspace the Lorentz curvature and the $\mathrm{U}(1)$ field strength are redundant and the torsion tensor is the main object determining the geometry of superspace.

\subsubsection{Chiral and antichiral superfields}
In order to define chiral and antichiral superfields, the covariant derivative $\cD$ is used, where $\cD$ is covariant with respect to all gauge transformations under which the superfield transforms. A superfield is called chiral, if the component $\cD^\sd{\alpha}$ of the covariant derivative vanishes, and it is called antichiral, if the $\cD_\alpha$ component vanishes. For example, for a superfield $\Phi$, which is a singlet under $\SO(1,3)$ and has $\U(1)$ weight $w(\Phi)$, the chirality conditions read
\begin{align}
&\Phi\quad\text{chiral} & \leftrightarrow\hspace{1.8cm} \cD^\sd{\alpha}\Phi &= D^\sd{\alpha}\Phi + w(\Phi)A^\sd{\alpha}\Phi = 0\,, \label{eq:superspace_chiral_superfields_phi_1}\\
&\Phi\quad\text{antichiral} & \leftrightarrow\hspace{1.8cm} \cD_\alpha\Phi &= D_\alpha\Phi + w(\Phi)A_\alpha\Phi= 0\,. \label{eq:superspace_chiral_superfields_phi_2}
\end{align}
In addition, a superfield $\chi^B$ with an upper Lorentz index and $\U(1)$ weight $w(\chi^B)$ is chiral or antichiral, if
\begin{align}
&\chi^B\quad\text{chiral} &\leftrightarrow\hspace{1cm} \cD^\sd{\alpha}\chi^B &= D^\sd{\alpha}\chi^B + \tensor{\Omega}{^{\sd{\alpha}}_C^B}\chi^C + w(\chi^B)A^\sd{\alpha}\chi^B = 0\,, \label{eq:superspace_chiral_superfields_chi_1}\\
&\chi^B\quad\text{antichiral} &\leftrightarrow\hspace{1cm} \cD_\alpha\chi^B &= D_\alpha\chi^B + {\Omega_{\alpha C}}^B \chi^C + w(\chi^B)A_\alpha\chi^B = 0\,. \label{eq:superspace_chiral_superfields_chi_2}
\end{align}
The definition of chiral and antichiral superfields by using the covariant derivative has the advantage that it is compatible with gauge transformations. This property can be illustrated on the basis of Eqs.~\eqref{eq:superspace_chiral_superfields_phi_1}--\eqref{eq:superspace_chiral_superfields_chi_2}, namely:
\begin{align}
\Phi,\;\chi^B & \quad\text{chiral:} & \Phi,\;\chi^B & \quad\text{antichiral:} \nonumber\\
\cD^\sd{\alpha}(\uge^{w(\Phi)}\Phi) &= \uge^{w(\Phi)}(\cD^\sd{\alpha}\Phi) = 0\,, & \cD_\alpha(\uge^{w(\Phi)}\Phi) &= \uge^{w(\Phi)}(\cD_\alpha\Phi) = 0\,, \nonumber\\
\cD^\sd{\alpha}(\uge^{w(\chi^B)}\chi^C{\soge_C}^D) &= \uge^{w(\chi^B)}(\cD^\sd{\alpha}\chi^C){\soge_C}^D = 0\,, & \cD_\alpha(\uge^{w(\chi^B)}\chi^C{\soge_C}^D) &= \uge^{w(\chi^B)}(\cD_\alpha\chi^C){\soge_C}^D = 0\,.
\end{align}
In Section~\ref{sec:matter_and_yang_mills} matter superfields $\phi^k$ and $\phib^\kb$ will be introduced, which are singlets under the structure group $\SO(1,3)\times\U(1)$, but transform under a Yang-Mills gauge group $G$. The condition that $\phi^k$ is chiral and $\phib^\kb$ antichiral is given by $\cD^\sd{\alpha}\phi^k=0$ and $\cD_\alpha\phib^\kb=0$ respectively, where $\cD$ is covariant with respect to Yang-Mills transformations.

\section{Solution of the Bianchi identity in $\U(1)$ superspace}
\label{sec:bianchi_sp_solution_all}
\subsection{Torsion Constraints}
Although the supergeometrical formalism discussed in Section~\ref{sec:superspace} it quite elegant, it turns out to be too general to describe supergravity. A general superspace is specified by the supervielbein $E^A$ and the Lorentz connection $\Omega$, whose components are independent quantities. These are far too many degrees of freedom to describe the fields in the supergravity multiplet. In principle, the same problem appears in general relativity, where the primary objects are the vielbein $e^a$ and the spin connection $\omega$. The choice of a vanishing torsion tensor ${T_{cb}}^a = 0$ allows to express the spin connection uniquely in terms of the vielbein, which contains the degrees of freedom of the graviton. To reduce the degrees of freedom in superspace, it would be straightforward to use the same strategy by implementing the torsion-free condition ${T_{CB}}^A = 0$. However, this approach turns out to be problematic from two points of view. On the one hand flat superspace can not be described by such a geometry, since ${T_\gamma}^{\sd{\beta}a} = -2i{(\sig^a\eps)_\gamma}^\sd{\beta}$. On the other hand Eq.~\eqref{eq:superspace_torsion_lorentz_curvature_identities} implies ${R_{DCB}}^A = 0$, leading to a trivial geometry.
\\\\
In superspace there is no general recipe to find proper torsion constraints to reduce the number of degrees of freedom, and there is also no unique solution to this problem. However, once a set of constraints is chosen, the consistency and the implications can be studied by solving the Bianchi identity in Eq.~\eqref{eq:superspace_algebraic_bianchi_identity_components}. It turns out, that distinct choices lead to geometries which describe either the minimal, the non-minimal or the new minimal supergravity multiplet. In the following, the discussion is limited to the case of the minimal supergravity multiplet. The explicit calculations in the subsequent sections show that the following covariant torsion constraints are suitable to describe supergravity with a minimal supergravity multiplet:
\begin{align}
\begin{split}
{T_{\su{\gamma}\su{\beta}}}^{\su{\alpha}} &= 0\,,\\
{T_{\gamma\beta}}^a &= {T_{\sd{\gamma}\sd{\beta}}}^a = 0\,,\\
{T_{\gamma\sd{\beta}}}^a &= {T_{\sd{\beta}\gamma}}^a = 2i\sig_{\gamma\sd{\beta}}^a\,,\\
{T_{\su{\gamma}b}}^a &= {T_{c\su{\beta}}}^a = 0\,,\\
{T_{cb}}^a &= 0\,,
\end{split}
\label{eq:bianchi_sp_torsion_constraints}
\end{align}
where $\su{\alpha}$ labels both $\alpha$ and $\sd{\alpha}$. Although there is no rigorous derivation for these constraints, beside explicitly calculating their implications, some of the identities in Eq.~\eqref{eq:bianchi_sp_torsion_constraints} can be motivated by simple arguments. As discussed later in Section~\ref{sec:matter_and_yang_mills}, in superspace matter is described by chiral and antichiral superfields $\phi$ and $\phib$, which means $\cD^\sd{\alpha} \phi = 0$ and $\cD_\alpha \phib = 0$ respectively. Assuming that both superfields are inert under $\mathrm{U}(1)$ transformations, Eq.~\eqref{eq:superspace_double_covariant_derivative_function_components} implies
\begin{align}
\begin{split}
0 &= \acom{\cD^\sd{\gamma}}{\cD^\sd{\beta}}\phi = -T^{\sd{\gamma}\sd{\beta}a} \cD_a \phi - T^{\sd{\gamma}\sd{\beta}\alpha} \cD_\alpha \phi\,,\\
0 &= \acom{\cD_\beta}{\cD_\alpha}\phib = -{T_{\gamma\beta}}^a \cD_a \phib - T_{\gamma\beta\sd{\alpha}} \cD^\sd{\alpha} \phib\,,
\end{split}
\end{align}
which are solved by setting
\begin{gather}
\begin{aligned}
T^{\sd{\gamma}\sd{\beta}a} &= 0\,, &\hspace{4cm} T^{\sd{\gamma}\sd{\beta}\alpha} &= 0\,,\\
{T_{\gamma\beta}}^a &= 0\,, & T_{\gamma\beta\sd{\alpha}} &= 0\,.
\label{eq:bianchi_sp_torsion_constraint_representation_preserving}
\end{aligned}
\end{gather}
In addition, consistency with flat superspace and general relativity suggests to demand
\begin{align}
{T_\gamma}^{\sd{\beta}a} &= -2i{(\sig^a\eps)_\gamma}^\sd{\beta}\,, & {T_{cb}}^a &= 0\,.
\label{eq:bianchi_sp_torsion_constraint_conventional}
\end{align}
The identities in Eq.~\eqref{eq:bianchi_sp_torsion_constraint_representation_preserving} are called representation preserving constraints, whereas all other constraints, including the ones in Eq.~\eqref{eq:bianchi_sp_torsion_constraint_conventional}, are referred to as conventional constraints.

\subsection{Solution of the Bianchi identity}
\label{sec:bianchi_sp_solution}
In this section, the Bianchi identity in $\mathrm{U}(1)$ superspace in Eq.~\eqref{eq:superspace_algebraic_bianchi_identity_components} is solved, by taking account of the torsion constraints listed in Eq.~\eqref{eq:bianchi_sp_torsion_constraints}. For the sake a completeness, the two equations are stated here again: the Bianchi identity reads
\begin{align}
\oint_{DCB} \big( \cD_D {T_{CB}}^A + {T_{DC}}^F {T_{FB}}^A - {R_{DCB}}^A - \omega(E^A)F_{DC} \delta^A_B \big) = 0\,,
\label{eq:bianchi_sp_bianchi_identity_components}
\end{align}
where $\oint_{DCB} DCB = DCB + (-1)^{D(C+B)}\, CBD + (-1)^{(D+C)B}\, BDC$, and the set of covariant torsion constraints is given by
\begin{align}
{T_{\su{\gamma}\su{\beta}}}^{\su{\alpha}} &= 0\,,\label{eq:bianchi_sp_torsion_constraint_1}\\
{T_{\gamma\beta}}^a &= {T_{\sd{\gamma}\sd{\beta}}}^a = 0\,,\label{eq:bianchi_sp_torsion_constraint_2}\\
{T_{\gamma\sd{\beta}}}^a &= {T_{\sd{\beta}\gamma}}^a = 2i\sig_{\gamma\sd{\beta}}^a\,,\label{eq:bianchi_sp_torsion_constraint_3}\\
{T_{\su{\gamma}b}}^a &= {T_{c\su{\beta}}}^a = 0\,,\label{eq:bianchi_sp_torsion_constraint_4}\\
{T_{cb}}^a &= 0\,,\label{eq:bianchi_sp_torsion_constraint_5}
\end{align}
where $\su{\alpha}$ labels both $\alpha$ and $\sd{\alpha}$. Identities for the components of the tensors are obtained from Eq.~\eqref{eq:bianchi_sp_bianchi_identity_components} by substituting the superspace indices $D,C,B,A$ with vector and spinor indices. There are the following non-equivalent tuples of indices:
\begin{align}
\begin{split}
A = a:\quad (D,C,B)\,=\,&(d,c,b),\,(d,c,\beta),\,(d,\gamma,\beta),\,(d,\gamma,\sd{\beta}),\,\underbrace{(\delta,\gamma,\beta),\,(\delta,\gamma,\sd{\beta})}_{\text{trivial}}\\
A = \alpha:\quad (D,C,B)\,=\,&(d,c,b),\,(d,c,\beta),\,(d,c,\sd{\beta}),\,(d,\gamma,\beta),\,(d,\gamma,\sd{\beta}),\,(d,\sd{\gamma},\sd{\beta}),\,\\
&(\delta,\gamma,\beta),\,(\delta,\gamma,\sd{\beta}),\,(\delta,\sd{\gamma},\sd{\beta}),\,\underbrace{(\sd{\delta},\sd{\gamma},\sd{\beta})}_{\text{trivial}}
\end{split}
\end{align}
and the corresponding ``conjugate tuples", where all dotted indices are switched to undotted ones and vice versa. The label "trivial" indicates that the Bianchi identity is trivially satisfied for the corresponding tuple of indices. Applying the torsion constraints given in Eqs.~\eqref{eq:bianchi_sp_torsion_constraint_1}--\eqref{eq:bianchi_sp_torsion_constraint_5} and neglecting the trivial cases, the Bianchi identity of Eq.~\eqref{eq:bianchi_sp_bianchi_identity_components} delivers $13$ pairs of equations, where in each pair the two equations are the conjugate of each other:
\begin{fleqn}
\begin{align}
&(D,C,B,A) = (\delta,\gamma,\beta,\alpha) \text{ and } (\sd{\delta},\sd{\gamma},\sd{\beta},\sd{\alpha}): \tag{I} \label{eq:bianchi_sp_bianchi_identity_eq01}\\
(a)\quad&{R_{\delta\gamma\beta}}^\alpha + {R_{\beta\delta\gamma}}^\alpha + {R_{\gamma\beta\delta}}^\alpha + F_{\delta\gamma}\delta_\beta^\alpha + F_{\beta\delta}\delta_\gamma^\alpha + F_{\gamma\beta}\delta_\delta^\alpha = 0 \notag\\
(b)\quad&{R_{\sd{\delta}\sd{\gamma}\sd{\beta}}}^\sd{\alpha} + {R_{\sd{\beta}\sd{\delta}\sd{\gamma}}}^\sd{\alpha} + {R_{\sd{\gamma}\sd{\beta}\sd{\delta}}}^\sd{\alpha} + F_{\sd{\delta}\sd{\gamma}}\delta_{\sd{\beta}}^\sd{\alpha} + F_{\sd{\beta}\sd{\delta}}\delta_{\sd{\gamma}}^\sd{\alpha} + F_{\sd{\gamma}\sd{\beta}}\delta_{\sd{\delta}}^\sd{\alpha} = 0 \notag
\end{align}
\end{fleqn}
\begin{fleqn}
\begin{align}
&(D,C,B,A) = (\delta,\sd{\gamma},\beta,\alpha) \text{ and } (\sd{\delta},\gamma,\sd{\beta},\sd{\alpha}): \tag{II} \label{eq:bianchi_sp_bianchi_identity_eq02}\\
(a)\quad&{R_{\delta\sd{\gamma}\beta}}^\alpha + {R_{\sd{\gamma}\beta\delta}}^\alpha + F_{\delta\sd{\gamma}} \delta_\beta^\alpha + F_{\sd{\gamma}\beta} \delta_\delta^\alpha = -2i\sig_{\delta\sd{\gamma}}^f {T_{\beta f}}^\alpha - 2i\sig_{\beta\sd{\gamma}}^f {T_{\delta f}}^\alpha \notag\\
(b)\quad&{R_{\sd{\delta}\gamma\sd{\beta}}}^\sd{\alpha} + {R_{\gamma\sd{\beta}\sd{\delta}}}^\sd{\alpha} + F_{\sd{\delta}\gamma} \delta_{\sd{\beta}}^\sd{\alpha} + F_{\gamma\sd{\beta}} \delta_{\sd{\delta}}^\sd{\alpha} = -2i\sig_{\gamma\sd{\delta}}^f {T_{\sd{\beta} f}}^\sd{\alpha} - 2i\sig_{\gamma\sd{\beta}}^f {T_{\sd{\delta} f}}^\sd{\alpha} \notag
\end{align}
\end{fleqn}
\begin{fleqn}
\begin{align}
&(D,C,B,A) = (\delta,\sd{\gamma},\beta,\sd{\alpha}) \text{ and } (\sd{\delta},\gamma,\sd{\beta},\alpha): \tag{III} \label{eq:bianchi_sp_bianchi_identity_eq03}\\
(a)\quad&{R_{\beta\delta\sd{\gamma}}}^\sd{\alpha} + F_{\beta\delta} \delta_\sd{\gamma}^\sd{\alpha} = -2i\sig_{\delta\sd{\gamma}}^f {T_{\beta f}}^\sd{\alpha} - 2i\sig_{\beta\sd{\gamma}}^f {T_{\delta f}}^\sd{\alpha} \notag\\
(b)\quad&{R_{\sd{\beta}\sd{\delta}\gamma}}^\alpha + F_{\sd{\beta}\sd{\delta}} \delta_\gamma^\alpha = -2i\sig_{\gamma\sd{\delta}}^f {T_{\sd{\beta} f}}^\alpha - 2i\sig_{\gamma\sd{\beta}}^f {T_{\sd{\delta} f}}^\alpha \notag
\end{align}
\end{fleqn}
\begin{fleqn}
\begin{align}
&(D,C,B,A) = (\delta,\gamma,b,\alpha) \text{ and } (\sd{\delta},\sd{\gamma},b,\sd{\alpha}): \tag{IV} \label{eq:bianchi_sp_bianchi_identity_eq04}\\
(a)\quad&{R_{\delta b \gamma}}^\alpha + {R_{\gamma b \delta}}^\alpha + F_{\delta b} \delta_\gamma^\alpha + F_{\gamma b} \delta_\delta^\alpha = -\cD_\gamma {T_{\delta b}}^\alpha - \cD_\delta {T_{\gamma b}}^\alpha \notag\\
(b)\quad&{R_{\sd{\delta} b \sd{\gamma}}}^\sd{\alpha} + {R_{\sd{\gamma} b \sd{\delta}}}^\sd{\alpha} + F_{\sd{\delta} b} \delta_\sd{\gamma}^\sd{\alpha} + F_{\sd{\gamma} b} \delta_\sd{\delta}^\sd{\alpha} = -\cD_\sd{\gamma} {T_{\sd{\delta} b}}^\sd{\alpha} - \cD_\sd{\delta} {T_{\sd{\gamma} b}}^\sd{\alpha} \notag
\end{align}
\end{fleqn}
\begin{fleqn}
\begin{align}
&(D,C,B,A) = (\delta,\sd{\gamma},b,\sd{\alpha}) \text{ and } (\sd{\delta},\gamma,b,\alpha): \tag{V} \label{eq:bianchi_sp_bianchi_identity_eq05}\\
(a)\quad&{R_{b \delta\sd{\gamma}}}^\sd{\alpha} + F_{b \delta} \delta_\sd{\gamma}^\sd{\alpha} = \cD_\delta {T_{\sd{\gamma} b}}^\sd{\alpha} + \cD_\sd{\gamma} {T_{\delta b}}^\sd{\alpha} + 2i\sig_{\delta\sd{\gamma}}^f {T_{fb}}^\sd{\alpha} \notag\\
(b)\quad&{R_{b \sd{\delta}\gamma}}^\alpha + F_{b \sd{\delta}} \delta_\gamma^\alpha = \cD_\sd{\delta} {T_{\gamma b}}^\alpha + \cD_\gamma {T_{\sd{\delta} b}}^\alpha + 2i\sig_{\gamma\sd{\delta}}^f {T_{fb}}^\alpha \notag
\end{align}
\end{fleqn}
\begin{fleqn}
\begin{align}
&(D,C,B,A) = (\sd{\delta},\sd{\gamma},b,\alpha) \text{ and } (\delta,\gamma,b,\sd{\alpha}): \tag{VI} \label{eq:bianchi_sp_bianchi_identity_eq06}\\
(a)\quad&\cD_\sd{\gamma} {T_{\sd{\delta} b}}^\alpha + \cD_\sd{\delta} {T_{\sd{\gamma} b}}^\alpha = 0 \notag \\
(b)\quad&\cD_\gamma {T_{\delta b}}^\sd{\alpha} + \cD_\delta {T_{\gamma b}}^\sd{\alpha} = 0 \notag
\end{align}
\end{fleqn}
\begin{fleqn}
\begin{align}
&(D,C,B,A) = (\sd{\delta},c,\sd{\beta},a) \text{ and } (\delta,c,\beta,a): \tag{VII} \label{eq:bianchi_sp_bianchi_identity_eq07}\\
(a)\quad&{R_{\sd{\beta}\sd{\delta} c}}^a = -2i\sig_{\varphi\sd{\delta}}^a {T_{\sd{\beta} c}}^\varphi - 2i\sig_{\varphi\sd{\beta}}^a {T_{\sd{\delta} c}}^\varphi \notag\\
(b)\quad&{R_{\beta\delta c}}^a = +2i\sig_{\delta\sd{\varphi}}^a {T_{\beta c}}^\sd{\varphi} + 2i\sig_{\beta\sd{\varphi}}^a {T_{\delta c}}^\sd{\varphi} \notag
\end{align}
\end{fleqn}
\begin{fleqn}
\begin{align}
&(D,C,B,A) = (\sd{\delta},c,\beta,a) \text{ and } (\delta,c,\sd{\beta},a): \tag{VIII} \label{eq:bianchi_sp_bianchi_identity_eq08}\\
(a)\quad&{R_{\beta\sd{\delta} c}}^a = -2i\sig_{\varphi\sd{\delta}}^a {T_{\beta c}}^\varphi + 2i\sig_{\beta\sd{\varphi}}^a {T_{\sd{\delta} c}}^\sd{\varphi} \notag\\
(b)\quad&{R_{\sd{\beta}\delta c}}^a = +2i\sig_{\delta\sd{\varphi}}^a {T_{\sd{\beta} c}}^\sd{\varphi} - 2i\sig_{\varphi\sd{\beta}}^a {T_{\delta c}}^\varphi \notag
\end{align}
\end{fleqn}
\begin{fleqn}
\begin{align}
&(D,C,B,A) = (d,\sd{\gamma},b,\sd{\alpha}) \text{ and } (d,\gamma,b,\alpha): \tag{IX} \label{eq:bianchi_sp_bianchi_identity_eq09}\\
(a)\quad&{R_{bd \sd{\gamma}}}^\sd{\alpha} + F_{bd} \delta_\sd{\gamma}^\sd{\alpha} = \cD_b {T_{d \sd{\gamma}}}^\sd{\alpha} + \cD_d {T_{\sd{\gamma} b}}^\sd{\alpha} + \cD_\sd{\gamma} {T_{bd}}^\sd{\alpha} + {T_{d \sd{\gamma}}}^{\su{\varphi}} {T_{\su{\varphi} b}}^\sd{\alpha} + {T_{\sd{\gamma} b}}^{\su{\varphi}} {T_{\su{\varphi} d}}^\sd{\alpha} \notag\\
(b)\quad&{R_{bd \gamma}}^\alpha + F_{bd} \delta_\gamma^\alpha = \cD_b {T_{d \gamma}}^\alpha + \cD_d {T_{\gamma b}}^\alpha + \cD_\gamma {T_{bd}}^\alpha + {T_{d \gamma}}^{\su{\varphi}} {T_{\su{\varphi} b}}^\alpha + {T_{\gamma b}}^{\su{\varphi}} {T_{\su{\varphi} d}}^\alpha \notag
\end{align}
\end{fleqn}
\begin{fleqn}
\begin{align}
&(D,C,B,A) = (d,\sd{\gamma},b,\alpha) \text{ and } (d,\gamma,b,\sd{\alpha}): \tag{X} \label{eq:bianchi_sp_bianchi_identity_eq10}\\
(a)\quad&\cD_b {T_{d \sd{\gamma}}}^\alpha + \cD_d {T_{\sd{\gamma} b}}^\alpha + \cD_\sd{\gamma} {T_{bd}}^\alpha + {T_{d \sd{\gamma}}}^{\su{\varphi}} {T_{\su{\varphi} b}}^\alpha + {T_{\sd{\gamma} b}}^{\su{\varphi}} {T_{\su{\varphi} d}}^\alpha = 0 \notag\\
(b)\quad&\cD_b {T_{d \gamma}}^\sd{\alpha} + \cD_d {T_{\gamma b}}^\sd{\alpha} + \cD_\gamma {T_{bd}}^\sd{\alpha} + {T_{d \gamma}}^{\su{\varphi}} {T_{\su{\varphi} b}}^\sd{\alpha} + {T_{\gamma b}}^{\su{\varphi}} {T_{\su{\varphi} d}}^\sd{\alpha} = 0 \notag
\end{align}
\end{fleqn}
\begin{fleqn}
\begin{align}
&(D,C,B,A) = (d,c,\beta,a) \text{ and } (d,c,\sd{\beta},a): \tag{XI} \label{eq:bianchi_sp_bianchi_identity_eq11}\\
(a)\quad&{R_{\beta dc}}^a + {R_{c \beta d}}^a = -2i\sig_{\beta\sd{\varphi}}^a {T_{dc}}^\sd{\varphi} \notag\\
(b)\quad&{R_{\sd{\beta} dc}}^a + {R_{c \sd{\beta} d}}^a = +2i\sig_{\varphi\sd{\beta}}^a {T_{dc}}^\varphi \notag
\end{align}
\end{fleqn}
\begin{fleqn}
\begin{align}
&(D,C,B,A) = (d,c,b,\alpha) \text{ and } (d,c,b,\sd{\alpha}): \tag{XII} \label{eq:bianchi_sp_bianchi_identity_eq12}\\
(a)\quad&\cD_b {T_{dc}}^\alpha + \cD_d {T_{cb}}^\alpha + \cD_c {T_{bd}}^\alpha + {T_{bd}}^{\su{\varphi}} {T_{\su{\varphi} c}}^\alpha + {T_{dc}}^{\su{\varphi}} {T_{\su{\varphi} b}}^\alpha + {T_{cb}}^{\su{\varphi}} {T_{\su{\varphi} d}}^\alpha = 0 \notag\\
(b)\quad&\cD_b {T_{dc}}^\sd{\alpha} + \cD_d {T_{cb}}^\sd{\alpha} + \cD_c {T_{bd}}^\sd{\alpha} + {T_{bd}}^{\su{\varphi}} {T_{\su{\varphi} c}}^\sd{\alpha} + {T_{dc}}^{\su{\varphi}} {T_{\su{\varphi} b}}^\sd{\alpha} + {T_{cb}}^{\su{\varphi}} {T_{\su{\varphi} d}}^\sd{\alpha} = 0 \notag
\end{align}
\end{fleqn}
\begin{fleqn}
\begin{align}
&(D,C,B,A) = (d,c,b,a): \tag{XIII} \label{eq:bianchi_sp_bianchi_identity_eq13}\\
(a)\quad&{R_{bdc}}^a + {R_{dcb}}^a + {R_{cbd}}^a = 0 \notag
\end{align}
\end{fleqn}
The Lorentz curvature takes values in the vector and in the spinor representations of the Lie algebra of the Lorentz group. Calculations are often much simpler if the vector components are expressed in terms of the spinor components. According to Eq.~\eqref{eq:superspace_lorentz_curvature_components_relation_4}, they are related in the following way
\begin{align}
{R_{DCb}}^a \quad\leftrightarrow\quad R_{DC\s\beta\sd{\beta}\s\alpha\sd{\alpha}} = 2\epsilon_{\sd{\beta}\sd{\alpha}} R_{DC\sym{2}{\beta\alpha}} - 2\epsilon_{\beta\alpha} R_{DC\sym{2}{\sd{\beta}\sd{\alpha}}}\,. \label{eq:bianchi_sp_lorentz_algebra_spinor_vector_identification}
\end{align}
In general, a vector index $a$ is turned into a pair of spinor indices $\alpha\sd{\alpha}$ via the contraction with the Pauli matrix $\sig^a_{\alpha\sd{\alpha}}$. The vector index $a$ is restored again by contracting $\alpha\sd{\alpha}$ with $-\frac{1}{2}\sigb_a^{\sd{\alpha}\alpha}$. Under conjugation $\alpha\sd{\alpha}'$ is treated as a vector index and is mapped to $\alpha'\sd{\alpha}$, because the Pauli matrices are Hermitian. The prime is written to explicitly distinguish the two spinor indices.
\\\\
First, the identities which are linear and do not contain derivatives, namely \eqref{eq:bianchi_sp_bianchi_identity_eq01}, \eqref{eq:bianchi_sp_bianchi_identity_eq02}, \eqref{eq:bianchi_sp_bianchi_identity_eq03}, \eqref{eq:bianchi_sp_bianchi_identity_eq07}, \eqref{eq:bianchi_sp_bianchi_identity_eq08}, \eqref{eq:bianchi_sp_bianchi_identity_eq11} and \eqref{eq:bianchi_sp_bianchi_identity_eq13}, are solved.
\begin{itemize}
\item \textbf{Identity \eqref{eq:bianchi_sp_bianchi_identity_eq07}:}\\
In (\ref{eq:bianchi_sp_bianchi_identity_eq07}a) it is convenient to express the vector indices in terms of spinor indices. The equation is then written as
\begin{align}
R_{\sd{\beta}\sd{\delta}\s\gamma\sd{\gamma}\s\alpha\sd{\alpha}} = 4i(\epsilon_{\sd{\alpha}\sd{\beta}} T_{\sd{\delta}\s\gamma\sd{\gamma}\s\alpha} + \epsilon_{\sd{\alpha}\sd{\delta}} T_{\sd{\beta}\s\gamma\sd{\gamma}\s\alpha})\,,
\end{align}
and by applying Eq.~\eqref{eq:bianchi_sp_lorentz_algebra_spinor_vector_identification}, it reads
\begin{align}
\epsilon_{\sd{\gamma}\sd{\alpha}} R_{\sd{\beta}\sd{\delta}\sym{2}{\gamma\alpha}} - \epsilon_{\gamma\alpha} R_{\sd{\beta}\sd{\delta}\sym{2}{\sd{\gamma}\sd{\alpha}}} = 2i(\epsilon_{\sd{\alpha}\sd{\beta}} T_{\sd{\delta}\s\gamma\sd{\gamma}\s\alpha} + \epsilon_{\sd{\alpha}\sd{\delta}} T_{\sd{\beta}\s\gamma\sd{\gamma}\s\alpha})\,. \label{eq:bianchi_sp_sol_eq07_1}
\end{align}
The tensor $T_{\sd{\delta}\s\gamma\sd{\gamma}\s\alpha}$ can be decomposed into components with definite symmetry properties:
\begin{align}
T_{\sd{\delta}\s\gamma\sd{\gamma}\s\alpha} = \epsilon_{\sd{\delta}\sd{\gamma}} \epsilon_{\gamma\alpha} T + \epsilon_{\sd{\delta}\sd{\gamma}} T_{\sym{2}{\gamma\alpha}} + \epsilon_{\gamma\alpha} T_{\sym{2}{\sd{\delta}\sd{\gamma}}} + T_{\sym{2}{\gamma\alpha}\sym{2}{\sd{\delta}\sd{\gamma}}}\,. \label{eq:bianchi_sp_sol_eq07_2}
\end{align}
The components are then determined by Eq.~\eqref{eq:bianchi_sp_sol_eq07_1} using the symmetry properties of the different terms. For example, if both sides of Eq.~\eqref{eq:bianchi_sp_sol_eq07_1} are contracted with $\epsilon^{\alpha\gamma}\epsilon^{\sd{\alpha}\sd{\gamma}}$, the Lorentz curvature tensor drops out and it follows immediately that
\begin{align}
T_{\sym{2}{\sd{\delta}\sd{\gamma}}} = 0\,.
\end{align}
In a similar manner follows, by taking into account that $R_{\sd{\beta}\sd{\delta}\sym{2}{\gamma\alpha}}$ and $R_{\sd{\beta}\sd{\delta}\sym{2}{\sd{\gamma}\sd{\alpha}}}$ are symmetric in~$\sd{\beta}\sd{\delta}$,
\begin{align}
T_{\sym{2}{\gamma\alpha}} &= 0\,, & T_{\sym{2}{\gamma\alpha}\sym{2}{\sd{\delta}\sd{\gamma}}} &= 0\,.
\end{align}
Thus, only the first term in the decomposition of Eq.~\eqref{eq:bianchi_sp_sol_eq07_2} remains:
\begin{align}
T_{\sd{\delta}\s\gamma\sd{\gamma}\s\alpha} = -2i\epsilon_{\sd{\delta}\sd{\gamma}} \epsilon_{\gamma\alpha} R\,, \label{eq:bianchi_sp_id_eq07_1}
\end{align}
with the definition $T=-2iR$. With these identities, Eq.~\eqref{eq:bianchi_sp_sol_eq07_1} implies
\begin{align}
R_{\sd{\beta}\sd{\delta}\sym{2}{\gamma\alpha}} &= 0\,, \label{eq:bianchi_sp_id_eq07_2}\\
R_{\sd{\beta}\sd{\delta}\sym{2}{\sd{\gamma}\sd{\alpha}}} &= 4(\epsilon_{\sd{\delta}\sd{\alpha}} \epsilon_{\sd{\beta}\sd{\gamma}} + \epsilon_{\sd{\beta}\sd{\alpha}} \epsilon_{\sd{\delta}\sd{\gamma}}) R\,. \label{eq:bianchi_sp_id_eq07_3}
\end{align}
In the same way, the identities
\begin{align}
T_{\delta\s\gamma\sd{\gamma}\s\sd{\alpha}} &= -2i\epsilon_{\delta\gamma} \epsilon_{\sd{\gamma}\sd{\alpha}} \Rb\,, \label{eq:bianchi_sp_id_eq07_4}\\
R_{\beta\delta\sym{2}{\sd{\gamma}\sd{\alpha}}} &= 0\,, \label{eq:bianchi_sp_id_eq07_5}\\
R_{\beta\delta\sym{2}{\gamma\alpha}} &= 4(\epsilon_{\delta\alpha} \epsilon_{\beta\gamma} + \epsilon_{\beta\alpha} \epsilon_{\delta\gamma}) \Rb\, \label{eq:bianchi_sp_id_eq07_6}
\end{align}
are derived from (\ref{eq:bianchi_sp_bianchi_identity_eq07}b). From Eq.~\eqref{eq:bianchi_sp_id_eq07_1} and \eqref{eq:bianchi_sp_id_eq07_4} follows, by considering the conjugation property of the torsion tensor components in Eq.~\eqref{eq:superspace_torsion_components_conjugate}, that the superfields $R$ and $\Rb$ are the conjugate of each other, i.e.\ $\Rb = R^\cc$. Furthermore, their weights with respect to the $\U(1)$ group are
\begin{align}
w(R) &= +2\,, & w(\Rb) &= -2\,.
\end{align}
It is straightforward to check that Eqs.~\eqref{eq:bianchi_sp_id_eq07_1}--\eqref{eq:bianchi_sp_id_eq07_6} solve \eqref{eq:bianchi_sp_bianchi_identity_eq07}.
\item \textbf{Identity \eqref{eq:bianchi_sp_bianchi_identity_eq01} and \eqref{eq:bianchi_sp_bianchi_identity_eq03}:}\\
Since the components of the torsion and the Lorentz curvature tensor which appear in \eqref{eq:bianchi_sp_bianchi_identity_eq01} and \eqref{eq:bianchi_sp_bianchi_identity_eq03} are already specified by \eqref{eq:bianchi_sp_bianchi_identity_eq07}, these two identities are solved if and only if the following components of the $\mathrm{U}(1)$ field strength vanish:
\begin{align}
F_{\delta\gamma} &= 0\,, \label{eq:bianchi_sp_id_eq01_1}\\
F_{\sd{\delta}\sd{\gamma}} &= 0\,. \label{eq:bianchi_sp_id_eq01_2}
\end{align}
\item \textbf{Identity \eqref{eq:bianchi_sp_bianchi_identity_eq02} and \eqref{eq:bianchi_sp_bianchi_identity_eq08}:}\\
In order to solve \eqref{eq:bianchi_sp_bianchi_identity_eq02} and \eqref{eq:bianchi_sp_bianchi_identity_eq08}, again all vector indices are turned into spinor indices, and the relation of Eq.~\eqref{eq:bianchi_sp_lorentz_algebra_spinor_vector_identification} is used to express the components of the curvature tensor in \eqref{eq:bianchi_sp_bianchi_identity_eq08}. In terms of components with definite symmetry properties the components of the torsion tensor are written as
\begin{align}
T_{\beta\s\gamma\sd{\gamma}\s\alpha} &= \epsilon_{\beta\gamma} T_{\alpha\sd{\gamma}} + \epsilon_{\beta\alpha} \tilde{T}_{\gamma\sd{\gamma}} + T_{\sym{3}{\beta\gamma\alpha}\sd{\gamma}}\,,\\
T_{\sd{\beta}\s\gamma\sd{\gamma}\s\sd{\alpha}} &= \epsilon_{\sd{\beta}\sd{\gamma}} T_{\sd{\alpha}\gamma} + \epsilon_{\sd{\beta}\sd{\alpha}} \tilde{T}_{\sd{\gamma}\gamma} + T_{\sym{3}{\sd{\beta}\sd{\gamma}\sd{\alpha}}\gamma}\,.
\end{align}
The identities \eqref{eq:bianchi_sp_bianchi_identity_eq02} and \eqref{eq:bianchi_sp_bianchi_identity_eq08} imply
\begin{align}
T_{\sym{3}{\beta\gamma\alpha}\sd{\gamma}} &= 0\,, & T_{\sym{3}{\sd{\beta}\sd{\gamma}\sd{\alpha}}\gamma} &= 0\,,\\
T_{\alpha\sd{\gamma}} &= T_{\sd{\gamma}\alpha}\,, & \tilde{T}_{\alpha\sd{\gamma}} &= \tilde{T}_{\sd{\gamma}\alpha}\,,
\end{align}
which in addition leads to
\begin{align}
T_{\beta\s\gamma\sd{\gamma}\s\alpha} &= +\epsilon_{\beta\gamma} T_{\alpha\sd{\gamma}} + \epsilon_{\beta\alpha} \tilde{T}_{\gamma\sd{\gamma}}\,, \label{eq:bianchi_sp_sol_eq02_1}\\
T_{\sd{\beta}\s\gamma\sd{\gamma}\s\sd{\alpha}} &= +\epsilon_{\sd{\beta}\sd{\gamma}} T_{\gamma\sd{\alpha}} + \epsilon_{\sd{\beta}\sd{\alpha}} \tilde{T}_{\gamma\sd{\gamma}}\,,\\
R_{\beta\sd{\delta}\sym{2}{\gamma\alpha}} &= -i(\epsilon_{\beta\gamma} T_{\alpha\sd{\delta}} + \epsilon_{\beta\alpha} T_{\gamma\sd{\delta}})\,,\\
R_{\beta\sd{\delta}\sym{2}{\sd{\gamma}\sd{\alpha}}} &= -i(\epsilon_{\sd{\delta}\sd{\gamma}} T_{\beta\sd{\alpha}} + \epsilon_{\sd{\delta}\sd{\alpha}} T_{\beta\sd{\gamma}})\,,\\
F_{\delta\sd{\gamma}} &= -i(T_{\delta\sd{\gamma}} - 2\tilde{T}_{\delta\sd{\gamma}}) \label{eq:bianchi_sp_sol_eq02_2}\,.
\end{align}
This is the general solution of \eqref{eq:bianchi_sp_bianchi_identity_eq02} and \eqref{eq:bianchi_sp_bianchi_identity_eq08}. In order to reduce the degrees of freedom further, the following additional restriction is assumed
\begin{align}
T_{\alpha\sd{\gamma}} = -\tilde{T}_{\alpha\sd{\gamma}}\,. \label{eq:bianchi_sp_sol_eq02_3}
\end{align}
Using the definition $T_{\alpha\sd{\gamma}} = -i G_{\alpha\sd{\gamma}}$, the tensor components in Eqs.~\eqref{eq:bianchi_sp_sol_eq02_1}--\eqref{eq:bianchi_sp_sol_eq02_2} are written as
\begin{align}
T_{\beta\s\gamma\sd{\gamma}\s\alpha} &= i\epsilon_{\gamma\alpha} G_{\beta\sd{\gamma}}\,, \label{eq:bianchi_sp_id_eq02_1}\\
T_{\sd{\beta}\s\gamma\sd{\gamma}\s\sd{\alpha}} &= i\epsilon_{\sd{\gamma}\sd{\alpha}} G_{\gamma\sd{\beta}}\,, \label{eq:bianchi_sp_id_eq02_2}\\
R_{\beta\sd{\delta}\sym{2}{\gamma\alpha}} &= -\epsilon_{\beta\alpha} G_{\gamma\sd{\delta}} - \epsilon_{\beta\gamma} G_{\alpha\sd{\delta}}\,, \label{eq:bianchi_sp_id_eq02_3}\\
R_{\delta\sd{\beta}\sym{2}{\sd{\gamma}\sd{\alpha}}} &= -\epsilon_{\sd{\beta}\sd{\alpha}} G_{\delta\sd{\gamma}} - \epsilon_{\sd{\beta}\sd{\gamma}} G_{\delta\sd{\alpha}}\,, \label{eq:bianchi_sp_id_eq02_4}\\
F_{\delta\sd{\gamma}} &= -3 G_{\delta\sd{\gamma}}  \label{eq:bianchi_sp_id_eq02_5}\,.
\end{align}
From Eq.~\eqref{eq:bianchi_sp_id_eq02_1} and \eqref{eq:bianchi_sp_id_eq02_2} follows that $(G_{\beta\sd{\gamma}})^\cc = G_{\gamma\sd{\beta}}$. Furthermore, the $\U(1)$ weight of $G_{\sd{\alpha}\gamma}$ vanishes:
\begin{align}
w(G_{\sd{\alpha}\gamma})=0\,.
\end{align}
\item \textbf{Identity \eqref{eq:bianchi_sp_bianchi_identity_eq11}:}\\
In (\ref{eq:bianchi_sp_bianchi_identity_eq11}a), the component $R_{\beta dca}$ is antisymmetric in $\beta d$, which are the two indices of the $2$-form, and also in $ca$, which are the two indices of the Lie algebra of the Lorentz group. Thus, the equation implies
\begin{align}
R_{\beta dca} = +i(\sig_{d\beta\sd{\varphi}} {T_{ca}}^\sd{\varphi} - \sig_{a\beta\sd{\varphi}} {T_{dc}}^\sd{\varphi} - \sig_{c\beta\sd{\varphi}} {T_{ad}}^\sd{\varphi}) \label{eq:bianchi_sp_id_eq11_1}\,.
\end{align}
With this identity, (\ref{eq:bianchi_sp_bianchi_identity_eq11}a) is solved. In the same way, (\ref{eq:bianchi_sp_bianchi_identity_eq11}b) leads to
\begin{align}
R_{\sd{\beta} dca} = -i(\sig_{d\varphi\sd{\beta}} {T_{ca}}^\varphi - \sig_{a\varphi\sd{\beta}} {T_{dc}}^\varphi - \sig_{c\varphi\sd{\beta}} {T_{ad}}^\varphi) \label{eq:bianchi_sp_id_eq11_2}\,.
\end{align}
\item \textbf{Identity \eqref{eq:bianchi_sp_bianchi_identity_eq13}:}\\
The identity \eqref{eq:bianchi_sp_bianchi_identity_eq13} represents the first Bianchi identity of the Lorentz curvature tensor in ordinary spacetime with the Levi-Civita connection. The general solution of this identity is derived by using the spinor notation. In terms of components with definite symmetry properties, the components of the Lorentz curvature tensor are written as
\begin{align}
\begin{split}
R_{\delta\sd{\delta}\s\gamma\sd{\gamma}\s\beta\sd{\beta}\s\alpha\sd{\alpha}} =& +4\epsilon_{\sd{\delta}\sd{\gamma}}\epsilon_{\sd{\beta}\sd{\alpha}} \chi_{\sym{2}{\delta\gamma}\sym{2}{\beta\alpha}} - 4\epsilon_{\sd{\delta}\sd{\gamma}}\epsilon_{\beta\alpha} \psi_{\sym{2}{\delta\gamma}\sym{2}{\sd{\beta}\sd{\alpha}}}\\
& - 4\epsilon_{\delta\gamma}\epsilon_{\sd{\beta}\sd{\alpha}} \psi_{\sym{2}{\sd{\delta}\sd{\gamma}}\sym{2}{\beta\alpha}} + 4\epsilon_{\delta\gamma}\epsilon_{\beta\alpha} \chi_{\sym{2}{\sd{\delta}\sd{\gamma}}\sym{2}{\sd{\beta}\sd{\alpha}}}\,.
\end{split}
\end{align}
From (\ref{eq:bianchi_sp_bianchi_identity_eq13}a) follows that
\begin{align}
\psi_{\sym{2}{\sd{\delta}\sd{\gamma}}\sym{2}{\beta\alpha}} &= \psi_{\sym{2}{\beta\alpha}\sym{2}{\sd{\delta}\sd{\gamma}}}\,,\\
{\chi^\alpha}_{\gamma\beta\alpha} &= 3\epsilon_{\gamma\beta} \chi\,,\\
{\chi^\sd{\alpha}}_{\sd{\gamma}\sd{\beta}\sd{\alpha}} &= 3\epsilon_{\sd{\gamma}\sd{\beta}} \chi^\ct\,,\\
\chi &= \chi^\ct\,,
\end{align}
which leads to
\begin{align}
\begin{split}
R_{\delta\sd{\delta}\s\gamma\sd{\gamma}\s\beta\sd{\beta}\s\alpha\sd{\alpha}} =& +4\epsilon_{\sd{\delta}\sd{\gamma}}\big(\epsilon_{\sd{\beta}\sd{\alpha}} \chi_{\sym{2}{\delta\gamma}\sym{2}{\beta\alpha}} - \epsilon_{\beta\alpha} \psi_{\sym{2}{\delta\gamma}\sym{2}{\sd{\beta}\sd{\alpha}}}\big)\\
& + 4\epsilon_{\delta\gamma}\big(\epsilon_{\beta\alpha} \chi_{\sym{2}{\sd{\delta}\sd{\gamma}}\sym{2}{\sd{\beta}\sd{\alpha}}} - \epsilon_{\sd{\beta}\sd{\alpha}} \psi_{\sym{2}{\beta\alpha}\sym{2}{\sd{\delta}\sd{\gamma}}}\big) \label{eq:bianchi_sp_id_eq13_1}\,,
\end{split}
\end{align}
and
\begin{align}
\chi_{\sym{2}{\delta\gamma}\sym{2}{\beta\alpha}} &= \chi_{\sym{4}{\delta\gamma\beta\alpha}} + (\epsilon_{\delta\beta}\epsilon_{\gamma\alpha} + \epsilon_{\delta\alpha}\epsilon_{\gamma\beta}) \chi\,, \label{eq:bianchi_sp_id_eq13_2}\\
\chi_{\sym{2}{\sd{\delta}\sd{\gamma}}\sym{2}{\sd{\beta}\sd{\alpha}}} &= \chi_{\sym{4}{\sd{\delta}\sd{\gamma}\sd{\beta}\sd{\alpha}}} + (\epsilon_{\sd{\delta}\sd{\beta}}\epsilon_{\sd{\gamma}\sd{\alpha}} + \epsilon_{\sd{\delta}\sd{\alpha}}\epsilon_{\sd{\gamma}\sd{\beta}}) \chi\,. \label{eq:bianchi_sp_id_eq13_3}
\end{align}
Note that $\chi$ is real. In addition, it is convenient for solving Eq.~\eqref{eq:bianchi_sp_bianchi_identity_eq09} below to write \eqref{eq:bianchi_sp_id_eq13_1} in the following form, by using Eq.~\eqref{eq:bianchi_sp_lorentz_algebra_spinor_vector_identification}:
\begin{align}
R_{\delta\sd{\delta}\s\gamma\sd{\gamma}\s\sym{2}{\beta\alpha}} &= +2(\epsilon_{\sd{\delta}\sd{\gamma}} \chi_{\sym{2}{\delta\gamma}\sym{2}{\beta\alpha}} - \epsilon_{\delta\gamma} \psi_{\sym{2}{\beta\alpha}\sym{2}{\sd{\delta}\sd{\gamma}}})\,,\\
R_{\delta\sd{\delta}\s\gamma\sd{\gamma}\s\sym{2}{\sd{\beta}\sd{\alpha}}} &= -2(\epsilon_{\delta\gamma} \chi_{\sym{2}{\sd{\delta}\sd{\gamma}}\sym{2}{\sd{\beta}\sd{\alpha}}} - \epsilon_{\sd{\delta}\sd{\gamma}} \psi_{\sym{2}{\delta\gamma}\sym{2}{\sd{\beta}\sd{\alpha}}})\,.
\end{align}
\end{itemize}
In a next step \eqref{eq:bianchi_sp_bianchi_identity_eq04}, \eqref{eq:bianchi_sp_bianchi_identity_eq05} and \eqref{eq:bianchi_sp_bianchi_identity_eq06} are solved, which contain derivative terms but are still linear.
\begin{itemize}
\item \textbf{Identity \eqref{eq:bianchi_sp_bianchi_identity_eq06}:}\\
The identity \eqref{eq:bianchi_sp_bianchi_identity_eq06} is solved if and only if
\begin{align}
\cD_\sd{\gamma} R &= 0\,, & \cD_\gamma \Rb &= 0\,, \label{eq:bianchi_sp_id_eq06_1}
\end{align}
which means that $R$ and $\Rb$ are a chiral and an antichiral superfield, respectively.
\item \textbf{Identity \eqref{eq:bianchi_sp_bianchi_identity_eq04} and \eqref{eq:bianchi_sp_bianchi_identity_eq05}:}\\
To get the solutions of \eqref{eq:bianchi_sp_bianchi_identity_eq04} and \eqref{eq:bianchi_sp_bianchi_identity_eq05}, Eq.~\eqref{eq:bianchi_sp_id_eq11_1} and \eqref{eq:bianchi_sp_id_eq11_2} are written in terms of spinor indices
\begin{align}
\epsilon_{\sd{\gamma}\sd{\alpha}} R_{\beta\s\delta\sd{\delta}\s\sym{2}{\gamma\alpha}} - \epsilon_{\gamma\alpha} R_{\beta\s\delta\sd{\delta}\s\sym{2}{\sd{\gamma}\sd{\alpha}}} = -i(\epsilon_{\delta\beta} T_{\gamma\sd{\gamma}\s\alpha\sd{\alpha}\s\sd{\delta}} - \epsilon_{\alpha\beta} T_{\delta\sd{\delta}\s\gamma\sd{\gamma}\s\sd{\alpha}} - \epsilon_{\gamma\beta} T_{\alpha\sd{\alpha}\s\delta\sd{\delta}\s\sd{\gamma}})\,, \label{eq:bianchi_sp_sol_eq04_1}\\
\epsilon_{\sd{\gamma}\sd{\alpha}} R_{\sd{\beta}\s\delta\sd{\delta}\s\sym{2}{\gamma\alpha}} - \epsilon_{\gamma\alpha} R_{\sd{\beta}\s\delta\sd{\delta}\s\sym{2}{\sd{\gamma}\sd{\alpha}}} = +i(\epsilon_{\sd{\delta}\sd{\beta}} T_{\gamma\sd{\gamma}\s\alpha\sd{\alpha}\s\delta} - \epsilon_{\sd{\alpha}\sd{\beta}} T_{\delta\sd{\delta}\s\gamma\sd{\gamma}\s\alpha} - \epsilon_{\sd{\gamma}\sd{\beta}} T_{\alpha\sd{\alpha}\s\delta\sd{\delta}\s\gamma})\,, \label{eq:bianchi_sp_sol_eq04_2}
\end{align}
where Eq.~\eqref{eq:bianchi_sp_lorentz_algebra_spinor_vector_identification} is used. In addition, the components of the torsion tensor appearing in these two equations is written in terms of components with definite symmetry properties
\begin{align}
T_{\gamma\sd{\gamma}\s\alpha\sd{\alpha}\s\su{\delta}} = 2\epsilon_{\sd{\gamma}\sd{\alpha}} T_{\sym{2}{\gamma\alpha}\su{\delta}} - 2\epsilon_{\gamma\alpha} T_{\sym{2}{\sd{\gamma}\sd{\alpha}}\su{\delta}}\,, \label{eq:bianchi_sp_sol_eq04_3}
\end{align}
with the further decomposition
\begin{align}
T_{\sym{2}{\gamma\alpha}\delta} = W_{\sym{3}{\gamma\alpha\delta}} + \frac{1}{3}(\epsilon_{\delta\gamma} S_\alpha + \epsilon_{\delta\alpha} S_\gamma)\,,
\label{eq:bianchi_sp_sol_eq04_4}\\
T_{\sym{2}{\sd{\gamma}\sd{\alpha}}\sd{\delta}} = W_{\sym{3}{\sd{\gamma}\sd{\alpha}\sd{\delta}}} + \frac{1}{3}(\epsilon_{\sd{\delta}\sd{\gamma}} S_\sd{\alpha} + \epsilon_{\sd{\delta}\sd{\alpha}} S_\sd{\gamma})\,. \label{eq:bianchi_sp_sol_eq04_5}
\end{align}
With these definitions, Eq.~\eqref{eq:bianchi_sp_sol_eq04_1} and \eqref{eq:bianchi_sp_sol_eq04_2} imply
\begin{align}
R_{\beta\s\delta\sd{\delta}\s\sym{2}{\gamma\alpha}} &= +i\sum_{\gamma\alpha}(\epsilon_{\beta\delta} T_{\sym{2}{\gamma\alpha}\sd{\delta}} + \epsilon_{\beta\alpha} T_{\sym{2}{\delta\gamma}\sd{\delta}} - \epsilon_{\beta\gamma} \epsilon_{\delta\alpha} S_\sd{\delta})\,, \label{eq:bianchi_sp_sol_eq04_6}\\
R_{\beta\s\delta\sd{\delta}\s\sym{2}{\sd{\gamma}\sd{\alpha}}} &= +4i\epsilon_{\beta\delta} W_{\sym{3}{\sd{\delta}\sd{\gamma}\sd{\alpha}}} + i\sum_{\sd{\gamma}\sd{\alpha}} \epsilon_{\sd{\delta}\sd{\gamma}} (T_{\sym{2}{\beta\delta}\sd{\alpha}} + \frac{1}{3}\epsilon_{\beta\delta} S_\sd{\alpha})\,, \label{eq:bianchi_sp_sol_eq04_7}\\
R_{\sd{\beta}\s\delta\sd{\delta}\s\sym{2}{\gamma\alpha}} &= -4i\epsilon_{\sd{\beta}\sd{\delta}} W_{\sym{3}{\delta\gamma\alpha}} - i\sum_{\gamma\alpha} \epsilon_{\delta\gamma} (T_{\sym{2}{\sd{\beta}\sd{\delta}}\alpha} + \frac{1}{3}\epsilon_{\sd{\beta}\sd{\delta}} S_\alpha)\,, \label{eq:bianchi_sp_sol_eq04_8}\\
R_{\sd{\beta}\s\delta\sd{\delta}\s\sym{2}{\sd{\gamma}\sd{\alpha}}} &= -i\sum_{\sd{\gamma}\sd{\alpha}}(\epsilon_{\sd{\beta}\sd{\delta}} T_{\sym{2}{\sd{\gamma}\sd{\alpha}}\delta} + \epsilon_{\sd{\beta}\sd{\alpha}} T_{\sym{2}{\sd{\delta}\sd{\gamma}}\delta} - \epsilon_{\sd{\beta}\sd{\gamma}} \epsilon_{\sd{\delta}\sd{\alpha}} S_\delta)\,. \label{eq:bianchi_sp_sol_eq04_9}
\end{align}
Using these identities, \eqref{eq:bianchi_sp_bianchi_identity_eq04} and \eqref{eq:bianchi_sp_bianchi_identity_eq05} are solved if and only if the following quantities are expressed in terms of derivatives of the superfields $R$, $\Rb$ and $G_{\alpha\sd{\alpha}}$:
\begin{align}
T_{\sym{2}{\delta\beta}\sd{\beta}} &= +\frac{1}{4} (\cD_\delta G_{\beta\sd{\beta}} + \cD_\beta G_{\delta\sd{\beta}})\,, \label{eq:bianchi_sp_id_eq04_1}\\
T_{\sym{2}{\sd{\delta}\sd{\beta}}\beta} &= -\frac{1}{4} (\cD_\sd{\delta} G_{\beta\sd{\beta}} + \cD_\sd{\beta} G_{\beta\sd{\delta}})\,, \label{eq:bianchi_sp_id_eq04_2}\\
S_\beta &= +\frac{1}{4} \cD^\sd{\beta} G_{\beta\sd{\beta}} - \cD_\beta R\,, \label{eq:bianchi_sp_id_eq04_3}\\
S_\sd{\beta} &= -\frac{1}{4} \cD^\beta G_{\beta\sd{\beta}} + \cD_\sd{\beta} \Rb\,, \label{eq:bianchi_sp_id_eq04_4}\\
F_{\delta\s\beta\sd{\beta}} &= \frac{3i}{2} \cD_\delta G_{\beta\sd{\beta}} + i\epsilon_{\delta\beta} X_\sd{\beta}\,, \label{eq:bianchi_sp_id_eq04_5}\\
F_{\sd{\delta}\s\beta\sd{\beta}} &= \frac{3i}{2} \cD_\sd{\delta} G_{\beta\sd{\beta}} + i\epsilon_{\sd{\delta}\sd{\beta}} X_\beta\,, \label{eq:bianchi_sp_id_eq04_6}
\end{align}
with
\begin{align}
X_\beta &= \cD_\beta R - \cD^\sd{\beta} G_{\beta\sd{\beta}}\,, \label{eq:bianchi_sp_id_eq04_7}\\
X_\sd{\beta} &= \cD_\sd{\beta} \Rb - \cD^\beta G_{\beta\sd{\beta}}\,. \label{eq:bianchi_sp_id_eq04_8}
\end{align}
From the definitions of $W_{\sym{3}{\gamma\beta\alpha}}$ and $W_{\sym{3}{\sd{\gamma}\sd{\beta}\sd{\alpha}}}$ in Eq.~\eqref{eq:bianchi_sp_sol_eq04_4} and \eqref{eq:bianchi_sp_sol_eq04_5} follows that $(W_{\sym{3}{\gamma\beta\alpha}})^\cc = -W_{\sym{3}{\sd{\gamma}\sd{\beta}\sd{\alpha}}}$, and the respective $\U(1)$ weights can be read off:
\begin{align}
w(W_{\sym{3}{\gamma\beta\alpha}}) &= +1\,, & w(W_{\sym{3}{\sd{\gamma}\sd{\beta}\sd{\alpha}}}) &= -1\,.
\end{align}
\end{itemize}
The remaining, non-linear identities are \eqref{eq:bianchi_sp_bianchi_identity_eq09}, \eqref{eq:bianchi_sp_bianchi_identity_eq10} and \eqref{eq:bianchi_sp_bianchi_identity_eq12}. In the derivation of the solution of these identities the following relations are used
\begin{align}
\acom{\cD_\sd{\beta}}{\cD_\beta} R &= -2i\cD_{\beta\sd{\beta}} R - 6 G_{\beta\sd{\beta}} R\,, \label{eq:bianchi_sp_sol_eq09_1}\\
\acom{\cD_\beta}{\cD_\sd{\beta}}  \Rb &= -2i\cD_{\beta\sd{\beta}} \Rb - 6 G_{\beta\sd{\beta}} \Rb\,, \label{eq:bianchi_sp_sol_eq09_2}\\
\acom{\cD_\beta}{\cD_\sd{\beta}} G_{\alpha\sd{\alpha}} &= -2i\cD_{\beta\sd{\beta}} G_{\alpha\sd{\alpha}}\,. \label{eq:bianchi_sp_sol_eq09_3}
\end{align}
They are derived from Eq.~\eqref{eq:superspace_double_covariant_derivative_1form_components} and \eqref{eq:superspace_double_covariant_derivative_function_components} by taking into account the chirality conditions of $R$ and $\Rb$ (see Eq.~\eqref{eq:bianchi_sp_id_eq06_1}), and the $\U(1)$ weights $w(R)=+2$, $w(\Rb)=-2$ and $w(G_{\alpha\sd{\alpha}})=0$.
\begin{itemize}
\item \textbf{Identity \eqref{eq:bianchi_sp_bianchi_identity_eq09}:}\\
Since the two vector indices $bd$ in the $\mathrm{U}(1)$ field strength $F_{bd}$ in \eqref{eq:bianchi_sp_bianchi_identity_eq09} are antisymmetric, in spinor notation the decomposition into components with definite symmetry properties is 
\begin{align}
F_{\beta\sd{\beta}\s\delta\sd{\delta}} &= 2\epsilon_{\sd{\beta}\sd{\delta}} F_{\sym{2}{\beta\delta}} - 2\epsilon_{\beta\delta} F_{\sym{2}{\sd{\beta}\sd{\delta}}}\,, \label{eq:bianchi_sp_sol_eq09_4}
\end{align}
where the components are determined by \eqref{eq:bianchi_sp_bianchi_identity_eq09}:
\begin{align}
F_{\sym{2}{\beta\delta}} &= +\frac{1}{8} \sum_{\beta\delta} (\cD_\beta \cD^\sd{\delta} G_{\delta\sd{\delta}} + 3i{\cD_\beta}^\sd{\delta} G_{\delta\sd{\delta}})\,, \label{eq:bianchi_sp_id_eq09_1}\\
F_{\sym{2}{\sd{\beta}\sd{\delta}}} &= -\frac{1}{8} \sum_{\sd{\beta}\sd{\delta}} (\cD_\sd{\beta} \cD^\delta G_{\delta\sd{\delta}} + 3i{\cD^\delta}_\sd{\beta} G_{\delta\sd{\delta}})\,. \label{eq:bianchi_sp_id_eq09_2}
\end{align}
In addition, the superfields which have been introduced in the solution of \eqref{eq:bianchi_sp_bianchi_identity_eq13} (see Eq.~\eqref{eq:bianchi_sp_id_eq13_1}) are now expressed in terms of the superfields $W_{\sym{3}{\gamma\beta\alpha}}$, $W_{\sym{3}{\sd{\gamma}\sd{\beta}\sd{\alpha}}}$ and $G_{\alpha\sd{\alpha}}$, and their derivatives:
\begin{align}
\psi_{\sym{2}{\beta\delta}\sym{2}{\sd{\gamma}\sd{\alpha}}} &= \frac{1}{8} \sum_{\sd{\gamma}\sd{\alpha}} \sum_{\beta\delta} \Big(G_{\beta\sd{\alpha}} G_{\delta\sd{\gamma}} -\frac{1}{2}[\cD_\beta,\cD_\sd{\alpha}] G_{\delta\sd{\gamma}} \Big)\,, \label{eq:bianchi_sp_id_eq09_3}\\
\chi_{\sym{4}{\beta\delta\gamma\alpha}} &= \frac{1}{4}(\cD_\beta W_{\sym{3}{\delta\gamma\alpha}} + \cD_\delta W_{\sym{3}{\gamma\alpha\beta}} + \cD_\gamma W_{\sym{3}{\alpha\beta\delta}} + \cD_\alpha W_{\sym{3}{\beta\delta\gamma}})\,, \label{eq:bianchi_sp_id_eq09_4}\\
\chi_{\sym{4}{\sd{\beta}\sd{\delta}\sd{\gamma}\sd{\alpha}}} &= \frac{1}{4}(\cD_\sd{\beta} W_{\sym{3}{\sd{\delta}\sd{\gamma}\sd{\alpha}}} + \cD_\sd{\delta} W_{\sym{3}{\sd{\gamma}\sd{\alpha}\sd{\beta}}} + \cD_\sd{\gamma} W_{\sym{3}{\sd{\alpha}\sd{\beta}\sd{\delta}}} + \cD_\sd{\alpha} W_{\sym{3}{\sd{\beta}\sd{\delta}\sd{\gamma}}})\,, \label{eq:bianchi_sp_id_eq09_5}
\end{align}
and
\begin{align}
\chi &= -\frac{1}{12}(\cD^\beta \cD_\beta R + \cD_\sd{\beta} \cD^\sd{\beta} \Rb) + \frac{1}{48}[\cD^\beta,\cD^\sd{\beta}] G_{\beta\sd{\beta}} - \frac{1}{8} G^{\beta\sd{\beta}} G_{\beta\sd{\beta}} + 2R\Rb\,. \label{eq:bianchi_sp_id_eq09_6}
\end{align}
The remaining identities, which are needed to solve \eqref{eq:bianchi_sp_bianchi_identity_eq09}, are
\begin{align}
\cD^\beta \cD_\beta R - \cD_\sd{\beta} \cD^\sd{\beta} \Rb &= -2i \cD^{\beta\sd{\beta}} G_{\beta\sd{\beta}}\,, \label{eq:bianchi_sp_id_eq09_7}
\end{align}
and
\begin{align}
\cD_\gamma W_{\sym{3}{\beta\delta\alpha}} &= \chi_{\sym{4}{\gamma\beta\delta\alpha}} - \frac{1}{3}(\epsilon_{\gamma\beta} F_{\sym{2}{\delta\alpha}} + \epsilon_{\gamma\delta} F_{\sym{2}{\alpha\beta}} + \epsilon_{\gamma\alpha} F_{\sym{2}{\beta\delta}})\,, \label{eq:bianchi_sp_id_eq09_8}\\
\cD_\sd{\gamma} W_{\sym{3}{\sd{\beta}\sd{\delta}\sd{\alpha}}} &= \chi_{\sym{4}{\sd{\gamma}\sd{\beta}\sd{\delta}\sd{\alpha}}} - \frac{1}{4}(\epsilon_{\sd{\gamma}\sd{\beta}} F_{\sym{2}{\sd{\delta}\sd{\alpha}}} + \epsilon_{\sd{\gamma}\sd{\delta}} F_{\sym{2}{\sd{\alpha}\sd{\beta}}} + \epsilon_{\sd{\gamma}\sd{\alpha}} F_{\sym{2}{\sd{\beta}\sd{\delta}}})\,. \label{eq:bianchi_sp_id_eq09_9}
\end{align}
\item \textbf{Identity \eqref{eq:bianchi_sp_bianchi_identity_eq10}:}\\
The identity \eqref{eq:bianchi_sp_bianchi_identity_eq10} is solved by the chirality constraints
\begin{align}
\cD_\sd{\delta} W_{\sym{3}{\gamma\beta\alpha}} &= 0\,, & \cD_\delta W_{\sym{3}{\sd{\gamma}\sd{\beta}\sd{\alpha}}} &= 0\,, \label{eq:bianchi_sp_id_eq10_1}\\
\cD_\sd{\beta} X_\alpha &= 0\,, & \cD_\beta X_\sd{\alpha} &= 0\,, \label{eq:bianchi_sp_id_eq10_2}
\end{align}
and the relations
\begin{align}
\cD^\varphi \cD_\varphi G_{\alpha\sd{\alpha}} &= +4i \cD_{\alpha\sd{\alpha}} \Rb\,, \label{eq:bianchi_sp_id_eq10_3}\\
\cD_\sd{\varphi} \cD^\sd{\varphi} G_{\alpha\sd{\alpha}} &= -4i \cD_{\alpha\sd{\alpha}} R\,. \label{eq:bianchi_sp_id_eq10_4}
\end{align}
\item \textbf{Identity \eqref{eq:bianchi_sp_bianchi_identity_eq12}:}\\
The identity \eqref{eq:bianchi_sp_bianchi_identity_eq12} implies derivative relations of the fields $R$, $\Rb$, $G_a$, $W_{\sym{3}{\gamma\beta\alpha}}$ and $W_{\sym{3}{\sd{\gamma}\sd{\beta}\sd{\alpha}}}$ which are not used in subsequent calculations.\footnote{Identity \eqref{eq:bianchi_sp_bianchi_identity_eq12} is the only identity which has the maximal mass dimension $5/2$ (see Section~\ref{sec:summary_tensor_components_further_relations} for the definition of the mass dimension). Thus, the derivative relations of the superfields arising from this identity have also mass dimension $5/2$.}
\end{itemize}

\subsection{Summary of the tensor components and further relations}
\label{sec:summary_tensor_components_further_relations}
This section provides a summary of the identities in Section~\ref{sec:bianchi_sp_solution}, which were obtained by solving the Bianchi identity and taking account of the torsion constraints in Eq.~\eqref{eq:bianchi_sp_torsion_constraints}.\footnote{In fact, the additional constraints $T_{\gamma\s\beta\sd{\beta}\s\alpha} \propto \eps_{\beta\alpha}$ and $T_{\sd{\gamma}\s\beta\sd{\beta}\s\sd{\alpha}} \propto \eps_{\sd{\beta}\sd{\alpha}}$ are used in the solution of the Bianchi identity to reduce the degrees of freedom (cf.~Eq.~\eqref{eq:bianchi_sp_sol_eq02_3}).} In particular, Section~\ref{sec:torsion_tensor_components} and \ref{sec:lorentz_curvature_u1_fieldstrength_components} provide a complete list of the components of the torsion tensor, the Lorentz curvature tensor and of the $\mathrm{U}(1)$ field strength in $\mathrm{U}(1)$ superspace. They are ordered according to their mass dimension, where $T$, ${R_B}^A$ and $F$ have mass dimension zero and the mass dimension of the supervielbein and the inverse supervielbein is given by $[E^a]=-1$, $[E^\su{\alpha}]=-\frac{1}{2}$ and $[E_a]=+1$, $[E^\su{\alpha}]=+\frac{1}{2}$, respectively. It turns out that all the components can be expressed in terms of the following few superfields and their derivatives:
\begin{align}
R\,,\quad \Rb\,,\quad G_a\,,\quad W_{\sym{3}{\gamma\beta\alpha}}\,,\quad W_{\sym{3}{\sd{\gamma}\sd{\beta}\sd{\alpha}}}\,,
\end{align}
where $G_a$ in terms of spinor indices is written as $G_{\alpha\sd{\alpha}} = \sig^a_{\alpha\sd{\alpha}} G_a$. Furthermore, the following chirality conditions apply
\begin{align}
\cD^\sd{\alpha} R &= 0\,, &  \cD_\alpha \Rb &= 0\,,\\
\cD^\sd{\delta} W_{\sym{3}{\gamma\beta\alpha}} &= 0\,, & \cD_\delta W_{\sym{3}{\sd{\gamma}\sd{\beta}\sd{\alpha}}} &= 0\,.
\end{align}
According to the appearance of these superfields in the torsion tensor, their $\U(1)$ weights are given by
\begin{gather}
w(R) = +2\,,\hspace{3cm} w(\Rb) = -2\,,\hspace{3cm} w(G_a) = 0\,,\\
w(W_{\sym{3}{\gamma\beta\alpha}}) = +1\,,\hspace{2.5cm} w(W_{\sym{3}{\sd{\gamma}\sd{\beta}\sd{\alpha}}}) = -1\,,
\end{gather}
and they transform under conjugation as follows
\begin{align}
R^\cc &= \Rb\,, & (G_a)^\cc &= G_a\,, & (W_{\sym{3}{\gamma\beta\alpha}})^\cc &= -W_{\sym{3}{\sd{\gamma}\sd{\beta}\sd{\alpha}}}\,.
\end{align}
In addition, in Section~\ref{sec:derivative_relations} the derivative relations of the components in $\mathrm{U}(1)$ superspace are listed. Many results are written in vector and in spinor notation, since both forms are used in further calculations. Moreover, the shorthand notation $\su{\alpha}$ is used, which labels both $\alpha$ and $\sd{\alpha}$.
\subsubsection{Torsion tensor components}
\label{sec:torsion_tensor_components}
Since the torsion tensor has two lower and one upper superspace index, the mass dimensions of the components range from $0$ to $\frac{3}{2}$.
\begin{itemize}
\item \textbf{Dimension $0$:}\\
The components with dimension $0$ are fixed by the torsion constraints in Eq.~\eqref{eq:bianchi_sp_torsion_constraint_2} and \eqref{eq:bianchi_sp_torsion_constraint_3}:
\begin{gather}
{T_{\gamma\beta}}^a = 0\,,\hspace{4cm} T^{\sd{\gamma}\sd{\beta}a} = 0\,, \label{eq:bianchi_sp_components_torsion_dim0_1}\\
{T_{\gamma}}^{\sd{\beta}a} = -2i{(\sig^a\eps)_\gamma}^\sd{\beta}\,. \label{eq:bianchi_sp_components_torsion_dim0_2}
\end{gather}
\item \textbf{Dimension $\frac{1}{2}$:}\\
According to the torsion constraints in Eq.~\eqref{eq:bianchi_sp_torsion_constraint_1} and \eqref{eq:bianchi_sp_torsion_constraint_4}, all components with dimension $\frac{1}{2}$ vanish:
\begin{align}
{T_{\su{\gamma}\su{\beta}}}^\su{\alpha} &= 0\,, & {T_{\su{\gamma}b}}^a &= 0\,. \label{eq:bianchi_sp_components_torsion_dim1/2_1}
\end{align}
\item \textbf{Dimension $1$:}\\
At dimension $1$, the components are specified by the solution of the Bianchi identity in Eq.~\eqref{eq:bianchi_sp_id_eq07_1}, \eqref{eq:bianchi_sp_id_eq07_4}, \eqref{eq:bianchi_sp_id_eq02_1} and \eqref{eq:bianchi_sp_id_eq02_2}, and by the torsion constraint in Eq.~\eqref{eq:bianchi_sp_torsion_constraint_5}:
\begin{align}
{T_{\gamma b}}^\alpha = +\frac{i}{2}{(\sig_c \sigb_b)_\gamma}^\alpha G^c &\quad\leftrightarrow\quad T_{\gamma\s\beta\sd{\beta}\s\alpha} = +i\eps_{\beta\alpha} G_{\gamma\sd{\beta}}\,, \label{eq:bianchi_sp_components_torsion_dim1_1}\\
{T^\sd{\gamma}}_{b\sd{\alpha}} = -\frac{i}{2}{(\sigb_c \sig_b)^\sd{\gamma}}_\sd{\alpha} G^c &\quad\leftrightarrow\quad T_{\sd{\gamma}\s\beta\sd{\beta}\s\sd{\alpha}} = +i\eps_{\sd{\beta}\sd{\alpha}} G_{\beta\sd{\gamma}}\,, \label{eq:bianchi_sp_components_torsion_dim1_2}\\
T_{\gamma b\sd{\alpha}} = -i\sig_{b\gamma\sd{\alpha}} \Rb &\quad\leftrightarrow\quad T_{\gamma\s\beta\sd{\beta}\s\sd{\alpha}} = -2i\eps_{\gamma\beta}\eps_{\sd{\beta}\sd{\alpha}} \Rb\,, \label{eq:bianchi_sp_components_torsion_dim1_3}\\
\tensor{T}{^{\sd{\gamma}}_b^\alpha} = -i\sigb_b^{\sd{\gamma}\alpha} R &\quad\leftrightarrow\quad T_{\sd{\gamma}\s\beta\sd{\beta}\s\alpha} = -2i \eps_{\sd{\gamma}\sd{\beta}} \eps_{\beta\alpha} R\,, \label{eq:bianchi_sp_components_torsion_dim1_4}\\
{T_{cb}}^a = 0 &\quad\leftrightarrow\quad T_{\gamma\sd{\gamma}\s\beta\sd{\beta}\s\alpha\sd{\alpha}} = 0\,. \label{eq:bianchi_sp_components_torsion_dim1_5}
\end{align}
\item \textbf{Dimension $\frac{3}{2}$:}\\
The components with dimension $\frac{3}{2}$ are conveniently written in spinor notation. In terms of components with definite symmetry properties, they are written as
\begin{align}
{T_{cb}}^\su{\alpha} \quad\leftrightarrow\quad T_{\gamma\sd{\gamma}\s\beta\sd{\beta}\s\su{\alpha}} = 2\eps_{\sd{\gamma}\sd{\beta}} T_{\sym{2}{\gamma\beta}\su{\alpha}} - 2\eps_{\gamma\beta} T_{\sym{2}{\sd{\gamma}\sd{\beta}}\su{\alpha}}\,, \label{eq:bianchi_sp_components_torsion_dim3/2_1}
\end{align}
with
\begin{align}
T_{\sym{2}{\gamma\beta}\alpha} &= W_{\sym{3}{\gamma\beta\alpha}} + \frac{1}{3}(\eps_{\alpha\gamma}S_\beta + \eps_{\alpha\beta}S_\gamma)\,, \label{eq:bianchi_sp_components_torsion_dim3/2_2}\\
T_{\sym{2}{\sd{\gamma}\sd{\beta}}\sd{\alpha}} &= W_{\sym{3}{\sd{\gamma}\sd{\beta}\sd{\alpha}}} + \frac{1}{3}(\eps_{\sd{\alpha}\sd{\gamma}}S_\sd{\beta} + \eps_{\sd{\alpha}\sd{\beta}}S_\sd{\gamma})\,. \label{eq:bianchi_sp_components_torsion_dim3/2_3}
\end{align}
The particular components are calculated in Eq.~\eqref{eq:bianchi_sp_id_eq04_3} and \eqref{eq:bianchi_sp_id_eq04_4},
\begin{align}
S_\beta = {T_{\sym{2}{\gamma\beta}}}^\gamma &= +\frac{1}{4} \cD^\sd{\beta} G_{\beta\sd{\beta}} - \cD_\beta R = +\frac{1}{2}{T_{cb}}^\alpha(\sig^{cb}\eps)_{\alpha\beta}\,, \label{eq:bianchi_sp_components_torsion_dim3/2_4}\\
S_\sd{\beta} = {T_{\sym{2}{\sd{\gamma}\sd{\beta}}}}^\sd{\gamma} &= -\frac{1}{4} \cD^\beta G_{\beta\sd{\beta}} + \cD_\sd{\beta} \Rb = +\frac{1}{2}T_{cb\sd{\alpha}}{(\sigb^{cb})^\sd{\alpha}}_\sd{\beta}\,, \label{eq:bianchi_sp_components_torsion_dim3/2_5}
\end{align}
and in Eq.~\eqref{eq:bianchi_sp_id_eq04_1} and \eqref{eq:bianchi_sp_id_eq04_2},
\begin{align}
T_{\sym{2}{\sd{\gamma}\sd{\beta}}\alpha} &= -\frac{1}{4}(\cD_\sd{\gamma} G_{\alpha\sd{\beta}} + \cD_\sd{\beta} G_{\alpha\sd{\gamma}})\,, \label{eq:bianchi_sp_components_torsion_dim3/2_6}\\
T_{\sym{2}{\gamma\beta}\sd{\alpha}} &= +\frac{1}{4}(\cD_\gamma G_{\beta\sd{\alpha}} + \cD_\beta G_{\gamma\sd{\alpha}})\,. \label{eq:bianchi_sp_components_torsion_dim3/2_7}
\end{align}
\end{itemize}

\subsubsection{Lorentz curvature and $\mathrm{U}(1)$ field strength components}
\label{sec:lorentz_curvature_u1_fieldstrength_components}
The Lorentz curvature tensor takes values in the Lie algebra of the Lorentz group. As stated in Eq.~\eqref{eq:bianchi_sp_lorentz_algebra_spinor_vector_identification}, the vector components are expressed in terms of the spinor components as follows:
\begin{align}
{R_{DCb}}^a \quad\leftrightarrow\quad R_{DC\s\beta\sd{\beta}\s\alpha\sd{\alpha}} = 2\epsilon_{\sd{\beta}\sd{\alpha}} R_{DC\sym{2}{\beta\alpha}} - 2\epsilon_{\beta\alpha} R_{DC\sym{2}{\sd{\beta}\sd{\alpha}}}\,. \label{eq:bianchi_sp_lorentz_algebra_spinor_vector_identification_1}
\end{align}
Since the Lorentz curvature and the $\mathrm{U}(1)$ field strength are both $2$-forms, the mass dimensions of their components reach from $1$ to $2$.
\begin{itemize}
\item \textbf{Dimension $1$:}\\
The components of the Lorentz curvature tensor with dimension $1$ are calculated in Eq.~\eqref{eq:bianchi_sp_id_eq07_2}, \eqref{eq:bianchi_sp_id_eq07_3}, \eqref{eq:bianchi_sp_id_eq07_5}, \eqref{eq:bianchi_sp_id_eq07_6}, \eqref{eq:bianchi_sp_id_eq02_3} and \eqref{eq:bianchi_sp_id_eq02_4}. Using vector indices they read
\begin{align}
R_{\delta\gamma ba} &= 8(\sig_{ba}\eps)_{\delta\gamma} \Rb\,, \label{eq:bianchi_sp_components_lorentz_curvature_dim1_1}\\
{R^{\sd{\delta}\sd{\gamma}}}_{ba} &= 8(\sigb_{ba}\eps)^{\sd{\delta}\sd{\gamma}} R\,, \label{eq:bianchi_sp_components_lorentz_curvature_dim1_2}\\
\tensor{R}{_\delta^{\sd{\gamma}}_{ba}} &= 2i G^c {(\sig^d\eps)_\delta}^\sd{\gamma} \eps_{dcba}\,, \label{eq:bianchi_sp_components_lorentz_curvature_dim1_3}
\end{align}
and in spinor notation they are written as
\begin{align}
R_{\delta\gamma\sym{2}{\beta\alpha}} &= 4(\eps_{\delta\beta}\eps_{\gamma\alpha} + \eps_{\delta\alpha}\eps_{\gamma\beta}) \Rb\,, \label{eq:bianchi_sp_components_lorentz_curvature_dim1_4}\\
R_{\delta\gamma\sym{2}{\sd{\beta}\sd{\alpha}}} &= 0\,, \label{eq:bianchi_sp_components_lorentz_curvature_dim1_5}\\
R_{\sd{\delta}\sd{\gamma}\sym{2}{\beta\alpha}} &= 0\,, \label{eq:bianchi_sp_components_lorentz_curvature_dim1_6}\\
R_{\sd{\delta}\sd{\gamma}\sym{2}{\sd{\beta}\sd{\alpha}}} &= 4(\eps_{\sd{\delta}\sd{\beta}}\eps_{\sd{\gamma}\sd{\alpha}} + \eps_{\sd{\delta}\sd{\alpha}}\eps_{\sd{\gamma}\sd{\beta}}) R\,, \label{eq:bianchi_sp_components_lorentz_curvature_dim1_7}\\
R_{\delta\sd{\gamma}\sym{2}{\beta\alpha}} &= -\eps_{\delta\beta}G_{\alpha\sd{\gamma}} - \eps_{\delta\alpha}G_{\beta\sd{\gamma}}\,, \label{eq:bianchi_sp_components_lorentz_curvature_dim1_8}\\
R_{\delta\sd{\gamma}\sym{2}{\sd{\beta}\sd{\alpha}}} &= -\eps_{\sd{\gamma}\sd{\beta}}G_{\delta\sd{\alpha}} - \eps_{\sd{\gamma}\sd{\alpha}}G_{\delta\sd{\beta}}\,. \label{eq:bianchi_sp_components_lorentz_curvature_dim1_9}
\end{align}
The components of the $\mathrm{U}(1)$ field strength are taken from Eq.~\eqref{eq:bianchi_sp_id_eq01_1}, \eqref{eq:bianchi_sp_id_eq01_2} and \eqref{eq:bianchi_sp_id_eq02_5}:
\begin{gather}
F_{\delta\gamma} = 0\,,\hspace{4cm} F^{\sd{\delta}\sd{\gamma}} = 0\,, \label{eq:bianchi_sp_components_u1_fieldstrength_dim1_1}\\
{F_\delta}^\sd{\gamma} = 3{(\sig^a\eps)_\delta}^\sd{\gamma} G_a\,. \label{eq:bianchi_sp_components_u1_fieldstrength_dim1_2}
\end{gather}
\item \textbf{Dimension $\frac{3}{2}$:}\\
According to Eq.~\eqref{eq:bianchi_sp_id_eq11_1} and \eqref{eq:bianchi_sp_id_eq11_2}, the components of the Lorentz curvature tensor with dimension $\frac{3}{2}$ are expressed in terms of the torsion tensor as:
\begin{align}
R_{\delta cba} &= i\sig_{c\delta\sd{\delta}} {T_{ba}}^\sd{\delta} - i\sig_{b\delta\sd{\delta}} {T_{ac}}^\sd{\delta} - i\sig_{a\delta\sd{\delta}} {T_{cb}}^\sd{\delta}\,, \label{eq:bianchi_sp_components_lorentz_curvature_dim3/2_1}\\
{R^\sd{\delta}}_{cba} &= i\sigb_c^{\sd{\delta\delta}} T_{ba\delta} - i\sigb_b^{\sd{\delta\delta}} T_{ac\delta} - i\sigb_a^{\sd{\delta\delta}} T_{cb\delta}\,, \label{eq:bianchi_sp_components_lorentz_curvature_dim3/2_2}
\end{align}
which takes the following form in spinor notation, as stated in Eqs.~\eqref{eq:bianchi_sp_sol_eq04_6}--\eqref{eq:bianchi_sp_sol_eq04_9}:
\begin{align}
R_{\delta\s\gamma\sd{\gamma}\s\sym{2}{\beta\alpha}} &= +i\sum_{\beta\alpha}\big(\eps_{\delta\gamma} T_{\sym{2}{\beta\alpha}\sd{\gamma}} + \eps_{\delta\alpha} T_{\sym{2}{\gamma\beta}\sd{\gamma}} - \eps_{\delta\beta}\eps_{\gamma\alpha}S_\sd{\gamma}\big)\,, \label{eq:bianchi_sp_components_lorentz_curvature_dim3/2_3}\\
R_{\delta\s\gamma\sd{\gamma}\s\sym{2}{\sd{\beta}\sd{\alpha}}} &= +4i\eps_{\delta\gamma} W_{\sym{3}{\sd{\gamma}\sd{\beta}\sd{\alpha}}} + i\sum_{\sd{\beta}\sd{\alpha}} \eps_{\sd{\gamma}\sd{\beta}} \Big(T_{\sym{2}{\delta\gamma}\sd{\alpha}} + \frac{1}{3}\eps_{\delta\gamma}S_\sd{\alpha}\Big)\,, \label{eq:bianchi_sp_components_lorentz_curvature_dim3/2_4}\\
R_{\sd{\delta}\s\gamma\sd{\gamma}\s\sym{2}{\beta\alpha}} &= -4i\eps_{\sd{\delta}\sd{\gamma}} W_{\sym{3}{\gamma\beta\alpha}} - i\sum_{\beta\alpha} \eps_{\gamma\beta} \Big(T_{\sym{2}{\sd{\delta}\sd{\gamma}}\alpha} + \frac{1}{3}\eps_{\sd{\delta}\sd{\gamma}}S_\alpha\Big)\,, \label{eq:bianchi_sp_components_lorentz_curvature_dim3/2_5}\\
R_{\sd{\delta}\s\gamma\sd{\gamma}\s\sym{2}{\sd{\beta}\sd{\alpha}}} &= -i\sum_{\sd{\beta}\sd{\alpha}} \big(\eps_{\sd{\delta}\sd{\alpha}} T_{\sym{2}{\sd{\gamma}\sd{\beta}}\gamma} + \eps_{\sd{\delta}\sd{\gamma}} T_{\sym{2}{\sd{\beta}\sd{\alpha}}\gamma} - \eps_{\sd{\delta}\sd{\beta}}\eps_{\sd{\gamma}\sd{\alpha}}S_\gamma\big)\,. \label{eq:bianchi_sp_components_lorentz_curvature_dim3/2_6}
\end{align}
Plugging in the identities from Eqs.~\eqref{eq:bianchi_sp_id_eq04_1}--\eqref{eq:bianchi_sp_id_eq04_4}, the components of the Lorentz curvature are finally written in terms of the superfields $R$, $\Rb$ and $G_a$:
\begin{align}
R_{\delta\s\gamma\sd{\gamma}\s\sym{2}{\beta\alpha}} &= +i\sum_{\beta\alpha} \Big(\frac{1}{2}\eps_{\delta\gamma}\cD_\beta G_{\alpha\sd{\gamma}} + \frac{1}{2}\eps_{\delta\beta} \cD_{\gamma} G_{\alpha\sd{\gamma}} - \eps_{\delta\beta}\eps_{\gamma\alpha} \cD_\sd{\gamma} \Rb\Big)\,, \label{eq:bianchi_sp_components_lorentz_curvature_dim3/2_7}\\
R_{\delta\s\gamma\sd{\gamma}\s\sym{2}{\sd{\beta}\sd{\alpha}}} &= +4i\eps_{\delta\gamma} W_{\sym{3}{\sd{\gamma}\sd{\beta}\sd{\alpha}}} + i\sum_{\sd{\beta}\sd{\alpha}} \eps_{\sd{\gamma}\sd{\alpha}} \Big(\frac{1}{3}\eps_{\delta\gamma}X_\sd{\beta} + \frac{1}{2}\cD_\delta G_{\gamma\sd{\beta}}\Big)\,, \label{eq:bianchi_sp_components_lorentz_curvature_dim3/2_8}\\
R_{\sd{\delta}\s\gamma\sd{\gamma}\s\sym{2}{\beta\alpha}} &= -4i\eps_{\sd{\delta}\sd{\gamma}} W_{\sym{3}{\gamma\beta\alpha}} + i\sum_{\beta\alpha} \eps_{\gamma\alpha} \Big(\frac{1}{3}\eps_{\sd{\delta}\sd{\gamma}}X_\beta + \frac{1}{2}\cD_\sd{\delta} G_{\beta\sd{\gamma}}\Big)\,, \label{eq:bianchi_sp_components_lorentz_curvature_dim3/2_9}\\
R_{\sd{\delta}\s\gamma\sd{\gamma}\s\sym{2}{\sd{\beta}\sd{\alpha}}} &= +i\sum_{\sd{\beta}\sd{\alpha}} \Big(\frac{1}{2}\eps_{\sd{\delta}\sd{\gamma}} \cD_\sd{\beta} G_{\gamma\sd{\alpha}} + \frac{1}{2}\eps_{\sd{\delta}\sd{\beta}} \cD_\sd{\gamma} G_{\gamma\sd{\alpha}} - \eps_{\sd{\delta}\sd{\beta}}\eps_{\sd{\gamma}\sd{\alpha}}\cD_\gamma R\Big)\,, \label{eq:bianchi_sp_components_lorentz_curvature_dim3/2_10}
\end{align}
with the definition
\begin{align}
X_\beta &= \cD_\beta R - \cD^\sd{\beta} G_{\beta\sd{\beta}}\,, \label{eq:bianchi_sp_components_lorentz_curvature_dim3/2_11}\\
X_\sd{\beta} &= \cD_\sd{\beta} \Rb - \cD^\beta G_{\beta\sd{\beta}}\,, \label{eq:bianchi_sp_components_lorentz_curvature_dim3/2_12}
\end{align}
where $(X_\beta)^\cc=X_\sd{\beta}$. The components of the $\mathrm{U}(1)$ field strength are calculated in Eq.~\eqref{eq:bianchi_sp_id_eq04_5} and \eqref{eq:bianchi_sp_id_eq04_6}. In vector and in spinor notation they read
\begin{align}
F_{\delta c} = \frac{3i}{2}\cD_\delta G_c + \frac{i}{2}\sig_{c\delta\sd{\delta}}X^\sd{\delta} \quad\leftrightarrow\quad F_{\delta\s\gamma\sd{\gamma}} = \frac{3i}{2}\cD_\delta G_{\gamma\sd{\gamma}} + i\eps_{\delta\gamma} X_\sd{\gamma}\,, \label{eq:bianchi_sp_components_u1_fieldstrength_dim3/2_1}\\
{F^\sd{\delta}}_c = \frac{3i}{2}\cD^\sd{\delta} G_c - \frac{i}{2}\sigb_c^{\sd{\delta}\delta}X_\delta \quad\leftrightarrow\quad F_{\sd{\delta}\s\gamma\sd{\gamma}} = \frac{3i}{2}\cD_\sd{\delta} G_{\gamma\sd{\gamma}} + i\eps_{\sd{\delta}\sd{\gamma}} X_\gamma\,. \label{eq:bianchi_sp_components_u1_fieldstrength_dim3/2_2}
\end{align}
\item \textbf{Dimension $2$:}\\
As in Eqs.~\eqref{eq:bianchi_sp_id_eq13_1}--\eqref{eq:bianchi_sp_id_eq13_3}, it is most convenient to write the components of the Lorentz curvature tensor with dimension $2$ in terms of spinor indices
\begin{align}
R_{\delta\sd{\delta}\s\gamma\sd{\gamma}\s\beta\sd{\beta}\s\alpha\sd{\alpha}} = 4\epsilon_{\sd{\delta}\sd{\gamma}}\big(\epsilon_{\sd{\beta}\sd{\alpha}} \chi_{\sym{2}{\delta\gamma}\sym{2}{\beta\alpha}} - \epsilon_{\beta\alpha} \psi_{\sym{2}{\delta\gamma}\sym{2}{\sd{\beta}\sd{\alpha}}}\big) + 4\epsilon_{\delta\gamma}\big(\epsilon_{\beta\alpha} \chi_{\sym{2}{\sd{\delta}\sd{\gamma}}\sym{2}{\sd{\beta}\sd{\alpha}}} - \epsilon_{\sd{\beta}\sd{\alpha}} \psi_{\sym{2}{\beta\alpha}\sym{2}{\sd{\delta}\sd{\gamma}}}\big)\,, \label{eq:bianchi_sp_components_lorentz_curvature_dim2_1}
\end{align}
with
\begin{align}
\chi_{\sym{2}{\delta\gamma}\sym{2}{\beta\alpha}} &= \chi_{\sym{4}{\delta\gamma\beta\alpha}} + (\epsilon_{\delta\beta}\epsilon_{\gamma\alpha} + \epsilon_{\delta\alpha}\epsilon_{\gamma\beta}) \chi\,,\label{eq:bianchi_sp_components_lorentz_curvature_dim2_2}\\
\chi_{\sym{2}{\sd{\delta}\sd{\gamma}}\sym{2}{\sd{\beta}\sd{\alpha}}} &= \chi_{\sym{4}{\sd{\delta}\sd{\gamma}\sd{\beta}\sd{\alpha}}} + (\epsilon_{\sd{\delta}\sd{\beta}}\epsilon_{\sd{\gamma}\sd{\alpha}} + \epsilon_{\sd{\delta}\sd{\alpha}}\epsilon_{\sd{\gamma}\sd{\beta}}) \chi\,. \label{eq:bianchi_sp_components_lorentz_curvature_dim2_3}
\end{align}
According to Eqs.~\eqref{eq:bianchi_sp_id_eq09_3}--\eqref{eq:bianchi_sp_id_eq09_6}, the dependence on the basic superfields is
\begin{align}
\psi_{\sym{2}{\delta\gamma}\sym{2}{\sd{\beta}\sd{\alpha}}} &= \frac{1}{8} \sum_{\delta\gamma} \sum_{\sd{\beta}\sd{\alpha}} \Big(G_{\delta\sd{\beta}} G_{\gamma\sd{\alpha}} -\frac{1}{2}[\cD_\delta,\cD_\sd{\beta}] G_{\gamma\sd{\alpha}} \Big)\,, \label{eq:bianchi_sp_components_lorentz_curvature_dim2_4}\\
\chi_{\sym{4}{\delta\gamma\beta\alpha}} &= \frac{1}{4}(\cD_\delta W_{\sym{3}{\gamma\beta\alpha}} + \cD_\gamma W_{\sym{3}{\beta\alpha\delta}} + \cD_\beta W_{\sym{3}{\alpha\delta\gamma}} + \cD_\alpha W_{\sym{3}{\delta\gamma\beta}})\,, \label{eq:bianchi_sp_components_lorentz_curvature_dim2_5}\\
\chi_{\sym{4}{\sd{\delta}\sd{\gamma}\sd{\beta}\sd{\alpha}}} &= \frac{1}{4}(\cD_\sd{\delta} W_{\sym{3}{\sd{\gamma}\sd{\beta}\sd{\alpha}}} + \cD_\sd{\gamma} W_{\sym{3}{\sd{\beta}\sd{\alpha}\sd{\delta}}} + \cD_\sd{\beta} W_{\sym{3}{\sd{\alpha}\sd{\delta}\sd{\gamma}}} + \cD_\sd{\alpha} W_{\sym{3}{\sd{\delta}\sd{\gamma}\sd{\beta}}})\,, \label{eq:bianchi_sp_components_lorentz_curvature_dim2_6}
\end{align}
and
\begin{align}
\chi &= -\frac{1}{12}(\cD^\alpha \cD_\alpha R + \cD_\sd{\alpha} \cD^\sd{\alpha} \Rb) + \frac{1}{48}[\cD^\alpha,\cD^\sd{\alpha}] G_{\alpha\sd{\alpha}} - \frac{1}{8} G^{\alpha\sd{\alpha}} G_{\alpha\sd{\alpha}} + 2R\Rb\,. \label{eq:bianchi_sp_components_lorentz_curvature_dim2_7}
\end{align}
Note that $\chi$ is real and it can also be written as
\begin{align}
\chi &= \frac{1}{24} {R_{ba}}^{ba}\,. \label{eq:bianchi_sp_components_lorentz_curvature_dim2_8}
\end{align}
In terms of components with definite symmetry properties, the components of the $\mathrm{U}(1)$ field strength are written in spinor notion as
\begin{align}
F_{dc} \quad\leftrightarrow\quad F_{\delta\sd{\delta}\s\gamma\sd{\gamma}} = 2\eps_{\sd{\delta}\sd{\gamma}} F_{\sym{2}{\delta\gamma}} - 2\eps_{\delta\gamma} F_{\sym{2}{\sd{\delta}\sd{\gamma}}}\,, \label{eq:bianchi_sp_components_u1_fieldstrength_dim2_1}
\end{align}
were the two symmetric tensors are calculated in Eq.~\eqref{eq:bianchi_sp_id_eq09_1} and \eqref{eq:bianchi_sp_id_eq09_2},
\begin{align}
F_{\sym{2}{\delta\gamma}} &= +\frac{1}{8} \sum_{\delta\gamma} (\cD_\delta \cD^\sd{\delta} G_{\gamma\sd{\delta}} + 3i{\cD_\delta}^\sd{\delta} G_{\gamma\sd{\delta}})\,, \label{eq:bianchi_sp_components_u1_fieldstrength_dim2_2}\\
F_{\sym{2}{\sd{\delta}\sd{\gamma}}} &= -\frac{1}{8} \sum_{\sd{\delta}\sd{\gamma}} (\cD_\sd{\delta} \cD^\delta G_{\delta\sd{\gamma}} + 3i{\cD^\delta}_\sd{\delta} G_{\delta\sd{\gamma}})\,. \label{eq:bianchi_sp_components_u1_fieldstrength_dim2_3}
\end{align}
\end{itemize}

\subsubsection{Derivative relations}
\label{sec:derivative_relations}
In this section the derivative relations of the component fields up to dimension $2$, which follow from the solution of the Bianchi identity, are listed. In addition, some reformulations of these identities are given, which are conveniently used in further calculation.
\\\\
Eq.~\eqref{eq:bianchi_sp_id_eq10_2} states the chirality conditions
\begin{align}
\cD^\sd{\beta} X_\alpha &= 0\,, & \cD_\beta X_\sd{\alpha} &= 0\,. \label{eq:bianchi_sp_x_chirality_condition}
\end{align}
Plugging in the definitions of $X_\alpha$ and $X^\sd{\alpha}$, the two relations
\begin{align}
\cD^\alpha X_\alpha &= \cD_\sd{\alpha}X^\sd{\alpha}\,, \label{eq:bianchi_sp_x_derivative_relation_1}\\
\cD^2 R - \cDb^2 \Rb &= 4i \cD_a G^a\,, \label{eq:bianchi_sp_x_derivative_relation_2}
\end{align}
with $\cD^2 = \cD^\alpha\cD_\alpha$ and $\cDb^2 = \cD_\sd{\alpha} \cD^\sd{\alpha}$, are equivalent and originate from Eq.~\eqref{eq:bianchi_sp_id_eq09_7}. Furthermore, the combination of Eq.~\eqref{eq:bianchi_sp_components_lorentz_curvature_dim2_7} and \eqref{eq:bianchi_sp_components_lorentz_curvature_dim2_8} leads to
\begin{align}
\cD^2 R + \cDb^2 \Rb &= -\frac{2}{3} {R_{ba}}^{ba} - \frac{2}{3} \cD^\alpha X_\alpha + 4G^a G_a + 32R \Rb\,, \label{eq:bianchi_sp_derivative_relation_id_1}
\end{align}
and the relations in Eq.~\eqref{eq:bianchi_sp_id_eq10_3} and \eqref{eq:bianchi_sp_id_eq10_4} are
\begin{align}
\cD^2 G_a &= +4i \cD_a \Rb\,, & \cDb^2 G_a &= -4i \cD_a R\,. \label{eq:bianchi_sp_derivative_relation_id_2}
\end{align}
Using the expressions of the different components, an explicit calculation delivers the relations
\begin{align}
\cD_\alpha R &= -\frac{1}{3}X_\alpha - \frac{4}{3} S_\alpha\,, \label{eq:bianchi_sp_derivative_relation_id_3}\\
\cD^\sd{\alpha} \Rb &= -\frac{1}{3}X^\sd{\alpha} + \frac{4}{3} S^\sd{\alpha}\,, \label{eq:bianchi_sp_derivative_relation_id_4}
\end{align}
and
\begin{align}
\cD_\beta G_{\alpha\sd{\alpha}} &= +2T_{\sym{2}{\beta\alpha}\sd{\alpha}} + \frac{2}{3}\eps_{\beta\alpha}S_\sd{\alpha} - \frac{2}{3}\eps_{\beta\alpha}X_\sd{\alpha}\,, \label{eq:bianchi_sp_derivative_relation_id_5}\\
\cD_\sd{\beta} G_{\alpha\sd{\alpha}} &= -2T_{\sym{2}{\sd{\beta}\sd{\alpha}}\alpha} - \frac{2}{3}\eps_{\sd{\beta}\sd{\alpha}}S_\alpha - \frac{2}{3}\eps_{\sd{\beta}\sd{\alpha}}X_\alpha\,. \label{eq:bianchi_sp_derivative_relation_id_6}
\end{align}
In addition, the commutator of mixed spinor derivatives applied on $G_{\alpha\sd{\alpha}}$ takes the form
\begin{align}
\begin{split}
\com{\cD_\beta}{\cD_\sd{\beta}}G_{\alpha\sd{\alpha}} =& -4\psi_{\sym{2}{\beta\alpha}\sym{2}{\sd{\beta}\sd{\alpha}}} + 2G_{\beta\sd{\beta}}G_{\alpha\sd{\alpha}} + 4(\eps_{\beta\alpha}F_{\sym{2}{\sd{\beta}\sd{\alpha}}} + \eps_{\sd{\beta}\sd{\alpha}}F_{\sym{2}{\beta\alpha}})\\
&+ 2i\eps_{\beta\alpha}{\cD^\varphi}_\sd{\beta} G_{\varphi\sd{\alpha}} - 2i\eps_{\sd{\beta}\sd{\alpha}}{\cD_\beta}^\sd{\varphi} G_{\alpha\sd{\varphi}}\\
&+ \eps_{\beta\alpha} \eps_{\sd{\beta}\sd{\alpha}}\Big(8R\Rb + 2G^c G_c - \frac{2}{3}\cD^\varphi X_\varphi - 4\chi\Big)\,.
\end{split} \label{eq:bianchi_sp_derivative_relation_id_7}
\end{align}
From Eq.~\eqref{eq:bianchi_sp_id_eq09_8} and \eqref{eq:bianchi_sp_id_eq09_9} follows
\begin{align}
\cD_\delta W_{\sym{3}{\gamma\beta\alpha}} &= \chi_{\sym{4}{\delta\gamma\beta\alpha}} + \frac{1}{4}\big(\eps_{\delta\gamma} \cD^\varphi W_{\sym{3}{\varphi\beta\alpha}} + \eps_{\delta\beta} \cD^\varphi W_{\sym{3}{\varphi\alpha\gamma}} + \eps_{\delta\alpha} \cD^\varphi W_{\sym{3}{\varphi\gamma\beta}}\big)\,, \label{eq:bianchi_sp_derivative_relation_id_8}\\
\cD_\sd{\delta} W_{\sym{3}{\sd{\gamma}\sd{\beta}\sd{\alpha}}} &= \chi_{\sym{4}{\sd{\delta}\sd{\gamma}\sd{\beta}\sd{\alpha}}} + \frac{1}{4}\big(\eps_{\sd{\delta}\sd{\gamma}} \cD^\sd{\varphi} W_{\sym{3}{\sd{\varphi}\sd{\beta}\sd{\alpha}}} + \eps_{\sd{\delta}\sd{\beta}} \cD^\sd{\varphi} W_{\sym{3}{\sd{\varphi}\sd{\alpha}\sd{\gamma}}} + \eps_{\sd{\delta}\sd{\alpha}} \cD^\sd{\varphi} W_{\sym{3}{\sd{\varphi}\sd{\gamma}\sd{\beta}}}\big)\,, \label{eq:bianchi_sp_derivative_relation_id_9}
\end{align}
and in combination with the identities in Eq.~\eqref{eq:bianchi_sp_id_eq09_1} and \eqref{eq:bianchi_sp_id_eq09_2} they imply
\begin{align}
\cD^\varphi W_{\sym{3}{\varphi\beta\alpha}} &= -\frac{1}{6} \sum_{\beta\alpha} (\cD_\beta \cD^\sd{\varphi} G_{\alpha\sd{\varphi}} + 3i{\cD_\beta}^\sd{\varphi} G_{\alpha\sd{\varphi}}) = -\frac{4}{3} F_{\sym{2}{\beta\alpha}}\,, \label{eq:bianchi_sp_derivative_relation_id_10}\\
\cD^\sd{\varphi} W_{\sym{3}{\sd{\varphi}\sd{\beta}\sd{\alpha}}} &= +\frac{1}{6} \sum_{\sd{\beta}\sd{\alpha}} (\cD_\sd{\beta} \cD^\varphi G_{\varphi\sd{\alpha}} + 3i{\cD^\varphi}_\sd{\beta} G_{\varphi\sd{\alpha}}) = -\frac{4}{3} F_{\sym{2}{\sd{\beta}\sd{\alpha}}}\,. \label{eq:bianchi_sp_derivative_relation_id_11}
\end{align}

\subsection{Integration by parts}
\label{sec:integration_by_parts}
Having solved the Bianchi identity by taking into account the torsion constraints, it is convenient at this point to discuss integration by parts in $\U(1)$ superspace, where the explicit expressions of the torsion tensor components are used. Consider an even supervector field $\chi=\chi^M \del_M = \chi^A E_A$. Since in the following equation the integrand is a total derivative, the integral is equal to a surface term and therefore vanishes:
\begin{align}
\int_*(-1)^M\del_M(E\,\chi^M) = 0\,, \label{eq:kahler_superspace_integration_parts_identity_1}
\end{align}
where $\int_*$ indicates integration over all coordinates $z^M$. The integrand can be rewritten in the form
\begin{align}
\begin{split}
(-1)^M\del_M(E\,\chi^M) &= (-1)^M \del_M(E\,\chi^A{E_A}^M) \\
&= (-1)^M\big(\del_M E\,\chi^M + E\,\del_M \chi^A {E_A}^M + (-1)^{MA}E\,\chi^A \del_M {E_A}^M\big) \\
&= (-1)^M E\big(\del_M \chi^A + \chi^N(\del_N{E_M}^A - (-1)^{MN}\del_M{E_N}^A)\big){E_A}^M \\
&= (-1)^A E\,D_A \chi^A + (-1)^A E\,\chi^B ({T_{BA}}^A - {\Omega_{BA}}^A) + (-1)^{A(B+1)} E\,\chi^B {\Omega_{AB}}^A \\
&\quad - (-1)^A w(E^A)\delta^A_A E\,\chi^B A_B + E\,w(E^A)\chi^A A_A \\
&= (-1)^A E\,\cD_A \chi^A + (-1)^A E\,\chi^B {T_{BA}}^A + (-1)^A E\big(w(E^A)-w(\chi^A)\big)A_A \chi^A\,,
\end{split} \label{eq:kahler_superspace_integration_parts_identity_2}
\end{align}
where the identities ${\Omega_{BA}}^A = 0$ and $(-1)^Aw(E^A)\delta_A^A=0$ are used. The only non-vanishing components of ${T_{BA}}^A$ are
\begin{align}
{T_{b\alpha}}^\alpha &= +iG_b\,, & \tensor{T}{_b^{\sd{\alpha}}_{\sd{\alpha}}} &= -iG_b\,,
\end{align}
thus
\begin{align}
(-1)^A \chi^B {T_{BA}}^A = -\chi^b({T_{b\alpha}}^\alpha + \tensor{T}{_b^{\sd{\alpha}}_{\sd{\alpha}}}) = -\chi^b(iG_b-iG_b) = 0\,. 
\end{align}
Moreover, if the $\U(1)$ weight of $\chi$ vanishes, i.e.\ $w(\chi^A)=w(E^A)$, Eq.~\eqref{eq:kahler_superspace_integration_parts_identity_2} takes the form
\begin{align}
(-1)^M\del_M(E\,\chi^M) &= (-1)^A E\,\cD_A \chi^A\,. \label{eq:kahler_superspace_integration_parts_identity_3}
\end{align}
According to Eq.~\eqref{eq:kahler_superspace_integration_parts_identity_1}, the following integral is just a surface term, which is equal to zero
\begin{align}
\int_* (-1)^A E\,\cD_A \chi^A = 0\,. \label{eq:kahler_superspace_integration_parts_identity_4}
\end{align}
In particular, for each component $A=(a,\alpha,\sd{\alpha})$ the integral vanishes separately:
\begin{align}
\int_* E\,\cD_a \chi^a &= 0\,, & \int_* E\,\cD_\alpha \chi^\alpha &= 0\,, & \int_* E\,\cD^\sd{\alpha} \chi_\sd{\alpha} &= 0\,. \label{eq:kahler_superspace_integration_parts_identity_5}
\end{align}
The identities in Eq.~\eqref{eq:kahler_superspace_integration_parts_identity_4} and \eqref{eq:kahler_superspace_integration_parts_identity_5} represent integration by parts in $\U(1)$ superspace for a supervector field with vanishing chiral weight, and they are frequently used in the derivation of the superfield equations of motion in Section~\ref{sec:equations_of_motion}.

\section{Matter superfields and Yang-Mills gauge group}
\label{sec:matter_and_yang_mills}
Matter in ordinary spacetime is described by the sigma model, that is a diffeomorphism $\Phi$ which goes from spacetime to a target manifold. Scalar fields $\varphi^k(x)$, with spacetime coordinates $x^m$, are then given by the pullback of the coordinate functions $z^k$ on the target manifold via $\Phi$, i.e.\ $\varphi^k=z^k\circ\Phi$ (see e.g.~\cite{Percacci:1998ag,Abdalla:1985nm,Bagger:1985pw}). It is often convenient to view the fields $\varphi^k$ simply as coordinates of the target manifold. The kinetic terms of the scalar fields, which must be positive definite, are formed by (the pullback of) the metric on the target manifold; thus the target manifold is Riemannian. The sigma model is called linear, if the Riemannian metric is constant, and it is called non-linear, if the metric is coordinate dependent, i.e.\ a function of the scalar fields. For supersymmetric theories, considered at the component field level, the target manifold is complex; thus the complex coordinates are represented by the complex scalar fields $\varphi^k(x)$ and $\varphib^\kb(x)$ and the metric $g_{k\kb}$ is Hermitian. When determining the supersymmetric Lagrangian of the component fields (cf.\ Section~\ref{sec:component_fields}), it turns out that for $N=1$ supersymmetry in $d=4$ dimensions the target manifold is K\"ahler,\footnote{Other structures on the target manifold arise in different dimensions and for a different number of supersymmetry generators (see e.g.~\cite{deWit:1995tf,Lindstrom:2012ci}).} hence the Hermitian metric is locally specified by the K\"ahler potential $K(\varphi^k,\varphib^\kb)$ (see e.g.~\cite{Moroianu:2007}).
\\\\
The superpotential $W(\varphi^k)$ is used, among others, to specify the scalar potential of the supersymmetric theory. In particular, the superpotential is a holomorphic section of a holomorphic line bundle over the target manifold, which carries locally the Hermitian metric $e^K$, where both quantities are locally defined with respect to some holomorphic section $\sig$ which forms a basis of the line bundle. Thus, the target manifold is a K\"ahler-Hodge manifold, which is also referred to as K\"ahler manifold of restricted type (see e.g.\ \cite{Freedman:2012zz,Fre:1995dw}). Since the line bundle is $1$-dimensional, the norm of the superpotential is just given by
\begin{align}
||W(\varphi^k)||^2 &= e^{K(\varphi^k,\varphib^\kb)} W(\varphi^k)\Wb(\varphib^\kb)\,. \label{eq:yang_mills_introduction_superpotential_norm}
\end{align}
Under a K\"ahler transformation, parametrized by a holomorphic function $F(\varphi^k)$, the K\"ahler potential transforms as
\begin{align}
K(\varphi^k,\varphib^\kb) &\mapsto K(\varphi^k,\varphib^\kb) + F(\varphi^k) + \Fb(\varphib^\kb)\,, \label{eq:yang_mills_introduction_kahlerpotential_transformation}
\end{align}
which does not affect the K\"ahler metric. Since the metric on the line bundle is invariant under K\"ahler transformations as well, Eq.~\eqref{eq:yang_mills_introduction_kahlerpotential_transformation} implies the basis change $\sigma\mapsto\sigma e^F$. Hence, the transfomations of $W(\varphi^k)$ and its conjugate $\Wb(\varphib^\kb)$ are given by:
\begin{align}
W(\varphi^k) &\mapsto W(\varphi^k) e^{-F(\varphi^k)}\,, & \Wb(\varphib^\kb) &\mapsto \Wb(\varphib^\kb) e^{-\Fb(\varphib^\kb)}\,. \label{eq:yang_mills_introduction_superpotential_transformation}
\end{align}
There is a unique connection, namely the Chern connection, on the line bundle associated with the Hermitian metric $e^K$ and the holomorphic structure $\delb$. The corresponding connection $1$-form with respect to the section $\sigma$ and its conjugate are given by $\del K$ and $\delb K$, where $\del\equiv\exd\varphi^k\frac{\del}{\del\varphi^k}$ and $\delb\equiv\exd\varphib^\kb\frac{\del}{\del\varphib^\kb}$. Under K\"ahler transformations, the two $1$-forms transform as
\begin{align}
\del K &\mapsto \del K + \del F\,, & \delb K &\mapsto \delb K + \delb\Fb\,.
\end{align}
Moreover, the covariant derivatives of $W$ and $\Wb$ with respect to K\"ahler transformations are thus given by
\begin{align}
D W &= \exd W + W\del K\,, & D\Wb &= \exd \Wb + \Wb\delb K\,,
\end{align}
which, in terms of components, take the form
\begin{align}
D_k W &= W_k + W K_k\,, & D_\kb\Wb &= \Wb_\kb + \Wb K_\kb\,.
\end{align}
The line bundle is mapped to a $\U(1)$ bundle by the multiplication with $e^{K/2}$, and the corresponding $\U(1)$ connection $\At$ has the form
\begin{align}
\At &= \frac{1}{4}(\del K - \delb K)\,. \label{eq:yang_mills_introduction_u1_connection}
\end{align}
In particular, under K\"ahler transformations $\At$, $e^{K/2}W$ and $e^{K/2}\Wb$ transform as
\begin{gather}
\At \mapsto \At + \frac{i}{2}\exd\im{F} \,, \\
e^{K/2}W \mapsto e^{-i\im{F}}e^{K/2}W\,, \hspace{2cm} e^{K/2}\Wb \mapsto e^{+i\im{F}}e^{K/2}\Wb\,,
\end{gather}
showing that the corresponding $\U(1)$ transformation is given by $-\frac{i}{2}\im{F}$, and that $e^{K/2}W$ and $e^{K/2}\Wb$ have weights $+2$ and $-2$, respectively. This consideration is applied in the construction of the Lagrangian at the superfield level in Section~\ref{sec:action_superfield_level}.
\\\\
The fermionic superpartners of the scalar fields are described by the section $\chi$ of a bundle over spacetime, which is the tensor product of a spinor bundle and the pullback via $\Phi$ of the holomorphic tangent bundle of the target manifold.\footnote{The Grassmann property of the fermions can be accounted for by considering the exterior algebra of the tensor product bundle.} In particular, $\chi_\alpha$ is a Weyl spinor with spinor index $\alpha$, and with respect to the canonical basis $\del_k\equiv\frac{\del}{\del\varphi^k}$ of the holomorphic tangent bundle it is written as $\chi_\alpha={\chi^k}_\alpha\del_k$. The field ${\chi^k}_\alpha$ is then the fermionic superpartner of the scalar field $\varphi^k$. Similarly, the components ${\chib^\kb}_\sd{\alpha}$ of the conjugated section $\chib_\sd{\alpha}$ with respect to the basis $\del_\kb\equiv\frac{\del}{\del\varphib^\kb}$ are the superpartners of $\varphib^\kb$. Moreover, the tangent bundle of the target manifold is equipped with the Levi-Civita connection of the K\"ahler metric $g_{k\kb}$, which is used below to define covariant derivatives of the fermions.
\\\\
A model is invariant under a compact Lie group $G$, if the group elements act as isometries on the target manifold. The action is parametrized by the Killing vector field $-\lac{r}(\cK_\g{r}+\cKb_\g{r})$ concerning the K\"ahler metric, where $\lac{r}(x)$ are real functions on spacetime, and $\cK_\g{r}=\cK_\g{r}^k(\varphi)\del_k$ and $\cKb_\g{r}=\cKb_\g{r}^\kb(\varphib)\del_\kb$ are holomorphic and anti-holomorphic vector fields, respectively, which obey the commutation relations
\begin{align}
\com{\cK_\g{p}}{\cK_\g{q}} &= \sco{p}{q}{r}\cK_\g{r}\,, & \com{\cKb_\g{p}}{\cKb_\g{q}} &= \sco{p}{q}{r}\cKb_\g{r}\,, & \com{\cK_\g{p}}{\cKb_\g{q}} &= 0\,.
\end{align}
The factors $\sco{p}{q}{r}$ are the real structure constants with respect to a set of Hermitian generators $\gen{r}$ of $G$. For infinitesimal $\lac{r}$, the changes of the scalars and the fermions under the group action read \footnote{The infinitesimal change of a tensor $T$ on the target manifold under the induced diffeomorphisms of the Killing vector fields is calculated by using the Lie derivative, namely $\delta T = -\lac{r}(\lied_{\cK_\g{r}}T+\lied_{\cKb_\g{r}}T)$. On the other hand, $\chi$ and $\chib$ are sections of a bundle over spacetime, thus $\delta(\chi^k\del_k)=0$, $\delta(\chib^\kb\del_\kb)=0$. These transformation rules imply Eq.~\eqref{eq:yang_mills_introduction_delta_phi} and Eq.~\eqref{eq:yang_mills_introduction_delta_chi}, which basically represent infinitesimal coordinate transformations.}
\begin{align}
\delta\varphi^k &= -\lac{r}\cK_\g{r}^k\,, & \delta\varphib^\kb &= -\lac{r}\cKb_\g{r}^\kb\,, \label{eq:yang_mills_introduction_delta_phi}\\
\delta\chi^k &= -\lac{r}(\del_j\cK_\g{r}^k)\chi^j\,, & \delta\chib^\kb &= -\lac{r}(\del_\jb\cKb_\g{r}^\kb)\chib^\jb\,. \label{eq:yang_mills_introduction_delta_chi}
\end{align}
Since the parameter $\lac{r}$ is spacetime-dependent, these transformations are called gauged isometries. In order to formulate covariant derivatives a connection $1$-form $\ca=\ca^\g{r}\gen{r}=\exd x^m\ca_m^\g{r}\gen{r}$ with values in the Lie algebra $\mathfrak{g}$ of $G$ is introduced. Under the infinitesimal group action $\ca^\g{r}$ transforms as
\begin{align}
\delta\ca^\g{r} &= \lac{p}\ca^\g{q}\sco{p}{q}{r} - \exd\lac{r}\,.
\end{align}
Thus, the covariant derivatives of the scalar and the fermion fields have the form \footnote{Note, compared to Eq.~\eqref{eq:yang_mills_introduction_cd_chi} the covariant derivatives in Eq.~\eqref{eq:component_fields_definition_matter_cd_fermion_1} and \eqref{eq:component_fields_definition_matter_cd_fermion_2} have an additional term which contains the $\U(1)$ connection $A_m$.}
\begin{align}
\begin{split}
\cD_m \varphi^k &= \del_m \varphi^k - {\ca_m}^\g{r}\cK_\g{r}^k\,, \\
\cD_m \varphib^\kb &= \del_m \varphib^\kb - {\ca_m}^\g{r}\cKb_\g{r}^\kb\,,
\end{split} \label{eq:yang_mills_introduction_cd_phi}
\end{align}
\begin{align}
\begin{split}
\cDt_m{\chi^k}_\alpha &= \del_m{\chi^k}_\alpha - {\omega_{m\alpha}}^\beta{\chi^k}_\beta - {\ca_m}^\g{r}(\del_j\cK_\g{r}^k){\chi^j}_\alpha + {\Gamma^k}_{ij}{\chi^i}_\alpha\cD_m \varphi^j\,, \\
\cDt_m\chib^{\kb\sd{\alpha}} &= \del_m\chib^{\kb\sd{\alpha}} - \tensor{\omega}{_m^{\sd{\alpha}}_{\sd{\beta}}}\chib^{\kb\sd{\beta}} - {\ca_m}^\g{r}(\del_\jb\cKb_\g{r}^\kb)\chi^{\jb\sd{\alpha}} + {\Gamma^\kb}_{\ib\jb}\chib^{\ib\sd{\alpha}}\cD_m \varphib^\jb\,,
\end{split} \label{eq:yang_mills_introduction_cd_chi}
\end{align}
where $\omega$ is the spin connection associated with local Lorentz transformations in spacetime. Furthermore, ${\Gamma^k}_{ij}$ and ${\Gamma^\kb}_{\ib\jb}$ are the non-vanishing Christoffel symbols of the Levi-Civita connection of $g_{k\kb}$. The derivatives in Eq.~\eqref{eq:yang_mills_introduction_cd_phi} and \eqref{eq:yang_mills_introduction_cd_chi} transform covariantly under local transformations of the group $G$ and under local Lorentz transformations. Moreover, the covariant derivatives of the fermions are covariant with respect to ungauged isometries of the K\"ahler metric, which basically represent coordinate transformations on the target manifold. The kinetic terms of the scalar fields and their fermionic superpartners are then written as
\begin{align}
\cL_\text{kin} &= - g_{k\kb}\,g^{mn}\cD_m\varphi^k\cD_n\varphib^\kb - \frac{i}{2}g_{k\kb}(\chi^k\sig^m\cDt_m\chib^\kb + \chib^\kb\sigb^m\cDt_m\chi^k)\,, \label{eq:yang_mills_introduction_kinetic_terms}
\end{align}
where $g^{mn}$ is the inverse spacetime metric.
\\\\
Ordinary Yang-Mills transformations are described, if $G$ acts via a linear representation on the scalar fields, i.e.\ the Killing vector fields have the following form: $\cK_\g{r}=+i(\gen{r}\varphi)^k\del_k$ and $\cKb_\g{r}=-i(\varphib\gen{r})^\kb\del_\kb$, where the commutation relations of the Hermitian generators are $\com{\gen{p}}{\gen{q}}=i\sco{p}{q}{r}\gen{r}$. In this case, Eq.~\eqref{eq:yang_mills_introduction_delta_phi} and \eqref{eq:yang_mills_introduction_delta_chi} read
\begin{align}
\delta\varphi^k &= -i\lac{r}(\gen{r}\varphi)^k\,, & \delta\varphib^\kb &= +i\lac{r}(\varphib\gen{r})^\kb\,, \label{eq:yang_mills_introduction_delta_phi_ym}\\
\delta\chi^k &= -i\lac{r}(\gen{r}\chi)^k\,, & \delta\chib^\kb &= +i\lac{r}(\chib\gen{r})^\kb\,. \label{eq:yang_mills_introduction_delta_chi_ym}
\end{align}
The sigma model in superspace is formulated in a straightforward way from the above discussion by promoting the scalar fields $\varphi^k,\varphib^\kb$ to $c$-type matter superfields $\phi^k,\phib^\kb$ and the connection $i\exd x^m{\ca_m}^\g{r}$ to a $c$-type $1$-superform ${\exd z^M\cA_M}^\g{r}$ (where the factor $i$ is convention). The Killing vector fields $\cK_\g{r},\cKb_\g{r}$ are now even supervector fields, and the metric $g_{k\kb}$, the K\"ahler potential $K$ and the superpotential $W$ are even functions of the superfields $\phi^k,\phib^\kb$. The matter superfields do not transform under the structure group of superspace, thus in $\U(1)$ superspace they have vanishing $\U(1)$ weights: $w(\phi^k)=0$, $w(\phib^\kb)=0$. In addition, the chirality conditions $\cD^\sd{\alpha}\phi^k=0$ and $\cD_\alpha\phib^k=0$ apply, where the derivatives are covariant with respect to the gauged isometries of the sigma model. The matter superfields contain the scalar and the fermion fields as component fields. The kinetic terms in Eq.~\eqref{eq:yang_mills_introduction_kinetic_terms} are part of the general supersymmetric Lagrangian at the component field level given in Section~\ref{sec:component_fields_actions_summary}, which is derived with superfield techniques.
\\\\\
In the following sections, gauged isometries at the superfield level are discussed and the corresponding Bianchi identity is solved in the framework of $\U(1)$ superspace. The considerations are restricted to Yang-Mills transformations.

\subsection{Yang-Mills connection and field strength}
In this section gauged isometries of the sigma model in superspace are discussed, where only Yang-Mills transformations are considered, i.e.\ the Lie group $G$ with Lie algebra $\mathfrak{g}$ acts via a linear representation on the matter superfields $\phi^k$ and $\phib^k$. Furthermore it is assumed that $G$ is compact. Generic elements $\lae\in\mathfrak{g}$ and $\lge\in G$ have the form
\begin{align}
\lae &= i \alpha^\g{r} \gen{r}\,, \\
\lge &= \exp(i \lac{r} \gen{r})\,,
\end{align}
where the coefficients $\lac{r}$ are real, even superfields and the $\gen{r}$ form a set of generators of the Lie algebra ($i\gen{r}\in\mathfrak{g}$). It is assumed that the generators are Hermitian ($\gen{r}^\ct = \gen{r}$). Thus, $\lge^\ct = \lge^{-1}$ and the structure constants $\sco{p}{q}{r}$, which are implicitly defined by 
\begin{align}
\com{\gen{p}}{\gen{q}} = i\sco{p}{q}{r} \gen{r}\,,
\end{align}
are real.
\begin{itemize}
\item\textbf{Yang-Mills connection:}\\
The Yang-Mills connection $\cA$ of the gauge group is an even $\mathfrak{g}$-valued $1$-superform:
\begin{align}
\cA = \cA^\g{r} \gen{r} = E^A {\cA_A}^\g{r} \gen{r}\,.
\end{align}
From this definition follows that the $1$-superform $\cA^\g{r}$ is imaginary:
\begin{align}
(\cA^\g{r})^\cc &= -\cA^\g{r}\,, & ({\cA_A}^\g{r})^\cc &= -{\cA_\sd{A}}^\g{r}\,.
\end{align}
Under transformations $\lge$ and infinitesimal transformations $\lge\approx\mathbb{1}+\lae$ of the Yang-Mills gauge group the connection transforms as
\begin{align}
\cA &\mapsto \lge^{-1}\cA\lge - \lge^{-1}\exd\lge\,, \label{eq:yang_mills_connection_gauge_transformation}\\
\cA &\mapsto \cA + \com{\cA}{\lae} - \exd\lae\,, \label{eq:yang_mills_connection_gauge_transformation_infinitesimal}
\end{align}
where in terms of components Eq.~\eqref{eq:yang_mills_connection_gauge_transformation_infinitesimal} is written as
\begin{align}
\cA^\g{r} &\mapsto \cA^\g{r} + \alpha^\g{p} \cA^\g{q} \sco{p}{q}{r} -i\exd \lac{r}\,. \label{eq:yang_mills_connection_gauge_transformation_infinitesimal_component}
\end{align}
Note that $\cA$ does not transform under the structure group $\SO(1,3)\times\U(1)$.
\item\textbf{Yang-Mills field strength:}\\
The curvature tensor $\cF$, also called field strength, with respect to the connection $\cA$ is defined by
\begin{align}
\cF &= \exd\cA + \cA\wedge\cA\,, \label{eq:yang_mills_curvature_definition}
\end{align}
which, with respect to the generators, is written as
\begin{align}
\cF = \cF^\g{r}\gen{r} \quad\text{with}\quad \cF^\g{r} &= \exd\cA^\g{r} + \frac{i}{2} \cA^\g{p}\wedge\cA^\g{q} \sco{p}{q}{r}\,. \label{eq:yang_mills_curvature_components_lie_algebra}
\end{align}
On the other hand, the field strength is a $2$-superform
\begin{align}
\cF &= \frac{1}{2}E^A\wedge E^B \cF_{BA}\,, \label{eq:yang_mills_curvature_form}
\end{align}
with
\begin{align}
\label{eq:yang_mills_curvature_components_form}
\cF_{BA} &= \cD_B \cA_A - (-1)^{BA} \cD_A \cA_B + {T_{BA}}^C \cA_C + \mcom{\cA_B}{\cA_A}\,,
\end{align}
where $\cA_A$ is considered as the component of a $1$-superform in the covariant derivative, namely
\begin{align}
\cD_B \cA_A &= D_B \cA_A - {\Omega_{BA}}^C\cA_C - w(E^A)A_B\cA_A\,,
\end{align}
using the notation $\cD=E^B\cD_B$ and $\exd=E^B D_B$. According to the definition in Eq.~\eqref{eq:yang_mills_curvature_components_lie_algebra}, $\cF^\g{r}$ is imaginary:
\begin{align}
(\cF^\g{r})^\cc &= -\cF^\g{r}\,, & ({\cF_{BA}}^\g{r})^\cc &= -(-1)^{BA} {\cF_{\sd{B}\sd{A}}}^\g{r}\,. \label{eq:yang_mills_curvature_conjugation}
\end{align}
The transformations in Eq.~\eqref{eq:yang_mills_connection_gauge_transformation} and \eqref{eq:yang_mills_connection_gauge_transformation_infinitesimal} imply that the field strength transforms in the adjoint representation of the Yang-Mills group, namely
\begin{align}
\begin{split}
\cF &= \exd\cA + \cA\wedge\cA\ \\
&\mapsto \exd(\lge^{-1}\cA\lge - \lge^{-1}\exd\lge) + (\lge^{-1}\cA\lge - \lge^{-1}\exd\lge)\wedge(\lge^{-1}\cA\lge - \lge^{-1}\exd\lge) \\
&= \lge^{-1}\cA\wedge\exd\lge + \lge^{-1}\exd\cA\,\lge - \exd\lge^{-1}\wedge\cA\lge - \exd\lge^{-1}\wedge\exd\lge \\
&\quad + \lge^{-1}\cA\wedge\cA\lge - \lge^{-1}\cA\wedge\exd\lge - \lge^{-1}\exd\lge\wedge\lge^{-1}\exd\cA\lge + \lge^{-1}\exd\lge\wedge\lge^{-1}\exd\lge \\
&= \lge^{-1}\exd\cA\,\lge + \lge^{-1}\cA\wedge\cA\lge \\
&= \lge^{-1}\cF\lge\,,
\end{split} \label{eq:yang_mills_curvature_gauge_transformation}
\end{align}
where $\exd\lge^{-1}=-\lge^{-1}\exd\lge\,\lge^{-1}$ is used. In case of infinitesimal gauge transformations Eq.~\eqref{eq:yang_mills_curvature_gauge_transformation} takes the form
\begin{align}
\cF &\mapsto \cF + \com{\cF}{\lae}\,, \label{eq:yang_mills_curvature_gauge_transformation_infinitesimal}
\end{align}
which in terms of components is written as
\begin{align}
\cF^\g{r} &\mapsto \cF^\g{r} + \lac{p} \cF^\g{q} \sco{p}{q}{r}\,. \label{eq:yang_mills_curvature_gauge_transformation_infinitesimal_component}
\end{align}
\end{itemize}

\subsubsection{Bianchi Identity}
The Bianchi identity of the Yang-Mills connection states that the covariant derivative of the field strength vanishes:
\begin{align}
\begin{split}
\cD\cF &= \exd\cF + \com{\cF}{\cA} \\
&= \exd\exd\cA + \exd(\cA\wedge\cA) + \com{\exd\cA}{\cA} + \com{\cA\wedge\cA}{\cA} \\
&= \cA\wedge\exd\cA - \exd\cA\wedge\cA + \com{\exd\cA}{\cA} \\
&= 0\,,
\end{split} \label{eq:yang_mills_bianchi_identity}
\end{align}
where $\com{\cA\wedge\cA}{\cA} = 0$ is used.
In terms of the Lie algebra components, Eq.~\eqref{eq:yang_mills_bianchi_identity} is written as
\begin{align}
\cD\cF^\g{r} &= \exd\cF^\g{r} + i \cF^\g{p}\wedge\cA^\g{q} \sco{p}{q}{r} = 0\,.
\end{align}
Furthermore, since $\cF$ is a $2$-superform, the covariant derivative has the form
\begin{align}
\begin{split}
\cD\cF &= \frac{1}{2}E^A\wedge E^B\wedge \cD\cF_{BA} + \frac{1}{2}E^A\wedge T^B\wedge \cF_{BA} - \frac{1}{2}T^A\wedge E^B\wedge \cF_{BA} \\
&= \frac{1}{2} E^A\wedge E^B\wedge E^C (\cD_C\cF_{BA} + {T_{CB}}^D\cF_{DA})\,.
\end{split}
\end{align}
Thus, in terms of the superspace components the Bianchi-identity in Eq.~\eqref{eq:yang_mills_bianchi_identity} takes the form
\begin{align}
\label{eq:yang_mills_bianchi_identity_components}
\oint_{CBA} \big( \cD_C \cF_{BA} + {T_{CB}}^D \cF_{DA}  \big) = 0\,,
\end{align}
with $\oint_{CBA} CBA = CBA + (-1)^{A(C+B)} ACB + (-1)^{(B+A)C} BAC$.

\subsubsection{Covariant derivative of matter superfields}
For the following discussion, it is convenient to write the matter superfields as a column and a row vector $\phi$ and $\phib$, respectively, neglecting the field index, where $\phib$ is the conjugate of $\phi$. The action of a gauge transformation $\lge$ on $\phi$ and $\phib$ is then given by
\begin{align}
\phi &\mapsto \lge^{-1}\phi\,, & \phib &\mapsto \phib\lge\,, \label{eq:yang_mills_field_transformation}
\end{align}
such that the term $\phib\phi$ is a gauge singlet. If the gauge transformation $\lge\approx\mathbb{1}+\lae$ is infinitesimal, Eq.~\eqref{eq:yang_mills_field_transformation} reads
\begin{align}
\phi &\mapsto \phi - \lae\phi\,, & \phib &\mapsto \phib + \phib\lae\,. \label{eq:yang_mills_field_transformation_infinitesimal}
\end{align}
In terms of components, Eq.~\eqref{eq:yang_mills_field_transformation_infinitesimal} is written as
\begin{align}
\phi^k &\mapsto \phi^k - i\lac{r}(\gen{r}\phi)^k\,, & \phib^\kb &\mapsto \phib^\kb + i\lac{r}(\phib\gen{r})^\kb\,. \label{eq:yang_mills_field_transformation_infinitesimal_components}
\end{align}
The covariant derivatives of $\phi$ and $\phib$ are thus given by
\begin{align}
\cD\phi &= d\phi - \cA\phi\,, \label{eq:yang_mills_phi_covariant_derivative}\\
\cD\phib &= d\phib + \phib\cA\,, \label{eq:yang_mills_conj_phi_covariant_derivative}
\end{align}
which, in terms of components, reads
\begin{align}
\cD_A\phi^k &= D_A\phi^k - {\cA_A}^\g{r}(\gen{r}\phi)^k\,, \label{eq:yang_mills_phi_covariant_derivative_components}\\
\cD_A\phib^\kb &= D_A\phib^\kb + {\cA_A}^\g{r}(\phib\gen{r})^\kb\,, \label{eq:yang_mills_conj_phi_covariant_derivative_components}
\end{align}
with the notations $\cD=E^A\cD_A$ and $\exd=E^A D_A$. Note, if $\phi$ and $\phib$ were $p$-superforms, the covariant derivatives would have the form $\cD\phi = d\phi - \phi\wedge\cA^\tr = d\phi -(-1)^p \cA\wedge\phi$ and $\cD\phib = d\phib + \phib\wedge\cA$. By using Eq.~\eqref{eq:yang_mills_connection_gauge_transformation}, it is straightforward to check that $\cD\phi$ and $\cD\phib$ from Eq.~\eqref{eq:yang_mills_phi_covariant_derivative} and \eqref{eq:yang_mills_conj_phi_covariant_derivative} transform the same way under gauge transformations as $\phi$ and $\phib$, respectively:
\begin{align}
\begin{split}
\cD\phi &= \exd\phi - \cA\phi \\
&\mapsto \exd(\lge^{-1}\phi) - (\lge^{-1}\cA\lge - \lge^{-1}\exd\lge)(\lge^{-1}\phi) \\
&= \lge^{-1}\exd\phi + \exd\lge^{-1}\phi - \lge^{-1}\cA\phi + \lge^{-1}\exd\lge\,\lge^{-1}\phi \\
&= \lge^{-1}\exd\phi - \lge^{-1}\cA\phi \\
&= \lge^{-1}(\cD\phi)\,, \label{eq:yang_mills_gauge_derivative_phi}
\end{split}
\end{align}
and, with a similar calculation,
\begin{align}
\cD\phib \mapsto (\cD\phib)\lge\,. \label{eq:yang_mills_gauge_derivative_phibar}
\end{align}
A second covariant derivative of $\phi$ and $\phib$ is the same as acting with the field strength $\cF$ on $\phi$ and $\phib$, namely:
\begin{align}
\cD\cD \phi &= \exd\exd\phi - \exd(\cA\phi) + \cA\wedge(\exd\phi) - \cA\wedge\cA\phi = -\cF\phi\,, \\
\cD\cD \phib &= \exd\exd\phib + \exd(\phib\cA) + (\exd\phib)\wedge\cA + \phib\cA\wedge\cA = +\phib\cF\,,
\end{align}
which leads to
\begin{align}
\mcom{\cD_B}{\cD_A}\phi &= -{T_{BA}}^C \cD_C \phi - \cF_{BA}\phi\,, \label{eq:yang_mills_double_derivative_phi}\\
\mcom{\cD_B}{\cD_A}\phib &= -{T_{BA}}^C \cD_C \phib + \phib\cF_{BA}\,. \label{eq:yang_mills_double_derivative_phibar}
\end{align}
where $\cD E^A = T^A$ is used, and $\mcom{\cD_B}{\cD_A} = \cD_B\cD_A -(-1)^{BA}\cD_A\cD_B$ is the supercommutator.
\\\\
As stated above $\phi^k$ and $\phib^\kb$ are a chiral and an antichiral superfield, respectively:
\begin{align}
\cD^\sd{\alpha}\phi^k &= 0\,, & \cD_\alpha\phib^\kb &= 0\,.
\end{align}
Because covariant derivatives are used to define the chirality conditions, they are compatible with Yang-Mills transformations (cf.\ Eq.~\eqref{eq:yang_mills_gauge_derivative_phi} and Eq.~\eqref{eq:yang_mills_gauge_derivative_phibar}):
\begin{align}
\cD^\sd{\alpha}(\lge^{-1}\phi) &= \lge^{-1}(\cD^\sd{\alpha}\phi) = 0\,, & \cD_\alpha(\phib\lge) &= (\cD_\alpha\phib)\lge = 0\,.
\end{align}

\subsection{Solution of the Yang-Mills Bianchi identity}
\subsubsection{Field strength constraints}
Since $\phi$ and $\phib$ are a chiral and an antichiral superfield respectively, i.e.\ $\cD^\sd{\alpha}\phi=0$ and $\cD_\alpha\phib=0$, Eq.~\eqref{eq:yang_mills_double_derivative_phi} and \eqref{eq:yang_mills_double_derivative_phibar} imply
\begin{align}
\acom{\cD^\sd{\beta}}{\cD^\sd{\alpha}} \phi &= -\cF^{\sd{\beta}\sd{\alpha}}\phi = 0\,, & \acom{\cD_\beta}{\cD_\alpha} \phib &= +\phib\cF_{\beta\alpha} = 0\,,
\end{align}
where the torsion constraints ${T_{\beta\alpha}}^C=0$ and $T^{\sd{\beta}\sd{\alpha}C}=0$ from Eq.~\eqref{eq:bianchi_sp_torsion_constraints} are used. In view of these identities, the following covariant constraints, the so called representation preserving constraints, for the components of the fields strength are introduced:
\begin{align}
\label{eq:yang_mills_constraints_1}
\cF^{\sd{\beta}\sd{\alpha}} &= 0\,, & \cF_{\beta\alpha} &= 0\,.
\end{align}
In addition, using the explicit form for the components of the field strength in Eq.~\eqref{eq:yang_mills_curvature_components_form}, from the identity
\begin{align}
{\cF_\beta}^\sd{\alpha} = \cD_\beta \cA^\sd{\alpha} + \cD^\sd{\alpha} \cA_\beta - 2i{(\sig^c\eps)_{\beta}}^\sd{\alpha} \cA_c - \acom{\cA_\beta}{\cA^\sd{\alpha}}
\end{align}
follows that the components ${\cF_\beta}^\sd{\alpha}$ can be absorbed into a linear covariant redefinition of the vector component $\cA_a$ of the connection. Thus, the so called covariant conventional constraint
\begin{align}
\label{eq:yang_mills_constraints_2}
{\cF_\beta}^\sd{\alpha} &= 0
\end{align}
is introduced. The representation preserving and the conventional constraints of the Yang-Mills field strength reduce the degrees of freedom, and their implications can be studied by solving the Bianchi identity in Eq.~\eqref{eq:yang_mills_bianchi_identity_components}.

\subsubsection{Solution of the Bianchi identity}
\label{sec:yang_mills_bianchi_identity_solution}
In this section, the Bianchi identity of the Yang-Mills connection in Eq.~\eqref{eq:yang_mills_bianchi_identity_components} is solved, by applying the torsion constraints given in Eq.~\eqref{eq:bianchi_sp_torsion_constraints} and the representation preserving and the conventional constraints for the Yang-Mills field strength in Eq.~\eqref{eq:yang_mills_constraints_1} and Eq.~\eqref{eq:yang_mills_constraints_2}. For the sake of completeness, the Bianchi identity
\begin{align}
\label{eq:yang_mills_bianchi_identity_components_solution}
\oint_{CBA} \big( \cD_C \cF_{BA} + {T_{CB}}^D \cF_{DA}  \big) = 0\,,
\end{align}
with $\oint_{CBA} CBA = CBA + (-1)^{A(C+B)} ACB + (-1)^{(B+A)C} BAC$, and the covariant constraints of the field strength
\begin{align}
\label{eq:yang_mills_constraints_compact}
\cF_{\su{\beta}\su{\alpha}} = 0\,,
\end{align}
where $\underline{\alpha}$ labels both $\alpha$ and $\sd{\alpha}$, are stated here again.
Using vector and spinor indices, there are the following non-equivalent tuples of indices:
\begin{align}
\begin{split}
(C,B,A)\,=\,&(c,b,a),\,(c,b,\alpha),\,(c,b,\sd{\alpha}),\,(c,\beta,\alpha),\,(c,\beta,\sd{\alpha}),\,(c,\sd{\beta},\sd{\alpha}),\\
&\underbrace{(\gamma,\beta,\alpha)}_{\text{trivial}},\,(\gamma,\beta,\sd{\alpha}),\,(\gamma,\sd{\beta},\sd{\alpha}),\,\underbrace{(\sd{\gamma},\sd{\beta},\sd{\alpha})}_{\text{trivial}}
\end{split}
\end{align}
The label "trivial" indicates that the Bianchi identity is trivially satisfied for the corresponding tuples of indices. Thus, the Bianchi identity delivers $8$ equations:
\begin{fleqn}
\begin{align}
&(C,B,A) = (\gamma,\sd{\beta},\alpha): \tag{I} \label{eq:yang_mills_eq01}\\
&2i\sig^d_{\gamma\sd{\beta}} \cF_{d\alpha} + 2i\sig^d_{\alpha\sd{\beta}} \cF_{d\gamma} = 0 \notag
\end{align}
\begin{align}
&(C,B,A) = (\gamma,\sd{\beta},\sd{\alpha}): \tag{II} \label{eq:yang_mills_eq02}\\
&2i\sig^d_{\gamma\sd{\beta}} \cF_{d\sd{\alpha}} + 2i\sig^d_{\gamma\sd{\alpha}} \cF_{d\sd{\beta}} = 0 \notag
\end{align}
\begin{align}
&(C,B,A) = (\gamma,\sd{\beta},a): \tag{III} \label{eq:yang_mills_eq03}\\
&\cD_\gamma \cF_{\sd{\beta}a} - \cD_\sd{\beta} \cF_{a\gamma} + 2i\sig^d_{\gamma\sd{\beta}} \cF_{da} = 0 \notag
\end{align}
\begin{align}
&(C,B,A) = (\gamma,\beta,a): \tag{IV} \label{eq:yang_mills_eq04}\\
&\cD_\gamma \cF_{\beta a} - \cD_{\beta} \cF_{a\gamma} = 0 \notag
\end{align}
\begin{align}
&(C,B,A) = (\sd{\gamma},\sd{\beta},a): \tag{V} \label{eq:yang_mills_eq05}\\
&\cD_\sd{\gamma} \cF_{\sd{\beta}a} - \cD_\sd{\beta} \cF_{a\sd{\gamma}} = 0 \notag
\end{align}
\begin{align}
&(C,B,A) = (\gamma,b,a): \tag{VI} \label{eq:yang_mills_eq06}\\
&\cD_\gamma \cF_{ba} + \cD_a \cF_{\gamma b} + \cD_b \cF_{a\gamma} + {T_{\gamma b}}^\su{\delta} \cF_{\su{\delta}a} + {T_{a\gamma}}^\su{\delta} \cF_{\su{\delta}b} = 0 \notag
\end{align}
\begin{align}
&(C,B,A) = (\sd{\gamma},b,a): \tag{VII} \label{eq:yang_mills_eq07}\\
&\cD_\sd{\gamma} \cF_{ba} + \cD_a \cF_{\sd{\gamma}b} + \cD_b \cF_{a\sd{\gamma}} + {T_{\sd{\gamma}b}}^\su{\delta} \cF_{\su{\delta}a} + {T_{a\sd{\gamma}}}^\su{\delta} \cF_{\su{\delta}b} = 0 \notag
\end{align}
\begin{align}
&(C,B,A) = (c,b,a): \tag{VIII} \label{eq:yang_mills_eq08}\\
&\cD_c \cF_{ba} + \cD_a \cF_{cb} + \cD_b \cF_{ac} + {T_{cb}}^\su{\delta} \cF_{\su{\delta}a} + {T_{ac}}^\su{\delta} \cF_{\su{\delta}b} + {T_{ba}}^\su{\delta} \cF_{\su{\delta}c} = 0 \notag
\end{align}
\end{fleqn}
In order to solve the above equations, it is convenient to express the vector indices in terms of spinor indices. The solutions of \eqref{eq:yang_mills_eq01}--\eqref{eq:yang_mills_eq08} are then the following:
\begin{itemize}
\item \textbf{Identity \eqref{eq:yang_mills_eq01} and \eqref{eq:yang_mills_eq02}:}\\
The identities \eqref{eq:yang_mills_eq01} and \eqref{eq:yang_mills_eq02} are written as
\begin{align}
\cF_{\alpha\s\gamma\sd{\beta}} + \cF_{\gamma\s\alpha\sd{\beta}} &= 0\,,\\
\cF_{\sd{\alpha}\s\gamma\sd{\beta}} + \cF_{\sd{\beta}\s\gamma\sd{\alpha}} &= 0\,,
\end{align}
and they have the respective solutions
\begin{align}
\label{eq:yang_mills_sol_eq01_1}
\cF_{\beta\s\alpha\sd{\alpha}} &= 2i\eps_{\beta\alpha} \cW_\sd{\alpha}\,,\\
\label{eq:yang_mills_sol_eq02_1}
\cF_{\sd{\beta}\s\alpha\sd{\alpha}} &= 2i\eps_{\sd{\beta}\sd{\alpha}} \cW_\alpha\,.
\end{align}
The superfields $\cW_\alpha$ and $\cW_\sd{\alpha}$ are Lie algebra valued, i.e.\
\begin{align}
\cW_\alpha &= {\cW_\alpha}^\g{r} \gen{r}\,, & \cW_\sd{\alpha} &= {\cW_\sd{\alpha}}^\g{r} \gen{r}\,,
\end{align}
and, according to the appearance of $\cW_\alpha$ and $\cW_\sd{\alpha}$ in Eq.~\eqref{eq:yang_mills_sol_eq01_1} and \eqref{eq:yang_mills_sol_eq02_1}, it follows that
\begin{align}
w(\cW_\alpha) &= +1\,, & w(\cW_\sd{\alpha}) &= -1\,,
\end{align}
and $({\cW_\alpha}^\g{r})^\cc = {\cW_\sd{\alpha}}^\g{r}$.
\item \textbf{Identity \eqref{eq:yang_mills_eq03}:}\\
Using the expressions for $\cF_{\beta\s\alpha\sd{\alpha}}$ and $\cF_{\sd{\beta}\s\alpha\sd{\alpha}}$ from Eq.~\eqref{eq:yang_mills_sol_eq01_1} and \eqref{eq:yang_mills_sol_eq02_1}, \eqref{eq:yang_mills_eq03} has the form
\begin{align}
\cF_{\beta\sd{\beta}\s\alpha\sd{\alpha}} &= -\eps_{\sd{\beta}\sd{\alpha}}\cD_\beta \cW_\alpha - \eps_{\beta\alpha}\cD_\sd{\beta} \cW_\sd{\alpha}\,.
\end{align}
From the antisymmetry $\cF_{ba} = -\cF_{ab}$ follows
\begin{align}
\label{eq:yang_mills_sol_eq03_1}
\cD^\alpha \cW_\alpha = \cD_\sd{\alpha} \cW^\sd{\alpha}\,,
\end{align}
which motivates the definition of the real superfield $\Db$ as follows
\begin{align}
\Db := -\frac{1}{2} \cD^\alpha \cW_\alpha\,.
\end{align}
In terms of components with definite symmetry properties, $\cF_{\beta\sd{\beta}\s\alpha\sd{\alpha}}$ is written as
\begin{align}
\label{eq:yang_mills_sol_eq03_2}
\cF_{\beta\sd{\beta}\s\alpha\sd{\alpha}} &= 2\eps_{\sd{\beta}\sd{\alpha}} \cF_{\sym{2}{\beta\alpha}} - 2\eps_{\beta\alpha} \cF_{\sym{2}{\sd{\beta}\sd{\alpha}}}\,,
\end{align}
with
\begin{align}
\cF_{\sym{2}{\beta\alpha}} &= -\frac{1}{4}(\cD_\beta \cW_\alpha + \cD_\alpha \cW_\beta)\,,\\
\cF_{\sym{2}{\sd{\beta}\sd{\alpha}}} &= +\frac{1}{4}(\cD_\sd{\beta} \cW_\sd{\alpha} + \cD_\sd{\alpha} \cW_\sd{\beta})\,,
\end{align}
and \eqref{eq:yang_mills_eq03} is then solved by Eq.~\eqref{eq:yang_mills_sol_eq03_1} and \eqref{eq:yang_mills_sol_eq03_2}.
\item \textbf{Identity \eqref{eq:yang_mills_eq04} and \eqref{eq:yang_mills_eq05}:}\\
Again using the identities from Eq.~\eqref{eq:yang_mills_sol_eq01_1} and \eqref{eq:yang_mills_sol_eq02_1}, \eqref{eq:yang_mills_eq04} and \eqref{eq:yang_mills_eq05} are fulfilled if
\begin{align}
\cD_\alpha \cW_\sd{\alpha} &= 0\,, & \cD_\sd{\alpha} \cW_\alpha &= 0\,.
\end{align}
This means $\cW_\alpha$ and $\cW_\sd{\alpha}$ are a chiral and an antichiral superfield, respectively.
\item \textbf{Identity \eqref{eq:yang_mills_eq06} and \eqref{eq:yang_mills_eq07}:}\\
From \eqref{eq:yang_mills_eq06} and \eqref{eq:yang_mills_eq07} the identities
\begin{align}
\cD_\alpha\Db &= iD_{\alpha\sd{\beta}} \cW^\sd{\beta}\,, \label{eq:yang_mills_sol_eq06_1}\\
\cD_\sd{\alpha}\Db &= i{D^\beta}_\sd{\alpha} \cW_\beta\,, \label{eq:yang_mills_sol_eq06_2}\\
\cD^2 \cW_\alpha &= 4i \cD_{\alpha\sd{\beta}} \cW^\sd{\beta} + 12\cW_\alpha \Rb\,, \label{eq:yang_mills_sol_eq06_3}\\
\cDb^2 \cW_\sd{\alpha} &= 4i {\cD^\beta}_\sd{\alpha} \cW_\beta + 12\cW_\sd{\alpha} R\,, \label{eq:yang_mills_sol_eq06_4}
\end{align}
with $\cD^2 = \cD^\alpha\cD_\alpha$ and $\cDb^2 = \cD_\sd{\alpha} \cD^\sd{\alpha}$, are obtained.
\item \textbf{Identity \eqref{eq:yang_mills_eq08}}: \\
The identity \eqref{eq:yang_mills_eq08} implies derivative relations of the field strength components $\cF_{ba}$ which are not used in subsequent calculations.\footnote{Identity \eqref{eq:yang_mills_eq08} is the only identity with mass dimension $3$. Thus, the derivative relations are of the same dimension.}
\end{itemize}

\subsubsection{Summary of the Yang-Mills field strength components and derivative relations}
\label{sec:summary_ym_fieldstrength_components_derivative_relations}
This section provides a summary of the identities from Section~\ref{sec:yang_mills_bianchi_identity_solution}, which were obtained by solving the Bianchi identity for the Yang-Mills connection, by taking the representation preserving and the conventional constraints in Eq.~\eqref{eq:yang_mills_constraints_compact} and the torsion constraints in Eq.~\eqref{eq:bianchi_sp_torsion_constraints} into account. It turned out that all components of the Yang-Mills field strength are expressed in terms of the following superfields and their derivatives:
\begin{align}
\cW_\alpha &= {\cW_\alpha}^\g{r} \gen{r}\,, & \cW^\sd{\alpha} &= \cW^{\sd{\alpha}\g{r}} \gen{r}\,,
\end{align}
where $\gen{r}$ are the generators of the Lie algebra $\mathfrak{g}$ ($i\gen{r}\in\mathfrak{g}$). Furthermore, $\cW_\sd{\alpha}$ and $\cW^\sd{\alpha}$ are a chiral and an antichiral superfield, respectively,
\begin{align}
\cD^\sd{\beta} \cW_\alpha &= 0\,, & \cD_\beta \cW^\sd{\alpha} &= 0\,,
\end{align}
and under conjugation the components transform as:
\begin{align}
({\cW_\alpha}^\g{r})^\cc = {\cW_\sd{\alpha}}^\g{r}\,.
\end{align}
According to the appearance of $\cW_\alpha$ and $\cW^\sd{\alpha}$ in the components of the Yang-Mills field strength, they have the $\U(1)$ weights
\begin{align}
w(\cW_\alpha) &= +1\,, & w(\cW_\sd{\alpha}) &= -1\,.
\end{align}
In the following, the components of the Yang-Mills field strength are listed, ordered according to their mass dimension. Since $\cF$ is a $2$-superform, the mass dimension of the components range from $1$ to $2$. All results are written in vector and spinor notation.
\begin{itemize}
\item \textbf{Dimension $1$:}\\
According to the constraints in Eq.~\eqref{eq:yang_mills_constraints_compact}, all components with dimension~$1$ vanish
\begin{align}
\cF_{\su{\beta}\su{\alpha}} = 0\,, \label{eq:yang_mills_components_curvature_dim1_1}
\end{align}
where $\su{\alpha}$ labels both $\alpha$ and $\sd{\alpha}$.
\item \textbf{Dimension $\frac{3}{2}$:}\\
The components with dimension~$\frac{3}{2}$ are stated in Eq.~\eqref{eq:yang_mills_sol_eq01_1} and \eqref{eq:yang_mills_sol_eq02_1}. They are given by
\begin{align}
\cF_{\beta a} &= +i\sig_{a\beta\sd{\beta}}\cW^\sd{\beta}\,, \label{eq:yang_mills_components_curvature_dim3/2_1}\\
{\cF^\sd{\beta}}_a &= -i\sigb_a^{\sd{\beta}\beta}\cW_\beta\,, \label{eq:yang_mills_components_curvature_dim3/2_2}
\end{align}
which, written in terms of spinor indices, read
\begin{align}
\cF_{\beta\s\alpha\sd{\alpha}} &= +2i\eps_{\beta\alpha} \cW_\sd{\alpha}\,, \label{eq:yang_mills_components_curvature_dim3/2_3}\\
\cF_{\sd{\beta}\s\alpha\sd{\alpha}} &= +2i\eps_{\sd{\beta}\sd{\alpha}} \cW_\alpha\,. \label{eq:yang_mills_components_curvature_dim3/2_4}
\end{align}
\item \textbf{Dimension $2$:}\\
It follows from Eq.~\eqref{eq:yang_mills_sol_eq03_2} that the components with dimension~$2$ have the form
\begin{align}
\cF_{ba} &= +\frac{1}{2}(\eps\sig_{ba})^{\beta\alpha}\cD_\alpha\cW_\beta + \frac{1}{2}(\sigb_{ba}\eps)^{\sd{\beta}\sd{\alpha}}\cD_\sd{\alpha}\cW_\sd{\beta}\,. \label{eq:yang_mills_components_curvature_dim2_1}
\end{align}
Using spinor indices the components are written as
\begin{align}
\cF_{\beta\sd{\beta}\s\alpha\sd{\alpha}} &= 2\eps_{\sd{\beta}\sd{\alpha}} \cF_{\sym{2}{\beta\alpha}} - 2\eps_{\beta\alpha} \cF_{\sym{2}{\sd{\beta}\sd{\alpha}}}\,, \label{eq:yang_mills_components_curvature_dim2_2}
\end{align}
with
\begin{align}
\cF_{\sym{2}{\beta\alpha}} &= -\frac{1}{4}(\cD_\beta \cW_\alpha + \cD_\alpha \cW_\beta)\,,\\
\cF_{\sym{2}{\sd{\beta}\sd{\alpha}}} &= +\frac{1}{4}(\cD_\sd{\beta} \cW_\sd{\alpha} + \cD_\sd{\alpha} \cW_\sd{\beta})\,.
\end{align}
\end{itemize}
The derivative relations of the components, up to mass dimension $\frac{5}{2}$, are summarized in the following. The relation from Eq.~\eqref{eq:yang_mills_sol_eq03_1},
\begin{align}
\cD^\alpha \cW_\alpha = \cD_\sd{\alpha} \cW^\sd{\alpha}\,, \label{eq:yang_mills_derivative_relations_1}
\end{align}
induces the definition of the real superfield $\Db$:
\begin{align}
\Db := -\frac{1}{2} \cD^\alpha \cW_\alpha\,, \label{eq:yang_mills_definition_dterm}
\end{align}
which has vanishing $\U(1)$ weight $w(\Db)=0$. The following useful identities are derived directly from Eq.~\eqref{eq:yang_mills_components_curvature_dim2_1} and Eq.~\eqref{eq:yang_mills_definition_dterm}:
\begin{align}
\cD_\beta\cW_\alpha + \cD_\alpha\cW_\beta &= -2(\sig^{ba}\eps)_{\beta\alpha}\cF_{ba}\,, \label{eq:yang_mills_derivative_relations_2}\\
\cD_\beta\cW_\alpha - \cD_\alpha\cW_\beta &= +\eps_{\beta\alpha}\cD^\varphi\cW_\varphi = -2\eps_{\beta\alpha}\Db\,, \label{eq:yang_mills_derivative_relations_3}\\
\cD_\sd{\beta}\cW_\sd{\alpha} + \cD_\sd{\alpha}\cW_\sd{\beta} &= -2(\eps\sigb^{ba})_{\sd{\beta}\sd{\alpha}}\cF_{ba}\,, \label{eq:yang_mills_derivative_relations_4}\\
\cD_\sd{\beta}\cW_\sd{\alpha} - \cD_\sd{\alpha}\cW_\sd{\beta} &= -\eps_{\sd{\beta\sd{\alpha}}}\cD_\sd{\varphi}\cW^\sd{\varphi} = +2\eps_{\sd{\beta\sd{\alpha}}}\Db\,. \label{eq:yang_mills_derivative_relations_5}
\end{align}
In addition, the relations stated in Eqs.~\eqref{eq:yang_mills_sol_eq06_1}--\eqref{eq:yang_mills_sol_eq06_4}, which contain a double covariant derivative, are given by
\begin{align}
\cD_\alpha\Db &= iD_{\alpha\sd{\beta}} \cW^\sd{\beta}\,, \label{eq:yang_mills_derivative_relations_6}\\
\cD_\sd{\alpha}\Db &= i{D^\beta}_\sd{\alpha} \cW_\beta\,, \label{eq:yang_mills_derivative_relations_7}\\
\cD^2 \cW_\alpha &= 4i \cD_{\alpha\sd{\beta}} \cW^\sd{\beta} + 12\cW_\alpha \Rb\,, \label{eq:yang_mills_derivative_relations_8}\\
\cDb^2 \cW_\sd{\alpha} &= 4i {\cD^\beta}_\sd{\alpha} \cW_\beta + 12\cW_\sd{\alpha} R\,, \label{eq:yang_mills_derivative_relations_9}
\end{align}
where $\cD^2 = \cD^\alpha\cD_\alpha$ and $\cDb^2 = \cD_\sd{\alpha} \cD^\sd{\alpha}$.

\subsection{Chiral projection operators}
\label{sec:chiral projection operators}
In this section the chiral and antichiral projection operators in $\U(1)$ superspace are introduced, by taking into account the constraints of the torsion tensor and the Yang-Mills field strength. The two operators will appear several times in calculations in following sections. Consider the superfield $\Phi$ with $\U(1)$ weight $w(\Phi)$, and which transforms in some representation of the Yang-Mills gauge group. The identities
\begin{gather}
\acom{\cD^\sd{\beta}}{\cD_\sd{\alpha}}\cD^\sd{\alpha}\Phi = 12R\,\cD^\sd{\beta}\Phi\,, \\
\cD^\sd{\beta}\cD^\sd{\alpha}\Phi = -\cD^\sd{\alpha}\cD^\sd{\beta}\Phi = \frac{1}{2}\eps^{\sd{\beta}\sd{\alpha}}\cD_\sd{\gamma}\cD^\sd{\gamma}\Phi\,,
\end{gather}
imply
\begin{align}
\cD^\sd{\beta}\cD_\sd{\alpha}\cD^\sd{\alpha}\Phi &= 8R\,\cD^\sd{\beta}\Phi\,.
\end{align}
With the chirality condition $\cD^\sd{\beta}R=0$ follows
\begin{align}
\cD^\sd{\beta}(\cDb^2 - 8R)\Phi = 0\,,
\end{align}
where $\cDb^2=\cD_\sd{\alpha}\cD^\sd{\alpha}$. Thus, $(\cDb^2 - 8R)\Phi$ is a chiral superfield. Note that this consideration holds true, if $\Phi$ has an additional undotted (upper or lower) spinor index $\gamma$, i.e.\ $\cD_\beta(\cDb^2 - 8R)\Phi_\gamma = 0$. A similar calculation leads to
\begin{align}
\cD_\beta(\cD^2 - 8\Rb)\Phi = 0\,,
\end{align}
with $\cD^2 = \cD^\alpha\cD_\alpha$, where the chirality condition $\cD_\beta\Rb=0$ is used. Thus, $(\cD^2 - 8\Rb)\Phi$ is an antichiral superfield. Again, this consideration holds true, if $\Phi$ has an additional dotted (upper or lower) spinor index $\sd{\gamma}$, i.e.\  $\cD_\beta(\cD^2 - 8\Rb)\Phi_\sd{\gamma} = 0$.
\\\\
If $\Phi$ and $\Phib$ are already a chiral and an antichiral superfield, respectively, then
\begin{align}
-\frac{1}{8R}(\cDb^2 - 8R)\Phi &= \Phi\,, \label{eq:kahler_superspace_chiral_projection_operator}\\
-\frac{1}{8\Rb}(\cD^2 - 8\Rb)\Phib &= \Phib\,. \label{eq:kahler_superspace_antichiral_projection_operator}
\end{align}
The operators in Eq.~\eqref{eq:kahler_superspace_chiral_projection_operator} and \eqref{eq:kahler_superspace_antichiral_projection_operator} are therefore the correctly normalized chiral and antichiral projection operators.

\section{K\"ahler superspace}
\label{sec:kahler_superspace}
In the previous sections the structure of $\U(1)$ superspace has been investigated by solving the Bianchi identities for certain torsion constraints, and matter superfields have been defined including Yang-Mills gauge transformations. Although, the correct geometric framework to describe the coupling of matter to supergravity is not $\U(1)$ superspace, but rather K\"ahler ($\UK(1)$) superspace, it is convenient to study first $\U(1)$ superspace and then use the results from there in $\UK(1)$ superspace. As mentioned earlier, the two superspaces are closely related, namely, $\UK(1)$ superspace is obtained from $\U(1)$ superspace by a suitable identification of the $\U(1)$ pre-potential and chiral pre-gauge transformations with the K\"ahler potential $K(\phi,\phib)$ and K\"ahler transformations $F(\phi)$, respectively, which are functions of the matter superfields. In the context of $\UK(1)$ superspace the $\U(1)$ weights $w$ defined in $\U(1)$ superspace are referred to as chiral weights and the $\U(1)$ connection $A$ is called K\"ahler connection.
\\\\
In Section~\ref{sec:u1_prepotentials} the $\U(1)$ pre-potential is introduced and K\"ahler superspace is defined. It is shown that the results in $\U(1)$ superspace from the previous sections hold true in $\UK(1)$ superspace. Furthermore, in Section~\ref{sec:sugra_transformations_superfield_level} supergravity transformations at the superfield level are introduced, and in Section~\ref{sec:action_superfield_level} the general action of the supergravity/matter/Yang-Mills system in K\"ahler superspace is given.

\subsection{$\U(1)$ pre-potential}
\label{sec:u1_prepotentials}
The solution of the Bianchi identities in $\mathrm{U}(1)$ superspace in Section~\ref{sec:bianchi_sp_solution_all} shows that the following components of the $\U(1)$ field strength vanish (cf.\ Eq.~\eqref{eq:bianchi_sp_components_u1_fieldstrength_dim1_1}):
\begin{align}
F_{\beta\alpha} = 0\,,\quad F^{\sd{\beta}\sd{\alpha}} = 0\,. \label{eq:kahler_superspace_curvature_components_vanish}
\end{align}
On the other hand, the explicit expression of these components in terms of the $\U(1)$ connection in Eq.~\eqref{eq:superspace_u1_field_strength_components} implies
\begin{align}
F_{\beta\alpha} = \cD_\beta A_\alpha + \cD_\gamma A_\alpha\,,\quad F^{\sd{\beta}\sd{\alpha}} = \cD^\sd{\beta} A^\sd{\alpha} + \cD^\sd{\alpha} A^\sd{\beta}\,, \label{eq:kahler_superspace_curvature_components_explicit}
\end{align}
where the torsion constraints ${T_{\beta\alpha}}^C = 0$ and $T^{\sd{\beta}\sd{\alpha}C} = 0$ are used. Thus, Eq.~\eqref{eq:kahler_superspace_curvature_components_vanish} is fulfilled, if $A_\alpha$ and $A^\sd{\alpha}$ are written in terms of the even superfield $T$ and its conjugate $\conj{T}=T^\cc$, respectively:
\begin{align}
\label{eq:kahler_superspace_connection_prepotential_1}
A_\alpha &= -T^{-1} D_\alpha T = -D_\alpha \log{T}\,,\\
\label{eq:kahler_superspace_connection_prepotential_2}
A^\sd{\alpha} &= +\conj{T}{}^{-1} D^\sd{\alpha} \conj{T} = +D^\sd{\alpha} \log{\conj{T}}\,,
\end{align}
which is consistent with the identity $(A_\alpha)^\cc = -A_\sd{\alpha}$ from Eq.~\eqref{eq:superspace_u1_connection_components_relation_1}. The superfield $T$ is called pre-potential and it is inert under $\mathrm{U}(1)$ transformations, thus the covariant derivative is just given by $\cD T = \exd T$. In order to restore the transformation property of the connection under $\U(1)$ transformations $g$ in Eq.~\eqref{eq:superspace_u1_connection_transformation_u1}, the pre-potential has to transform in the following way
\begin{align}
T &\mapsto \Pb\,T\,g\,, \\
\Tb &\mapsto P\,\Tb\,g^{-1}\,,
\end{align}
where $P$ is a chiral superfield, i.e.\ \hbox{$\cD^\sd{\alpha}P=0$}, and is called pre-gauge transformation. Note that $\Pb = P^\cc$ is an antichiral superfield, and that $g^\cc = g^{-1}$. This is an extra transformation, beside the gauge transformation $g$, at the level of the pre-potential, which does not appear in the gauge transformations of $A_\alpha$ and $A^\sd{\alpha}$. The product $(T^a\,\Tb^{-b})$ with arbitrary powers $a,b\in\mathbb{R}$ transforms under gauge and pre-gauge transformations as follows
\begin{align}
(T^a \Tb^{-b}) &\mapsto (T^a \Tb^{-b})\,\Pb^a\,P^{-b}\,g^{a+b}\,. \label{eq:kahler_superspace_t_term_transformation}
\end{align}
For a superfield $\Phi$ with $\U(1)$ weight $w(\Phi)$ the new superfield $\Phi(a,b)$ in the so called $(a,b)$-basis is defined as
\begin{align}
\label{eq:kahler_superspace_phi_ab_basis}
\Phi(a,b) &:= (T^a\,\Tb^{-b})^{-w(\Phi)} \Phi\,.
\end{align}
Note, if $w(\Phi)=0$ then $\Phi(a,b)=\Phi$, and the label $(a,b)$ is usually neglected. According to Eq.~\eqref{eq:kahler_superspace_t_term_transformation}, the transformation of $\Phi(a,b)$ under gauge and pre-gauge transformations is given by
\begin{align}
\label{eq:kahler_superspace_phi_transformation_ab_basis}
\Phi(a,b) &\mapsto \big(g^{(a+b-1)}\,\Pb^a\,P^{-b}\big)^{-w(\Phi)} \Phi(a,b)\,.
\end{align}
Furthermore, the $\U(1)$ connection and the field strength in the $(a,b)$-basis have the form
\begin{align}
\label{eq:kahler_superspace_connection_ab_basis}
A(a,b) &:= A + (T^a\,\Tb^{-b})^{-1} \exd(T^a\,\Tb^{-b}) = A + a\,\exd\log{T} - b\,\exd\log{\Tb}\,,\\
\label{eq:kahler_superspace_curvature_ab_basis}
F(a,b) &:= \exd A(a,b) = \exd A = F\,,
\end{align}
which shows that the field strength is independent of the chosen basis. Under gauge and pre-gauge transformations the connection transforms as follows
\begin{align}
\label{eq:kahler_superspace_connection_transformation_ab_basis}
A(a,b) &\mapsto A(a,b) + (a+b-1)\,\exd\log{g} + a\,\exd\log{\Pb} - b\,\exd\log{P}\,,
\end{align}
whereas $F(a,b)$ does not transform. The covariant derivative of $\Phi(a,b)$ contains the connection in the $(a,b)$-basis and is defined as
\begin{align}
\cD\Phi(a,b) &:= \exd\Phi(a,b) + w(\Phi)\Phi(a,b)A(a,b) = (T^a\,\Tb^{-b})^{-w(\Phi)} \cD\Phi\,. \label{eq:kahler_superspace_phi_covariant_derivative_ab_basis}
\end{align}
The second identity shows, that this definition is consistent with the fact that $\cD\Phi$ has the same $\U(1)$ weight as $\Phi$. Since the supervielbein $E^A$ and its inverse $E_A$ have $\U(1)$ weights $w(E^A)$ and $w(E_A)$ respectively, in the $(a,b)$-basis they are given by
\begin{align}
E^A(a,b) &= (T^a\,\Tb^{-b})^{-w(E^A)} E^A\,, \label{eq:kahler_superspace_vielbein_ab_basis}\\
E_A(a,b) &= (T^a\,\Tb^{-b})^{-w(E_A)} E_A\,, \label{eq:kahler_superspace_inverse_vielbein_ab_basis}
\end{align}
according to Eq.~\eqref{eq:kahler_superspace_phi_ab_basis}. Tensors are then written with respect the basis $E^A(a,b)$ and $E_A(a,b)$. In particular, the components of a $1$-superform $\omega$ and a supervector field $\chi$ (with $w(\omega)=w(\chi)=0$) are given by
\begin{align}
\omega &= E^A(a,b)\omega_A(a,b)\,, & \text{with}\quad\quad\omega_A(a,b) &= (T^a\,\Tb^{-b})^{+w(E^A)} \omega_A\,, \label{eq:kahler_superspace_form_ab_basis}\\
\chi &= E_A(a,b)\chi^A(a,b)\,, & \text{with}\quad\quad\chi^A(a,b) &= (T^a\,\Tb^{-b})^{+w(E_A)} \chi^A\,. \label{eq:kahler_superspace_vector_ab_basis}
\end{align}
In particular, Eq.~\eqref{eq:kahler_superspace_form_ab_basis} indicates the following notation for the components of the covariant derivative $\cD$ in the $(a,b)$-basis:
\begin{align}
\cD &= E^A(a,b)\cD_A(a,b)\,, & \text{with}\quad\quad\cD_A(a,b) &= (T^a\,\Tb^{-b})^{+w(E^A)} \cD_A\,.
\end{align}
With this notation, the covariant derivative of $\Phi$ in Eq.~\eqref{eq:kahler_superspace_phi_covariant_derivative_ab_basis} is written as
\begin{align}
\cD_A(a,b)\Phi(a,b) &:= (\cD_A\Phi)(a,b) = (T^a\,\Tb^{-b})^{+w(E^A)-w(\Phi)}\cD_A\Phi\,, \label{eq:kahler_superspace_phi_covariant_derivative_components_ab_basis}
\end{align}
which, for a double covariant derivative, implies 
\begin{align}
\cD_B(a,b)\cD_A(a,b)\Phi(a,b) &= (\cD_B\cD_A\Phi)(a,b) = (T^a\,\Tb^{-b})^{+w(E^B)+w(E^A)-w(\Phi)}\cD_A\Phi\,. \label{eq:kahler_superspace_phi_double_covariant_derivative_components_ab_basis}
\end{align}
For later use, it is convenient to write the spinor components of the $\U(1)$ connection in the $(a,b)$-basis (see Eq.~\eqref{eq:kahler_superspace_connection_ab_basis}) as
\begin{align}
\label{eq:kahler_superspace_connection_components_ab_basis_1}
A_\alpha(a,b) &= (a - \frac{1}{2})\big(D_\alpha\log{T}\big)(a,b) - (b - \frac{1}{2})\big(D_\alpha\log{\Tb}\big)(a,b) - \frac{1}{2}\big(D_\alpha\log{W}\big)(a,b)\,,\\
\label{eq:kahler_superspace_connection_components_ab_basis_2}
A^\sd{\alpha}(a,b) &= (a - \frac{1}{2})\big(D^\sd{\alpha}\log{T}\big)(a,b) - (b - \frac{1}{2})\big(D^\sd{\alpha}\log{\Tb}\big)(a,b) + \frac{1}{2}\big(D^\sd{\alpha}\log{W}\big)(a,b)\,.
\end{align}
The real superfield $W$ is given by
\begin{align}
W &= T\,\Tb\,, \label{eq:kahler_superspace_w_definition}
\end{align}
thus, it is inert under gauge transformations, and under pre-gauge transformations it transforms as follows
\begin{align}
W &\mapsto \Pb\,W\,P\,. \label{eq:kahler_superspace_w_transformation}
\end{align}
In the case of $a+b=1$, the gauge transformation $g$ is absent in the transformations of $\Phi(a,b)$ and $A(a,b)$, i.e.\ Eq.~\eqref{eq:kahler_superspace_phi_transformation_ab_basis} and \eqref{eq:kahler_superspace_connection_transformation_ab_basis} have the simplified form
\begin{align}
\Phi(a,b) &\mapsto (\Pb^a\,P^{-b})^{-w(\Phi)} \Phi(a,b)\,, \label{eq:kahler_superspace_phi_transformation_ab_basis_1}\\
A(a,b) &\mapsto A(a,b) + a\,\exd\log{\Pb} - b\,\exd\log{P}\,. \label{eq:kahler_superspace_connection_transformation_ab_basis_1}
\end{align}
For $a+b=1$ it is convenient to label the basis only by $a$, namely \hbox{$\Phi(a)\equiv\Phi(a,b)$} and \hbox{$A(a)\equiv A(a,b)$}.\footnote{In the basis $(0,1)$ the chirality condition \hbox{$\cD^\sd{\alpha}\Phi=0$} takes the form \hbox{$D^\sd{\alpha}(0)\Phi(0)=0$}, since \hbox{$A^\sd{\alpha}(0)=0$}, and $\Phi(0)$ transforms under chiral pre-gauge transformations: \hbox{$\Phi(0)\mapsto P^{w(\Phi)}\Phi(0)$}. Analogously, in the basis $(1,0)$ the antichiral condition reads $D_\alpha(1)\conj{\Phi}(1)=0$, since $A_\alpha(1)=0$, and the transformation is antichiral: \hbox{$\conj{\Phi}(1)\mapsto\Pb^{w(\Phi)}\conj{\Phi}(1)$} with \hbox{$w(\conj{\Phi})=-w(\Phi)$}. Thus, the traditional chirality conditions and gauge transformations are recovered. The gauge invariant product is then written as \hbox{$\conj{\Phi}\Phi=\conj{\Phi}(1)W^{-w(\Phi)}\Phi(0)=\conj{\Phi}(1)e^{2w(\Phi)V}\Phi(0)$} with \hbox{$W=e^{-2V}$}, recovering the traditional form. Similar considerations apply to general Yang-Mills gauge groups.}

\subsubsection{Definition of K\"ahler superspace}
In the previous section the $(a,b)$-basis has been introduced with respect to the $\U(1)$ pre-potential $T$ for a superfield $\Phi$ with weight $w(\Phi)$ and the $\U(1)$ connection $A$. K\"ahler superspace is formulated by choosing the basis $(a,b)=(\frac{1}{2},\frac{1}{2})\equiv(\frac{1}{2})$. In this basis $\Phi$ takes the form (see Eq.~\eqref{eq:kahler_superspace_phi_ab_basis})
\begin{align}
\Phi(\tfrac{1}{2}) &= (T\,\Tb^{-1})^{-\frac{1}{2}w(\Phi)} \Phi\,, \label{eq:kahler_superspace_1/2_phi}
\end{align}
and from Eq.~\eqref{eq:kahler_superspace_connection_components_ab_basis_1} and \eqref{eq:kahler_superspace_connection_components_ab_basis_2} follows that $A_\alpha(\frac{1}{2})$ and $A^\sd{\alpha}(\frac{1}{2})$ are functions of the superfield $W$ only, defined in Eq.~\eqref{eq:kahler_superspace_w_definition}:
\begin{align}
A_\alpha(\tfrac{1}{2}) &= - \frac{1}{2}\big(D_\alpha\log{W}\big)(\tfrac{1}{2})\,, \label{eq:kahler_superspace_1/2_connection_w_identification_1}\\
A^\sd{\alpha}(\tfrac{1}{2}) &= + \frac{1}{2}\big(D^\sd{\alpha}\log{W}\big)(\tfrac{1}{2})\,. \label{eq:kahler_superspace_1/2_connection_w_identification_2}
\end{align}
Furthermore, the supervielbein $E^A(\tfrac{1}{2})$ and its inverse $E_A(\tfrac{1}{2})$ are given by
\begin{align}
E^A(\tfrac{1}{2}) &= (T\,\Tb^{-1})^{-\frac{1}{2}w(E^A)} E^A\,, & E_A(\tfrac{1}{2}) &= (T\,\Tb^{-1})^{-\frac{1}{2}w(E_A)} E_A\,, \label{eq:kahler_superspace_1/2_vielbein}
\end{align}
and the basic superfields of the torsion tensor in $\U(1)$ superspace take the form, using the respective chiral weights,
\begin{align}
\begin{split}
R(\tfrac{1}{2}) = (T^{-1}\,\Tb)R\,, \quad \Rb(\tfrac{1}{2}) = (T\,\Tb^{-1})\Rb\,,\quad G_a(\tfrac{1}{2}) = G_a\,, \\
W_{\sym{3}{\gamma\beta\alpha}}(\tfrac{1}{2}) = (T^{-1}\,\Tb)^\frac{1}{2} W_{\sym{3}{\gamma\beta\alpha}}\,, \quad W_{\sym{3}{\sd{\gamma}\sd{\beta}\sd{\alpha}}}(\tfrac{1}{2}) = (T\,\Tb^{-1})^\frac{1}{2} W_{\sym{3}{\sd{\gamma}\sd{\beta}\sd{\alpha}}}\,.
\end{split} \label{eq:kahler_superspace_1/2_torion_fields_1}
\end{align}
The matter superfields and the basic superfields of the Yang-Mills field strength in the $(\tfrac{1}{2},\tfrac{1}{2})$-basis read
\begin{align}
\phi^k(\tfrac{1}{2}) &= \phi^k\,, & \phib^\kb(\tfrac{1}{2}) &= \phib^\kb\,, \label{eq:kahler_superspace_1/2_matter_fields}\\
\cW_\alpha(\tfrac{1}{2}) &= (T^{-1}\,\Tb)^\frac{1}{2}\cW_\alpha\,, & \cW^\sd{\alpha}(\tfrac{1}{2}) &= (T\,\Tb^{-1})^\frac{1}{2}\cW^\sd{\alpha}\,. \label{eq:kahler_superspace_1/2_ym_fieldstrength_fields}
\end{align}
In $\UK(1)$ superspace, the real superfield $W$ and the pre-gauge transformation $P$, which is a chiral superfield, are identified with the K\"ahler potential $K(\phi,\phib)$ and the K\"ahler transformation $F(\phi)$ in the following way
\begin{align}
W &= e^{-\frac{1}{2}K(\phi,\phib)}\,, \label{eq:kahler_superspace_1/2_kahler_potential_identification}\\
P &= e^{-\frac{1}{2}F(\phi)}\,, \label{eq:kahler_superspace_1/2_pregauge_transformation_identification_1}\\
\Pb &= e^{-\frac{1}{2}\Fb(\phib)}\,. \label{eq:kahler_superspace_1/2_pregauge_transformation_identification_2}
\end{align}
Note that the K\"ahler potential and the K\"ahler transformation are functions of the matter superfields, where $K(\phi,\phib)$ is real and $F(\phi)$ is holomorphic in $\phi$. Furthermore, $\Fb(\phib)$ is the conjugate of $F(\phi)$. With the notation
\begin{align}
K_k &= \frac{\del K(\phi,\phib)}{\del\phi^k}\,, & K_\kb &= \frac{\del K(\phi,\phib)}{\del\phib^\kb}\,,
\end{align}
the metric $g_{k\kb}$ of the K\"ahler manifold is given by
\begin{align}
g_{k\kb} &\equiv K_{k\kb} = \frac{\del^2 K(\phi,\phib)}{\del\phi^k\del\phib^\kb}\,.
\end{align}
The inverse of the metric is written as $g^{\kb k}$. According to the transformation property of $W$ under pre-gauge transformations in Eq.~\eqref{eq:kahler_superspace_w_transformation}, the K\"ahler potential transforms under a K\"ahler transformation
\begin{align}
K(\phi,\phib) \mapsto K(\phi,\phib) + F(\phi) + \Fb(\phib)\,, \label{eq:kahler_superspace_kahler_transformation}
\end{align}
leaving the metric $g_{k\kb}$ unchanged. Furthermore, because the superfields $W$ and $P$ are singlets under Yang-Mills transformations, Eq.~\eqref{eq:kahler_superspace_1/2_kahler_potential_identification} and \eqref{eq:kahler_superspace_1/2_pregauge_transformation_identification_1} imply that  $K(\phi,\phib)$ and $F(\phi)$ are singlets too. In particular, the invariance of the K\"ahler potential leads to the identity
\begin{align}
K_k(\gen{r}\phi)^k = K_\kb(\phib\gen{r})^\kb\,, \label{eq:kahler_superspace_1/2_invariance_kahler_potential}
\end{align}
and it also guarantees that the Yang-Mills transformations are isometries with respect to the K\"ahler metric $g_{k\kb}$. The Killing potential $G_\g{r}$ thus has the form
\begin{align}
G_\g{r} &= -\frac{1}{2}\big(K_k(\gen{r}\phi)^k + K_\kb(\phib\gen{r})^\kb\big)\,, \label{eq:kahler_superspace_1/2_killing_potential}
\end{align}
i.e.\ the isometry (Yang-Mills) transformations are obtained by taking a derivative of $G_\g{r}$ with respect to the matter superfields, namely
\begin{align}
\frac{\del G_\g{r}}{\del\phi^k} &= -g_{k\kb}(\phib\gen{r})^\kb\,, & \frac{\del G_\g{r}}{\del\phib^\kb} &= -g_{k\kb}(\gen{r}\phi)^k\,.
\end{align}
Furthermore, Eq.~\eqref{eq:kahler_superspace_1/2_invariance_kahler_potential} and the chirality conditions for $\phi^k$ and $\phib^\kb$ imply
\begin{align}
\begin{split}
D_\alpha K &= K_k D_\alpha\phi^k + K_\kb D_\alpha\phib^\kb \\
&= K_k \cD_\alpha\phi^k + {\cA_\alpha}^\g{r}K_k(\gen{r}\phi)^k + K_\kb \cD_\alpha\phib^\kb - {\cA_\alpha}^\g{r}K_\kb(\phib\gen{r})^\kb \\
&= K_k \cD_\alpha\phi^k\,,
\end{split} \label{eq:kahler_superspace_1/2_derivative_k_1}\\
\nonumber\\
\begin{split}
D^\sd{\alpha} K &= K_k D^\sd{\alpha}\phi^k + K_\kb D^\sd{\alpha}\phib^\kb \\
&= K_k \cD^\sd{\alpha}\phi^k + \cA^{\sd{\alpha}\g{r}}K_k(\gen{r}\phi)^k + K_\kb \cD^\sd{\alpha}\phib^\kb - \cA^{\sd{\alpha}\g{r}}K_\kb(\phib\gen{r})^\kb \\
&= K_\kb \cD^\sd{\alpha}\phib^\kb\,.
\end{split} \label{eq:kahler_superspace_1/2_derivative_k_2}
\end{align}
In K\"ahler superspace the former $\U(1)$ connection $A$ is now called K\"ahler connection and the former $U(1)$ weights $w$ are referred to as chiral weights, based on the chiral transformation $F(\phi)$. Note, since the K\"ahler potential has no well defined weight the covariant derivative $\cD_A K$ is not defined.
\\\\
By plugging Eq.~\eqref{eq:kahler_superspace_1/2_kahler_potential_identification} into Eq.~\eqref{eq:kahler_superspace_1/2_connection_w_identification_1} and \eqref{eq:kahler_superspace_1/2_connection_w_identification_2} the spinor components of the connection are expressed in terms of the K\"ahler potential
\begin{align}
A_\alpha (\tfrac{1}{2}) &= + \frac{1}{4}\big(D_\alpha K\big)(\tfrac{1}{2})\,, \label{eq:kahler_superspace_1/2_connection_component_1}\\
A^\sd{\alpha} (\tfrac{1}{2}) &= - \frac{1}{4}\big(D^\sd{\alpha}K\big)(\tfrac{1}{2})\,. \label{eq:kahler_superspace_1/2_connection_component_2}
\end{align}
With the identities \footnote{Because the covariant derivative $\cD_B A_\alpha$ is well defined, namely $A_\alpha$ is considered as the component of a $1$-superform, the covariant derivative $\cD_B(\frac{1}{2})\big(D_\alpha K\big)(\frac{1}{2})$ is defined as well, according to the identification in Eq.~\eqref{eq:kahler_superspace_1/2_connection_component_1}. In order to write $\cD_B(\frac{1}{2})\big(D_\alpha K\big)(\frac{1}{2})$ in terms of covariant derivatives of $\phi^k$ and $\phib^\kb$, as for example in Eq.~\eqref{eq:kahler_superspace_1/2_double_covariant_derivative_kahler_potential_1}, it is convenient to assume that $\cD_B K$ is well defined, where $K$ has chiral weight $0$. This calculation is valid since $K$ is invariant under Yang-Mills transformations (cf.\ Eq.~\eqref{eq:kahler_superspace_1/2_invariance_kahler_potential}). The same considerations apply to $D^\sd{\alpha}K$.}
\begin{align}
\cD_\sd{\alpha}(\tfrac{1}{2})\big(D_\alpha K\big)(\tfrac{1}{2}) &= g_{k\kb} \cD_\sd{\alpha}(\tfrac{1}{2})\phib^\kb \cD_\alpha(\tfrac{1}{2})\phi^k - 2i\sig^a_{\alpha\sd{\alpha}}K_k\cD_a\phi^k\,, \label{eq:kahler_superspace_1/2_double_covariant_derivative_kahler_potential_1}\\
\cD_\alpha(\tfrac{1}{2})\big(D_\sd{\alpha}K\big)(\tfrac{1}{2}) &= g_{k\kb} \cD_\alpha(\tfrac{1}{2})\phi^k \cD_\sd{\alpha}(\tfrac{1}{2})\phib^\kb - 2i\sig^a_{\alpha\sd{\alpha}}K_\kb\cD_a\phib^\kb\,, \label{eq:kahler_superspace_1/2_double_covariant_derivative_kahler_potential_2}
\end{align}
follows from Eq.~\eqref{eq:kahler_superspace_1/2_connection_component_1}, \eqref{eq:kahler_superspace_1/2_connection_component_2}, \eqref{eq:superspace_u1_field_strength_components} and \eqref{eq:bianchi_sp_components_u1_fieldstrength_dim1_2} that
\begin{align}
A_a(\tfrac{1}{2}) &= \frac{1}{4}(K_k\cD_a\phi^k - K_\kb\cD_a\phib^\kb) + \frac{3i}{2}G_a + \frac{i}{8} \sigb_a^{\sd{\alpha}\alpha} g_{k\kb} \cD_\alpha(\tfrac{1}{2})\phi^k \cD_\sd{\alpha}(\tfrac{1}{2})\phib^\kb\,. \label{eq:kahler_superspace_1/2_connection_component_3}
\end{align}
Note that in Eq.~\eqref{eq:kahler_superspace_1/2_double_covariant_derivative_kahler_potential_1} the notation $\cD_\sd{\alpha}(\tfrac{1}{2})\cD_\alpha(\tfrac{1}{2})K=(\cD_\sd{\alpha}\cD_\alpha K)(\tfrac{1}{2})$ is used, and that $\cD_a(\tfrac{1}{2})=\cD_a$ and $G_a(\tfrac{1}{2})=G_a$. The K\"ahler connection in the $(\frac{1}{2},\frac{1}{2})$ basis is then written as \footnote{The connection $A(\frac{1}{2})$ corresponds to the connection $\At$ in Eq.~\eqref{eq:yang_mills_introduction_u1_connection} at the superfield level and with covariant derivatives with respect to the Yang-Mills gauge group. The extra term in Eq.~\eqref{eq:kahler_superspace_1/2_u1_connection} may be absorbed by a redefinition of the vector component $A_a(\frac{1}{2})$.}
\begin{align}
A(\tfrac{1}{2}) &= \frac{1}{4}(K_k\cD\phi^k - K_\kb\cD\phib^\kb) + \frac{i}{8} E^a\big(12G_a + \sigb_a^{\sd{\alpha}\alpha} g_{k\kb} \cD_\alpha(\tfrac{1}{2})\phi^k \cD_\sd{\alpha}(\tfrac{1}{2})\phib^\kb\big)\,. \label{eq:kahler_superspace_1/2_u1_connection}
\end{align}
Furthermore, according to Eq.~\eqref{eq:kahler_superspace_phi_transformation_ab_basis_1} and \eqref{eq:kahler_superspace_connection_transformation_ab_basis_1} the superfield $\Phi(\frac{1}{2})$ and the K\"ahler connection $A(\frac{1}{2})$ transform under K\"ahler transformations as
\begin{align}
\Phi(\tfrac{1}{2}) &\mapsto e^{-\frac{i}{2}w(\Phi)\im{F}} \Phi(\tfrac{1}{2})\,, \label{eq:kahler_superspace_1/2_phi_u1_transformation}\\
A(\tfrac{1}{2}) &\mapsto A(\tfrac{1}{2}) + \frac{i}{2}\exd \im{F}\,, \label{eq:kahler_superspace_1/2_connection_u1_transformation}
\end{align}
where $\im{F}=\frac{1}{2i}(F-\Fb)$ is the imaginary part of $F$. Finally, Eq.~\eqref{eq:kahler_superspace_1/2_phi_u1_transformation} implies the transformation properties under K\"ahler transformations of the quantities listed in Eqs.~\eqref{eq:kahler_superspace_1/2_vielbein}--\eqref{eq:kahler_superspace_1/2_ym_fieldstrength_fields}.
\\\\
In the remainder of this section it is argued that all the identities stated in Section~\ref{sec:bianchi_sp_solution_all} and \ref{sec:matter_and_yang_mills} in $\U(1)$ superspace, which were derived by solving the Bianchi identities in $\U(1)$ superspace and in the Yang-Mills sector, can be taken up literally in $\UK(1)$ superspace. First, it is shown that the torsion constraints in Eq.~\eqref{eq:bianchi_sp_torsion_constraints} and the constraints of the Yang-Mill field strength in Eq.~\eqref{eq:yang_mills_constraints_compact} have the same form in K\"ahler superspace. Because the components of the torsion tensor have the chiral weights $w({T_{CB}}^A)=w(E^A)-w(E^B)-w(E^C)$, in the $(\frac{1}{2},\frac{1}{2})$-basis they are written as
\begin{align}
{T_{CB}}^A(\tfrac{1}{2}) &= (T\,\Tb^{-1})^{-\frac{1}{2}(w(E^A)-w(E^B)-w(E^C))} {T_{CB}}^A\,. \label{eq:kahler_superspace_1/2_torsion_components}
\end{align}
Thus, all constraints of the form ${T_{CB}}^A=0$ for some $A,B,C$ are equivalent to ${T_{CB}}^A(\tfrac{1}{2})=0$, and since $w({T_{\gamma\sd{\beta}}}^a)=0$, Eq.~\eqref{eq:kahler_superspace_1/2_torsion_components} implies \hbox{${T_{\gamma\sd{\beta}}}^a(\tfrac{1}{2}) = {T_{\gamma\sd{\beta}}}^a=2i\sig^a_{\gamma\sd{\beta}}$}. A similar consideration applies to the Yang-Mills field strength. The chiral weights of the components are given by $w(\cF_{BA})=-w(E^A)-w(E^B)$, thus
\begin{align}
\cF_{BA}(\tfrac{1}{2}) &= (T\,\Tb^{-1})^{+\frac{1}{2}(w(E^A)+w(E^B))} \cF_{BA}\,, \label{eq:kahler_superspace_1/2_ym_connection_components}
\end{align}
which implies that the constraint $\cF_{\su{\beta}\su{\alpha}}=0$ holds true in K\"ahler superspace. Because the covariant derivative is compatible with the $(\frac{1}{2},\frac{1}{2})$-basis as shown in Eq.~\eqref{eq:kahler_superspace_phi_covariant_derivative_components_ab_basis} and \eqref{eq:kahler_superspace_phi_double_covariant_derivative_components_ab_basis}, the Bianchi identities stated in Eq.~\eqref{eq:bianchi_sp_bianchi_identity_components} and \eqref{eq:yang_mills_bianchi_identity_components_solution} have the same form in K\"ahler superspace. In addition, since the constraints of the torsion tensor and the Yang-Mills field strength have the same form in K\"ahler superspace too, as stated above, the Bianchi identities lead to the same identities as listed in Section~\ref{sec:summary_tensor_components_further_relations} and \ref{sec:summary_ym_fieldstrength_components_derivative_relations}.
\\\\
In the following sections $\UK(1)$ superspace is considered. In order to avoid clutter in the calculations the label $(\frac{1}{2})$ is not written.

\subsubsection{Further definitions and identities in K\"ahler superspace}
Having defined K\"ahler superspace, it is appropriate at this point to express the superfields $X_\alpha$, $X^\sd{\alpha}$ and $-\frac{1}{2}\cD^\alpha X_\alpha$ in terms of the matter superfields in an appropriate way. These expressions are used later in Section~\ref{sec:component_fields} in the derivation of the Lagrangian of the supergravity/matter/Yang-Mills system at the component field level. Note that the label $(\tfrac{1}{2})$ is neglected in the following calculations.
\\\\
In the calculation of $-\frac{1}{2}\cD^\alpha X_\alpha$ the superfields $F^k$ and $\Fb^\kb$ will appear, which are defined as
\begin{align}
F^k &:= -\frac{1}{4}\cDt^\alpha\cD_\alpha\phi^k\,, \label{eq:kahler_superspace_1/2_fterm_1}\\
\Fb^\kb &:= -\frac{1}{4}\cDt_\sd{\alpha}\cD^\sd{\alpha}\phib^\kb\,, \label{eq:kahler_superspace_1/2_fterm_2}
\end{align}
where the covariant derivatives with the tilde are given by
\begin{align}
\cDt_B\cD_A\phi^k &:= \cD_B\cD_A\phi^k + {\Gamma^k}_{ij} \cD_B \phi^i \cD_A \phi^j\,, \label{eq:kahler_superspace_1/2_covariant_derivative_isometry_1}\\
\cDt_B\cD_A\phib^\kb &:= \cD_B\cD_A\phib^\kb + {\Gamma^\kb}_{\ib\jb} \cD_B \phib^\ib \cD_A \phib^\jb\,. \label{eq:kahler_superspace_1/2_covariant_derivative_isometry_2}
\end{align}
The superfields
\begin{align}
{\Gamma^k}_{ij} &= g^{\kb k}g_{i\kb,j}\,, \\
{\Gamma^\kb}_{\ib\jb} &= g^{\kb k}g_{k\ib,\jb}\,,
\end{align}
are the (non-vanishing) Christoffel symbols of the Levi-Civita connection of the metric $g_{k\kb}$. This means, the derivative $\cDt_B$ of $\cD_A\phi^k$ and $\cD_A\phib^\kb$ is covariant with respect to the gauge symmetries, as well as with respect to the (ungauged) Levi-Civita connection of the K\"ahler manifold. Furthermore, the corresponding Riemann tensor will be used, which is given by
\begin{align}
R_{k\kb j\jb} = g_{k\kb,j\jb} - g^{\lb l}\,g_{k\lb,j}\,g_{l\kb,\jb}\,.
\end{align}
\begin{itemize}
\item\textbf{Superfields $X_\alpha$ and $X^\sd{\alpha}$:}\\
Plugging Eq.~\eqref{eq:bianchi_sp_components_u1_fieldstrength_dim1_2} into \eqref{eq:superspace_u1_field_strength_components}, it follows that
\begin{align}
A_{\alpha\sd{\alpha}} &= \frac{3i}{2}G_{\alpha\sd{\alpha}} + \frac{i}{2}(\cD_\alpha A_\sd{\alpha} + \cD_\sd{\alpha} A_\alpha)\,,
\end{align}
which, again by using Eq.~\eqref{eq:superspace_u1_field_strength_components}, implies
\begin{align}
\begin{split}
F_{\sd{\beta}\s\alpha\sd{\alpha}} &= \frac{i}{2}(\cD_\sd{\beta}\cD_\alpha A_\sd{\alpha} + \cD_\sd{\beta}\cD_\sd{\alpha} A_\alpha) - \frac{i}{2}\acom{\cD_\alpha}{\cD_\sd{\alpha}} A_\sd{\beta} + 2i\eps_{\sd{\beta}\sd{\alpha}} {G_\alpha}^\sd{\delta} A_\sd{\delta} \\
&\quad + \frac{3i}{2} \cD_\sd{\beta} G_{\alpha\sd{\alpha}} + 2i\eps_{\sd{\beta}\sd{\alpha}} R\,A_\alpha\,,
\end{split} \\
\nonumber\\
\begin{split}
F_{\beta\s\alpha\sd{\alpha}} &= \frac{i}{2}(\cD_\beta\cD_\alpha A_\sd{\alpha} + \cD_\beta\cD_\sd{\alpha} A_\alpha) - \frac{i}{2}\acom{\cD_\alpha}{\cD_\sd{\alpha}} A_\beta - 2i\eps_{\beta\alpha} {G^\delta}_\sd{\alpha} A_\delta \\
&\quad + \frac{3i}{2} \cD_\beta G_{\alpha\sd{\alpha}} - 2i\eps_{\beta\alpha} \Rb A_\sd{\alpha}\,.
\end{split}
\end{align}
On the other hand, according to Eq.~\eqref{eq:bianchi_sp_components_u1_fieldstrength_dim3/2_1} and \eqref{eq:bianchi_sp_components_u1_fieldstrength_dim3/2_2} these components of the K\"ahler field strength have the form
\begin{align}
F_{\sd{\beta}\s\alpha\sd{\alpha}} &= \frac{3i}{2} \cD_\sd{\beta} G_{\alpha\sd{\alpha}} + i\eps_{\sd{\beta}\sd{\alpha}} X_\alpha\,, \\
F_{\beta\s\alpha\sd{\alpha}} &= \frac{3i}{2} \cD_\beta G_{\alpha\sd{\alpha}} + i\eps_{\beta\alpha} X_\sd{\alpha}\,,
\end{align}
thus the superfields $X_\alpha$ and $X^\sd{\alpha}$ are expressed as
\begin{align}
X_\alpha &= -\frac{1}{4}\cDb^2 A_\alpha + \frac{1}{4}\cD_\alpha\cD^\sd{\alpha} A_\sd{\alpha} + \frac{1}{2}\cD^\sd{\alpha}\cD_\alpha A_\sd{\alpha} + 2{G_\alpha}^\sd{\alpha} A_\sd{\alpha} + 2R\,A_\alpha\,, \label{eq:kahler_superspace_1/2_x_1}\\
X^\sd{\alpha} &= +\frac{1}{4}\cD^2 A^\sd{\alpha} + \frac{1}{4}\cD^\sd{\alpha}\cD^\alpha A_\alpha + \frac{1}{2}\cD^\alpha\cD^\sd{\alpha} A_\alpha - 2G^{\alpha\sd{\alpha}} A_\alpha - 2\Rb A^\sd{\alpha}\,. \label{eq:kahler_superspace_1/2_x_2}
\end{align}
Writing the spinor components of the K\"ahler connection in terms of the K\"ahler potential as in Eq.~\eqref{eq:kahler_superspace_1/2_connection_component_1} and \eqref{eq:kahler_superspace_1/2_connection_component_2}, the two superfields have the simple form.
\begin{align}
X_\alpha &= -\frac{1}{8}(\cDb^2-8R) D_\alpha K\,, \label{eq:kahler_superspace_1/2_x_3}\\
X^\sd{\alpha} &= -\frac{1}{8}(\cD^2-8\Rb) D^\sd{\alpha}K\,, \label{eq:kahler_superspace_1/2_x_4}
\end{align}
where $(\cDb^2-8R)$ is the chiral and $(\cD^2-8\Rb)$ the antichiral projection operator, as in Eq.~\eqref{eq:kahler_superspace_chiral_projection_operator} and \eqref{eq:kahler_superspace_antichiral_projection_operator}. This is consistent with $\cD^\sd{\alpha}X_\alpha=0$ and $\cD_\alpha X^\sd{\alpha}=0$ in Eq.~\eqref{eq:bianchi_sp_x_chirality_condition}. An expansion of the terms in Eq.~\eqref{eq:kahler_superspace_1/2_x_3} and \eqref{eq:kahler_superspace_1/2_x_4} finally leads to the desired expressions for $X_\alpha$ and $X^\sd{\alpha}$:
\begin{align}
X_\alpha &= -\frac{i}{2} g_{k\kb} \sig^a_{\alpha\sd{\alpha}} \cD_a\phi^k \cD^\sd{\alpha} \phib^\kb + \frac{1}{2} g_{k\kb} \cD_\alpha\phi^k\Fb^\kb + \cW_\alpha^\g{r}G_\g{r}\,, \label{eq:kahler_superspace_1/2_x_5}\\
X^\sd{\alpha} &= -\frac{i}{2} g_{k\kb} \sigb^{a\sd{\alpha}\alpha} \cD_a\phib^\kb \cD_\alpha \phi^k + \frac{1}{2} g_{k\kb} \cD^\sd{\alpha}\phib^\kb F^k + \cW^{\g{r}\sd{\alpha}}G_\g{r}\,, \label{eq:kahler_superspace_1/2_x_6}
\end{align}
where $G_\g{r}$ is the Killing potential, as defined in Eq.~\eqref{eq:kahler_superspace_1/2_killing_potential}.
\item\textbf{Superfield $-\frac{1}{2}\cD^\alpha X_\alpha$:}\\
For a superfield $\Phi$ with $\U(1)$ weight $w(\Phi)$ and which transform in some representation of the Yang-Mills gauge group the identities
\begin{align}
\acom{\cD_\alpha}{\cDb^2}\Phi &= -2\cD_\sd{\alpha}\cD_\alpha\cD^\sd{\alpha}\Phi - 8G_{\alpha\sd{\alpha}}\cD^\sd{\alpha}\Phi + 8R\,\cD_\alpha\Phi - 4w(\Phi)X_\alpha\Phi - 8(\Phi\cdot\cW_\alpha)\,, \label{eq:kahler_superspace_1/2_identity_general_fields_1}\\
\acom{\cD^\sd{\alpha}}{\cD^2}\Phi &= -2\cD^\alpha\cD^\sd{\alpha}\cD_\alpha\Phi + 8G^{\alpha\sd{\alpha}}\cD_\alpha\Phi + 8\Rb\,\cD^\sd{\alpha}\Phi + 4w(\Phi)X^\sd{\alpha}\Phi + 8(\Phi\cdot\cW^\sd{\alpha})\,. \label{eq:kahler_superspace_1/2_identity_general_fields_2}
\end{align}
are used in the calculation of $-\frac{1}{2}\cD^\alpha X_\alpha$. If, for example, $\Phi=\phi,\phib$ represents a matter superfield, the two identities get simplified because of the chirality conditions and the vanishing $\U(1)$ weight, and the term $(\Phi\cdot\cW_\su{\alpha})$ takes the form
\begin{align}
(\phi\cdot\cW_\su{\alpha}) &= -({\cW_\su{\alpha}}^\g{r}\phi) = -(\gen{r}\phi){\cW_\su{\alpha}}^\g{r}\,, \\
(\phib\cdot\cW_\su{\alpha}) &= +(\phib{\cW_\su{\alpha}}^\g{r}) = +(\phib\gen{r}){\cW_\su{\alpha}}^\g{r}\,.
\end{align}
These considerations finally lead to the expression 
\begin{align}
\begin{split}
-\frac{1}{2}\cD^\alpha X_\alpha &= -g_{k\kb} \cD_a\phi^k \cD^a\phib^\kb - \frac{i}{4}g_{k\kb} \cD^\alpha\phi^k \cDt_{\alpha\sd{\alpha}} \cD^\sd{\alpha}\phib^\kb -\frac{i}{4}g_{k\kb} \cD^\sd{\alpha}\phib^\kb \cDt_{\alpha\sd{\alpha}} \cD^\alpha\phi^k \\
&\quad + g_{k\kb} F^k \Fb^\kb + \frac{1}{16} R_{k\kb j\jb} \cD^\alpha\phi^k \cD_\alpha\phi^j \cD_\sd{\alpha}\phib^\kb \cD^\sd{\alpha}\phib^\jb \\
&\quad + g_{k\kb} (\phib\gen{r})^\kb \cW^{\alpha\g{r}}\cD_\alpha\phi^k + g_{k\kb} (\gen{r}\phi)^k {\cW_\sd{\alpha}}^\g{r}\cD^\sd{\alpha}\phib^\kb \\
&\quad - \frac{1}{2} \cD^\alpha {\cW_\alpha}^\g{r} G_\g{r}\,.
\end{split} \label{eq:kahler_superspace_1/2_x_7}
\end{align}
Note that the term in Eq.~\eqref{eq:kahler_superspace_1/2_x_7} is real, and that $-\frac{1}{2}\cD^\alpha X_\alpha = -\frac{1}{2}\cD_\sd{\alpha} X^\sd{\alpha}$, as expected from Eq.~\eqref{eq:bianchi_sp_x_derivative_relation_1}.
\end{itemize}

\subsection{Supergravity transformations}
\label{sec:sugra_transformations_superfield_level}
Supergravity transformations are a subset of (infinitesimal) coordinate dependent transformations
\renewcommand{\arraystretch}{1.3}
\begin{flalign}
\quad & \begin{array}{lll}
\bullet & \xi^M & \text{(local) superspace diffeomorphism,} \\
\bullet & {\soae_B}^A & \text{Lorentz transformation,} \\
\bullet & F(\phi) & \text{K\"ahler transformation,} \\
\bullet & \lac{r} & \text{Yang-Mills transformation,}
\end{array} & \label{eq:sugra_transf_transformations}
\end{flalign}\\
\renewcommand{\arraystretch}{1}
in $\UK(1)$ superspace, where $\xi=\xi^M\del_M$ is a real and even supervector field, ${\soae_B}^A\in\so(1,3)$ and $i\lac{r}\gen{r}\in\mathfrak{g}$. Note that the infinitesimal action of the local diffeomorphism is given by the Lie derivative $\lied_\xi$. Under these transformations, the infinitesimal change of a $p$-superform $\chi^A$ with Lorentz index $A$, chiral weight $w(\chi^A)$, and which transforms in any representation of the Yang-Mills gauge group, has the form
\begin{align}
\chi^A &\mapsto \chi^A + \delta\chi^A\,, \label{eq:sugra_transf_general_delta_pform_short}
\end{align}
with
\begin{align}
\begin{split}
\delta\chi^A &= \lied_\xi\chi^A - \frac{i}{2}w(\chi^A)\chi^A\im{F} + \chi^B{\soae_B}^A + i\lac{r}(\chi^A\cdot\gen{r}) \\
&= (\intp_\xi\cD + \cD\intp_\xi)\chi^A + \chi^B({\soae_B}^A - \intp_\xi{\Omega_B}^A) - w(\chi^A)\chi^A(\intp_\xi A + \frac{i}{2}\im F) \\
&\quad + i(\lac{r} + i\intp_\xi\cA^\g{r})(\chi^A\cdot\gen{r})\,,
\end{split} \label{eq:sugra_transf_general_delta_pform}
\end{align}
where
\begin{align}
\intp_\xi A = \frac{1}{4}(K_k\intp_\xi\cD\phi^k - K_\kb\intp_\xi\cD\phib^\kb) + \frac{i}{8} \xi^a(12G_a + \sigb_a^{\sd{\alpha}\alpha} g_{k\kb} \cD_\alpha\phi^k \cD_\sd{\alpha}\phib^\kb)\,. \label{eq:sugra_transf_interior_product_u1connection}
\end{align}
Supergravity transformations $\dsg$ are those transformations, where the terms containing the Lorentz and the Yang-Mills connection in Eq.~\eqref{eq:sugra_transf_general_delta_pform} are cancelled by a compensating Lorentz and Yang-Mills transformation, namely
\begin{align}
{\soae_B}^A &= \intp_\xi{\Omega_B}^A\,, \label{eq:sugra_transf_compensating_lorentz}\\
\alpha^\g{r} &= -i\intp_\xi \cA^\g{r}\,. \label{eq:sugra_transf_compensating_ym}
\end{align}
Furthermore, in a supergravity transformation the K\"ahler transformation is equal to zero:
\begin{align}
F(\phi) &= 0\,. \label{eq:sugra_transf_compensating_kahler}
\end{align}
The components of the superspace diffeomorphism are conveniently written in the form $\xi^A=(\xi^a,\xi^\alpha,\xib_\sd{\alpha})$ to explicitly indicate that $(\xi^\alpha)^\cc = \xib^\sd{\alpha}$. Strictly speaking the definition of a supergravity transformation contains the additional condition that the lowest superspace component of $\xi^a$  vanishes, i.e.\ $\xi^a\p=0$ (cf.\ Section~\ref{sec:component_fields_definition_projection}). If $\xi$ is an arbitrary real and even supervector field, the transformation is called supergauge transformation. This distinction only becomes important in Section~\ref{sec:component_fields_sugra_transformations}, where the supergravity transformation at the component field level is considered. Because the K\"ahler transformation $F$ is a holomorphic function of the matter superfields $\phi^k$, in general there exists no K\"ahler transformation, such that the term $\intp_\xi A$ gets cancelled by $\frac{i}{2}\im F$. The supergravity transformation of $\chi^A$ is then given by
\begin{align}
\dsg_\xi\chi^A &= \clied_\xi\chi^A - w(\chi^A)\chi^A\intp_\xi A\,, \label{eq:sugra_transf_delta_pform}
\end{align}
where the covariant Lie derivative $\clied_\xi$ is defined as
\begin{align}
\clied_\xi &= \intp_\xi\cD + \cD\intp_\xi\,. \label{eq:sugra_transf_covariant_liederivative}
\end{align}
The identity in Eq.~\eqref{eq:sugra_transf_delta_pform} shows that the supergravity transformation is completely specified by the superspace diffeomorphism $\xi$. Moreover, the supergravity transformation is covariant with respect to the Lorentz and the Yang-Mills gauge group. Hence, with an appropriate definition of the component fields of the supermultiplets (cf.\ Section~\ref{sec:component_fields}), the Wess-Zumino gauge is implemented in a geometric manner. Supergravity transformations therefore correspond to local supersymmetry transformations adapted to the Wess-Zumino gauge and are also called Wess-Zumino transformations.\footnote{Pure local supersymmetry transformations correspond to local superspace diffeomorphisms, where the infinitesimal action is given by the Lie derivative $\lied_\xi$.} Note, Eq.~\eqref{eq:sugra_transf_delta_pform} has the same form for superforms with an arbitrary number of upper and lower Lorentz indices.
\\\\
Particular cases of Eq.~\eqref{eq:sugra_transf_delta_pform} are the supergravity transformations of the supervielbein and the matter superfields:
\begin{align}
\dsg_\xi E^A &= \cD\xi^A + \intp_\xi T^A - w(E^A)E^A\intp_\xi A\,, \label{eq:sugra_transf_delta_vielbein}\\
\nonumber\\
\dsg_\xi \phi^k &= \intp_\xi \cD\phi^k\,, \label{eq:sugra_transf_delta_phi}\\
\nonumber\\
\dsg_\xi \phib^\kb &= \intp_\xi \cD\phib^\kb\,. \label{eq:sugra_transf_delta_phib}
\end{align}
The supergravity transformations of the Lorentz and the Yang-Mills connection are calculated in a similar way. The changes under the general transformations of Eq.~\eqref{eq:sugra_transf_transformations} are given by
\begin{align}
\begin{split}
\delta{\Omega_B}^A &= \lied_\xi{\Omega_B}^A + {\com{\Omega}{\soae}_B}^A - \exd{\soae_B}^A \\
&= \intp_\xi {R_B}^A + {\com{\Omega}{\soae-\intp_\xi\Omega}_B}^A - \exd({\soae_B}^A-{\intp_\xi\Omega_B}^A)\,,
\end{split} \label{eq:sugra_transf_general_delta_lorentzconnection}
\\\nonumber\\
\begin{split}
\delta\cA^\g{r} &= \lied_\xi \cA^\g{r} + \lac{p}\cA^\g{q} \sco{p}{q}{r} - i\exd\alpha^\g{r} \\
&= \intp_\xi \cF^\g{r} + (\lac{p} + i\intp_\xi\cA^\g{p})\cA^\g{q}\sco{p}{q}{r} - i\exd(\lac{r} + i\intp_\xi\cA^\g{r})\,.
\end{split} \label{eq:sugra_transf_general_delta_ymconnection}
\end{align}
Therefore, using Eqs.~\eqref{eq:sugra_transf_compensating_lorentz}--\eqref{eq:sugra_transf_compensating_kahler}, the supergravity transformations of the connections are given by
\begin{align}
\dsg_\xi{\Omega_B}^A &= \intp_\xi {R_B}^A\,, \label{eq:sugra_transf_delta_lorentzconnection}
\\\nonumber\\
\dsg_\xi\cA^\g{r} &= \intp_\xi \cF^\g{r}\,. \label{eq:sugra_transf_delta_ymconnection}
\end{align}
Again, these transformations are covariant with respect to the Lorentz and the Yang-Mills gauge group.
\\\\
For two infinitesimal superspace diffeomorphisms $\xi_1$ and $\xi_2$ the commutation relation of the corresponding covariant Lie derivatives has the form
\begin{align}
\com{\clied_{\xi_1}}{\clied_{\xi_2}}\chi^A &= \clied_{\com{\xi_1}{\xi_2}}\chi^A + \chi^B\intp_{\xi_2}\intp_{\xi_1}{R_B}^A + w(\chi^A)\chi^A\intp_{\xi_2}\intp_{\xi_1}\exd A + \intp_{\xi_2}\intp_{\xi_1}\cF^\g{r}(\chi^A\cdot\gen{r})\,.
\end{align}
This implies that the commutator of two supergravity transformations is given by
\begin{align}
\com{\dsg_{\xi_1}}{\dsg_{\xi_2}}\chi^A &= \dsg_{\com{\xi_1}{\xi_2}}\chi^A + \chi^B\intp_{\xi_2}\intp_{\xi_1}{R_B}^A + \intp_{\xi_2}\intp_{\xi_1}\cF^\g{r}(\chi^A\cdot\gen{r})\,. \label{eq:sugra_transf_delta_pform_commutator}
\end{align}
Thus, the commutator of two supergravity transformations is again a supergravity transformation plus a Lorentz transformation ${\soae_B}^A=\intp_{\xi_2}\intp_{\xi_1}{R_B}^A$ and a Yang-Mills transformations $\lac{r}=-i\intp_{\xi_2}\intp_{\xi_1}\cF^\g{r}$, which are both field dependent. The additional gauge transformations are present, because supergravity transformations correspond to local supersymmetry transformations, adapted to the Wess-Zumino gauge. The same conclusion applies to the Lorentz and the Yang-Mills connection:
\begin{align}
\com{\dsg_{\xi_1}}{\dsg_{\xi_2}}{\Omega_B}^A &= \dsg_{\com{\xi_1}{\xi_2}}{\Omega_B}^A + {\com{\Omega}{\intp_{\xi_2}\intp_{\xi_1}R}_B}^A - \exd\intp_{\xi_2}\intp_{\xi_1}{R_B}^A\,, \\
\nonumber\\
\com{\dsg_{\xi_1}}{\dsg_{\xi_2}}\cA^\g{r} &= \dsg_{\com{\xi_1}{\xi_2}}\cA^\g{r} - i\intp_{\xi_2}\intp_{\xi_1}\cF^\g{p}\cA^\g{q}\sco{p}{q}{r} - \exd\intp_{\xi_2}\intp_{\xi_1}\cF^\g{r}\,.
\end{align}
Because of the covariant definition of the supergravity transformation, it is straight forward to show that the commutator of a supergravity transformation $\dsg_\xi$ with a Lorentz transformation $\delta_L$ and a Yang-Mills transformation $\delta_\alpha$, respectively, vanishes:
\begin{align}
\com{\dsg_\xi}{\delta_L}\chi^A &= 0\,, \label{eq:sugra_transf_delta_pform_commutator_lorentz}\\
\com{\dsg_\xi}{\delta_\alpha}\chi^A &= 0\,. \label{eq:sugra_transf_delta_pform_commutator_yangmills}
\end{align}
Furthermore, the commutator of a supergravity transformation $\dsg_\xi$ and a K\"ahler transformation $\delta_F$ is given by
\begin{align}
\com{\dsg_\xi}{\delta_F}\chi^A &= +\frac{i}{2} w(\chi^A)\chi^A\intp_\xi\exd\im{F}\,, \label{eq:sugra_transf_delta_pform_commutator_kahler}
\end{align}
where $F(\phi^k) + \intp_\xi\exd F(\phi^k)=F(\phi^k+\dsg_\xi\phi^k)$ is the same K\"ahler transformation as a function of the transformed matter superfields. Together with Eq.~\eqref{eq:sugra_transf_delta_pform_commutator} this shows that the algebra spanned by supergravity, Lorentz, K\"ahler and Yang-Mills transformations closes. Note, if $\xi$ is not infinitesimal, the corresponding finite supergravity transformation has the form $\Dsg_\xi=\exp(\dsg_\xi)$, and the commutation relation in Eq.~\eqref{eq:sugra_transf_delta_pform_commutator} is then used to calculate compositions of finite supergravity transformations.
\\\\
Chiral and antichiral superfields retain their property under transformations as in Eq.~\eqref{eq:sugra_transf_transformations}, and consequently also under supergravity transformations. This applies because the chirality conditions are specified with respect to the basis of the supervielbein, which transforms as well. In particular, the covariant derivative of a superfield $\Phi$ transforms as follows
\begin{align}
\begin{split}
E^A\cD_A\Phi &\mapsto E^A\cD_A\Phi + \delta(E^A\cD_A\Phi) \\
&= E^A\cD_A\Phi + (\delta E^A)\cD_A\Phi + E^A(\delta\cD_A\Phi) \\
&= (E^A + \delta E^A)(\cD_A\Phi + \delta\cD_A\Phi)\,,
\end{split}
\end{align}
up to first order in $\delta$.
The chirality conditions for the transformed field $\Phi+\delta\Phi$ are specified with respect to the transformed supervielbein $E^A+\delta E^A$, and are thus given by
\begin{align}
\cD^\sd{\alpha}\Phi + \delta\cD^\sd{\alpha}\Phi &= 0\,, \\
\cD_\alpha\Phi + \delta\cD_\alpha\Phi &= 0\,.
\end{align}
The two conditions are obviously fulfilled for a chiral superfield, i.e.\ $\cD^\sd{\alpha}\Phi=0$, and an antichiral superfield, i.e.\ $\cD_\alpha\Phi=0$, respectively.

\subsubsection{Further identities}
\label{sec:sugra_transformations_superfield_level_identities}
As a preparation for the calculation of the supergravity transformations at the component field level in Section~\ref{sec:component_fields_sugra_transformations} and \ref{sec:component_fields_actions}, Eqs.~\eqref{eq:sugra_transf_delta_vielbein}--\eqref{eq:sugra_transf_delta_phib} and \eqref{eq:sugra_transf_delta_ymconnection} are stated in a more explicit form and the supergravity transformations for further quantities are determined in the following.
\begin{itemize}
\item \textbf{Supergravity sector:}\\
In terms of superspace indices Eq.~\eqref{eq:sugra_transf_delta_vielbein} is written as
\begin{align}
\dsg_\xi{E_M}^A &= \cD_M\xi^A + {E_M}^B\xi^C {T_{CB}}^A - w(E^A){E_M}^A\intp_\xi A\,. \label{eq:sugra_transf_delta_vielbein_explicit}
\end{align}
Furthermore, under a transformation as stated in Eq.~\eqref{eq:sugra_transf_transformations} the superfields $R$ and $\Rb$ transform as
\begin{align}
\begin{split}
\delta R &= \lied_\xi R - iR\im{F} \\
&= \intp_\xi\cD R - 2R(\intp_\xi A + \frac{i}{2}\im{F})\,,
\end{split} \\
\nonumber\\
\begin{split}
\delta \Rb &= \lied_\xi\Rb + i\Rb\im{F} \\
&= \intp_\xi\cD\Rb + 2\Rb(\intp_\xi A + \frac{i}{2}\im{F})\,.
\end{split}
\end{align}
With the identities (cf.\ Eq.~\eqref{eq:bianchi_sp_components_torsion_dim3/2_4}, \eqref{eq:bianchi_sp_components_torsion_dim3/2_5}, \eqref{eq:bianchi_sp_derivative_relation_id_3} and \eqref{eq:bianchi_sp_derivative_relation_id_4})
\begin{align}
\cD_\alpha R &= -\frac{1}{3}X_\alpha - \frac{2}{3}{(\sig^{cb})_\alpha}^\varphi T_{cb\varphi}\,, \\
\cD^\sd{\alpha}\Rb &= -\frac{1}{3}X^\sd{\alpha} - \frac{2}{3}{(\sigb^{cb})^\sd{\alpha}}_\sd{\varphi}{T_{cb}}^\sd{\varphi}\,,
\end{align}
and the chirality conditions for $R$ and $\Rb$ follows, that the supergravity transformations have the form
\begin{align}
\dsg_\xi R &= \xi^a\cD_a R - \frac{1}{3}\xi^\alpha X_\alpha - \frac{2}{3}(\xi\sig^{cb})^\alpha T_{cb\alpha}\,, \label{eq:sugra_transf_delta_r_explicit}\\
\dsg_\xi \Rb &= \xi^a\cD_a\Rb - \frac{1}{3}\xib_\sd{\alpha}X^\sd{\alpha} - \frac{2}{3}(\xi\sigb^{cb})_\sd{\alpha}{T_{cb}}^\sd{\alpha}\,. \label{eq:sugra_transf_delta_rb_explicit}
\end{align}
For the superfield $G_a$, the general transformation of Eq.~\eqref{eq:sugra_transf_transformations} reads
\begin{align}
\begin{split}
\delta G_a &= \lied_\xi G_a - {\soae_a}^b G_b \\
&= \intp_\xi\cD G_a - ({\soae_a}^b - \intp_\xi{\Omega_a}^b)\,.
\end{split}
\end{align}
Using the identities (cf.\ Eq.~\eqref{eq:bianchi_sp_components_torsion_dim3/2_4}, \eqref{eq:bianchi_sp_components_torsion_dim3/2_5}, \eqref{eq:bianchi_sp_derivative_relation_id_5} and \eqref{eq:bianchi_sp_derivative_relation_id_6})
\begin{align}
\cD_\beta G_a &= +\frac{1}{2}(\sig^{cb}\sig_a)_{\beta\sd{\alpha}}{T_{cb}}^\sd{\alpha} - \frac{1}{6}(\sig_a\sigb^{cb})_{\beta\sd{\alpha}}{T_{cb}}^\sd{\alpha} - \frac{1}{3}\sig_{a\beta\sd{\alpha}}X^\sd{\alpha}\,, \\
\cD^\sd{\beta}G_a &= -\frac{1}{2}(\sigb^{cb}\sigb_a)^{\sd{\beta}\alpha}T_{cb\alpha} + \frac{1}{6}(\sigb_a\sig^{cb})^{\sd{\beta}\alpha}T_{cb\alpha} + \frac{1}{3}\sigb_a^{\sd{\beta}\alpha}X_\alpha\,,
\end{align}
implies, that the supergravity transformation is given by
\begin{align}
\begin{split}
\dsg_\xi G_a &= \xi^b\cD_b G_a \\
&\quad +\frac{1}{2}(\xi\sig^{cb}\sig_a)_\sd{\alpha}{T_{cb}}^\sd{\alpha} - \frac{1}{6}(\xi\sig_a\sigb^{cb})_\sd{\alpha}{T_{cb}}^\sd{\alpha} - \frac{1}{3}(\xi\sig_a)_\sd{\alpha}X^\sd{\alpha} \\
&\quad -\frac{1}{2}(\xib\sigb^{cb}\sigb_a)^\alpha T_{cb\alpha} + \frac{1}{6}(\xib\sigb_a\sig^{cb})^\alpha T_{cb\alpha} + \frac{1}{3}(\xib\sigb_a)^\alpha X_\alpha\,.
\end{split} \label{eq:sugra_transf_delta_ga_explicit}
\end{align}
\item \textbf{Matter sector:}\\
In a more explicit form Eq.~\eqref{eq:sugra_transf_delta_phi} and \eqref{eq:sugra_transf_delta_phib} are written as
\begin{align}
\dsg_\xi\phi^k &= \xi^b\cD_b\phi^k + \xi^\beta\cD_\beta\phi^k\,, \label{eq:sugra_transf_delta_phi_explicit}\\
\dsg_\xi\phib^\kb &= \xi^b\cD_b\phib^\kb + \xib_\sd{\beta}\cD^\sd{\beta}\phib^\kb\,, \label{eq:sugra_transf_delta_phib_explicit}
\end{align}
taking into account the chirality conditions for the matter superfields. Under an infinitesimal transformation, as stated in Eq.~\eqref{eq:sugra_transf_transformations}, the terms $\cD_\alpha\phi^k$ and $\cD^\sd{\alpha}\phib^\kb$ change as
\begin{align}
\begin{split}
\delta\cD_\alpha\phi^k &= \lied_\xi\cD_\alpha\phi^k - {\soae_\alpha}^\beta\cD_\beta\phi^k + \frac{i}{2}\cD_\alpha\phi^k\im{F} - i\lac{r}(\gen{r}\cD_\alpha\phi)^k \\
&= \intp_\xi\cD\cD_\alpha\phi^k - ({\soae_\alpha}^\beta - \intp_\xi{\Omega_\alpha}^\beta)\cD_\beta\phi^k + \cD_\alpha\phi^k(\intp_\xi A + \frac{i}{2}\im{F}) \\
&\quad - i(\lac{r}+i\intp_\xi\cA^\g{r})(\gen{r}\cD_\alpha\phi)^k\,,
\end{split} \\
\nonumber\\
\begin{split}
\delta\cD^\sd{\alpha}\phib^\kb &= \lied_\xi\cD^\sd{\alpha}\phib^\kb - {\soae^\sd{\alpha}}_\sd{\beta}\cD^\sd{\beta}\phib^\kb - \frac{i}{2}\cD^\sd{\alpha}\phib^\kb\im{F} + i\lac{r}(\cD^\sd{\alpha}\phib\gen{r})^\kb \\
&= \intp_\xi\cD\cD^\sd{\alpha}\phib^\kb - ({\soae^\sd{\alpha}}_\sd{\beta} - \intp_\xi{\Omega^\sd{\alpha}}_\sd{\beta})\cD^\sd{\beta}\phib^\kb - \cD^\sd{\alpha}\phib^\kb(\intp_\xi A + \frac{i}{2}\im{F}) \\
&\quad + i(\lac{r}+i\intp_\xi\cA^\g{r})(\cD^\sd{\alpha}\phib\gen{r})^\kb\,.
\end{split}
\end{align}
By applying the identities
\begin{align}
\cD_\beta\cD_\alpha\phi^k &= \frac{1}{2}\eps_{\beta\alpha}\cD^2\phi^k\,,\\
\cD^\sd{\beta}\cD_\alpha\phi^k &= 2i{(\sigb^a\eps)^\sd{\beta}}_\alpha\cD_a\phi^k\,,\\
\cD^\sd{\beta}\cD^\sd{\alpha}\phib^\kb &= \frac{1}{2}\eps^{\sd{\beta}\sd{\alpha}}\cDb^2\phib^\kb\,,\\
\cD_\beta\cD^\sd{\alpha}\phib^\kb &= 2i{(\sig^a\eps)_\beta}^\sd{\alpha}\cD_a\phib^\kb\,,
\end{align}
the supergravity transformations are given by
\begin{align}
\dsg_\xi\cD_\alpha\phi^k &= \xi^b\cD_b\cD_\alpha\phi^k + 2\xi_\alpha F^k + \frac{1}{2}\xi_\alpha{\Gamma^k}_{ij}\cD^\beta\phi^i\cD_\beta\phi^j + 2i(\xib\sigb^a\eps)_\alpha\cD_a\phi^k + \cD_\alpha\phi^k\intp_\xi A\,, \label{eq:sugra_transf_delta_dphi_explicit}\\
\dsg_\xi\cD^\sd{\alpha}\phib^\kb &= \xi^b\cD_b\cD^\sd{\alpha}\phib^\kb + 2\xib^\sd{\alpha} \Fb^\kb + \frac{1}{2}\xib^\sd{\alpha}{\Gamma^\kb}_{\ib\jb}\cD_\sd{\beta}\phib^\ib\cD^\sd{\beta}\phib^\jb + 2i(\xi\sig^a\eps)^\sd{\alpha}\cD_a\phib^\kb - \cD^\sd{\alpha}\phib^\kb\intp_\xi A\,. \label{eq:sugra_transf_delta_dphib_explicit}
\end{align}
Furthermore, the infinitesimal transformations of the terms $\cDt^\alpha\cD_\alpha\phi^k$ and $\cDt_\sd{\alpha}\cD^\sd{\alpha}\phib^\kb$ are
\begin{align}
\begin{split}
\delta\cDt^\alpha\cD_\alpha\phi^k &= \lied_\xi\cDt^\alpha\cD_\alpha\phi^k + i\cDt^\alpha\cD_\alpha\phi^k\im{F} - i\lac{r}(\gen{r}\cDt^\alpha\cD_\alpha\phi)^k\,, \\
&= \intp_\xi\cD\cDt^\alpha\cD_\alpha\phi^k + 2\cDt^\alpha\cD_\alpha\phi^k(\intp_\xi A +\frac{i}{2}\im{F}) - i(\lac{r}+i\intp_\xi\cA^\g{r})(\gen{r}\cDt^\alpha\cD_\alpha\phi)^k\,,
\end{split} \\
\nonumber\\
\begin{split}
\delta\cDt_\sd{\alpha}\cD^\sd{\alpha}\phib^\kb &= \lied_\xi\cDt_\sd{\alpha}\cD^\sd{\alpha}\phib^\kb - i\cDt_\sd{\alpha}\cD^\sd{\alpha}\phib^\kb\im{F} + i\lac{r}(\cDt_\sd{\alpha}\cD^\sd{\alpha}\phib\gen{r})^\kb\,, \\
&= \intp_\xi\cD\cDt_\sd{\alpha}\cD^\sd{\alpha}\phib^\kb - 2\cDt_\sd{\alpha}\cD^\sd{\alpha}\phib^\kb(\intp_\xi A +\frac{i}{2}\im{F}) + i(\lac{r}+i\intp_\xi\cA^\g{r})(\cDt_\sd{\alpha}\cD^\sd{\alpha}\phib\gen{r})^\kb\,.
\end{split}
\end{align}
With the identities
\begin{align}
\cD_\beta\cDt^\alpha\cD_\alpha\phi^k &= +8\Rb\,\cD_\beta\phi^k + 4{\Gamma^k}_{ij}\cD_\beta\phi^i F^j\,, \\
\begin{split}
\cD^\sd{\beta}\cDt^\alpha\cD_\alpha\phi^k &= -4i\sigb^{a\sd{\beta}\alpha}\cDt_a\cD_\alpha\phi^k + 4G^{\alpha\sd{\beta}}\cD_\alpha\phi^k + 8\cW^{\sd{\beta}\g{r}}(\gen{r}\phi)^k \\
&\quad + g^{\kb k}R_{i\kb j\lb}\cD^\sd{\beta}\phib^\lb\cD^\alpha\phi^i\cD_\alpha\phi^j\,,
\end{split} \\
\cD^\sd{\beta}\cDt_\sd{\alpha}\cD^\sd{\alpha}\phib^\kb &= +8R\,\cD^\sd{\beta}\phib^\kb + 4{\Gamma^\kb}_{\ib\jb}\cD^\sd{\beta}\phib^\ib \Fb^\jb\,, \\
\begin{split}
\cD_\beta\cDt_\sd{\alpha}\cD^\sd{\alpha}\phib^\kb &= -4i\sig^a_{\beta\sd{\alpha}}\cDt_a\cD^\sd{\alpha}\phib^\kb - 4G_{\beta\sd{\alpha}}\cD^\sd{\alpha}\phib^\kb + 8{\cW_\beta}^\g{r}(\phib\gen{r})^\kb \\
&\quad + g^{\kb k}R_{k\ib l\jb}\cD_\beta\phi^l\cD_\sd{\alpha}\phib^\ib\cD^\sd{\alpha}\phib^\jb\,,
\end{split}
\end{align}
the supergravity transformations are written in the explicit form
\begin{align}
\begin{split}
\dsg_\xi\cDt^\alpha\cD_\alpha\phi^k &= \xi^b\cD_b\cDt^\alpha\cD_\alpha\phi^k + 8\Rb\,\xi^\beta\cD_\beta\phi^k - 4i(\xib\sigb^a)^\alpha\cDt_a\cD_\alpha\phi^k \\
&\quad + 4(\xib\sigb^a)^\alpha\cD_\alpha\phi^k G_a + 8(\xib\cWb^\g{r})(\gen{r}\phi)^k + 4{\Gamma^k}_{ij}\xi^\beta\cD_\beta\phi^i F^j \\
&\quad + g^{\kb k}R_{i\kb j\lb}\xib_\sd{\beta}\cD^\sd{\beta}\phib^\lb\cD^\alpha\phi^i\cD_\alpha\phi^j - 8F^k\intp_\xi A\,,
\end{split} \label{eq:sugra_transf_delta_ddphi_explicit}\\\nonumber\\
\begin{split}
\dsg_\xi\cDt_\sd{\alpha}\cD^\sd{\alpha}\phib^\kb &= \xi^b\cD_b\cDt_\sd{\alpha}\cD^\sd{\alpha}\phib^\kb + 8R\,\xib_\sd{\beta}\cD^\sd{\beta}\phib^\kb - 4i(\xi\sig^a)_\sd{\alpha}\cDt_a\cD^\sd{\alpha}\phib^\kb \\
&\quad - 4(\xi\sig^a)_\sd{\alpha}\cD^\sd{\alpha}\phib^\kb G_a + 8(\xi\cW^\g{r})(\phib\gen{r})^\kb + 4{\Gamma^\kb}_{\ib\jb}\xib_\sd{\beta}\cD^\sd{\beta}\phib^\ib \Fb^\jb \\
&\quad + g^{\kb k}R_{k\ib l\jb}\xi^\beta\cD_\beta\phi^l\cD_\sd{\alpha}\phib^\ib\cD^\sd{\alpha}\phib^\jb + 8\Fb^\kb\intp_\xi A\,. \label{eq:sugra_transf_delta_ddphib_explicit}
\end{split}
\end{align}
\item \textbf{Yang-Mills sector:}\\
In terms of superspace indices Eq.~\eqref{eq:sugra_transf_delta_ymconnection} has the form
\begin{align}
\dsg_\xi{\cA_M}^\g{r} &= {E_M}^A\xi^B{\cF_{BA}}^\g{r}\,. \label{eq:sugra_transf_delta_ymconnection_explicit}
\end{align}
Furthermore, the infinitesimal changes of the superfields ${\cW_\alpha}^\g{r}$ and $\cW^{\sd{\alpha}\g{r}}$, which appear in the Yang-Mills field strength, under transformations as in Eq.~\eqref{eq:sugra_transf_transformations} are given by
\begin{align}
\begin{split}
\delta{\cW_\alpha}^\g{r} &= \lied_\xi{\cW_\alpha}^\g{r} - {\soae_\alpha}^\beta{\cW_\beta}^\g{r} - \frac{i}{2}{\cW_\alpha}^\g{r}\im{F} + \lac{p}{\cW_\alpha}^\g{q}\sco{p}{q}{r} \\
&= \intp_\xi\cD{\cW_\alpha}^\g{r} -({\soae_\alpha}^\beta - \intp_\xi{\Omega_\alpha}^\beta){\cW_\beta}^\g{r} - {\cW_\alpha}^\g{r}(\intp_\xi A +\frac{i}{2}\im{F}) \\
&\quad + (\lac{p}+i\intp_\xi\cA^\g{p}){\cW_\alpha}^\g{q}\sco{p}{q}{r}\,,
\end{split} \\\nonumber\\
\begin{split}
\delta\cW^{\sd{\alpha}\g{r}} &= \lied_\xi\cW^{\sd{\alpha}\g{r}} - {\soae^\sd{\alpha}}_\sd{\beta}\cW^{\sd{\beta}\g{r}} + \frac{i}{2}\cW^{\sd{\alpha}\g{r}}\im{F} + \lac{p}\cW^{\sd{\alpha}\g{q}}\sco{p}{q}{r} \\
&= \intp_\xi\cD\cW^{\sd{\alpha}\g{r}} - ({\soae^\sd{\alpha}}_\sd{\beta} - \intp_\xi{\Omega^\sd{\alpha}}_\sd{\beta})\cW^{\sd{\beta}\g{r}} + \cW^{\sd{\alpha}\g{r}}(\intp_\xi A +\frac{i}{2}\im{F}) \\
&\quad + (\lac{p}+i\intp_\xi\cA^\g{p})\cW^{\sd{\alpha}\g{q}}\sco{p}{q}{r}\,.
\end{split}
\end{align}
The identities
\begin{align}
\cD_\beta{\cW_\alpha}^\g{r} &= -(\sig^{ba}\eps)_{\beta\alpha}{\cF_{ba}}^\g{r} - \eps_{\beta\alpha}\Db^\g{r}\,, \\
\cD^\sd{\beta}\cW^{\sd{\alpha}\g{r}} &= +(\sigb^{ba}\eps)^{\sd{\beta}\sd{\alpha}}{\cF_{ba}}^\g{r} - \eps^{\sd{\beta}\sd{\alpha}}\Db^\g{r}\,,
\end{align}
are then used to write the supergravity transformations as follows:
\begin{align}
\dsg_\xi{\cW_\alpha}^\g{r} &= \xi^a\cD_a{\cW_\alpha}^\g{r} - (\xi\sig^{ba}\eps)_\alpha{\cF_{ba}}^\g{r} + \xi_\alpha\Db^\g{r} - {\cW_\alpha}^\g{r}\intp_\xi A\,, \label{eq:sugra_transf_delta_w_explicit}\\
\dsg_\xi\cW^{\sd{\alpha}\g{r}} &= \xi^a\cD_a\cW^{\sd{\alpha}\g{r}} + (\xib\sigb^{ba}\eps)^\sd{\alpha}\cF_{\sd{\beta}\sd{\alpha}} + \xib^\sd{\alpha}\Db^\g{r} + \cW^{\sd{\alpha}\g{r}}\intp_\xi A\,. \label{eq:sugra_transf_delta_wb_explicit}
\end{align}
Finally, for the term $\cD^\alpha{\cW_\alpha}^\g{r}$ the infinitesimal transformation
\begin{align}
\begin{split}
\delta\cD^\alpha{\cW_\alpha}^\g{r} &= \lied_\xi\cD^\alpha{\cW_\alpha}^\g{r} + \lac{p}\cD^\alpha{\cW_\alpha}^\g{q}\sco{p}{q}{r} \\
&= \intp_\xi\cD\cD^\alpha{\cW_\alpha}^\g{r} + (\lac{p}+i\intp_\xi\cA^\g{p})\cD^\alpha{\cW_\alpha}^\g{q}\sco{p}{q}{r}\,.
\end{split}
\end{align}
and the identities
\begin{align}
\cD_\beta\cD^\alpha{\cW_\alpha}^\g{r} &= -2i\sig^a_{\beta\sd{\alpha}}\cD_a\cW^{\sd{\alpha}\g{r}}\,, \\
\cD^\sd{\beta}\cD^\alpha{\cW_\alpha}^\g{r} &= -2i\sigb^{a\sd{\beta}\alpha}\cD_a{\cW_\alpha}^\g{r}\,,
\end{align}
are used to calculate the supergravity transformation
\begin{align}
\dsg_\xi\cD^\alpha{\cW_\alpha}^\g{r} &= \xi^a\cD_a\cD^\alpha{\cW_\alpha}^\g{r} - 2i(\xi\sig^a)_\sd{\alpha}\cD_a\cW^{\sd{\alpha}\g{r}} - 2i(\xib\sigb^a)^\alpha\cD_a{\cW_\alpha}^\g{r}\,. \label{eq:sugra_transf_delta_dw_explicit}
\end{align}
\item \textbf{Chiral superfield:}\\
Under infinitesimal transformations as in Eq.~\eqref{eq:sugra_transf_transformations} a chiral superfield $\Phi$ with chiral weight $w(\Phi)$ and an antichiral superfield $\Phib$ with chiral weight $w(\Phib)$, both singlets with respect to the Lorentz and the Yang-Mills group, transform as
\begin{align}
\begin{split}
\delta\Phi &= \lied_\xi\Phi - \frac{i}{2}w(\Phi)\Phi\im{F} \\
&= \intp_\xi\cD\Phi - w(\Phi)\Phi(\intp_\xi A + \frac{i}{2}\im{F})\,,
\end{split} \\
\nonumber\\
\begin{split}
\delta\Phib &= \lied_\xi\Phib - \frac{i}{2}w(\Phib)\Phib\im{F} \\
&= \intp_\xi\cD\Phib - w(\Phib)\Phib(\intp_\xi A + \frac{i}{2}\im{F})\,.
\end{split}
\end{align}
Thus the supergravity transformations of $\Phi$ and $\Phib$ have the form
\begin{align}
\dsg_\xi\Phi &= \xi^b\cD_b\Phi + \xi^\beta\cD_\beta\Phi - w(\Phi)\Phi\,\intp_\xi A\,, \label{eq:sugra_transf_delta_general_phi_explicit}\\
\dsg_\xi\Phib &= \xi^b\cD_b\Phib + \xib_\sd{\beta}\cD^\sd{\beta}\Phi - w(\Phib)\Phib\,\intp_\xi A\,. \label{eq:sugra_transf_delta_general_phib_explicit}
\end{align}
Furthermore, the infinitesimal transformations of the terms $\cD_\alpha\Phi$ and $\cD^\sd{\alpha}\Phib$ are given by
\begin{align}
\begin{split}
\delta\cD_\alpha\Phi &= \lied_\xi\cD_\alpha\Phi - {\soae_\alpha}^\beta\cD_\beta\Phi - \frac{i}{2}\big(w(\Phi)-1\big)\cD_\alpha\Phi\im{F} \\
&= \intp_\xi\cD\cD_\alpha\Phi - ({\soae_\alpha}^\beta - \intp_\xi{\Omega_\alpha}^\beta)\cD_\beta\Phi - \big(w(\Phi)-1\big)\cD_\alpha\Phi(\intp_\xi A + \frac{i}{2}\im{F})\,,
\end{split} \\
\nonumber\\
\begin{split}
\delta\cD^\sd{\alpha}\Phib &= \lied_\xi\cD^\sd{\alpha}\Phib - {\soae^\sd{\alpha}}_\sd{\beta}\cD^\sd{\beta}\Phib - \frac{i}{2}\big(w(\Phib)-1\big)\cD^\sd{\alpha}\Phib\im{F} \\
&= \intp_\xi\cD\cD^\sd{\alpha}\Phib - ({\soae^\sd{\alpha}}_\sd{\beta} - \intp_\xi{\Omega^\sd{\alpha}}_\sd{\beta})\cD^\sd{\beta}\Phib - \big(w(\Phib)+1\big)\cD^\sd{\alpha}\Phib(\intp_\xi A + \frac{i}{2}\im{F})\,.
\end{split}
\end{align}
With the identities
\begin{align}
\cD_\beta\cD_\alpha\Phi &= +\frac{1}{2}\eps_{\beta\alpha}\cD^2\Phi\,, \\
\cD^\sd{\beta}\cD_\alpha\Phi &= +2i{(\sigb^a\eps)^\sd{\beta}}_\alpha\cD_a\Phi + 3w(\Phi){(\sigb^a\eps)^\sd{\beta}}_\alpha G_a\Phi\,, \\
\cD^\sd{\beta}\cD^\sd{\alpha}\Phib &= +\frac{1}{2}\eps^{\sd{\beta}\sd{\alpha}}\cDb^2\Phib\,, \\
\cD_\beta\cD^\sd{\alpha}\Phib &= +2i{(\sig^a\eps)_\beta}^\sd{\alpha}\cD_a\Phib + 3w(\Phib){(\sig^a\eps)_\beta}^\sd{\alpha}G_a\Phib\,,
\end{align}
the supergravity transformations read
\begin{align}
\begin{split}
\dsg_\xi\cD_\alpha\Phi &= \xi^b\cD_b\cD_\alpha\Phi - \frac{1}{2}\xi_\alpha\cD^2\Phi + 2i(\xib\sigb^a\eps)_\alpha\cD_a\Phi + 3w(\Phi)(\xib\sigb^a\eps)_\alpha G_a\Phi \\
&\quad - \big(w(\Phi)-1\big)\cD_\alpha\Phi\,\intp_\xi A\,,
\end{split} \label{eq:sugra_transf_delta_general_dphi_explicit}\\
\nonumber\\
\begin{split}
\dsg_\xi\cD^\sd{\alpha}\Phib &= \xi^b\cD_b\cD^\sd{\alpha}\Phib - \frac{1}{2}\xi^\sd{\alpha}\cDb^2\Phib + 2i(\xi\sig^a\eps)^\sd{\alpha}\cD_a\Phib + 3w(\Phib)(\xi\sig^a\eps)^\sd{\alpha}G_a\Phib \\
&\quad - \big(w(\Phib)+1\big)\cD^\sd{\alpha}\Phib\,\intp_\xi A\,.
\end{split} \label{eq:sugra_transf_delta_general_dphib_explicit}
\end{align}
Finally, for the terms $\cD^2\Phi$ and $\cDb^2\Phib$ the infinitesimal transformations
\begin{align}
\begin{split}
\delta\cD^2\Phi &= \lied_\xi\cD^2\Phi - \frac{i}{2}\big(w(\Phi)-2\big)\cD^2\Phi\im{F} \\
&= \intp_\xi\cD\cD^2\Phi - \big(w(\Phi)-2\big)\cD^2\Phi(\intp_\xi A + \frac{i}{2}\im{F})\,,
\end{split} \\
\nonumber\\
\begin{split}
\delta\cDb^2\Phib &= \lied_\xi\cDb^2\Phib - \frac{i}{2}\big(w(\Phib)+2\big)\cDb^2\Phib\im{F} \\
&= \intp_\xi\cD\cDb^2\Phib - \big(w(\Phib)+2\big)\cDb^2\Phib(\intp_\xi A + \frac{i}{2}\im{F})\,,
\end{split}
\end{align}
and the identities
\begin{align}
\cD_\beta\cD^2\Phi &= +8\Rb\,\cD_\beta\Phi\,, \\
\cD^\sd{\beta}\cD^2\Phi &= -4i\sigb^{a\sd{\beta}\alpha}\cD_a\cD_\alpha\Phi - 2\big(3w(\Phi)-2\big)\sigb^{a\sd{\beta}\alpha}G_a\cD_\alpha\Phi - 4w(\Phi)\Phi X^\sd{\beta}\,, \\
\cD^\sd{\beta}\cDb^2\Phib &= +8R\,\cD^\sd{\beta}\Phib\,, \\
\cD_\beta\cD^2\Phib &= -4i\sig^a_{\beta\sd{\alpha}}\cD_a\cD^\sd{\alpha}\Phib - 2\big(3w(\Phib)+2\big)\sig^a_{\beta\sd{\alpha}}G_a\cD^\sd{\alpha}\Phib - 4w(\Phib)\Phib X_\beta\,,
\end{align}
imply, that the supergravity transformations have the following form:
\begin{align}
\begin{split}
\dsg_\xi\cD^2\Phi &= \xi^b\cD_b\cD^2\Phi + 8\Rb\,\xi^\beta\cD_\beta\Phi - 4i(\xib\sigb^a)^\alpha\cD_a\cD_\alpha\Phi \\
&\quad - 2\big(3w(\Phi)-2\big)(\xib\sigb^a)^\alpha G_a\cD_\alpha\Phi - 4w(\Phi)\Phi\,\xib_\sd{\beta}X^\sd{\beta} - \big(w(\Phi)-2\big)\cD^2\Phi\,\intp_\xi A\,,
\end{split} \label{eq:sugra_transf_delta_general_ddphi_explicit}\\
\nonumber\\
\begin{split}
\dsg_\xi\cDb^2\Phib &= \xi^b\cD_b\cDb^2\Phib + 8R\,\xi_\sd{\beta}\cD^\sd{\beta}\Phib - 4i(\xi\sig^a)_\sd{\alpha}\cD_a\cD^\sd{\alpha}\Phib \\
&\quad - 2\big(3w(\Phib)+2\big)(\xi\sig^a)_\sd{\alpha}G_a\cD^\sd{\alpha}\Phib - 4w(\Phib)\Phib\,\xi^\beta X_\beta - \big(w(\Phib)+2\big)\cDb^2\Phib\,\intp_\xi A\,.
\end{split} \label{eq:sugra_transf_delta_general_ddphib_explicit}
\end{align}
\end{itemize}

\subsection{Invariant actions at the superfield level}
\label{sec:action_superfield_level}
An action $\cS$ in K\"ahler superspace is constructed by using the real density $E$, which is the superdeterminant of the supervielbein components ${E_M}^A$ as defined in Eq.~\eqref{eq:superspace_vielbein_sdet}. In general $\cS$ has the form
\begin{align}
\cS &= \int_*E\,\Theta\,, \label{eq:superfield_actions_general_construction_1}
\end{align}
where $\Theta$ is an even, real superfield with vanishing chiral weight $w(\Theta)=0$ and where $\int_*$ indicates integration over the whole superspace (cf.\ Section~\ref{sec:integration}). Note that $\Theta$ is also inert under Lorentz and Yang-Mills transformations. Actions written in the form of Eq.~\eqref{eq:superfield_actions_general_construction_1} are manifestly invariant under Lorentz, Yang-Mills and K\"ahler transformations, as well as under superspace diffeomorphisms. In particular, they are invariant under supergravity transformations $\dsg$ at the superfield level.
\\\\
Sometimes it is convenient to write the action in a different way, in terms of the so called chiral density $\frac{E}{R}$ and an even, chiral superfield $\Xi$ with $w(\Xi)=+2$, compensating the chiral weight $w(R^{-1})=-2$.\footnote{This construction is also referred to as generalized $F$-term construction.} In that case, the action is written as
\begin{align}
\cS &= \frac{1}{2}\int_*\frac{E}{R}\,\Xi + \frac{1}{2}\int_*\frac{E}{\Rb}\,\Xib\,, \label{eq:superfield_actions_general_construction_2}
\end{align}
where $\Xib=\Xi^\cc$. The two constructions in Eq.~\eqref{eq:superfield_actions_general_construction_1} and \eqref{eq:superfield_actions_general_construction_2} are equivalent. This is evident, if $\Xi$ and $\Xib$ are written in terms of the (unconstrained) superfields $\Sigma$ and $\Sigmab$ by applying the projection operators $(\cDb^2-8R)$ and $(\cD^2-8\Rb)$ form Section~\eqref{sec:chiral projection operators}:
\begin{align}
\Xi &= (\cDb^2-8R)\Sigma\,, & \Xib &= -8(\cD^2-8\Rb)\Sigmab\,.
\end{align}
The terms in Eq.~\eqref{eq:superfield_actions_general_construction_2} thus take the form
\begin{align}
\int_*\frac{E}{R}\,\Xi &= \int_*E\,\cDb^2\Big(\frac{\Sigma}{R}\Big) - 8\int_*E\,\Sigma = - 8\int_*E\,\Sigma\,, \\
\int_*\frac{E}{\Rb}\,\Xib &= \int_*E\,\cD^2\Big(\frac{\Sigmab}{\Rb}\Big) - 8\int_*E\,\Sigma = - 8\int_*E\,\Sigmab\,.
\end{align}
According to Eq.~\eqref{eq:kahler_superspace_integration_parts_identity_5}, the parts of the integrand which contain $\cDb^2$ and $\cD^2$ represent a total derivative, therefore the integral of these terms vanishes. The real superfield in Eq.~\eqref{eq:superfield_actions_general_construction_1} is then given by $\Theta=-4(\Sigma+\Sigmab)$.
\\\\
The general constructions in Eq.~\eqref{eq:superfield_actions_general_construction_1} and \eqref{eq:superfield_actions_general_construction_2} are used in the following to write down the action of the supergravity/matter/Yang-Mills system in K\"ahler superspace at the superfield level. The action decomposes into a supergravity+matter part, a superpotential part and a Yang-Mills part, namely
\begin{align}
\cS = \cS_\text{supergravity+matter} + \cS_\text{superpotential} + \cS_\text{Yang-Mills}\,. \label{eq:superfield_actions_all}
\end{align}
\begin{itemize}
\item \textbf{Supergravity+matter:}\\
The supergravity+matter part contains, among others, the kinetic terms of the supergravity+matter system, and is given by the volume element
\begin{align}
\cS_\text{supergravity+matter} &= -3\int_* E\,, \label{eq:superfield_actions_sugra_matter}
\end{align}
thus the real superfield
\begin{align}
\Theta_\text{supergravity+matter}=-3 \label{eq:superfield_actions_sugra_matter_real_field}
\end{align}
is just a constant.
\item \textbf{Superpotential:}\\
The coupling of the superpotential to supergravity is described by
\begin{align}
\cS_\text{superpotential} = \frac{1}{2}\int_* \frac{E}{R}\,e^{\frac{1}{2}K(\phi,\phib)} W(\phi) + \frac{1}{2}\int_* \frac{E}{\Rb}\,e^{\frac{1}{2}K(\phi,\phib)} \Wb(\phib)\,, \label{eq:superfield_actions_superpotential}
\end{align}
using the chiral density. The superpotential $W(\phi)$ and its conjugate $\Wb(\phib) = \big(W(\phi)\big)^*$ are holomorphic and anti-holomorphic functions of the matter superfields $\phi^k$ and $\phib^\kb$, respectively. They are both singlets with respect to the Yang-Mills gauge group, hence
\begin{align}
W_k(\gen{r}\phi)^k &= 0\,, & \Wb_\kb(\phib\gen{r})^\kb &= 0\,. \label{eq:superfield_actions_superpotential_id_1}
\end{align}
From these identities follows
\begin{equation}
\begin{aligned}
D_\alpha W &= W_k\cD_\alpha\phi^k\,, &\hspace{3cm} D_\alpha\Wb &= 0\,, \\
D^\sd{\alpha}W &= 0\,, &\hspace{3cm} D^\sd{\alpha}\Wb &= \cD^\sd{\alpha}\phib^\kb\,.
\end{aligned}\label{eq:superfield_actions_superpotential_id_2}
\end{equation}
In order that the action is invariant under K\"ahler transformations, the superpotential has to transform as
\begin{align}
W &\mapsto e^{-F}W\,, & \Wb &\mapsto e^{-\Fb}\Wb\,, \label{eq:superfield_actions_superpotential_kahler_transformation_superpotential}
\end{align}
which implies
\begin{align}
e^{K/2}W &\mapsto e^{-i\im{F}}e^{K/2}W\,, & e^{K/2}\Wb &\mapsto e^{+i\im{F}}e^{K/2}\Wb\,, \label{eq:superfield_actions_u1_transformation_product_term}
\end{align}
by using the transformation property of the K\"ahler potential in Eq.~\eqref{eq:kahler_superspace_kahler_transformation} (cf. the discussion in Section~\ref{sec:matter_and_yang_mills}). Although neither the K\"ahler potential $K(\phi,\phib)$ nor the superpotential $W(\phi)$ has a well defined $\U(1)$ weight, the product terms have the weights
\begin{align}
w\big(e^{\frac{1}{2}K(\phi,\phib)} W(\phi)\big) &= +2\,, & w\big(e^{\frac{1}{2}K(\phi,\phib)} \Wb(\phib)\big) &= -2\,, \label{eq:superfield_actions_superpotential_chiral_field_weight}
\end{align}
which are derived by comparing the transformations in Eq.~\eqref{eq:superfield_actions_u1_transformation_product_term} with Eq.~\eqref{eq:kahler_superspace_1/2_phi_u1_transformation}. Thus, the chiral superfield and its conjugate are given by
\begin{align}
\Xi_\text{superpotential} &= e^{\frac{1}{2}K(\phi,\phib)} W(\phi)\,, & \Xib_\text{superpotential} &= e^{\frac{1}{2}K(\phi,\phib)} \Wb(\phib)\,. \label{eq:superfield_actions_superpotential_chiral_field}
\end{align}
In order to check that these two superfields are indeed chiral and antichiral respectively, the covariant derivatives are written explicitly in term of the K\"ahler connection, namely:
\begin{align}
\begin{split}
\cD^\sd{\alpha}\Xi_\text{superpotential} &= D^\sd{\alpha}\Xi_\text{superpotential} + 2A^\sd{\alpha}\Xi_\text{superpotential} \\
&= \frac{1}{2}e^{K/2}\big((D^\sd{\alpha}K)W + 2D^\sd{\alpha}W + 4A^\sd{\alpha}W\big) \\
&= 0\,,
\end{split} \\
\nonumber\\
\begin{split}
\cD_\alpha\Xib_\text{superpotential} &= D_\alpha\Xib_\text{superpotential} - 2A_\alpha\Xib_\text{superpotential} \\
&= \frac{1}{2}e^{K/2}\big((D_\alpha K)\Wb + 2D_\alpha\Wb - 4A_\alpha\Wb\big) \\
&= 0\,,
\end{split}
\end{align}
where the expressions $A_\alpha=+\frac{1}{4}D_\alpha K$ and $A^\sd{\alpha}=-\frac{1}{4}D^\sd{\alpha}K$, and the identities from Eq.~\eqref{eq:superfield_actions_superpotential_id_2} are used. Note that the covariant derivative for $K$, $W$ and $\Wb$ is not defined, because they do not have a well defined chiral weight. Moreover, if the K\"ahler transformation is chosen $F=\log{W}$ (provided that $\log{W}$ exists), then $K\mapsto\cG=K+\log{|W|^2}$ and $W\mapsto 1$, i.e.\ $\Xi_\text{superpotential}=e^{\cG/2}$. This means, the action only depends on the function $\cG$, and not on $K$ and $W$ separately.
\item \textbf{Yang-Mills:}\\
The coupling of the kinetic terms in the Yang-Mills sector to supergravity are contained in
\begin{align}
\cS_\text{Yang-Mills} &= \frac{1}{8}\int_* \frac{E}{R}\,\gkf{r}{s}(\phi) \cW^{\alpha\g{r}}{\cW_\alpha}^\g{s} + \frac{1}{8}\int_* \frac{E}{\Rb}\,\gkfb{r}{s}(\phib) {\cW_\sd{\alpha}}^\g{r}\cW^{\sd{\alpha}\g{s}}\,. \label{eq:superfield_actions_yang_mills}
\end{align}
The gauge kinetic function $\gkf{r}{s}(\phi)$ and its conjugate $\gkfb{r}{s}(\phib) = \big(\gkf{r}{s}(\phi)\big)^\cc$ are holomorphic and anti-holomorphic functions of the matter superfields $\phi^k$ and $\phib^\kb$, respectively. In addition, they are symmetric in their indices, i.e.\ $\gkf{r}{s}(\phi)=\gkf{s}{r}(\phi)$ and $\gkfb{r}{s}(\phib)=\gkfb{s}{r}(\phib)$. From Eq.~\eqref{eq:yang_mills_curvature_gauge_transformation_infinitesimal_component} follows that the gauge kinetic function has to transform as a tensor with two adjoint indices, in order that the action is invariant under Yang-Mills transformations. Therefore, $\gkf{r}{s}(\phi)$ and $\gkfb{r}{s}(\phib)$ fulfil the following identities
\begin{align}
+i\frac{\del\gkf{r}{s}}{\del\phi^k} (\gen{q}\phi)^k = \sco{q}{r}{p}\gkf{p}{s} + \sco{q}{s}{p}\gkf{r}{p}\,, \\
-i\frac{\del\gkfb{r}{s}}{\del\phib^\kb} (\phib\gen{q})^\kb = \sco{q}{r}{p}\gkfb{p}{s} + \sco{q}{s}{p}\gkfb{r}{p}\,.
\end{align}
Hence, the chiral superfield and its conjugate are given by
\begin{align}
\Xi_\text{Yang-Mills} &= \frac{1}{4}\gkf{r}{s}(\phi) \cW^{\alpha\g{r}}{\cW_\alpha}^\g{s}\,, & \Xib_\text{Yang-Mills} &= \frac{1}{4}\gkfb{r}{s}(\phib){\cW_\sd{\alpha}}^\g{r}\cW^{\sd{\alpha}\g{s}}\,. \label{eq:superfield_actions_yang_mills_chiral_field}
\end{align}
Note, because $\gkf{r}{s}(\phi)$ and ${\cW_\alpha}^\g{r}$ are chiral superfields, $\Xi_\text{Yang-Mills}$ is a chiral superfield as well.
\end{itemize}
In order to justify that the action $\cS$, as stated in Eq.~\eqref{eq:superfield_actions_all}, describes correctly the supergravity/\-matter/\-Yang-Mills system, either the superfield equations of motions can be calculated, or the Lagrangian at the component field level can be determined. This is done in Section~\ref{sec:equations_of_motion} and \ref{sec:component_fields}, respectively.

\subsection{$R$-symmetry}
\label{sec:r_symmetry_superfield_level}
In a supergravity theory, an $R$-symmetry $\UR$ is a special type of $\U(1)$ symmetry, whose transformations consist of a Yang-Mills gauge transformation and a simultaneous K\"ahler transformation. Analogous to ordinary $\U(1)$ Yang-Mills gauge symmetries, the matter superfields $\phi^k$ and $\phib^\kb$ transform under $\UR$ as
\begin{align}
\phi^k &\mapsto e^{+i\wR(\phi^k)\lacR}\phi^k\,, & \phib^\kb &\mapsto e^{-i\wR(\phi^k)\lacR}\phib^\kb\,, \label{eq:superfield_r_symmetry_matterfield}
\end{align}
where $\wR(\phi^k)$ is the corresponding weight of $\phi^k$, also called $R$-charge, and $\lacR$ is a real, even superfield. If it is assumed that the superpotential $W(\phi^k)$, as a function of the fields $\phi^k$, has a definite weight $\wR(W)$, $W$ and $\Wb$ have well defined transformation properties under $\UR$, namely \footnote{The assignment of an $R$-charge to the superpotential is compatible with derivatives with respect to the matter superfields, e.g.\ $\wR(D_k W)=\wR(W_k)=\wR(W)-\wR(\phi^k)$ where the first identity holds under the assumption of Eq.~\eqref{eq:superfield_r_symmetry_kahlerpotential_weight}.}
\begin{align}
W &\mapsto e^{+i\wR(W)\lacR}W\,, & \Wb &\mapsto e^{-i\wR(W)\lacR}\Wb\,. \label{eq:superfield_r_symmetry_superpotential}
\end{align}
In contrast to ordinary Yang-Mills gauge symmetries, the superpotential is not invariant with respect to an $R$-symmetry. However, it is still assumed that the K\"ahler potential $K(\phi^k,\phib^\kb)$ and the gauge kinetic function $\gkf{r}{s}(\phi^k)$ are a singlets under $\UR$ transformations, i.e.
\begin{gather}
\wR(K) = 0\,, \label{eq:superfield_r_symmetry_kahlerpotential_weight} \\
\wR(\gkf{r}{s}) = 0\,, \hspace{4cm} \wR(\gkfb{r}{s}) = 0\,.  \label{eq:superfield_r_symmetry_gaugekineticfunction_weight}
\end{gather}
As stated in Section~\ref{sec:action_superfield_level}, the superpotential appears in the action only in combination with the term $e^{K/2}$, such that a K\"ahler transformation $F(\phi^k)$ induces a $\U(1)$ transformation (see Eq.~\eqref{eq:superfield_actions_u1_transformation_product_term}):
\begin{align}
e^{K/2}W &\mapsto e^{-i\im{F}}e^{K/2}W\,, & e^{K/2}\Wb &\mapsto e^{+i\im{F}}e^{K/2}\Wb\,. \label{eq:superfield_r_symmetry_kahler_transformation}
\end{align}
If the $\UR$ transformation takes the value $\lacR=-\frac{1}{2}\im{F}$, i.e.\ it is parametrized by a K\"ahler transformation, and the superpotential has the weight
\begin{align}
\wR(W) &= +2\,, & \wR(\Wb) &= -2\,, \label{eq:superfield_r_symmetry_superpotential_weight}
\end{align}
the Yang-Mills transformation in Eq.~\eqref{eq:superfield_r_symmetry_matterfield} corresponds to the K\"ahler transformation in Eq.~\eqref{eq:superfield_r_symmetry_kahler_transformation}. Since K\"ahler and Yang-Mills transformations leave the action invariant, this holds true for $\UR$ transformations, if and only if Eq.~\eqref{eq:superfield_r_symmetry_kahlerpotential_weight}, \eqref{eq:superfield_r_symmetry_gaugekineticfunction_weight} and \eqref{eq:superfield_r_symmetry_superpotential_weight} apply. Note that chiral weights $w$ are added to the $R$-charge, e.g.
\begin{align}
\wR(\cD_A\phi^k) &= w(E_A) + \wR(\phi^k)\,, & \wR(\cD_A\phib^\kb) &= w(E_A) + \wR(\phib^\kb)\,,
\end{align}
where $\wR(\phib^\kb)=-\wR(\phi^k)$, and $w(E_A)$ is given in Eq.~\eqref{eq:superspace_inversevielbein_weight_u1_group}. Hence, different component fields of a supermultiplet usually have different $R$-charges (cf.\ Section~\ref{sec:component_fields_r_symmetry}). In general, multiple $\UR$ symmetries may be present in a supergravity theory.
\\\\
Since $\lacR$ is parametrized by a K\"ahler transformation $F$, the commutator of an infinitesimal supergravity transformation $\dsg_\xi$ and an infinitesimal $\UR$ transformation $\delta_\lacR$ is again a K\"ahler transformation (cf.\ Section~\ref{sec:sugra_transformations_superfield_level}), namely
\begin{align}
\com{\dsg_\xi}{\delta_\lacR}\chi^A &= +\frac{i}{2} w(\chi^A)\chi^A\intp_\xi\exd\im{F}\,,
\end{align}
where $\chi^A$ is a $p$-superform with Lorentz index $A$, chiral weight $w(\chi^A)$ (and $R$-charge $\wR(\chi^A)$), and which transforms in any representation of the Yang-Mills gauge group.

\section{Equations of motion in K\"ahler superspace}
\label{sec:equations_of_motion}
The equations of motion in K\"ahler superspace at the superfield level are obtained by applying the calculus of variation to the action $S$ stated in Section~\ref{sec:action_superfield_level}. The basic objects in K\"ahler superspace, which are varied, are the supervielbein, the Lorentz and the Yang-Mills connection, and the matter superfields. In fact, these variations have to be considered modulo compensating superspace diffeomorphisms, and modulo compensating Lorentz and Yang-Mills transformations. An extra complication arises, because the torsion tensor and the Yang-Mills field strength, which are derived quantities from the basic objects, as well as the matter superfields fulfil covariant constraints, which must be respected by the variations. However, these constraints are violated by the most general variations of $E^A$, $\Omega$, $\cA$, $\phi^k$ and $\phib^\kb$. Therefore, the variational version of the constraint equations have to be solved, which provides the expressions of the variations in terms of unconstrained entities. This calculation is carried out in the remainder of this section.

\subsection{Variation of the torsion tensor}
\label{sec:variation_torsion}
The definition of the torsion tensor $T^A$ in Eq.~\eqref{eq:superspace_structure_equation_torsion} contains the supervielbein $E^A$, the Lorentz connection ${\Omega_B}^A$ and the K\"ahler connection $A$. Although in K\"ahler superspace $A$ is not a basic object but rather a function of the matter superfields, as stated in Eq.~\eqref{eq:kahler_superspace_1/2_u1_connection}, it is convenient to treat $A$ as such in the following. In Section~\ref{sec:variation_kahler_connection} the variation of the K\"ahler connection is then put down to variations of the matter superfields. The infinitesimal variations of the supervielbein, the Lorentz and the K\"ahler connection have the general form
\begin{align}
\delta E^A &= \exd z^M \delta {E_M}^A\,, \label{eq:eom_superspace_variation_vielbein_1}\\
\delta {\Omega_B}^A &= \exd z^M \delta {\Omega_{MB}}^A\,, \label{eq:eom_superspace_variation_lorentz_connection_1}\\
\delta A &= \exd z^M \delta A_M\,. \label{eq:eom_superspace_variation_kahler_connection_1}
\end{align}
This can also be written as
\begin{align}
\delta E^A &= H^A\,, & \text{with} && \quad H^A &= E^B {H_B}^A\,, & {H_B}^A &= {E_B}^M\delta {E_M}^A\,, \label{eq:eom_superspace_variation_vielbein_2}\\
\delta {\Omega_B}^A &= {\phi_B}^A\,, & \text{with} && \quad {\phi_B}^A &= E^C {\phi_{CB}}^A\,, & {\phi_{CB}}^A &= {E_C}^M\delta {\Omega_{MB}}^A\,, \label{eq:eom_superspace_variation_lorentz_connection_2}\\
\delta A &= \omega\,, & \text{with} && \quad \omega &= E^A \omega_A\,, & \omega_A &= {E_A}^M\delta A_M\,, \label{eq:eom_superspace_variation_kahler_connection_2}
\end{align}
where the $1$-superforms $H^A$, ${\phi_B}^A$ and $\omega$ are introduced, which parametrize $\delta E^A$, $\delta{\Omega_B}^A$ and $\delta A$, respectively. Note that in leading order of the variation, the components of the infinitesimal superforms $H^A$, ${\phi_B}^A$ and $\omega$ are written with respect to the supervielbein $E^A$, and not with respect to the varied supervielbein $(E^A+\delta E^A)$. From Eq.~\eqref{eq:eom_superspace_variation_vielbein_1} and \eqref{eq:eom_superspace_variation_vielbein_2} follows that
\begin{align}
\delta{E_M}^A = +{E_M}^B{H_B}^A\,,\quad \delta{E_A}^M = -{H_A}^B{E_B}^M\,, \label{eq:eom_superspace_variation_vielbein_3}
\end{align}
where $\delta({E_A}^M{E_M}^B)=0$ is used to derive the second identity. By using the definition of the torsion tensor and by applying Eqs.~\eqref{eq:eom_superspace_variation_vielbein_2}--\eqref{eq:eom_superspace_variation_kahler_connection_2}, the variation of $T^A$ is given by
\begin{align}
\begin{split}
\delta T^A &= \delta\big(\exd E^A + E^B\wedge{\Omega_B}^A + w(E^A)E^A\wedge A\big) \\
&= \exd(\delta E^A) + (\delta E^B)\wedge{\Omega_B}^A + E^B\wedge(\delta{\Omega_B}^A) + w(E^A)(\delta E^A)\wedge A \\
&\quad + w(E^A)E^A\wedge (\delta A) \\
&= \cD H^A + E^B\wedge{\phi_B}^A + w(E^A) E^A\wedge\omega\,.
\end{split} \label{eq:eom_superspace_variation_torsion}
\end{align}
If the last three terms in Eq.~\eqref{eq:eom_superspace_variation_torsion} are written explicitly as $2$-superforms, namely
\begin{align}
\begin{split}
\cD H^A &= E^B\wedge\cD{H_B}^A + T^B{H_B}^A \\
&= \frac{1}{2}E^B\wedge E^C\big(\cD_C{H_B}^A - (-1)^{CB}\cD_B{H_C}^A + {T_{CB}}^D {H_D}^A\big)\,,
\end{split}\\\nonumber\\
E^B\wedge{\phi_B}^A &= \frac{1}{2}E^B\wedge E^C\big({\phi_{CB}}^A - (-1)^{CB}{\phi_{BC}}^A\big)\,,\\\nonumber\\
\begin{split}
w(E^A)E^A\wedge\omega &= w(E^A)E^A\wedge E^B\omega_B \\
&= w(E^A)\frac{1}{2}E^B\wedge E^C\big(\delta_B^A\,\omega_C - (-1)^{CB}\delta_C^A\,\omega_B\big)\,,
\end{split}
\end{align}
and the variation of the torsion tensor is expanded as
\begin{align}
\begin{split}
\delta T^A &= \delta\Big(\frac{1}{2}E^B\wedge E^C{T_{CB}}^A\Big) \\
&= \frac{1}{2}(H^B\wedge E^C + E^B\wedge H^C){T_{CB}}^A + \frac{1}{2}E^B\wedge E^C \delta{T_{CB}}^A \\
&= \frac{1}{2}E^B\wedge E^C\big({H_C}^D{T_{DB}}^A -(-1)^{CB}{H_B}^D{T_{DC}}^A + \delta{T_{CB}}^A\big)\,,
\end{split}
\end{align}
the variation of the torsion tensor components can be read off:
\begin{align}
\begin{split}
\delta{T_{CB}}^A &= \cD_C{H_B}^A - (-1)^{BC}\cD_B{H_C}^A + {T_{CB}}^D {H_D}^A \\
&\quad - {H_C}^D{T_{DB}}^A + (-1)^{BC}{H_B}^D{T_{DC}}^A \\
&\quad + {\phi_{CB}}^A - (-1)^{BC}{\phi_{BC}}^A \\
&\quad + w(E^A)\big(\delta_B^A\,\omega_C - (-1)^{BC}\delta_C^A\,\omega_B\big)\,.
\end{split} \label{eq:eom_superspace_variation_torsion_components}
\end{align}
Two variations for a certain object, stated in Eqs.~\eqref{eq:eom_superspace_variation_vielbein_1}--\eqref{eq:eom_superspace_variation_kahler_connection_1}, are equivalent, if the difference is just an infinitesimal transformation under a superspace diffeomorphism $\xi=E^A\xi_A$, with $\xi^A=(\xi^a,\xi^\alpha,\xib_\sd{\alpha})$, and a Lorentz transformation $\soae$, i.e.\ $\xi$ and $\soae$ can be used to compensate parts of the variation. Note, because the K\"ahler transformation $F(\phi)$ depends on the matter superfields, it can not be used to compensate variations. The transformation under a generic, infinitesimal superspace diffeomorphism and Lorentz transformation is labelled by $\deltat$. For the supervielbein, the Lorentz and the K\"ahler connection this is given by
\begin{align}
\begin{split}
\deltat E^A &= \lied_\xi E^A + E^B{\soae_B}^A \\
&= \intp_\xi T^A + \cD\xi^A + E^B({\soae_B}^A - \intp_\xi{\Omega_B}^A) - w(E^A)E^A\intp_\xi A\,,
\end{split} \label{eq:eom_superspace_transf_vielbein}\\
\nonumber\\
\begin{split}
\deltat{\Omega_B}^A &= \lied_\xi{\Omega_B}^A + {\com{\Omega}{\soae}_B}^A  - \exd{\soae_B}^A \\
&= \intp_\xi {R_B}^A - \cD({\soae_B}^A - \intp_\xi{\Omega_B}^A)\,,
\end{split} \label{eq:eom_superspace_transf_lorentz_connection}\\
\nonumber\\
\deltat A &= \lied_\xi A = \intp_\xi F\,. \label{eq:eom_superspace_transf_u1_connection}
\end{align}
Again, in leading order of the transformation $\deltat$ the components of the infinitesimal superforms $H^A$, ${\phi_B}^A$ and $\omega$ are written with respect to the supervielbein $E^A$, and not with respect to the transformed supervielbein $(E^A+\deltat E^A)$. Furthermore, with the definitions
\begin{align}
\Delta H^A &:= \deltat E^A\,, \label{eq:eom_superspace_transf_h}\\
\Delta {\phi_B}^A &:= \deltat{\Omega_B}^A\,, \label{eq:eom_superspace_transf_phi}\\
\Delta \omega &:= \deltat A\,, \label{eq:eom_superspace_transf_omega}
\end{align}
follows
\begin{align}
\Delta {H_B}^A &=  \xi^C{T_{CB}}^A + \cD_B\xi^A + {\soaet_B}^A - w(E^A)\delta^A_B\,\intp_\xi A\,, \label{eq:eom_superspace_transf_h_components}\\
\Delta {\phi_{CB}}^A &= \xi^D {R_{DCB}}^A - \cD_C{\soaet_B}^A\,, \label{eq:eom_superspace_transf_phi_components}\\
\Delta \omega_A &= \xi^B F_{BA}\,, \label{eq:eom_superspace_transf_omega_components}
\end{align}
where ${\soaet_B}^A := ({\soae_B}^A - \intp_\xi{\Omega_B}^A)$. The usage of the symbol $\Delta$ indicates, that the transformation may be of the same order as the object, namely $H^A$, ${\phi_A}^B$ and $\omega$, itself. The goal in the remainder of this section is to solve the variational version of the torsion constraints, stated in Eq.~\eqref{eq:bianchi_sp_torsion_constraints}, by taking into account compensating superspace diffeomorphisms and Lorentz transformations.

\subsubsection{Torsion constraints I}
According to Eq.~\eqref{eq:bianchi_sp_torsion_constraints} the following components of the torsion tensor have the constant values
\begin{gather}
{T_{\gamma\beta}}^a = 0\,,\hspace{4cm} T^{\sd{\gamma}\sd{\beta}a} = 0\,, \label{eq:eom_superspace_torsion_constraints_1_constr_torsion_comp_1}\\
T_{\gamma\beta\sd{\alpha}} = 0\,,\hspace{4cm} T^{\sd{\gamma}\sd{\beta}\alpha} = 0\,, \label{eq:eom_superspace_torsion_constraints_1_constr_torsion_comp_2}\\
{T_{\gamma}}^{\sd{\beta}a} = -2i{(\sig^a\eps)_\gamma}^\sd{\beta}\,. \label{eq:eom_superspace_torsion_constraints_1_constr_torsion_comp_3}
\end{gather}
Thus, the corresponding variations have to vanish. According to Eq.~\eqref{eq:eom_superspace_variation_torsion_components}, the variations of ${T_{\gamma\beta}}^a$ and $T_{\gamma\beta\sd{\alpha}}$ are written as
\begin{align}
0 = \delta {T_{\gamma\beta}}^a &= \sum_{\gamma\beta}\big(\cD_\gamma {H_\beta}^a - H_{\gamma\sd{\delta}}\tensor{T}{^{\sd{\delta}}_\beta^a}\big)\,, \label{eq:eom_superspace_torsion_constraints_1_var_torsion_comp_1}\\
0 = \delta T_{\gamma\beta\sd{\alpha}} &= \sum_{\gamma\beta}\big(\cD_\gamma H_{\beta\sd{\alpha}} - {H_\gamma}^d T_{d\beta\sd{\alpha}}\big)\,. \label{eq:eom_superspace_torsion_constraints_1_var_torsion_comp_2}
\end{align}
The solution for these two equations is given by
\begin{align}
{H_\beta}^a &= \cD_\beta\Xib^a + \Xib_\sd{\delta}\tensor{T}{^{\sd{\delta}}_\beta^a}\,, \label{eq:eom_superspace_torsion_constraints_1_solution_1}\\
H_{\beta\sd{\alpha}} &= \cD_\beta\Xib_\sd{\alpha} + \Xib^d T_{d\beta\sd{\alpha}}\,, \label{eq:eom_superspace_torsion_constraints_1_solution_2}
\end{align}
where the superfields $\Xib^a$ and $\Xib_\sd{\alpha}$ are introduced. A similar consideration holds true for the components $T^{\sd{\gamma}\sd{\beta}a}$ and $T^{\sd{\gamma}\sd{\beta}\alpha}$. The corresponding variations are given by
\begin{align}
0 = \delta T^{\sd{\gamma}\sd{\beta}a} &= \sum_{\sd{\gamma}\sd{\beta}}(\cD^\sd{\gamma}H^{\sd{\beta}a} - H^{\sd{\gamma}\delta}{T_\delta}^{\sd{\beta}a})\,, \label{eq:eom_superspace_torsion_constraints_1_var_torsion_comp_3}\\
0 = \delta T^{\sd{\gamma}\sd{\beta}\alpha} &= \sum_{\sd{\gamma}\sd{\beta}}(\cD^\sd{\gamma}H^{\sd{\beta}\alpha} - H^{\sd{\gamma}d}{T_d}^{\sd{\beta}\alpha})\,. \label{eq:eom_superspace_torsion_constraints_1_var_torsion_comp_4}
\end{align}
The two equations are solved by introducing the superfields $\Xi^a$ and $\Xi^\alpha$ as follows:
\begin{align}
H^{\sd{\beta}a} &= \cD^\sd{\beta}\Xi^a + \Xi^\delta {T_\delta}^{\sd{\beta}a}\,, \label{eq:eom_superspace_torsion_constraints_1_solution_3}\\
H^{\sd{\beta}\alpha} &= \cD^\sd{\beta}\Xi^\alpha + \Xi^d {T_d}^{\sd{\beta}\alpha}\,, \label{eq:eom_superspace_torsion_constraints_1_solution_4}
\end{align}
where $(\Xi^a)^\cc=\Xib^a$ and $(\Xi^\alpha)^\cc=\Xib^\sd{\alpha}$. Having solved the variational version of the torsion constraints in Eq.~\eqref{eq:eom_superspace_torsion_constraints_1_constr_torsion_comp_1} and \eqref{eq:eom_superspace_torsion_constraints_1_constr_torsion_comp_2}, there is still the freedom of using an infinitesimal superspace diffeomorphism $\xi$ to bring the variations in Eq.~\eqref{eq:eom_superspace_torsion_constraints_1_solution_1}, \eqref{eq:eom_superspace_torsion_constraints_1_solution_2}, \eqref{eq:eom_superspace_torsion_constraints_1_solution_3} and \eqref{eq:eom_superspace_torsion_constraints_1_solution_4} to a simpler form. In particular, according to Eq.~\eqref{eq:eom_superspace_transf_h_components}, the variations ${H_\beta}^a$ and $H^{\sd{\beta}a}$ transform as
\begin{align}
\Delta {H_\beta}^a &= \xib_\sd{\gamma}\tensor{T}{^{\sd{\gamma}}_\beta^a} + \cD_\beta\xi^a\,, \\
\Delta H^{\sd{\beta}a} &= \xi^\gamma {T_\gamma}^{\sd{\beta}a} + \cD^\sd{\beta}\xi^a\,,
\end{align}
under an infinitesimal superspace diffeomorphism, and they are inert under Lorentz transformations. Thus, the following choice for the components of $\xi$,
\begin{align}
\xi^a = -\frac{1}{2}(\Xib^a + \Xi^a)\,, \quad \xi^\alpha = -\Xi^\alpha\,, \quad \xib_\sd{\alpha} = -\Xib_\sd{\alpha}\,, \label{eq:eom_superspace_torsion_constraints_1_choice_xi}
\end{align}
leads to
\begin{align}
{H_\beta}^a &= +\cD_\beta\Nu^a\,, \label{eq:eom_superspace_torsion_constraints_1_sol_1}\\
H^{\sd{\beta}a} &= -\cD^\sd{\beta}\Nu^a\,, \label{eq:eom_superspace_torsion_constraints_1_sol_2}
\end{align}
where the imaginary superfield $\Nu^a := \frac{1}{2}(\Xib^a - \Xi^a)$ is introduced. Since the value of the infinitesimal superspace diffeomorphism is now fixed (cf.\ Eq.~\eqref{eq:eom_superspace_torsion_constraints_1_choice_xi}) the corresponding transformation has to be applied to the other variations as well. Under an infinitesimal superspace diffeomorphism the variations $H_{\beta\sd{\alpha}}$ and $H^{\sd{\beta}\alpha}$ transform as
\begin{align}
\Delta H_{\beta\sd{\alpha}} &= \xi^c T_{c\beta\sd{\alpha}} + \cD_\beta\xib_\sd{\alpha}\,, \\
\Delta H^{\sd{\beta}\alpha} &= \xi^c {T_c}^{\sd{\beta}\alpha} + \cD^\sd{\beta}\xi_\alpha\,.
\end{align}
which leads to
\begin{align}
H_{\beta\sd{\alpha}} &= +i\sig_{a\beta\sd{\alpha}}\Rb\,\Nu^a = +i\Rb\,\Nu_{\beta\sd{\alpha}}\,, \label{eq:eom_superspace_torsion_constraints_1_sol_3}\\
H^{\sd{\beta}\alpha} &= -i\sigb_a^{\sd{\beta}\alpha}R\,\Nu^a = -iR\,\Nu^{\alpha\sd{\beta}}\,, \label{eq:eom_superspace_torsion_constraints_1_sol_4}
\end{align}
where the explicit forms of $T_{c\beta\sd{\alpha}}$ and ${T_c}^{\sd{\beta}\alpha}$ are used. Beside superspace diffeomorphisms, the variations ${H_\beta}^\alpha$ and ${H^\sd{\beta}}_\sd{\alpha}$ transform also under Lorentz transformations:
\begin{align}
\Delta {H_\beta}^\alpha &= \cD_\beta\xi^\alpha + {\soaet_\beta}^\alpha - \delta_\beta^\alpha\intp_\xi A\,, \\
\Delta {H^\sd{\beta}}_\sd{\alpha} &= \cD^\sd{\beta}\xib_\sd{\alpha} + {\soaet^\sd{\beta}}_\sd{\alpha} + \delta^\sd{\beta}_\sd{\alpha}\intp_\xi A\,.
\end{align}
By the following choice of ${\soaet_\beta}^\alpha$ and ${\soaet^\sd{\beta}}_\sd{\alpha}$,
\begin{align}
\soaet_{\sym{2}{\beta\alpha}} &= -\frac{1}{2}\sum_{\beta\alpha}\big(H_{\beta\alpha} + \cD_\beta\xi_\alpha\big)\,,& \soaet_{\sym{2}{\sd{\beta}\sd{\alpha}}} &= -\frac{1}{2}\sum_{\sd{\beta}\sd{\alpha}}\big(H_{\sd{\beta}\sd{\alpha}} + \cD_\sd{\beta}\xib_\sd{\alpha}\big)\,, \label{eq:eom_superspace_torsion_constraints_1_choice_l}
\end{align}
the traceless parts of ${H_\beta}^\alpha$ and ${H^\sd{\beta}}_\sd{\alpha}$ are eliminated:
\begin{align}
{H_\beta}^\alpha &= \frac{1}{2}\delta_\beta^\alpha H\,, \label{eq:eom_superspace_torsion_constraints_1_sol_5}\\
{H^\sd{\beta}}_\sd{\alpha} &= \frac{1}{2}\delta^\sd{\beta}_\sd{\alpha} \Hb\,, \label{eq:eom_superspace_torsion_constraints_1_sol_6}
\end{align}
where
\begin{align}
H &:= -({H^\alpha}_\alpha + \cD^\alpha\xi_\alpha + 2\intp_\xi A)\,, \\
\Hb &:= -({H_\sd{\alpha}}^\sd{\alpha} + \cD_\sd{\alpha}\xib^\sd{\alpha} - 2\intp_\xi A)\,,
\end{align}
are the conjugate of each other. Note that Eq.~\eqref{eq:eom_superspace_torsion_constraints_1_choice_l} also fixes the component $\soaet_{ba}$. In a last step the variational form of the torsion constraint in Eq.~\eqref{eq:eom_superspace_torsion_constraints_1_constr_torsion_comp_3} is solved. The variation of ${T_\gamma}^{\sd{\beta}a}$ is given by
\begin{align}
\begin{split}
0 = \delta {T_\gamma}^{\sd{\beta}a} &= \cD_\gamma H^{\sd{\beta}a} + \cD^\sd{\beta}{H_\gamma}^a + {T_\gamma}^{\sd{\beta}d} {H_d}^a \\
&\quad - {H_\gamma}^\delta{T_\delta}^{\sd{\beta}a} - {H^\sd{\beta}}_\sd{\delta}\tensor{T}{^{\sd{\delta}}_\gamma^a} \\
&= -\cD_\gamma\cD^\sd{\beta}\Nu^a + \cD^\sd{\beta}\cD_\gamma\Nu^a + {T_\gamma}^{\sd{\beta}d}{H_d}^a\\
&\quad - \frac{1}{2}H\,{T_\gamma}^{\sd{\beta}a} - \frac{1}{2}\Hb\,\tensor{T}{^{\sd{\beta}}_\gamma^a}\,.
\end{split} \label{eq:eom_superspace_torsion_constraints_1_var_torsion_comp_5}
\end{align}
Here, ${H_b}^a$ represents the variation after the transformation under the infinitesimal superspace diffeomorphism and Lorentz transformation specified in Eq.~\eqref{eq:eom_superspace_torsion_constraints_1_choice_xi} and \eqref{eq:eom_superspace_torsion_constraints_1_choice_l}, respectively, i.e.\ ${H_b}^a\equiv{H_b}^a+\Delta{H_b}^a$. This can be done without loss of generality, because the variation ${H_b}^a$ is generic. From Eq.~\eqref{eq:eom_superspace_torsion_constraints_1_var_torsion_comp_5} follows
\begin{align}
{T_\gamma}^{\sd{\beta}d}{H_d}^a = \com{\cD_\gamma}{\cD^\sd{\beta}}\Nu^a + \frac{1}{2}(H+\Hb){T_\gamma}^{\sd{\beta}a}\,.
\end{align}
This identity is written efficiently in terms of spinor indices:
\begin{align}
H_{\beta\sd{\beta}\s\alpha\sd{\alpha}} &= -\frac{i}{2}\com{\cD_\beta}{\cD_\sd{\beta}}\Nu_{\alpha\sd{\alpha}} - \eps_{\beta\alpha}\eps_{\sd{\beta}\sd{\alpha}}(H+\Hb)\,. \label{eq:eom_superspace_torsion_constraints_1_sol_7}
\end{align}
As discussed in Section~\ref{sec:variation_action}, in the variation of the density $E$ the supertrace of the variation ${H_B}^A$ appears. The supertrace is calculated from Eq.~\eqref{eq:eom_superspace_torsion_constraints_1_sol_5}, \eqref{eq:eom_superspace_torsion_constraints_1_sol_6} and \eqref{eq:eom_superspace_torsion_constraints_1_sol_7}, and has the following value:
\begin{align}
(-1)^A {H_A}^A &= \frac{i}{4}\com{\cD^\alpha}{\cD^\sd{\alpha}}\Nu_{\alpha\sd{\alpha}} + H+\Hb\,. \label{eq:eom_superspace_torsion_constraints_1_trace_h}
\end{align}

\subsubsection{Torsion constraints II}
In this section, the following torsion constraints from Eq.~\eqref{eq:bianchi_sp_torsion_constraints} are considered:
\begin{align}
{T_{\gamma b}}^a &= 0\,, \label{eq:eom_superspace_torsion_constraints_2_constr_torsion_comp_1}\\
\tensor{T}{_\gamma^{\sd{\beta}}_{\sd{\alpha}}} &= 0\,, \label{eq:eom_superspace_torsion_constraints_2_constr_torsion_comp_2}\\
{T_{\gamma\beta}}^\alpha &= 0\,. \label{eq:eom_superspace_torsion_constraints_2_constr_torsion_comp_3}
\end{align}
To solve the variational versions of these constraints, it is convenient to define some new superfields:
\begin{align}
{X_b}^a &:= {H_b}^a - \cD_b\Nu^a\,, & X_{b\sd{\alpha}} &:= H_{b\sd{\alpha}} - \Nu^c T_{cb\sd{\alpha}}\,, \nonumber\\
{X_\beta}^\alpha &:= {H_\beta}^\alpha - \Nu^c{T_{c\beta}}^\alpha\,, & {X^\sd{\beta}}_\sd{\alpha} &:= {H^\sd{\beta}}_\sd{\alpha} - \Nu^c\tensor{T}{_c^{\sd{\beta}}_{\sd{\alpha}}}\,, \label{eq:eom_superspace_torsion_constraints_2_field_definitions}\\
{\Pi_{\gamma B}}^A &:= {\phi_{\gamma B}}^A - \Nu^d{R_{d\gamma B}}^A\,, & \Sigma_\gamma &:= \omega_\gamma - \Nu^d F_{d\gamma}\,. \nonumber
\end{align}
In terms of spinor indices ${X_b}^a$ and ${\Pi_{\gamma a}}^b$ are written as
\begin{align}
X_{\beta\sd{\beta}\,\alpha\sd{\alpha}} &= -i\cD_\beta\cD_\sd{\beta}\Nu_{\alpha\sd{\alpha}} - \eps_{\beta\alpha}\eps_{\sd{\beta}\sd{\alpha}}(H+\Hb) - i\eps_{\sd{\beta}\sd{\alpha}}\Nu_{\beta\sd{\gamma}}{G_\alpha}^\sd{\gamma} - i\eps_{\beta\alpha}{\Nu^\gamma}_\sd{\beta}G_{\gamma\sd{\alpha}}\,, \label{eq:eom_superspace_torsion_constraints_2_id_1}\\
\Pi_{\gamma\s\beta\sd{\beta}\s\alpha\sd{\alpha}} &= + 2\eps_{\sd{\beta}\sd{\alpha}}\Pi_{\gamma\sym{2}{\beta\alpha}} - 2\eps_{\beta\alpha}\Pi_{\gamma\sym{2}{\sd{\beta}\sd{\alpha}}}\,, \label{eq:eom_superspace_torsion_constraints_2_id_2}
\end{align}
where Eq.~\eqref{eq:eom_superspace_torsion_constraints_1_sol_7} is used to derive Eq.~\eqref{eq:eom_superspace_torsion_constraints_2_id_1}. According to Eq.~\eqref{eq:eom_superspace_variation_torsion_components}, the variation of ${T_{\gamma b}}^a$ is given by
\begin{align}
\begin{split}
0 = \delta {T_{\gamma b}}^a &= \cD_\gamma{H_b}^a - \cD_b{H_\gamma}^a + {T_{\gamma b}}^{\su{\delta}}{H_{\su{\delta}}}^a + H_{b\sd{\delta}}\tensor{T}{^{\sd{\delta}}_\gamma^a} + {\phi_{\gamma b}}^a \\
&= {\Pi_{\gamma b}}^a + \cD_\gamma{X_b}^a + X_{b\sd{\delta}}\tensor{T}{^{\sd{\delta}}_\gamma^a} - 2T_{\gamma b\sd{\delta}}\cD^\sd{\delta}\Nu^a\,,
\end{split}
\end{align}
with the solution
\begin{align}
X_{\beta\sd{\beta}\s\sd{\alpha}} &= +\frac{i}{2}\eps_{\sd{\beta}\sd{\alpha}}\cD_\beta(H+\Hb) + \frac{1}{4}\cD_\beta\cD^\varphi\cD_\sd{\alpha}\Nu_{\varphi\sd{\beta}} - \Rb\cD_\sd{\beta}\Nu_{\beta\sd{\alpha}}\,, \label{eq:eom_superspace_torsion_constraints_2_sol_1}\\
\Pi_{\gamma\sym{2}{\beta\alpha}} &= +\frac{i}{4}\cD_\gamma\sum_{\beta\alpha}\big({G_\beta}^\sd{\varphi}\Nu_{\alpha\sd{\varphi}}\big) - \sum_{\beta\alpha}\eps_{\gamma\beta}\cD_\alpha\Big(\frac{1}{2}(H+\Hb) + \frac{i}{4}\cD^\varphi\cD^\sd{\varphi}\Nu_{\varphi\sd{\varphi}}\Big)\,, \label{eq:eom_superspace_torsion_constraints_2_sol_2}\\
\Pi_{\gamma\sym{2}{\sd{\beta}\sd{\alpha}}} &= -\frac{i}{4}\cD_\gamma\sum_{\sd{\beta}\sd{\alpha}}\big(\cD^\varphi\cD_\sd{\beta}\Nu_{\varphi\sd{\alpha}} - {G^\varphi}_\sd{\beta}\Nu_{\varphi\sd{\alpha}}\big)\,. \label{eq:eom_superspace_torsion_constraints_2_sol_3}
\end{align}
Furthermore, the variation of $\tensor{T}{_\gamma^{\sd{\beta}}_{\sd{\alpha}}}$ has the form
\begin{align}
\begin{split}
0 = \delta \tensor{T}{_\gamma^{\sd{\beta}}_{\sd{\alpha}}} &= \cD_\gamma {H^\sd{\beta}}_\sd{\alpha} + \cD^\sd{\beta} H_{\gamma\sd{\alpha}} + {T_{\gamma}}^{\sd{\beta}d}H_{d\sd{\alpha}} - {H_\gamma}^d\tensor{T}{_d^{\sd{\beta}}_{\sd{\alpha}}} - H^{\sd{\beta}d}T_{d\gamma\sd{\alpha}} + \tensor{\phi}{_\gamma^{\sd{\beta}}_{\sd{\alpha}}} - \delta^\sd{\beta}_\sd{\alpha}\omega_\gamma \\
&= \tensor{\Pi}{_\gamma^{\sd{\beta}}_{\sd{\alpha}}} - \delta^\sd{\beta}_\sd{\alpha}\Sigma_\gamma + {T_\gamma}^{\sd{\beta}d}X_{d\sd{\alpha}} + \cD_\gamma{X^\sd{\beta}}_\sd{\alpha} - 2(\cD^\sd{\beta}\Nu^d)T_{\gamma d\sd{\alpha}}\,.
\end{split}
\end{align}
This equation is solved by
\begin{align}
\Sigma_\gamma &= -\cD_\gamma\Big(H + \frac{1}{2}\Hb + \frac{i}{4}\cD^\varphi\cD^\sd{\varphi}\Nu_{\varphi\sd{\varphi}} - \frac{i}{2}\Nu^a G_a\Big)\,. \label{eq:eom_superspace_torsion_constraints_2_sol_4}
\end{align}
An explicit calculation shows, that by using the expressions for ${\Pi_{\gamma\beta}}^\alpha$ and $\Sigma_\gamma$ stated in Eq.~\eqref{eq:eom_superspace_torsion_constraints_2_sol_2} and \eqref{eq:eom_superspace_torsion_constraints_2_sol_4}, respectively, the equation
\begin{align}
\begin{split}
0 = \delta {T_{\gamma\beta}}^\alpha &= \sum_{\gamma\beta}\big(\cD_\gamma{H_\beta}^\alpha - {H_\gamma}^d{T_{d\beta}}^\alpha + {\phi_{\gamma\beta}}^\alpha + \delta_\beta^\alpha\omega_\gamma\big) \\
&= \sum_{\gamma\beta}\big({\Pi_{\gamma\beta}}^\alpha + \delta_\beta^\alpha\Sigma_\gamma + \cD_\gamma{X_\beta}^\alpha\big)
\end{split}
\end{align}
is automatically fulfilled.

\subsubsection{Torsion constraints III}
The torsion constraints
\begin{align}
\tensor{T}{^{\sd{\gamma}}_b^a} &= 0\,, \\
\tensor{T}{^{\sd{\gamma}}_\beta^\alpha} &= 0\,, \\
{T^{\sd{\gamma}\sd{\beta}}}_\sd{\alpha} &= 0\,,
\end{align}
are the conjugate versions of the ones in Eqs.~\eqref{eq:eom_superspace_torsion_constraints_2_constr_torsion_comp_1}--\eqref{eq:eom_superspace_torsion_constraints_2_constr_torsion_comp_3}. Again, it is convenient to introduce some new superfields
\begin{align}
{\Xb_b}^a &:= {H_b}^a + \cD_b\Nu^a\,, & {\Xb_b}^\alpha &:= {H_b}^\alpha + \Nu^c {T_{cb}}^\alpha\,, \nonumber\\
{\Xb_\beta}^\alpha &:= {H_\beta}^\alpha + \Nu^c{T_{c\beta}}^\alpha\,, & {\Xb^\sd{\beta}}_\sd{\alpha} &:= {H^\sd{\beta}}_\sd{\alpha} + \Nu^c\tensor{T}{_c^{\sd{\beta}}_{\sd{\alpha}}}\,, \label{eq:eom_superspace_torsion_constraints_3_field_definitions}\\
\tensor{\Pib}{^{\sd{\gamma}}_B^A} &:= \tensor{\phi}{^{\sd{\gamma}}_B^A} + \Nu^d\tensor{R}{_d^{\sd{\gamma}}_B^A}\,, & \Sigmab^\sd{\gamma} &:= \omega^\sd{\gamma} + \Nu^d {F_d}^\sd{\gamma}\,, \nonumber
\end{align}
where, in terms of spinor indices, ${\Xb_b}^a$ and $\tensor{\phi}{^{\sd{\gamma}}_b^a}$ are written as
\begin{align}
\Xb_{\beta\sd{\beta}\,\alpha\sd{\alpha}} &= +i\cD_\sd{\beta}\cD_\beta\Nu_{\alpha\sd{\alpha}} - \eps_{\beta\alpha}\eps_{\sd{\beta}\sd{\alpha}}(H+\Hb) + i\eps_{\sd{\beta}\sd{\alpha}}\Nu_{\beta\sd{\gamma}}{G_\alpha}^\sd{\gamma} + i\eps_{\beta\alpha}{\Nu^\gamma}_\sd{\beta}G_{\gamma\sd{\alpha}}\,, \label{eq:eom_superspace_torsion_constraints_3_id_1}\\
{\Pib^\sd{\gamma}}_{\beta\sd{\beta}\s\alpha\sd{\alpha}} &= + 2\eps_{\sd{\beta}\sd{\alpha}}{\Pib^\sd{\gamma}}_{\sym{2}{\beta\alpha}} - 2\eps_{\beta\alpha}{\Pib^\sd{\gamma}}_{\sym{2}{\sd{\beta}\sd{\alpha}}}\,, \label{eq:eom_superspace_torsion_constraints_3_id_2}
\end{align}
by using Eq.~\eqref{eq:eom_superspace_torsion_constraints_1_sol_7} to derive \eqref{eq:eom_superspace_torsion_constraints_3_id_1}. The equation corresponding to the variation of $\tensor{T}{^{\sd{\gamma}}_b^a}$, namely
\begin{align}
\begin{split}
0 = \delta \tensor{T}{^{\sd{\gamma}}_b^a} &= \cD^\sd{\gamma}{H_b}^a - \cD_b H^{\sd{\gamma}a} + \tensor{T}{^{\sd{\gamma}}_b^{\su{\delta}}}{H_{\su{\delta}}}^a + {H_b}^\delta{T_\delta}^{\sd{\gamma}a} + \tensor{\phi}{^{\sd{\gamma}}_b^a} \\
&= \tensor{\Pib}{^{\sd{\gamma}}_b^a} + \cD^\sd{\gamma}{\Xb_b}^a + {\Xb_b}^\delta{T_\delta}^{\sd{\gamma}a} + 2\tensor{T}{^{\sd{\gamma}}_b^\delta}\cD_\delta\Nu^a\,,
\end{split}
\end{align}
is solved by
\begin{align}
X_{\beta\sd{\beta}\s\alpha} &= -\frac{i}{2}\eps_{\beta\alpha}\cD_\sd{\beta}(H+\Hb) + \frac{1}{4}\cD_\sd{\beta}\cD^\sd{\varphi}\cD_\alpha\Nu_{\beta\sd{\varphi}} + R\,\cD_\beta\Nu_{\alpha\sd{\beta}}\,, \label{eq:eom_superspace_torsion_constraints_3_sol_1}\\
\Pib_{\sd{\gamma}\sym{2}{\sd{\beta}\sd{\alpha}}} &= -\frac{i}{4}\cD_\sd{\gamma}\sum_{\sd{\beta}\sd{\alpha}}\big({G^\varphi}_\sd{\beta}\Nu_{\varphi\sd{\alpha}}\big) + \sum_{\sd{\beta}\sd{\alpha}}\eps_{\sd{\gamma}\sd{\beta}}\cD_\sd{\alpha}\Big(\frac{1}{2}(H+\Hb) - \frac{i}{4}\cD^\sd{\varphi}\cD^\varphi\Nu_{\varphi\sd{\varphi}}\Big)\,, \label{eq:eom_superspace_torsion_constraints_3_sol_2}\\
\Pib_{\sd{\gamma}\sym{2}{\beta\alpha}} &= -\frac{i}{4}\cD_\sd{\gamma}\sum_{\beta\alpha}\big(\cD^\sd{\varphi}\cD_\beta\Nu_{\alpha\sd{\varphi}} + {G_\beta}^\sd{\varphi}\Nu_{\alpha\sd{\varphi}}\big)\,. \label{eq:eom_superspace_torsion_constraints_3_sol_3}
\end{align}
In addition, the variation of $\tensor{T}{^{\sd{\gamma}}_\beta^\alpha}$ is given by
\begin{align}
\begin{split}
0 = \delta \tensor{T}{^{\sd{\gamma}}_\beta^\alpha} &= \cD^\sd{\gamma} {H_\beta}^\alpha + \cD_\beta H^{\sd{\gamma}\alpha} + \tensor{T}{^{\sd{\gamma}}_\beta^d}{H_d}^\alpha - H^{\sd{\gamma}d}{T_{d\beta}}^\alpha - {H_\beta}^d {T_d}^{\sd{\gamma}\alpha} + \tensor{\phi}{^{\sd{\gamma}}_\beta^\alpha} + \delta_\beta^\alpha\omega^\sd{\gamma} \\
&= \tensor{\Pib}{^{\sd{\gamma}}_\beta^\alpha} + \delta_\beta^\alpha\Sigmab^\sd{\gamma} + {T_\beta}^{\sd{\gamma}d}{\Xb_d}^\alpha + \cD^\sd{\gamma}{\Xb_\beta}^\alpha + 2(\cD_\beta\Nu^d)\tensor{T}{^{\sd{\gamma}}_d^\alpha}\,,
\end{split}
\end{align}
with the solution
\begin{align}
\Sigmab^\sd{\gamma} &= +\cD^\sd{\gamma}\Big(\Hb + \frac{1}{2}H - \frac{i}{4}\cD^\sd{\varphi}\cD^\varphi\Nu_{\varphi\sd{\varphi}} - \frac{i}{2}\Nu^a G_a\Big)\,. \label{eq:eom_superspace_torsion_constraints_3_sol_4}
\end{align}
Finally, the equation
\begin{align}
\begin{split}
0 = \delta {T^{\sd{\gamma}\sd{\beta}}}_\sd{\alpha} &= \sum_{\sd{\gamma}\sd{\beta}}\big(\cD^\sd{\gamma}{H^\sd{\beta}}_\sd{\alpha} - H^{\sd{\gamma}d}\tensor{T}{_d^{\sd{\beta}}_{\sd{\alpha}}} + {\phi^{\sd{\gamma}\sd{\beta}}}_\sd{\alpha} - \delta^\sd{\beta}_\sd{\alpha}\omega^\sd{\gamma}\big) \\
&= \sum_{\sd{\gamma}\sd{\beta}}\big({\Pib^{\sd{\gamma}\sd{\beta}}}_\sd{\alpha} - \delta^\sd{\beta}_\sd{\alpha}\Sigmab^\sd{\gamma} + \cD^\sd{\gamma}{\Xb^\sd{\beta}}_\sd{\alpha}\big)
\end{split}
\end{align}
is automatically fulfilled, by plugging in Eq.~\eqref{eq:eom_superspace_torsion_constraints_3_sol_2} and \eqref{eq:eom_superspace_torsion_constraints_3_sol_4} for ${\Pib^{\sd{\gamma}\sd{\beta}}}_\sd{\alpha}$ and $\Sigmab^\sd{\gamma}$, respectively.

\subsubsection{Variation of $R$ and $\Rb$}
In this section the variation of the superfields $R$ and $\Rb$ are derived, by using the expressions for ${H_B}^A$ from above. The superfield $R$ appears in the component $\tensor{T}{^{\sd{\gamma}}_b^\alpha}$ of the torsion tensor (cf.\ Eq.~\eqref{eq:bianchi_sp_components_torsion_dim1_4}) and can be expressed as
\begin{align}
R &= -\frac{i}{8}\sig^b_{\alpha\sd{\gamma}} \tensor{T}{^{\sd{\gamma}}_b^\alpha} = +\frac{i}{8}{T^{\sd{\beta}\alpha}}_{\sd{\beta}\alpha}\,. \label{eq:eom_superspace_var_r_id_1}
\end{align}
According to Eq.~\eqref{eq:eom_superspace_variation_torsion_components}, the variation of $\tensor{T}{^{\sd{\gamma}}_b^\alpha}$ reads
\begin{align}
\delta \tensor{T}{^{\sd{\gamma}}_b^\alpha} &= \cD^\sd{\gamma}{H_b}^\alpha - \cD_b H^{\sd{\gamma}\alpha} + \tensor{T}{^{\sd{\gamma}}_b^{\su{\delta}}}{H_\delta}^\alpha - H^{\sd{\gamma}d}{T_{db}}^\alpha - H^{\sd{\gamma}\su{\delta}}{T_{\su{\delta}b}}^\alpha + {H_b}^d{T_d}^{\sd{\gamma}\alpha}\,, \label{eq:eom_superspace_var_r_id_2}
\end{align}
which implies
\begin{align}
\delta {T^{\sd{\beta}\alpha}}_{\sd{\beta}\alpha} &= -4i\Nu^{\alpha\sd{\beta}}(\cD_{\alpha\sd{\beta}}R) + 8iR\,\Hb - 4R\,\Nu^{\alpha\sd{\beta}}G_{\alpha\sd{\beta}} + i\cDb^2(H+\Hb) + \frac{1}{2}\cDb^2\cD^\sd{\beta}\cD^\alpha\Nu_{\alpha\sd{\beta}}\,, \label{eq:eom_superspace_var_r_id_3}
\end{align}
by using spinor notation. Thus, from Eq.~\eqref{eq:eom_superspace_var_r_id_1} follows that the variation of $R$ is given by
\begin{align}
\delta R &= -(\Nu^a\cD_a + \Hb - i\Nu^a G_a)R - \frac{1}{8}\cDb^2\Big(H + \Hb - \frac{i}{2}\cD^\sd{\varphi}\cD^\varphi\Nu_{\varphi\sd{\varphi}}\Big)\,. \label{eq:eom_superspace_var_r_variation_r}
\end{align}
In a similar way the variation of the superfield $\Rb$ is derived, which is contained in the component $T_{\gamma b\sd{\alpha}}$ of the torsion tensor (cf.\ Eq.~\eqref{eq:bianchi_sp_components_torsion_dim1_3}). It can be written as
\begin{align}
\Rb &= -\frac{i}{8}\sigb^{b\sd{\alpha}\gamma}T_{\gamma b\sd{\alpha}} = +\frac{i}{8}\tensor{T}{^\beta_\beta^{\sd{\alpha}}_{\sd{\alpha}}}\,. \label{eq:eom_superspace_var_r_id_4}
\end{align}
The variation of $T_{\gamma b\sd{\alpha}}$ is given by
\begin{align}
\delta T_{\gamma b\sd{\alpha}} &= \cD_\gamma H_{b\sd{\alpha}} - \cD_b H_{\gamma\sd{\alpha}} + {T_{\gamma b}}^\su{\delta} H_{\su{\delta}\sd{\alpha}} - {H_\gamma}^d T_{db\sd{\alpha}} - {H_\gamma}^\su{\delta}T_{\su{\delta}b\sd{\alpha}} + {H_b}^dT_{d\gamma\sd{\alpha}}\,, \label{eq:eom_superspace_var_r_id_5}
\end{align}
from which follows that
\begin{align}
\delta \tensor{T}{^\beta_\beta^{\sd{\alpha}}_{\sd{\alpha}}} &= +4i\Nu^{\beta\sd{\alpha}}(\cD_{\beta\sd{\alpha}}\Rb) + 8i\Rb\,H - 4\Rb\,\Nu^{\beta\sd{\alpha}}G_{\beta\sd{\alpha}} + i\cD^2(H+\Hb) - \frac{1}{2}\cD^2\cD^\beta\cD^\sd{\alpha}\Nu_{\beta\sd{\alpha}}\,. \label{eq:eom_superspace_var_r_id_6}
\end{align}
Therefore, according to Eq.~\eqref{eq:eom_superspace_var_r_id_4}, the variation of $\Rb$ reads
\begin{align}
\delta \Rb &= +(\Nu^a\cD_a - H + i\Nu^a G_a)\Rb - \frac{1}{8}\cD^2\Big(H + \Hb + \frac{i}{2}\cD^\varphi\cD^\sd{\varphi}\Nu_{\varphi\sd{\varphi}}\Big)\,. \label{eq:eom_superspace_var_r_variation_rb}
\end{align}
The variations $\delta R$ and $\delta\Rb$ are used Section~\ref{sec:variation_action}, where the variation of the action is considered. In principle, the variation of the superfields $G^a$, $W_{\sym{3}{\gamma\beta\alpha}}$ and $W_{\sym{3}{\sd{\gamma}\sd{\beta}\sd{\alpha}}}$ can be derived from the expressions of $\delta{T_{\gamma b}}^\alpha$, $\delta{T_{cb}}^\alpha$ and $\delta T_{cb\sd{\alpha}}$. However, this calculation not done here, because there is no further use for these quantities.

\subsection{Variation of the Yang-Mills field strength}
\label{sec:variation_ym_fieldstrength}
Beside the supervielbein and the Lorentz connection, another basic object in K\"ahler superspace with a Yang-Mills gauge symmetry is the Yang-Mills connection $\cA = \cA^\g{r}\gen{r}$. The variations of $\cA$, which determine the variations of the Yang-Mills field strength $\cF$, must be compatible with the covariant constraints of $\cF_{BA}$. Thus, the variational version of the constraint equations have to be solved. The infinitesimal variation of the component $\cA^\g{r}$ has the general form
\begin{align}
\delta\cA^\g{r} = \exd z^M \delta{\cA_M}^\g{r}\,. \label{eq:eom_ym_variation_ym_connection_1}
\end{align}
This can also be written as
\begin{align}
\delta\cA^\g{r} &= \Gamma^\g{r}\,, & \text{with} && \quad \Gamma^\g{r} &= E^A {\Gamma_A}^\g{r}\,, & {\Gamma_A}^\g{r} &= {E_A}^M\delta{\cA_M}^\g{r}\,, \label{eq:eom_ym_variation_ym_connection_2}
\end{align}
where the $1$-superform $\Gamma^\g{r}$ is introduced, which parametrizes $\delta\cA^\g{r}$. By using the expression of the Yang-Mills field strength from Eq.~\eqref{eq:yang_mills_curvature_components_lie_algebra}, and by applying Eq.~\eqref{eq:eom_ym_variation_ym_connection_2}, the infinitesimal variation of $\cF^\g{r}$ is given by
\begin{align}
\begin{split}
\delta\cF^\g{r} &= \delta\big(\exd\cA^\g{r} + \frac{i}{2}\cA^\g{p}\wedge\cA^\g{q}\sco{p}{q}{r}\big) \\
&= \exd(\delta\cA^\g{r}) + \frac{i}{2}(\delta\cA^\g{p})\wedge\cA^\g{q}\sco{p}{q}{r} + \frac{i}{2}\cA^\g{p}\wedge(\delta\cA^\g{q})\sco{p}{q}{r} \\
&= \exd\Gamma^\g{r} + i\Gamma^p\wedge\cA^q\sco{p}{q}{r} \\
&= \cD\Gamma^\g{r}\,.
\end{split} \label{eq:eom_ym_variation_fieldstrength}
\end{align}
If the last term in Eq.~\eqref{eq:eom_ym_variation_fieldstrength} is written explicitly as a $2$-superform, namely
\begin{align}
\begin{split}
\cD\Gamma^\g{r} &= E^A\wedge\cD{\Gamma_A}^\g{r} + T^A{\Gamma_A}^\g{r} \\
&= \frac{1}{2}E^A\wedge E^B\big(\cD_B{\Gamma_A}^\g{r} -(-1)^{BA} \cD_A{\Gamma_B}^\g{r} + {T_{BA}}^C{\Gamma_C}^\g{r}\big)\,,
\end{split}
\end{align}
and the variation of the field strength is expanded as
\begin{align}
\begin{split}
\delta\cF^\g{r} &= \delta\Big(\frac{1}{2}E^A\wedge E^B{\cF_{BA}}^\g{r}\Big) \\
&= \frac{1}{2}(H^A\wedge E^B + E^A\wedge H^B){\cF_{BA}}^\g{r} + \frac{1}{2}E^A\wedge E^B\delta{\cF_{BA}}^\g{r} \\
&= \frac{1}{2}E^A\wedge E^B\big({H_B}^C{\cF_{CA}}^\g{r} - (-1)^{BA}{H_A}^C{\cF_{CB}}^\g{r} + \delta{\cF_{BA}}^\g{r}\big)\,,
\end{split}
\end{align}
the variation of the field strength components can be read off:
\begin{align}
\begin{split}
\delta{\cF_{BA}}^\g{r} &= \cD_B{\Gamma_A}^\g{r} -(-1)^{BA}\cD_A{\Gamma_B}^\g{r} + {T_{BA}}^C{\Gamma_C}^\g{r} \\
&\quad -{H_B}^C{\cF_{CA}}^\g{r} + (-1)^{BA}{H_A}^C{\cF_{CB}}^\g{r}\,.
\end{split} \label{eq:eom_ym_variation_fieldstrength_components}
\end{align}
There is the freedom to use a Yang-Mills gauge transformation to bring the variation to a more convenient form. Furthermore, the Yang-Mills connection is also affected by superspace diffeomorphisms. Under an infinitesimal Yang-Mills transformation, parametrized by $\lac{r}$ where $i\lac{r}\gen{r}\in\mathfrak{g}$, and an infinitesimal superspace diffeomorphism $\xi$ the connection transforms as
\begin{align}
\begin{split}
\deltat\cA^\g{r} &= \lied_\xi\cA^\g{r} + \lac{p}\cA^\g{q}\sco{p}{q}{r} - i\exd\lac{r} \\
&= \intp_\xi\cF^\g{r} - i\cD(\lac{r} + i\intp_\xi\cA^\g{r})\,.
\end{split}
\end{align}
With the definition
\begin{align}
\Delta\Gamma^\g{r} &:= \deltat\cA^\g{r}\,, \label{eq:eom_ym_transf_gamma}
\end{align}
follows
\begin{align}
\Delta{\Gamma_A}^\g{r} &= \xi^B{\cF_{BA}}^\g{r} - i\cD_A\lact{r}\,, \label{eq:eom_ym_transf_gamma_components_1}
\end{align}
where $\lact{r} := \lac{r} + i\intp_\xi\cA^\g{r}$. Since $\xi$ has already been fixed in Eq.~\eqref{eq:eom_superspace_torsion_constraints_1_choice_xi}, in the following calculations ${\Gamma_A}^\g{r}$ represents the generic variation after the transformation under the infinitesimal superspace diffeomorphism. Thus Eq.~\eqref{eq:eom_ym_transf_gamma_components_1} takes the form
\begin{align}
\Delta{\Gamma_A}^\g{r} &= - i\cD_A\lact{r}\,. \label{eq:eom_ym_transf_gamma_components_2}
\end{align}

\subsubsection{Field strength constraints}
According to Eq.~\eqref{eq:yang_mills_constraints_compact} the following field strength components vanish:
\begin{align}
{\cF_{\beta\alpha}}^\g{r} &= 0\,, \label{eq:eom_ym_fieldstrength_constraints_constr_fs_comp_1}\\
\cF^{\sd{\beta}\sd{\alpha}\g{r}} &= 0\,. \label{eq:eom_ym_fieldstrength_constraints_constr_fs_comp_2}
\end{align}
By using Eq.~\eqref{eq:eom_ym_variation_fieldstrength_components}, the variational versions of these equations are written as
\begin{align}
0 = \delta{\cF_{\beta\alpha}}^\g{r} &= \sum_{\beta\alpha}\big(\cD_\beta{\Gamma_\alpha}^\g{r} - {H_\beta}^d{\cF_{d\alpha}}^\g{r}\big)\,, \label{eq:eom_ym_fieldstrength_constraints_var_fs_comp_1}\\
0 = \delta\cF^{\sd{\beta}\sd{\alpha}\g{r}} &= \sum_{\sd{\beta}\sd{\alpha}}\big(\cD^\sd{\beta}\Gamma^{\sd{\alpha}\g{r}} - H^{\sd{\beta}d}{\cF_d}^{\sd{\alpha}\g{r}}\big)\,, \label{eq:eom_ym_fieldstrength_constraints_var_fs_comp_2}
\end{align}
and have the solutions
\begin{align}
{\Gamma_\alpha}^\g{r} &= +\cD_\alpha\Xi^\g{r} + \Nu^d{\cF_{d\alpha}}^\g{r}\,, \label{eq:eom_ym_fieldstrength_constraints_solution_1}\\
\Gamma^{\sd{\alpha}\g{r}} &= -\cD^\sd{\alpha}\Xib^\g{r} - \Nu^d{\cF_d}^{\sd{\alpha}\g{r}}\,, \label{eq:eom_ym_fieldstrength_constraints_solution_2}
\end{align}
where the superfields $\Xi^\g{r}$ and $\Xib^\g{r}$ are introduced ($(\Xi^\g{r})^\cc=\Xib^\g{r}$). According to Eq.~\eqref{eq:eom_ym_transf_gamma_components_2}, the variations ${\Gamma_\alpha}^\g{r}$ and $\Gamma^{\sd{\alpha}\g{r}}$ transform as
\begin{align}
\Delta{\Gamma_\alpha}^\g{r} &= - i\cD_\alpha\lact{r}\,, \label{eq:eom_ym_fieldstrength_constraints_identity_1}\\
\Delta\Gamma^{\sd{\alpha}\g{r}} &=  - i\cD^\sd{\alpha}\lact{r}\,, \label{eq:eom_ym_fieldstrength_constraints_identity_2}
\end{align}
under an infinitesimal Yang-Mills transformation. Thus, the following choice of $\lact{r}$,
\begin{align}
\lact{r}=-\frac{i}{2}(\Xi^\g{r}-\Xib^\g{r})\,, \label{eq:eom_ym_fieldstrength_constraints_choice_alpha}
\end{align}
leads to
\begin{align}
{\Gamma_\alpha}^\g{r} &= +\cD_\alpha\Sigma^\g{r} + \Nu^d{\cF_{d\alpha}}^\g{r}\,, \label{eq:eom_ym_fieldstrength_constraints_identity_3}\\
\Gamma^{\sd{\alpha}\g{r}} &= -\cD^\sd{\alpha}\Sigma^\g{r} - \Nu^d{\cF_d}^{\sd{\alpha}\g{r}}\,, \label{eq:eom_ym_fieldstrength_constraints_identity_4}
\end{align}
with the real superfield $\Sigma^\g{r}:=\frac{1}{2}(\Xi^\g{r}+\Xib^\g{r})$.
\\\\
The remaining constraint of the strength components stated in Eq.~\eqref{eq:yang_mills_constraints_compact}
is the following
\begin{align}
{\cF_\beta}^{\sd{\alpha}\g{r}} = 0\,, \label{eq:eom_ym_fieldstrength_constraints_constr_fs_comp_3}
\end{align}
which implies
\begin{align}
\begin{split}
0 = \delta{\cF_\beta}^{\sd{\alpha}\g{r}} &= \cD_\beta\Gamma^{\sd{\alpha}\g{r}} + \cD^\sd{\alpha}{\Gamma_\beta}^\g{r} + {T_\beta}^{\sd{\alpha}d}{\Gamma_d}^\g{r} - {H_\beta}^d{\cF_d}^{\sd{\alpha}\g{r}} - H^{\sd{\alpha}d}{\cF_{d\beta}}^\g{r} \\
&= \cD_\beta\Gamma^{\sd{\alpha}\g{r}} + \cD^\sd{\alpha}{\Gamma_\beta}^\g{r} + {T_\beta}^{\sd{\alpha}d}{\Gamma_d}^\g{r} - \cD_\beta\Nu^d{\cF_d}^{\sd{\alpha}\g{r}} + \cD^\sd{\alpha}\Nu^d{\cF_{d\beta}}^\g{r}\,.
\end{split} \label{eq:eom_ym_fieldstrength_constraints_var_fs_comp_3}
\end{align}
It is convenient for later calculations to write the solution of this equation in two different ways, by introducing the two superfields
\begin{align}
{\Lambda_a}^\g{r} &:= {\Gamma_a}^\g{r} - \cD_a\Sigma^\g{r} - \Nu^b{\cF_{ba}}^\g{r}\,, \label{eq:eom_ym_fieldstrength_constraints_identity_5}\\
{\Lambdat_a}^\g{r} &:= {\Gamma_a}^\g{r} + \cD_a\Sigma^\g{r} + \Nu^b{\cF_{ba}}^\g{r}\,, \label{eq:eom_ym_fieldstrength_constraints_identity_6}
\end{align}
where $({\Lambda_a}^\g{r})^\cc=-{\Lambdat_a}^\g{r}$. Using spinor notation, the solution is given by
\begin{align}
{\Lambda_{\alpha\sd{\alpha}}}^\g{r} &= i\cD_\alpha{\Gamma_\sd{\alpha}}^\g{r} + i(\cD_\sd{\alpha}\Nu^b){\cF_{b\alpha}}^\g{r}\,, \label{eq:eom_ym_fieldstrength_constraints_solution_3}
\end{align}
which is equivalent to
\begin{align}
{\Lambdat_{\alpha\sd{\alpha}}}^\g{r} &= i\cD_\sd{\alpha}{\Gamma_\alpha}^\g{r} - i(\cD_\alpha\Nu^b){\cF_{b\sd{\alpha}}}^\g{r}\,. \label{eq:eom_ym_fieldstrength_constraints_solution_4}
\end{align}

\subsubsection{Variation of $\cW_\alpha$ and $\cW^\sd{\alpha}$}
In this section the variations of the superfields ${\cW_\alpha}^\g{r}$ and $\cW^{\sd{\alpha}\g{r}}$ are derived, by using the expressions for ${\Gamma_\alpha}^\g{r}$ and $\Gamma^{\sd{\alpha}\g{r}}$ from the previous section. The superfield ${\cW_\alpha}^\g{r}$ appears in the component $\tensor{\cF}{^{\sd{\beta}}_a^{\g{r}}}$ of the Yang-Mills field strength (cf.\ Eq.~\eqref{eq:yang_mills_components_curvature_dim3/2_2}) and can be expressed as
\begin{align}
{\cW_\alpha}^\g{r} &= -\frac{i}{4}\sig^a_{\alpha\sd{\beta}}\tensor{\cF}{^{\sd{\beta}}_a^{\g{r}}} = -\frac{i}{4}\tensor{\cF}{^{\sd{\alpha}}_{\alpha\sd{\alpha}}^{\g{r}}}\,. \label{eq:eom_ym_var_w_id_1}
\end{align}
According to Eq.~\eqref{eq:eom_ym_variation_fieldstrength_components}, the variation of $\tensor{\cF}{^{\sd{\beta}}_a^{\g{r}}}$ is given by
\begin{align}
\begin{split}
\delta \tensor{\cF}{^{\sd{\beta}}_a^{\g{r}}} &= \cD^\sd{\beta}{\Gamma_a}^\g{r} - \cD_a\Gamma^{\sd{\beta}\g{r}} + \tensor{T}{^{\sd{\beta}}_a^\gamma}{\Gamma_\gamma}^\g{r} + {T^\sd{\beta}}_{a\sd{\gamma}}\Gamma^{\sd{\gamma}\g{r}} \\
&\quad - H^{\sd{\beta}c}{\cF_{ca}}^\g{r} - H^{\sd{\beta}\gamma}{\cF_{\gamma a}}^\g{r} - {H^\sd{\beta}}_\sd{\gamma}\tensor{\cF}{^{\sd{\gamma}}_a^{\g{r}}} + {H_a}^c{\cF_c}^{\sd{\beta}\g{r}}\,, \label{eq:eom_ym_var_w_id_2}
\end{split}
\end{align}
which implies
\begin{align}
\begin{split}
\delta \tensor{\cF}{^{\sd{\alpha}}_{\alpha\sd{\alpha}}^{\g{r}}} &= -4i\Nu^b\cD_b{\cW_\alpha}^\g{r} - 4i\big(\Hb+\frac{1}{2}H\big){\cW_\alpha}^\g{r} + 4\Sigma^p{\cW_\alpha}^\g{q}\sco{p}{q}{r} \\
&\quad - 2(\cD^\sd{\beta}\cD_\alpha\Nu_{\beta\sd{\beta}})\cW^{\beta\g{r}} + 2\Nu_{\alpha\sd{\beta}}G^{\beta\sd{\beta}}{\cW_\beta}^\g{r} - i(\cDb^2 - 8R){\Gamma_\alpha}^\g{r}\,, \label{eq:eom_ym_var_w_id_3}
\end{split}
\end{align}
using spinor notation. Thus, from Eq.~\eqref{eq:eom_ym_var_w_id_1} follows that the variation of ${\cW_\alpha}^\g{r}$ is given by
\begin{align}
\begin{split}
\delta {\cW_\alpha}^\g{r} &= -\Big(\Nu^b\cD_b + \Hb+\frac{1}{2}H\Big){\cW_\alpha}^\g{r} - i\Sigma^p{\cW_\alpha}^\g{q}\sco{p}{q}{r} \\
&\quad + \frac{i}{2}\big(\cD_\sd{\beta}\cD_\alpha\Nu^{\beta\sd{\beta}} - \Nu_{\alpha\sd{\beta}}G^{\beta\sd{\beta}}\big){\cW_\beta}^\g{r} - \frac{1}{4}(\cDb^2 - 8R){\Gamma_\alpha}^\g{r}\,.
\end{split} \label{eq:eom_ym_var_w_variation_w}
\end{align}
In a similar way the variation of the superfield $\cW^{\sd{\alpha}\g{r}}$ is derived, which is contained in the component ${\cF_{\beta a}}^\g{r}$ of the field strength (cf.\ Eq.~\eqref{eq:yang_mills_components_curvature_dim3/2_1}). It can be written as
\begin{align}
\cW^{\sd{\alpha}\g{r}} &= +\frac{i}{4}\sigb^{a\sd{\alpha}\beta}{\cF_{\beta a}}^\g{r} = +\frac{i}{4}{\cF_\alpha}^{\alpha\sd{\alpha}\s\g{r}}\,. \label{eq:eom_ym_var_w_id_4}
\end{align}
The variation of ${\cF_{\beta a}}^\g{r}$ is given by
\begin{align}
\begin{split}
\delta {\cF_{\beta a}}^\g{r} &= \cD_\beta{\Gamma_a}^\g{r} - \cD_a{\Gamma_\beta}^\g{r} + {T_{\beta a}}^\gamma{\Gamma_\gamma}^\g{r} + T_{\beta a\sd{\gamma}}\Gamma^{\sd{\gamma}\g{r}} \\
&\quad - {H_\beta}^c {\cF_{ca}}^\g{r} - {H_\beta}^\gamma {\cF_{\gamma a}}^\g{r} - H_{\beta\sd{\gamma}}\tensor{\cF}{^{\sd{\gamma}}_a^{\g{r}}} + {H_a}^c{\cF_{c\beta}}^\g{r}\,, \label{eq:eom_ym_var_w_id_5}
\end{split}
\end{align}
from which follows that
\begin{align}
\begin{split}
\delta {\cF_\alpha}^{\alpha\sd{\alpha}\s\g{r}} &= -4i\Nu^b\cD_b\cW^{\sd{\alpha}\g{r}} + 4i\big(H + \frac{1}{2}\Hb\big)\cW^{\sd{\alpha}\g{r}} + 4\Sigma^\g{p}\cW^{\sd{\alpha}\g{q}}\sco{p}{q}{r} \\
&\quad - 2(\cD^\beta\cD^\sd{\alpha}\Nu_{\beta\sd{\beta}})\cW^{\sd{\beta}\g{r}} - 2\Nu^{\beta\sd{\alpha}}G_{\beta\sd{\beta}}\cW^{\sd{\beta}\g{r}} - i(\cD^2 - 8\Rb)\Gamma^{\sd{\alpha}\g{r}}\,. \label{eq:eom_ym_var_w_id_6}
\end{split}
\end{align}
Therefore, according to Eq.~\eqref{eq:eom_ym_var_w_id_4}, the variation of $\cW^{\sd{\alpha}\g{r}}$ reads
\begin{align}
\begin{split}
\delta \cW^{\sd{\alpha}\g{r}} &= +\Big(\Nu^b\cD_b - H - \frac{1}{2}\Hb\Big)\cW^{\sd{\alpha}\g{r}} + i\Sigma^\g{p}\cW^{\sd{\alpha}\g{q}}\sco{p}{q}{r} \\
&\quad - \frac{i}{2}\big(\cD^\beta\cD^\sd{\alpha}\Nu_{\beta\sd{\beta}} + \Nu^{\beta\sd{\alpha}}G_{\beta\sd{\beta}}\big)\cW^{\sd{\beta}\g{r}} + \frac{1}{4}(\cD^2 - 8\Rb)\Gamma^{\sd{\alpha}\g{r}}\,.
\end{split} \label{eq:eom_ym_var_w_variation_wb}
\end{align}
The variations of ${\cW_\alpha}^\g{r}$ and $\cW^{\sd{\alpha}\g{r}}$ are used Section~\ref{sec:variation_action} where the variation of the action in the Yang-Mills sector is considered.

\subsection{Variation of the matter superfields}
\label{sec:variation_matter_superfields}
The variations of the matter superfields have to be compatible with the condition, that $\phi^k$ and $\phib^\kb$ are a chiral and an antichiral superfield, respectively. Thus, the variational versions of these constraints have to be considered. The general infinitesimal variations of the matter superfields are written as $\delta\phi^k$ and $\delta\phib^\kb$. The infinitesimal variations of their covariant derivatives are then given by
\begin{align}
\begin{split}
\delta\cD\phi^k &= \delta\big(\exd\phi^k - \cA^\g{r}(\gen{r}\phi)^k\big) \\
&= \exd(\delta\phi^k) - (\delta\cA^\g{r})(\gen{r}\phi)^k - \cA^\g{r}(\gen{r}\delta\phi)^k \\
&= \cD(\delta\phi^k) - \Gamma^\g{r}(\gen{r}\phi)^k\,,
\end{split} \label{eq:eom_matter_variation_covariant_derivative_phi_1}\\
\nonumber\\
\begin{split}
\delta\cD\phib^\kb &= \delta\big(\exd\phib^\kb + \cA^\g{r}(\phib\gen{r})^\kb\big) \\
&= \exd(\delta\phib^\kb) + (\delta\cA^\g{r})(\phib\gen{r})^\kb + \cA^\g{r}(\delta\phib\gen{r})^\kb \\
&= \cD(\delta\phib^\kb) + \Gamma^\g{r}(\phib\gen{r})^\kb\,.
\end{split} \label{eq:eom_matter_variation_covariant_derivative_phib_1}
\end{align}
On the other hand $\delta\cD\phi^k$ and $\delta\cD\phib^\kb$ can also be evaluated as follows
\begin{align}
\begin{split}
\delta\cD\phi^k &= (\delta E^A)\cD_A\phi^k + E^A(\delta\cD_A\phi^k) \\
&= E^A({H_A}^B\cD_B\phi^k + \delta\cD_A\phi^k)\,,
\end{split} \label{eq:eom_matter_variation_covariant_derivative_phi_2}\\
\nonumber\\
\begin{split}
\delta\cD\phib^\kb &= (\delta E^A)\cD_A\phib^\kb + E^A(\delta\cD_A\phib^\kb) \\
&= E^A({H_A}^B\cD_B\phib^\kb + \delta\cD_A\phib^\kb)\,,
\end{split} \label{eq:eom_matter_variation_covariant_derivative_phib_2}
\end{align}
which leads to expressions
\begin{align}
\delta\cD_A\phi^k &= \cD_A(\delta\phi^k) - {\Gamma_A}^\g{r}(\gen{r}\phi)^k - {H_A}^B\cD_B\phi^k\,, \label{eq:eom_matter_variation_covariant_derivative_phi_components}\\
\delta\cD_A\phib^\kb &= \cD_A(\delta\phib^\kb) + {\Gamma_A}^\g{r}(\phib\gen{r})^\kb - {H_A}^B\cD_B\phib^\kb\,. \label{eq:eom_matter_variation_covariant_derivative_phib_components}
\end{align}

\subsubsection{Chirality constraints}
The matter superfields $\phi^k$ and $\phib^\kb$ fulfil the chirality constraints
\begin{align}
\cD^\sd{\alpha}\phi^k &= 0\,, & \cD_\alpha\phib^\kb &= 0\,. \label{eq:eom_matter_chirality_constr_phi_phib_constr}
\end{align}
According to Eq.~\eqref{eq:eom_matter_variation_covariant_derivative_phi_components} and \eqref{eq:eom_matter_variation_covariant_derivative_phib_components}, the variational versions of these constraints are given by
\begin{align}
\begin{split}
0 = \delta \cD^\sd{\alpha}\phi^k &= \cD^\sd{\alpha}(\delta\phi^k) - \Gamma^{\sd{\alpha}\g{r}}(\gen{r}\phi)^k - H^{\sd{\alpha}B}\cD_B\phi^k \\
&= \cD^\sd{\alpha}\big(\delta\phi^k + \Sigma^\g{r}(\gen{r}\phi)^k + \Nu^b\cD_b\phi^k\big) \\
&= \cD^\sd{\alpha}\eta^k\,,
\end{split} \label{eq:eom_matter_chirality_constr_var_phi_constr}\\
\nonumber\\
\begin{split}
0 = \delta \cD_\alpha\phib^\kb &= \cD_\alpha(\delta\phib^\kb) + {\Gamma_\alpha}^\g{r}(\phib\gen{r})^\kb - {H_\alpha}^B\cD_B\phib^\kb \\
&= \cD_\alpha\big(\delta\phib^\kb + \Sigma^\g{r}(\phib\gen{r})^\kb - \Nu^b\cD_b\phib^\kb\big) \\
&= \cD_\alpha\etab^\kb\,,
\end{split} \label{eq:eom_matter_chirality_constr_var_phib_constr}
\end{align}
with the definitions
\begin{align}
\eta^k &:= \delta\phi^k + \Sigma^\g{r}(\gen{r}\phi)^k + \Nu^b\cD_b\phi^k\,, \label{eq:eom_matter_chirality_constr_def_eta}\\
\etab^\kb &:= \delta\phib^\kb + \Sigma^\g{r}(\phib\gen{r})^\kb - \Nu^b\cD_b\phib^\kb\,. \label{eq:eom_matter_chirality_constr_def_etab}
\end{align}
From Eq.~\eqref{eq:eom_matter_chirality_constr_var_phi_constr} and \eqref{eq:eom_matter_chirality_constr_var_phib_constr} follows that $\eta^k$ and $\etab^\kb$ are chiral and antichiral, respectively. Thus, they can be written in terms of the unconstrained superfields $\zeta^k$ and $\zetab^\kb$ by using the projection operators $(\cDb^2-8R)$ and $(\cD^2-8\Rb)$ from Section~\ref{sec:chiral projection operators}:
\begin{align}
\eta^k &= (\cDb^2-8R)\zeta^k\,, \label{eq:eom_matter_chirality_constr_id_1}\\
\etab^\kb &= (\cD^2-8\Rb)\zetab^\kb\,. \label{eq:eom_matter_chirality_constr_id_2}
\end{align}

\subsubsection{Variation of the K\"ahler connection}
\label{sec:variation_kahler_connection}
In K\"ahler superspace the K\"ahler connection $A$ is not a basic object, but rather a function of the matter superfields, and in particular of the K\"ahler potential. In this section, the variations of the spinor components $A_\alpha$ and $A^\sd{\alpha}$ are related to the variation of $K$, and finally to the variations of $\phi^k$ and $\phib^\kb$. According to Eq.~\eqref{eq:eom_superspace_variation_kahler_connection_2}, the general infinitesimal variation of $A$ is parametrized by the $1$-superform $\omega$, namely
\begin{align}
\delta A = \omega\,,\quad\text{with}\quad \omega=E^A\omega_A\,. \label{eq:eom_matter_kahler_connection_variation_a_1}
\end{align}
On the other hand, $\delta A$ can also be written as
\begin{align}
\delta A = \delta(E^A A_A) = (\delta E^A)A_A + E^A(\delta A_A) = E^A {H_A}^B A_B + E^A(\delta A_A)\,, \label{eq:eom_matter_kahler_connection_variation_a_2}
\end{align}
therefore
\begin{align}
0 = \omega_A - \delta A_A - {H_A}^B A_B\,. \label{eq:eom_matter_kahler_connection_id_1}
\end{align}
In K\"ahler superspace the components $A_\alpha$ and $A^\sd{\alpha}$ are identified with derivatives of the K\"ahler potential $K$, as stated in Eq.~\eqref{eq:kahler_superspace_1/2_connection_component_1} and \eqref{eq:kahler_superspace_1/2_connection_component_2}:
\begin{align}
A_\alpha &= +\frac{1}{4} D_\alpha K = +\frac{1}{4}{E_\alpha}^M\del_M K\,, \label{eq:eom_matter_kahler_connection_id_2}\\
\quad A^\sd{\alpha} &= -\frac{1}{4} D^\sd{\alpha}K = -\frac{1}{4}E^{\sd{\alpha}M}\del_M K\,. \label{eq:eom_matter_kahler_connection_id_3}
\end{align}
The variations of $A_\alpha$ and $A^\sd{\alpha}$ are thus determined by the variation of $E^A$ and $K$, and are given by
\begin{align}
\begin{split}
\delta A_\alpha &= +\frac{1}{4}(\delta{E_\alpha}^M)\del_M K + \frac{1}{4}{E_\alpha}^M\delta(\del_M K) \\
&= -\frac{1}{4}{H_\alpha}^B{E_B}^M\del_M K + \frac{1}{4}{E_\alpha}^M\del_M(\delta K)\,,
\end{split} \label{eq:eom_matter_kahler_connection_variation_spinor_component_1}\\
\nonumber\\
\begin{split}
\delta A^\sd{\alpha} &= -\frac{1}{4}(\delta E^{\sd{\alpha}M})\del_M K - \frac{1}{4}E^{\sd{\alpha}M}\delta(\del_M K) \\
&= +\frac{1}{4}H^{\sd{\alpha}B}{E_B}^M\del_M K - \frac{1}{4}E^{\sd{\alpha}M}\del_M(\delta K)\,.
\end{split} \label{eq:eom_matter_kahler_connection_variation_spinor_component_2}
\end{align}
Plugging these two identities into Eq.~\eqref{eq:eom_matter_kahler_connection_id_1} leads to
\begin{align}
0 &= \omega_\alpha - {H_\alpha}^B\Big(A_B - \frac{1}{4} D_B K\Big) - \frac{1}{4}\cD_\alpha(\delta K)\,, \label{eq:eom_matter_kahler_connection_id_4}\\
0 &= \omega^\sd{\alpha} - H^{\sd{\alpha}B}\Big(A_B + \frac{1}{4} D_B K\Big) + \frac{1}{4}\cD^\sd{\alpha}(\delta K)\,. \label{eq:eom_matter_kahler_connection_id_5}
\end{align}
With the expressions for ${H_\alpha}^B$ and $H^{\sd{\alpha}B}$ from the previous sections, and with $\omega_\alpha=\Sigma_\alpha+\Nu^bF_{b\alpha}$ and $\omega^\sd{\alpha}=\Sigmab^\sd{\alpha}-\Nu^b{F_b}^\sd{\alpha}$ (cf.\ Eq.~\eqref{eq:eom_superspace_torsion_constraints_2_field_definitions} and \eqref{eq:eom_superspace_torsion_constraints_3_field_definitions}), where the components of the K\"ahler field strength have the explicit form
\begin{align}
F_{b\alpha} &= \cD_b A_\alpha - \cD_\alpha A_b - {T_{\alpha b}}^\su{\gamma}A_\su{\gamma}\,, \label{eq:eom_matter_kahler_connection_id_6}\\
{F_b}^\sd{\alpha} &= \cD_b A^\sd{\alpha} - \cD^\sd{\alpha} A_b - \tensor{T}{^{\sd{\alpha}}_b^{\su{\gamma}}}A_\su{\gamma}\,, \label{eq:eom_matter_kahler_connection_id_7}
\end{align}
Eq.~\eqref{eq:eom_matter_kahler_connection_id_4} and \eqref{eq:eom_matter_kahler_connection_id_5} imply
\begin{align}
\Sigma_\alpha &= +\cD_\alpha\Big(\frac{1}{4}\delta K + \Nu^a A_a - \frac{1}{4}\Nu^a D_a K\Big)\,, \label{eq:eom_matter_kahler_connection_id_8}\\
\Sigmab^\sd{\alpha} &= -\cD^\sd{\alpha}\Big(\frac{1}{4}\delta K + \Nu^a A_a + \frac{1}{4}\Nu^a D_a K\Big)\,. \label{eq:eom_matter_kahler_connection_id_9}
\end{align}
On the other hand, Eq.~\eqref{eq:eom_superspace_torsion_constraints_2_sol_4} and \eqref{eq:eom_superspace_torsion_constraints_3_sol_4} state that
\begin{align}
\Sigma_\alpha &= -\cD_\alpha\Big(H + \frac{1}{2}\Hb + \frac{i}{4}\cD^\varphi\cD^\sd{\varphi}\Nu_{\varphi\sd{\varphi}} - \frac{i}{2}\Nu^a G_a\Big)\,, \label{eq:eom_matter_kahler_connection_id_10}\\
\Sigmab^\sd{\alpha} &= +\cD^\sd{\alpha}\Big(\Hb + \frac{1}{2}H - \frac{i}{4}\cD^\sd{\varphi}\cD^\varphi\Nu_{\varphi\sd{\varphi}} - \frac{i}{2}\Nu^a G_a\Big)\,, \label{eq:eom_matter_kahler_connection_id_11}
\end{align}
which implies the following identities
\begin{align}
0 &= \cD_\alpha\Big(H + \frac{1}{2}\Hb + \frac{1}{4}\delta K + \frac{i}{4}\cD^\varphi\cD^\sd{\varphi}\Nu_{\varphi\sd{\varphi}} + \Nu^a\big(A_a-\frac{i}{2}G_a\big) - \frac{1}{4}\Nu^a D_a K\Big)\,, \label{eq:eom_matter_kahler_connection_sol_1}\\
0 &= \cD^\sd{\alpha}\Big(\Hb + \frac{1}{2}H + \frac{1}{4}\delta K - \frac{i}{4}\cD^\sd{\varphi}\cD^\varphi\Nu_{\varphi\sd{\varphi}} + \Nu^a\big(A_a-\frac{i}{2}G_a\big) + \frac{1}{4}\Nu^a D_a K\Big)\,. \label{eq:eom_matter_kahler_connection_sol_2}
\end{align}
In order that the equations in Eq.~\eqref{eq:eom_matter_kahler_connection_sol_1} and \eqref{eq:eom_matter_kahler_connection_sol_2} are fulfilled, the terms in the corresponding brackets have to vanish up to an antichiral and a chiral superfield, respectively. The antichiral and the chiral superfield can be written in terms of unconstrained superfields $U$, $\Ub$, where $U^\cc = \Ub$, by applying the projection operators $(\cD^2-8\Rb)$ and $(\cDb^2-8R)$ from Section~\ref{sec:chiral projection operators}. The identities in Eq.~\eqref{eq:eom_matter_kahler_connection_sol_1} and \eqref{eq:eom_matter_kahler_connection_sol_2} are then equivalent to the expressions
\begin{align}
H + \frac{1}{2}\Hb &= -\frac{1}{4}\delta K - \frac{i}{4}\cD^\varphi\cD^\sd{\varphi}\Nu_{\varphi\sd{\varphi}} - \Nu^a\Big(A_a-\frac{i}{2}G_a\Big) + \frac{1}{4}\Nu^a D_a K + (\cD^2-8\Rb)U\,, \label{eq:eom_matter_kahler_connection_sol_3}\\
\Hb + \frac{1}{2}H &= -\frac{1}{4}\delta K +\frac{i}{4}\cD^\sd{\varphi}\cD^\varphi\Nu_{\varphi\sd{\varphi}} - \Nu^a\Big(A_a-\frac{i}{2}G_a\Big) - \frac{1}{4}\Nu^a D_a K + (\cDb^2-8R)\Ub\,. \label{eq:eom_matter_kahler_connection_sol_4}
\end{align}
Since the K\"ahler potential is a function of the matter superfields, the variation $\delta K$ can be expressed in terms of the variations $\delta\phi^k$ and $\delta\phib^\kb$, namely
\begin{align}
\delta K &= K_k\delta\phi^k + K_\kb\delta\phib^\kb\,.
\end{align}
Furthermore, according to Eq.~\eqref{eq:kahler_superspace_1/2_connection_component_3} the vector component of the K\"ahler connection is given by
\begin{align}
A_a &= \frac{1}{4}(K_k\cD_a\phi^k - K_\kb\cD_a\phib^\kb) + \frac{3i}{2}G_a + \frac{i}{8}g_{k\kb}\sigb_a^{\sd{\alpha}\alpha}\cD_\alpha\phi^k\cD_\sd{\alpha}\phib^\kb\,.
\end{align}
By using these to identities, Eq.~\eqref{eq:eom_matter_kahler_connection_sol_3} and \eqref{eq:eom_matter_kahler_connection_sol_4} take the form
\begin{align}
\begin{split}
H + \frac{1}{2}\Hb &= -\frac{i}{4}\cD^\varphi\cD^\sd{\varphi}\Nu_{\varphi\sd{\varphi}} - \frac{1}{4}(K_k\delta\phi^k - K_\kb\delta\phib^\kb + 2\Sigma^\g{r}G_\g{r}) \\
&\quad - i\Nu^aG_a - \frac{i}{8}\Nu^a\sigb_a^{\sd{\alpha}\alpha}g_{k\kb}\cD_\alpha\phi^k\cD_\sd{\alpha}\phib^\kb \\
&\quad - 2g_{k\kb}F^k\zetab^\kb + g_{k\kb}\cD^\varphi\phi^k\cD_\varphi\zetab^\kb + (\cD^2 - 8\Rb)Z\,,
\end{split} \label{eq:eom_matter_kahler_connection_id_12}\\
\nonumber\\
\begin{split}
\Hb + \frac{1}{2}H &= +\frac{i}{4}\cD^\sd{\varphi}\cD^\varphi\Nu_{\varphi\sd{\varphi}} + \frac{1}{4}(K_k\delta\phi^k - K_\kb\delta\phib^\kb - 2\Sigma^\g{r}G_\g{r}) \\
&\quad - i\Nu^aG_a - \frac{i}{8}\Nu^a\sigb_a^{\sd{\alpha}\alpha}g_{k\kb}\cD_\alpha\phi^k\cD_\sd{\alpha}\phib^\kb \\
&\quad - 2g_{k\kb}\Fb^\kb\zeta^k + g_{k\kb}\cD_\sd{\varphi}\phib^\kb\cD^\sd{\varphi}\zeta^k + (\cDb^2 - 8R)\Zb\,,
\end{split} \label{eq:eom_matter_kahler_connection_id_13}
\end{align}
where the superfields $Z:=U-\frac{1}{2}K_\kb\zetab^\kb$ and $\Zb:=\Ub-\frac{1}{2}K_k\zeta^k$ are introduced, and $G_\g{r}$ is the Killing potential defined in Eq.~\eqref{eq:kahler_superspace_1/2_killing_potential}. The terms $H + \frac{1}{2}\Hb$ and $\Hb + \frac{1}{2}H$ appear in the variation of the action in Section~\ref{sec:variation_action}.

\subsection{Variation of the action}
\label{sec:variation_action}
The action $\cS$ of the supergravity/matter/Yang-Mills system has the form
\begin{align}
\cS = \cS_\text{supergravity+matter} + \cS_\text{superpotential} + \cS_\text{Yang-Mills}\,,
\end{align}
where the individual parts are listed in Eq.~\eqref{eq:superfield_actions_sugra_matter}, \eqref{eq:superfield_actions_superpotential} and \eqref{eq:superfield_actions_yang_mills}. All these expressions contain the real density $E$, namely the superdeterminant of the components of the supervielbein. According to Eq.~\eqref{eq:superspace_vielbein_sdet_jacobis_formula}, the variation of $E$ reads
\begin{align}
\begin{split}
\delta E &= (-1)^M E(\delta{E_M}^A){E_A}^M \\
&= (-1)^M E\,{E_M}^B{H_B}^A{E_A}^M \\
&= (-1)^A E\,{E_A}^M{E_M}^B{H_B}^A \\
&= (-1)^A E\,{H_A}^A\,,
\end{split} \label{eq:eom_var_action_variation_supervielbein_determinant_1}
\end{align}
where the expression for the supertrace $(-1)^A {H_A}^A$ is stated in Eq.~\eqref{eq:eom_superspace_torsion_constraints_1_trace_h}, thus
\begin{align}
\delta E &= E\Big(\frac{i}{4}\com{\cD^\alpha}{\cD^\sd{\alpha}}\Nu_{\alpha\sd{\alpha}} + H+\Hb\Big)\,. \label{eq:eom_var_action_variation_supervielbein_determinant_2}
\end{align}
The variation of an action written in terms of the real density $E$ and a real superfield $\Theta$, as in Eq.~\eqref{eq:superfield_actions_general_construction_1}, is then given by
\begin{align}
\begin{split}
\delta\int_*E\,\Theta &= \int_*(\delta E)\Theta + \int_*E(\delta \Theta) \\
&= \int_*E(H+\Hb)\Theta + \int_*E(\delta\Theta)\,. \label{eq:eom_var_action_variation_action_variation_general_1}
\end{split}
\end{align}
Note that according to Eq.~\eqref{eq:kahler_superspace_integration_parts_identity_5} the first term in Eq.~\eqref{eq:eom_var_action_variation_supervielbein_determinant_2} vanishes if the integration over the whole superspace is performed. On the other hand, if the action is written in terms of the chiral density $\frac{E}{R}$ and a chiral superfield $\Xi$ with $w(\Xi)=+2$, as in Eq.~\eqref{eq:superfield_actions_general_construction_2}, then the variation has the form
\begin{align}
\begin{split}
\delta\int_*\frac{E}{R}\,\Xi &= \int_*\Big(\frac{\delta E}{R}\,\Xi - \frac{E}{R^2}(\delta R)\Xi + \frac{E}{R}\,\delta \Xi\Big) \\
&= \int_*\frac{E}{R}\Big((\delta \Xi + \Nu^a\cD_a \Xi) + (H + 2\Hb - i\Nu^a G_a)\Xi\Big)\,,
\end{split} \label{eq:eom_var_action_variation_action_variation_general_2}\\
\nonumber\\
\begin{split}
\delta\int_*\frac{E}{\Rb}\,\Xib &= \int_*\Big(\frac{\delta E}{\Rb}\,\Xib - \frac{E}{\Rb^2}(\delta \Rb)\Xib + \frac{E}{\Rb}\,\delta \Xib\Big) \\
&= \int_*\frac{E}{\Rb}\Big((\delta \Xib - \Nu^a\cD_a \Xib) + (\Hb + 2H - i\Nu^a G_a)\Xib\Big)\,,
\end{split} \label{eq:eom_var_action_variation_action_variation_general_3}
\end{align}
where the expressions for $\delta R$ and $\delta \Rb$ from Eq.~\eqref{eq:eom_superspace_var_r_variation_r} and \eqref{eq:eom_superspace_var_r_variation_rb} are used, and $\Xib=\Xi^\cc$.

\subsubsection{Supergravity+matter part}
According to Eq.~\eqref{eq:superfield_actions_sugra_matter}, the action of the supergravity+matter part is constructed by using the real density $E$ and the real superfield $\Theta$, which in this case is just a constant:
\begin{align}
\Theta_\text{supergravity+matter} &= -3\,. \label{eq:eom_var_action_sugra_matter_id_1}
\end{align}
Thus, from Eq.~\eqref{eq:eom_var_action_variation_action_variation_general_1} follows that the variation is given by
\begin{align}
\delta \cS_\text{supergravity+matter} &= -3\int_*E(H+\Hb)\,, \label{eq:eom_var_action_sugra_matter_id_2}
\end{align}
where the term $(H+\Hb)$ is evaluated by taking the sum of Eq~\eqref{eq:eom_matter_kahler_connection_id_12} and \eqref{eq:eom_matter_kahler_connection_id_13}:
\begin{align}
\begin{split}
\frac{3}{2}(H+\Hb) &= \frac{i}{4}\com{\cD^\sd{\varphi}}{\cD^\varphi}\Nu_{\varphi\sd{\varphi}} - \Sigma^\g{r}G_\g{r} - 2i\Nu^a G_a - \frac{i}{4}\Nu^a\sigb_a^{\sd{\alpha}\alpha}g_{k\kb} \cD_\alpha\phi^k\cD_\sd{\alpha}\phib^\kb \\
&\quad - 2g_{k\kb}F^k\zetab^\kb + g_{k\kb}\cD^\varphi\phi^k\cD_\varphi\zetab^\kb - 2g_{k\kb}\Fb^\kb\zeta^k + g_{k\kb}\cD_\sd{\varphi}\phib^\kb\cD^\sd{\varphi}\zeta^k \\
&\quad + (\cD^2-8\Rb)Z + (\cDb^2-8R)\Zb\,.
\end{split} \label{eq:eom_var_action_sugra_matter_id_3}
\end{align}
Neglecting total derivative terms, the variation of the supergravity+matter part reads
\begin{align}
\begin{split}
\delta \cS_\text{supergravity+matter} &= + 4i\int_*E\,\Nu^a\Big(G_a + \frac{1}{8}\sigb_a^{\sd{\alpha}\alpha}g_{k\kb} \cD_\alpha\phi^k\cD_\sd{\alpha}\phib^\kb\Big) \\
&\quad + 16\int_*E\,Z\,\Rb + 16\int_*E\,\Zb\,R \\
&\quad - 4\int_*E\,g_{k\kb}F^k\zetab^\kb - 4\int_*E\,g_{k\kb}\Fb^\kb\zeta^k \\
&\quad + 2\int_*E\,\Sigma^\g{r}G_\g{r}\,.
\end{split} \label{eq:eom_var_action_sugra_matter_variation}
\end{align}

\subsubsection{Superpotential part}
According to Eq.~\eqref{eq:superfield_actions_superpotential}, the superpotential part of the action is written in terms of the chiral density $\frac{E}{R}$, and the chiral superfield $\Xi$ and its conjugate $\Xib$, which are given by
\begin{align}
\Xi_\text{superpotential} &= e^{K/2}W\,, \\
\Xib_\text{superpotential} &= e^{K/2}\Wb\,,
\end{align}
with the variations
\begin{align}
\delta \Xi_\text{superpotential} &= \frac{1}{2}e^{K/2}\big((\delta K)W + 2\delta W\big)\,, \\
\delta \Xib_\text{superpotential} &= \frac{1}{2}e^{K/2}\big((\delta K)\Wb + 2\delta \Wb\big)\,.
\end{align}
Using the identities
\begin{gather}
\delta K = K_k \delta\phi^k + K_\kb \delta\phib^\kb\,, \\
\delta W = W_k\delta\phi^k\,,\hspace{4cm} \delta \Wb = \Wb_\kb \delta\phib^\kb\,,
\end{gather}
and neglecting total derivative terms, Eq.~\eqref{eq:eom_var_action_variation_action_variation_general_2} and \eqref{eq:eom_var_action_variation_action_variation_general_3} imply
\begin{align}
\delta \cS_\text{superpotential} &= \delta\,\frac{1}{2}\int_*\frac{E}{R}\,\Xi_\text{superpotential} + \delta\,\frac{1}{2}\int_*\frac{E}{\Rb}\,\Xib_\text{superpotential}\,, \label{eq:eom_var_action_superpotential_variation_1}
\end{align}
where
\begin{align}
\delta\,\frac{1}{2}\int_*\frac{E}{R}\,\Xi_\text{superpotential} &= -8\int_*E\,\Zb\,e^{K/2}W - 4\int_*E\,\zeta^k e^{K/2}(K_k W + W_k)\,, \label{eq:eom_var_action_superpotential_variation_2}\\
\nonumber\\
\delta\,\frac{1}{2}\int_*\frac{E}{\Rb}\Xib_\text{superpotential} &= -8\int_*E\,Z\,e^{K/2}\Wb - 4\int_*E\,\zetab^\kb e^{K/2}(K_\kb \Wb + \Wb_\kb)\,. \label{eq:eom_var_action_superpotential_variation_3}
\end{align}

\subsubsection{Yang-Mills part}
Like the superpotential part, the Yang-Mills part of the action is constructed by using the chiral density and the superfields $\Xi$ and $\Xib$, which have the following form (cf.\ Eq.~\eqref{eq:superfield_actions_yang_mills}):
\begin{align}
\Xi_\text{Yang-Mills} &= \frac{1}{4}\gkf{r}{s} \cW^{\alpha\g{r}}{\cW_\alpha}^\g{s}\,, \\
\Xib_\text{Yang-Mills} &= \frac{1}{4}\gkfb{r}{s} {\cW_\sd{\alpha}}^\g{r}\cW^{\sd{\alpha}\g{s}}\,.
\end{align}
The corresponding variations read
\begin{align}
\delta \Xi_\text{Yang-Mills} &= \frac{1}{4}(\delta\gkf{r}{s}) \cW^{\alpha\g{r}}{\cW_\alpha}^\g{s} + \frac{1}{2}\gkf{r}{s} \cW^{\alpha\g{r}}(\delta{\cW_\alpha}^\g{s})\,, \\
\delta \Xib_\text{Yang-Mills} &= \frac{1}{4}(\delta\gkfb{r}{s}) {\cW_\sd{\alpha}}^\g{r}\cW^{\sd{\alpha}\g{s}} + \frac{1}{2}\gkfb{r}{s} {\cW_\sd{\alpha}}^\g{r}(\delta\cW^{\sd{\alpha}\g{s}})\,,
\end{align}
where
\begin{align}
\delta\gkf{r}{s} &= \frac{\del\gkf{r}{s}}{\del\phi^k}\delta\phi^k\,, & \delta\gkfb{r}{s} &= \frac{\del\gkfb{r}{s}}{\del\phib^\kb}\delta\phib^\kb\,,
\end{align}
and the expressions for $\delta{\cW_\alpha}^\g{s}$ and $\delta\cW^{\sd{\alpha}\g{s}}$ are stated in Eq.~\eqref{eq:eom_ym_var_w_variation_w} and \eqref{eq:eom_ym_var_w_variation_wb}, respectively. Neglecting total derivative terms, the variation of the Yang-Mills part of the action is then given by
\begin{align}
\delta\cS_\text{Yang-Mills} &= \delta\,\frac{1}{2}\int_*\frac{E}{R}\,\Xi_\text{Yang-Mills} + \delta\,\frac{1}{2}\int_*\frac{E}{\Rb}\,\Xib_\text{Yang-Mills}\,, \label{eq:eom_var_action_yang_mills_variation_1}
\end{align}
with
\begin{align}
\begin{split}
\delta\,\frac{1}{2}\int_*\frac{E}{R}\,\Xi_\text{Yang-Mills} &= - \frac{1}{2}\int_*E\,\Sigma^\g{r}\Big(\gkf{r}{s}\cD^\alpha{\cW_\alpha}^\g{s} + \frac{\del\gkf{r}{s}}{\del\phi^k}\cD^\alpha\phi^k{\cW_\alpha}^\g{s}\Big) \\
&\quad - \frac{i}{2}\int_*E\,\gkf{r}{s}\Nu^a\sig_{a\alpha\sd{\alpha}}\cW^{\alpha\g{r}}\cW^{\sd{\alpha}\g{s}} \\
&\quad - \int_*E\,\zeta^k\frac{\del\gkf{r}{s}}{\del\phi^k}\cW^{\alpha\g{r}}{\cW_\alpha}^\g{s}\,,
\end{split} \label{eq:eom_var_action_yang_mills_variation_2}\\
\nonumber\\
\begin{split}
\delta\,\frac{1}{2}\int_*\frac{E}{\Rb}\,\Xib_\text{Yang-Mills} &= - \frac{1}{2}\int_*E\,\Sigma^\g{r}\Big(\gkfb{r}{s}\cD_\sd{\alpha}\cW^{\sd{\alpha}\g{s}} + \frac{\del\gkfb{r}{s}}{\del\phib^\kb}\cD_\sd{\alpha}\phib^\kb\cW^{\sd{\alpha}\g{s}}\Big) \\
&\quad - \frac{i}{2}\int_*E\,\gkfb{r}{s}\Nu^a\sig_{a\alpha\sd{\alpha}}\cW^{\alpha\g{r}}\cW^{\sd{\alpha}\g{s}} \\
&\quad - \int_*E\,\zetab^\kb\frac{\del\gkfb{r}{s}}{\del\phib^\kb}{\cW_\sd{\alpha}}^\g{r}\cW^{\sd{\alpha}\g{s}}\,.
\end{split} \label{eq:eom_var_action_yang_mills_variation_3}
\end{align}

\subsubsection{Superfield equations of motion}
The variation of the action $\cS$ is given by
\begin{align}
\delta\cS = \delta\cS_\text{supergravity+matter} + \delta\cS_\text{superpotential} + \delta\cS_\text{Yang-Mills}\,,
\end{align}
where the expressions for the variations of the supergravity+matter, the superpotential and the Yang-Mills part are stated in Eq.~\eqref{eq:eom_var_action_sugra_matter_variation}, \eqref{eq:eom_var_action_superpotential_variation_1} and \eqref{eq:eom_var_action_yang_mills_variation_1}, respectively. The equations of motion at the superfield level are obtained from the condition $\delta\cS=0$. The variation of $\cS$ vanishes if and only if the variation of the integrand of the superspace integral $\int_*$ vanishes up to total derivatives terms. The calculations in the previous sections showed that the variations of the supervielbein, the Lorentz and Yang-Mills connection, and the matter superfields are highly restricted by the covariant constraints in K\"ahler superspace. In addition, the degrees of freedom are further reduced by applying a compensating superspace diffeomorphism, and a compensating Lorentz and Yang-Mills transformation. It turned out that the variations can completely be parametrized by the (unconstrained) superfields $\Zb$, $Z$, $\Nu^a$, $\zetab^\kb$, $\zeta^k$, $\Sigma^\g{r}$. The variation of $\cS$ with respect to each of these superfields has to vanish separately. The corresponding equations of motion are listed in the following:
\begin{itemize}
\item Variation $\Zb$:
\begin{align}
R - \frac{1}{2}e^{K/2}W &= 0 \label{eq:eom_var_action_eom_1}
\end{align}
\item Variation $Z$:
\begin{align}
\Rb - \frac{1}{2}e^{K/2}\Wb &= 0 \label{eq:eom_var_action_eom_2}
\end{align}
\item Variation $\Nu^a$:
\begin{align}
G_a + \frac{1}{8}g_{k\kb}\sig_{a\alpha\sd{\alpha}}\cD^\alpha\phi^k\cD^\sd{\alpha}\phib^\kb - \frac{1}{4}\re{\gkf{r}{s}}\sig_{a\alpha\sd{\alpha}}\cW^{\alpha\g{r}}\cW^{\sd{\alpha}\g{s}} &= 0 \label{eq:eom_var_action_eom_3}
\end{align}
\item Variation $\zetab^\kb$:
\begin{align}
g_{k\kb}F^k + e^{K/2}D_\kb\Wb + \frac{1}{4}\frac{\del\gkfb{r}{s}}{\del\phib^\kb}{\cW_\sd{\alpha}}^\g{r}\cW^{\sd{\alpha}\g{s}} &= 0 \label{eq:eom_var_action_eom_4}
\end{align}
\item Variation $\zeta^k$:
\begin{align}
g_{k\kb}\Fb^\kb + e^{K/2}D_k W + \frac{1}{4}\frac{\del\gkf{r}{s}}{\del\phi^k}\cW^{\alpha\g{r}}{\cW_\alpha}^\g{s} &= 0 \label{eq:eom_var_action_eom_5}
\end{align}
\item Variation $\Sigma^\g{r}$:
\begin{align}
\re{\gkf{r}{s}}\Db^\g{s} - K_k(\gen{r}\phi)^k - \frac{1}{4}\Big(\frac{\del\gkf{r}{s}}{\del\phi^k}\cD^\alpha\phi^k{\cW_\alpha}^\g{s} + \frac{\del\gkfb{r}{s}}{\del\phib^\kb}\cD_\sd{\alpha}\phib^\kb\cW^{\sd{\alpha}\g{s}}\Big) = 0 \label{eq:eom_var_action_eom_6}
\end{align}
\end{itemize}
In Eq.~\eqref{eq:eom_var_action_eom_4} and \eqref{eq:eom_var_action_eom_5} the derivatives $D_\kb\Wb := \Wb_\kb + \Wb K_\kb$ and $D_k W := W_k + W K_k$ are used.

\section{Component field formalism}
\label{sec:component_fields}

\subsection{Definition of component fields}
\label{sec:component_fields_definition}
In supersymmetric theories, component fields come along in so called supermultiplets which form irreducible representations with respect to the super-Poincar\'e algebra (see e.g.~\cite{Wess:1992cp}). Supermultiplets are considered either on-shell or off-shell, depending on whether the component fields fulfil the equations of motion or not. A well known property of a supermultiplet, either on-shell or off-shell, is that it contains the same number of bosonic and fermionic degrees of freedom. On the other hand, the equations of motion eliminate a different number of degrees of freedom for fields with different spin. Hence, an off-shell supermultiplet contains not only the physical fields which are present in the on-shell formulation of the supermultiplet, but also additional fields, so called auxiliary fields. The equations of motion of these fields are purely algebraic, i.e.\ they have no kinetic term, such that the corresponding degrees of freedom vanish on-shell. In the present superspace formulation of supergravity, supersymmetry is implemented off-shell, which is appropriate if the theory gets quantized.
\\\\
The action at the component field level in Section~\ref{sec:component_fields_actions} contains three different supermultiplets, namely the minimal supergravity multiplet, the matter (chiral) multiplet, and the Yang-Mills (vector) multiplet in Wess-Zumino gauge.\footnote{The Wess-Zumino gauge of a vector multiplet is only present in gauge theories, where a particular chiral gauge transformation can be used to reduce the number of auxiliary fields.} An overview of these supermultiplets is given below. They are written in the form $\big(\,\text{\textit{physical}}\;\big\vert\,\text{\textit{auxiliary}}\,\big)$, where on the left-hand side of the bar the physical fields are listed, and on the right-hand side the auxiliary fields are specified. Furthermore, $\mathbf{b}$ and $\mathbf{f}$ stand for bosonic and fermionic (off-shell) real degrees of freedom, respectively.
\renewcommand{\arraystretch}{1.3}
\begin{itemize}
\item \textbf{Supergravity multiplet:}\\
The off-shell minimal supergravity multiplet consists of the graviton ${e_m}^a$, the gravitino ${\psi_m}^\alpha$ and its conjugate $\psib_{m\sd{\alpha}}$, and the auxiliary fields $M$, its conjugate $\Mb$, and $b_a$:
\begin{align}
\big({e_m}^a\,,\;{\psi_m}^\alpha\,(\psib_{m\sd{\alpha}})\,\big\vert\;M\,(\Mb)\,,\;b_a\big) \quad\left\{
\begin{array}{lll}
{e_m}^a & 6\,\mathbf{b} & \text{graviton} \\
{\psi_m}^\alpha\,(\psib_{m\sd{\alpha}}) & 12\,\mathbf{f} & \text{gravitino} \\
M\,(\Mb) & 2\,\mathbf{b} & \text{complex scalar} \\
b_a & 4\,\mathbf{b} & \text{real vector}
\end{array}\right.
\end{align}
Note that six degrees of freedom of the graviton are removed by local Lorentz transformations and four are removed by spacetime diffeomorphism transformations. In the case of the gravitino, four degrees of freedom are removed by supergravity transformations. Furthermore, other supergravity multiplets, like the new minimal or the non-minimal multiplet, contain the graviton and the gravitino as well, but a different set of auxiliary fields (see e.g.\ \cite{Binetruy:2000zx}). The supergravity multiplet is real.
\item \textbf{Matter multiplet:}\\
The off-shell matter (chiral) multiplet consists of a complex scalar $\varphi^k$, the corresponding Weyl fermion ${\chi^k}_\alpha$ and an auxiliary field $F^k$:
\begin{align}
\big(\varphi^k\,,\;{\chi^k}_\alpha\,\big\vert\;F^k\big) \quad\left\{
\begin{array}{lll}
\varphi^k & 2\,\mathbf{b} & \text{complex scalar} \\
{\chi^k}_\alpha & 4\,\mathbf{f} & \text{Weyl spinor} \\
F^k & 2\,\mathbf{b} & \text{complex scalar}
\end{array}\right.
\end{align}
The conjugate multiplet $\big(\varphib^\kb\,,\;\chib^{\kb\sd{\alpha}}\,\big\vert\;\Fb^\kb\big)$ contains the conjugated component fields. Note that the matter multiplet is complex.
\item \textbf{Yang-Mills multiplet:}\\
In Wess-Zumino gauge the off-shell Yang-Mills (vector) multiplet consists of a gauge boson ${\ca_m}^\g{r}$, the corresponding gaugino ${\cl_\alpha}^\g{r}$ and its conjugate $\clb^{\sd{\alpha}\g{r}}$, and an auxiliary field $\Db^\g{r}$:
\begin{align}
\big({\ca_m}^\g{r}\,,\;{\cl_\alpha}^\g{r}\,(\clb^{\sd{\alpha}\g{r}})\,\big\vert\;\Db^\g{r}\big) \quad\left\{
\begin{array}{lll}
{\ca_m}^\g{r} & 3\,\mathbf{b} & \text{vector boson} \\
{\cl_\alpha}^\g{r}\,(\clb^{\sd{\alpha}\g{r}}) & 4\,\mathbf{f} & \text{Weyl spinor} \\
\Db^\g{r} & 1\,\mathbf{b} & \text{real scalar}
\end{array}\right.
\end{align}
Note that one degree of freedom of the vector boson is removed by gauge transformations. The Yang-Mills multiplet is real.
\end{itemize}
In the context of superspace, a component field is identified as the lowest component of a superfield, i.e.\ the component which does not contain any anti-commuting coordinates $\theta^\mu$ or $\thetab_\sd{\mu}$. Supergravity transformations of the component fields are then defined by the supergravity transformations at the superfield level. Since the supergravity transformation contains only covariant objects, the component fields are present as lowest components in a chain of covariant derivatives, applied on the superfield which contains the lowest component of the supermultiplet. In Section~\ref{sec:component_fields_definition_projection} projection operators, which project to lowest superspace components, are introduced. These operators are then used in Section~\ref{sec:component_fields_definition_sugra} and \ref{sec:component_fields_definition_matter} to define the supergravity, the matter and the Yang-Mills multiplet.

\subsubsection{Projection to lowest superspace components}
\label{sec:component_fields_definition_projection}
For a superfield $\Phi(x^m,\theta^\mu,\thetab_\sd{\mu})=\phi_{00}(x^m)+\mathcal{O}(\theta^\mu,\thetab_\sd{\mu})$, where $\mathcal{O}(\theta^\mu,\thetab_\sd{\mu})$ indicates terms with at least one $\theta^\mu$ or $\thetab^\sd{\mu}$, the so called bar projection $\p$ is given by
\begin{align}
\Phi(x^m,\theta^\mu,\thetab_\sd{\mu})\p &:= \phi_{00}(x^m)\,. \label{eq:component_fields_definition_lowest_components_field}
\end{align}
If $\Psi$ is another superfield, the bar projection is compatible with multiplication:
\begin{align}
(\Phi\Psi)\p = \Phi\p\,\Psi\p\,. \label{eq:component_fields_definition_lowest_components_id_1}
\end{align}
Furthermore, for a $p$-superform $\omega$, i.e.\
\begin{align}
\omega &= \frac{1}{p!}\exd z^{M_1}\wedge...\wedge\exd z^{M_p}\omega_{M_p...M_1}\,,
\end{align}
the so called double bar projection $\pp$ is defined as
\begin{align}
\omega\pp &:= \exd x^{m_1}\wedge...\wedge\exd x^{m_p}\omega_{m_p...m_1}\p\,. \label{eq:component_fields_definition_lowest_components_form}
\end{align}
If $\omega$ and $\tau$ are two pure superforms, the double bar projection is compatible with the wedge product, namely
\begin{align}
(\omega\wedge\tau)\pp &= \omega\pp\wedge\tau\pp\,. \label{eq:component_fields_definition_lowest_components_id_2}
\end{align}

\subsubsection{Supergravity multiplet}
\label{sec:component_fields_definition_sugra}
The double bar projection of the vector component $E^a$ of the supervielbein identifies the vielbein $e^a$ of ordinary spacetime
\begin{align}
E^a\pp = \exd z^m {E_m}^a\p := \exd z^m {e_m}^a =: e^a\,, \label{eq:component_fields_definition_sugra_vielbein}
\end{align}
where ${e_m}^a$ is the graviton. Since ${E_M}^A$ is invertible (${E_M}^A{E_A}^N=\delta_M^N$), ${E_m}^a$ must be invertible too (${E_m}^a{E_a}^n=\delta_m^n$), thus
\begin{align}
{e_m}^a{e_a}^n = {E_m}^a\p{E_a}^n\p = ({E_m}^a{E_a}^n)\p = \delta_m^n\p = \delta_m^n\,. \label{eq:component_fields_definition_sugra_vielbein_inverse}
\end{align}
with ${e_a}^m={E_a}^m\p$. The metric $g_{nm}$ of ordinary spacetime is then given by
\begin{align}
g_{nm} := {e_n}^b{e_m}^a \rho_{ba}\p = {e_n}^b{e_m}^a \eta_{ba}\,. \label{eq:component_fields_definition_sugra_metric}
\end{align}
where $g_{nm}$ and $\eta_{ba}$, and their inverses, are used to raise and lower indices of the component fields.
Furthermore, for the Pauli matrices $\sig^a$ and $\sigb^a$ the corresponding matrices with a spacetime index are defined as
\begin{align}
\sig_{m\alpha\sd{\alpha}} &:= {e_m}^a\sig_{a\alpha\sd{\alpha}}\,, & \sigb_m^{\sd{\alpha}\alpha} &:= {e_m}^a\sigb_a^{\sd{\alpha}\alpha}\,. \label{eq:component_fields_definition_sugra_pauli_matrices}
\end{align}
The gravitino ${\psi_m}^\alpha$ and its conjugate $\psib_{m\sd{\alpha}}$ are contained in the double bar projection of the spinor components $E^\alpha$ and $E_\sd{\alpha}$ of the supervielbein, namely
\begin{align}
\begin{split}
E^\alpha\pp &= \exd z^m {E_m}^\alpha\p =: \frac{1}{2} \exd z^m {\psi_m}^\alpha =: e^\alpha\,, \\
E_\sd{\alpha}\pp &= \exd z^m E_{m\sd{\alpha}}\p =: \frac{1}{2} \exd z^m \psib_{m\sd{\alpha}} =: e_\sd{\alpha}\,.
\end{split} \label{eq:component_fields_definition_sugra_gravitino}
\end{align}
According to the appearance of ${e_m}^a$, ${\psi_m}^\alpha$ and $\psib_{m\sd{\alpha}}$ in the supervielbein, their chiral weights are
\begin{gather}
w({e_m}^a) = 0\,, \label{eq:component_fields_definition_sugra_vielbein_weight}\\
w({\psi_m}^\alpha) = +1\,,\hspace{4cm} w(\psib_{m\sd{\alpha}}) = -1\,. \label{eq:component_fields_definition_sugra_gravitino_weight}
\end{gather}
The auxiliary fields $M$, $\Mb$ and $b_a$ of the supergravity multiplet are defined as
\begin{gather}
R\p =: -\frac{1}{6}M\,,\hspace{4cm} \Rb\p =: -\frac{1}{6}\Mb\,, \label{eq:component_fields_definition_sugra_auxiliary_1}\\
G_a\p =: -\frac{1}{3}b_a\,, \label{eq:component_fields_definition_sugra_auxiliary_2}
\end{gather}
with the chiral weights
\begin{gather}
w(M) = +2\,,\hspace{4cm} w(\Mb) = -2\,, \label{eq:component_fields_definition_sugra_auxiliary_1_weight}\\
w(b_a) = 0\,. \label{eq:component_fields_definition_sugra_auxiliary_2_weight}
\end{gather}
Furthermore, from the double bar projection of the Lorentz connection ${\Omega_B}^A$ the spin connection ${\omega_B}^A$ is recovered, namely
\begin{align}
\begin{split}
{\Omega_b}^a\pp &= \exd z^m{\Omega_{mb}}^a\p =: \exd z^m{\omega_{mb}}^a =: {\omega_b}^a\,, \\
{\Omega_\beta}^\alpha\pp &= \exd z^m{\Omega_{m\beta}}^\alpha\p =: \exd z^m{\omega_{m\beta}}^\alpha =: {\omega_\beta}^\alpha\,, \\
{\Omega^\sd{\beta}}_\sd{\alpha}\pp &= \exd z^m\tensor{\Omega}{_m^{\sd{\beta}}_{\sd{\alpha}}}\p =: \exd z^m\tensor{\omega}{_m^{\sd{\beta}}_{\sd{\alpha}}} =: {\omega^\sd{\beta}}_\sd{\alpha}\,,
\end{split} \label{eq:component_fields_definition_sugra_spin_connection}
\end{align}
where ${\omega_{mB}}^A$ inherits the properties from ${\Omega_{mB}}^A$, in particular
\begin{align}
\begin{gathered}
\omega_{m\sym{2}{\beta\alpha}} = +\frac{1}{2}(\sig^{ba}\eps)_{\beta\alpha} \omega_{mba}\,,\hspace{3cm} \omega_{m\sym{2}{\sd{\beta}\sd{\alpha}}} = +\frac{1}{2}(\eps\sigb^{ba})_{\sd{\beta}\sd{\alpha}} \omega_{mba}\,, \\
\omega_{mba} = -(\eps\sig_{ba})^{\beta\alpha} \omega_{m\sym{2}{\beta\alpha}} + (\sigb_{ba}\eps)^{\sd{\beta}\sd{\alpha}} \omega_{m\sym{2}{\sd{\beta}\sd{\alpha}}}\,.
\end{gathered} \label{eq:component_fields_definition_sugra_spin_connection_id}
\end{align}
In the following it is shown, that the spin connection is not an independent quantity, but rather a function of the graviton and the gravitino. This is analogous to general relativity, where the torsion free condition implies that the spin connection can be expressed in terms of the vielbein. In order to prove the above statement, the double bar projection of the vector component $T^a$ of the torsion tensor is considered. By definition, the double bar projection of a $2$-superform is given by
\begin{align}
T^a\pp = \frac{1}{2}\exd z^m\wedge\exd z^n {T_{nm}}^a\p\,. \label{eq:component_fields_definition_sugra_torsion_vector_dbp_1}
\end{align}
On the other hand, if $T^a$ is written in terms of the supervielbein by using the covariant derivative, the double bar projection reads
\begin{align}
\begin{split}
T^a\pp &= \exd E^a\pp + E^b\wedge{\Omega_b}^a\pp \\
&= \exd z^M\wedge \exd z^N\del_N{E_M}^a\pp - \exd z^M\wedge \exd z^N{E_N}^b{\Omega_{Mb}}^a\pp \\
&= \exd z^m\wedge \exd z^n\del_n{e_m}^a - \exd z^m\wedge \exd z^n{e_n}^b{\omega_{mb}}^a \\
&= \frac{1}{2}\exd z^m\wedge \exd z^n(\cD_n{e_m}^a - \cD_m{e_n}^a)\,,
\end{split} \label{eq:component_fields_definition_sugra_torsion_vector_dbp_2}
\end{align}
with the covariant derivative of the vielbein components \footnote{Note that the definition of the covariant derivative of ${e_m}^a$ in Eq.~\eqref{eq:component_fields_definition_sugra_vielbein_cd} is not covariant with respect to the Levi-Civita connection of spacetime. The fully covariant derivative would also contain the term $-{\Gamma^p}_{nm}{e_p}^a$ with the corresponding Christoffel symbols ${\Gamma^p}_{nm}$. Since the Levi-Civita connection is torsion-free, the Christoffel symbols are symmetric in the indices $n,m$ and thus the term drops out in the antisymmetric combination $\cD_n{e_m}^a - \cD_m{e_n}^a$.}
\begin{align}
\cD_n{e_m}^a:=\del_n{e_m}^a + {e_m}^b{\omega_{nb}}^a\,. \label{eq:component_fields_definition_sugra_vielbein_cd}
\end{align}
A comparison of Eq.~\eqref{eq:component_fields_definition_sugra_torsion_vector_dbp_1} and \eqref{eq:component_fields_definition_sugra_torsion_vector_dbp_2} leads to
\begin{align}
{T_{nm}}^a\p &= \cD_n{e_m}^a - \cD_m{e_n}^a\,. \label{eq:component_fields_definition_sugra_id_0}
\end{align}
There is another way to evaluate $T^a\pp$ using the torsion constraints, namely
\begin{align}
\begin{split}
T^a\pp &= \frac{1}{2}E^B\wedge E^C {T_{CB}}^a\pp = E_\sd{\beta}\wedge E^\gamma{T_\gamma}^{\sd{\beta}a}\pp \\
&= e_\sd{\beta}\wedge e^\gamma{T_\gamma}^{\sd{\beta}a}\p = -\frac{1}{4}\exd z^m\wedge\exd z^n \psib_{n\sd{\beta}}{\psi_m}^\gamma {T_\gamma}^{\sd{\beta}a}\p \\
&= \frac{i}{2}\exd z^m\wedge\exd z^n \psib_{n\sd{\beta}}{\psi_m}^\gamma {(\sigma^a\eps)_\gamma}^\sd{\beta} \\
&= \frac{i}{4}\exd z^m\wedge\exd z^n (\psi_n\sigma^a\psib_m - \psi_m\sigma^a\psib_n)\,,
\end{split} \label{eq:component_fields_definition_sugra_torsion_vector_dbp_3}
\end{align}
thus
\begin{align}
\cD_n{e_m}^a - \cD_m{e_n}^a &= \frac{i}{2}(\psi_n\sigma^a\psib_m - \psi_m\sigma^a\psib_n)\,. \label{eq:component_fields_definition_sugra_id_1}
\end{align}
According to Eq.~\eqref{eq:component_fields_definition_sugra_vielbein_cd}, this identity is equivalent to
\begin{align}
\omega_{nmp} - \omega_{mnp} &= -{e_p}^a\del_n e_{ma} + {e_p}^a\del_m e_{na} + \frac{i}{2}(\psi_n\sig_p\psib_m - \psi_m\sig_p\psib_n)\,, \label{eq:component_fields_definition_sugra_id_2}
\end{align}
which implies
\begin{align}
\begin{split}
\omega_{mnp} &= +\frac{1}{2}({e_m}^a\del_n e_{pa} - {e_p}^a\del_m e_{na} - {e_n}^a\del_p e_{ma}) \\
&\quad -\frac{1}{2}({e_m}^a\del_p e_{na} - {e_n}^a\del_m e_{pa} - {e_p}^a\del_n e_{ma}) \\
&\quad +\frac{i}{4}(\psi_p\sig_m\psib_n - \psi_m\sig_n\psib_p - \psi_n\sig_p\psib_m) \\
&\quad -\frac{i}{4}(\psi_n\sig_m\psib_p - \psi_m\sig_p\psib_n - \psi_p\sig_n\psib_m)\,,
\end{split} \label{eq:component_fields_definition_sugra_spin_connection_identity}
\end{align}
where $\omega_{mnp}:={e_n}^b {e_p}^a\omega_{mba}$. Hence, the spin connection is expressed in terms of the graviton and the gravitino. A similar strategy for the spinor components $T^\alpha$ and $T_\sd{\alpha}$ of the torsion tensor is applied to write the components ${T_{cb}}^\alpha\p$ and $T_{cb\sd{\alpha}}\p$ in terms of the component fields of the supergravity multiplet. The following two different ways of evaluating the double bar projections $T^\alpha\pp$ and $T_\sd{\alpha}\pp$, namely
\begin{align}
T^\alpha\pp &= \frac{1}{2}\exd z^m\wedge\exd z^n{T_{nm}}^\alpha\p\,, \label{eq:component_fields_definition_sugra_torsion_spinor_dbp_1}\\
T_\sd{\alpha}\pp &= \frac{1}{2}\exd z^m\wedge\exd z^n T_{nm\sd{\alpha}}\p\,, \label{eq:component_fields_definition_sugra_torsion_spinorb_dbp_1}
\end{align}
and
\begin{align}
\begin{split}
T^\alpha\pp &= \exd E^\alpha\pp + E^\beta\wedge{\Omega_\beta}^\alpha\pp + E^\alpha\wedge A\pp \\
&= \frac{1}{4}\exd z^m\wedge \exd z^n(\cD_n{\psi_m}^\alpha - \cD_m{\psi_n}^\alpha)\,,
\end{split} \label{eq:component_fields_definition_sugra_torsion_spinor_dbp_2}\\
\nonumber\\
\begin{split}
T_\sd{\alpha}\pp &= \exd E_\sd{\alpha}\pp + E_\sd{\beta}\wedge{\Omega^\sd{\beta}}_\sd{\alpha}\pp - E_\sd{\alpha}\wedge A\pp \\
&= \frac{1}{4}\exd z^m\wedge \exd z^n(\cD_n\psib_{m\sd{\alpha}} - \cD_m\psib_{n\sd{\alpha}})\,,
\end{split} \label{eq:component_fields_definition_sugra_torsion_spinorb_dbp_2}
\end{align}
with the covariant derivative of the gravitino \footnote{As the covariant derivative of the graviton in Eq.~\eqref{eq:component_fields_definition_sugra_vielbein_cd}, the covariant derivative of the gravitino in Eq.~\eqref{eq:component_fields_definition_sugra_cd_gravitino_1} and \eqref{eq:component_fields_definition_sugra_cd_gravitino_2} is not covariant with respect to the Levi-Civita connection, since the terms containing the Christoffel symbols vanish in the antisymmetric combination of the derivatives.}
\begin{align}
\cD_n{\psi_m}^\alpha := \del_n{\psi_m}^\alpha + {\psi_m}^\beta{\omega_{n\beta}}^\alpha + {\psi_m}^\alpha A_n\,, \label{eq:component_fields_definition_sugra_cd_gravitino_1}\\
\cD_n\psib_{m\sd{\alpha}} := \del_n\psib_{m\sd{\alpha}} + \psi_{m\sd{\beta}}\tensor{\omega}{_n^{\sd{\beta}}_{\sd{\alpha}}} - \psi_{m\sd{\alpha}}A_n\,, \label{eq:component_fields_definition_sugra_cd_gravitino_2}
\end{align}
lead to
\begin{align}
{T_{nm}}^\alpha\p &= \frac{1}{2}(\cD_n{\psi_m}^\alpha - \cD_m{\psi_n}^\alpha)\,, \label{eq:component_fields_definition_sugra_id_3}\\
T_{nm\sd{\alpha}}\p &= \frac{1}{2}(\cD_n\psib_{m\sd{\alpha}} - \cD_m\psib_{n\sd{\alpha}})\,. \label{eq:component_fields_definition_sugra_id_4}
\end{align}
Since the K\"ahler connection $A$ is a function of the matter superfields $\phi^k$ and $\phib^\kb$, $A_n:=A_n\p$ can be expressed in terms of the component fields of the matter multiplet (cf.\ Eq.~\eqref{eq:component_fields_definition_matter_kahler_connection_component_1}). On the other hand, the double bar projection can also be evaluated as follows:
\begin{align}
T^\alpha\pp &= \frac{1}{2}e^b\wedge e^c {T_{cb}}^\alpha\p + e^b\wedge e^\gamma{T_{\gamma b}}^\alpha\p + e^b\wedge e_\sd{\gamma}\tensor{T}{^{\sd{\gamma}}_b^\alpha}\p\,, \label{eq:component_fields_definition_sugra_torsion_spinor_dbp_3}\\
T_\sd{\alpha}\pp &= \frac{1}{2}e^b\wedge e^c T_{cb\sd{\alpha}}\p + e^b\wedge e_\sd{\gamma}{T^\sd{\gamma}}_{b\sd{\alpha}}\p + e^b\wedge e^\gamma T_{\gamma b\sd{\alpha}}\p\,. \label{eq:component_fields_definition_sugra_torsion_spinorb_dbp_3}
\end{align}
By using the explicit expressions for ${T_{\su{\gamma}b}}^\su{\alpha}$ in terms of the superfields $R$, $\Rb$ and $G_a$, the corresponding bar projections read
\begin{align}
{T_{\gamma b}}^\alpha\p &= -\frac{i}{6}{(\sig_c\sigb_b)_\gamma}^\alpha b_c\,, & \tensor{T}{^{\sd{\gamma}}_b^\alpha}\p &= +\frac{i}{6}\sigb_b^{\sd{\gamma}\alpha} M\,, \label{eq:component_fields_definition_sugra_id_5}\\
{T^\sd{\gamma}}_{b\sd{\alpha}}\p &= +\frac{i}{6}{(\sigb_c\sig_b)^\sd{\gamma}}_\sd{\alpha} b^c\,, & T_{\gamma b\sd{\alpha}}\p &= +\frac{i}{6}\sig_{b\gamma\sd{\alpha}} \Mb\,, \label{eq:component_fields_definition_sugra_id_6}
\end{align}
which implies
\begin{align}
\begin{split}
{T_{cb}}^\alpha\p &= + \frac{1}{2}{e_c}^n{e_b}^m(\cD_n{\psi_m}^\alpha - \cD_m{\psi_n}^\alpha) \\
&\quad + \frac{i}{12}({e_c}^m\psi_m\sig_a\sigb_b - {e_b}^m\psi_m\sig_a\sigb_c)^\alpha b^a \\
&\quad - \frac{i}{12}({e_c}^m\psib_m\sigb_b - {e_b}^m\psib_m\sigb_c)^\alpha M\,,
\end{split} \label{eq:component_fields_definition_sugra_id_7}\\
\nonumber\\
\begin{split}
T_{cb\sd{\alpha}}\p &= + \frac{1}{2}{e_c}^n{e_b}^m(\cD_n\psib_{m\sd{\alpha}} - \cD_m\psib_{n\sd{\alpha}}) \\
&\quad - \frac{i}{12}({e_c}^m\psib_m\sigb_a\sig_b - {e_b}^m\psib_m\sigb_a\sig_c)_\sd{\alpha} b^a \\
&\quad - \frac{i}{12}({e_c}^m\psi_m\sig_b - {e_b}^m\psi_m\sig_c)_\sd{\alpha} \Mb\,.
\end{split} \label{eq:component_fields_definition_sugra_id_8}
\end{align}
These two quantities are used in subsequent calculations. In particular, they appear in the term ${R_{ba}}^{ba}\p$, which is also used in later calculations, where ${R_a}^b$ is the vector component of the Lorentz curvature. According to the definition of the double bar projection for a $2$-superform, ${R_b}^a\pp$ has the form
\begin{align}
{R_b}^a\pp &= \frac{1}{2}\exd z^m\wedge\exd z^n {R_{nmb}}^a\p\,, \label{eq:component_fields_definition_sugra_dbp_lorentz_curvature_1}
\end{align}
On the other hand, if the Lorentz curvature is expressed in terms the Lorentz connection, ${R_b}^a\pp$ reads
\begin{align}
\begin{split}
{R_b}^a\pp &= \exd{\Omega_b}^a\pp + {\Omega_b}^c{\Omega_c}^a\pp \\
&= \frac{1}{2}\exd z^m\wedge\exd z^n(\del_n{\omega_{mb}}^a - \del_m{\omega_{nb}}^a + {\com{\omega_m}{\omega_n}_b}^a)\,,
\end{split} \label{eq:component_fields_definition_sugra_dbp_lorentz_curvature_2}
\end{align}
thus
\begin{align}
{R_{nmb}}^a\p &= \del_n{\omega_{mb}}^a - \del_m{\omega_{nb}}^a + {\com{\omega_m}{\omega_n}_b}^a\,, \label{eq:component_fields_definition_sugra_id_9}
\end{align}
showing that ${R_{nmb}}^a\p$ is the curvature tensor associated to the spin connection ${\omega_{mb}}^a$. The corresponding curvature scalar $\cR$ is defined as
\begin{align}
\cR := {e_a}^n{e_b}^m {R_{nm}}^{ba}\p\,. \label{eq:component_fields_definition_sugra_curvature_scalar}
\end{align}
Another way to evaluate ${R_b}^a\pp$ is the following:
\begin{align}
{R_b}^a\pp &= \frac{1}{2}e^c\wedge e^d {R_{dcb}}^a\p + e^c\wedge e^\su{\delta} {R_{\su{\delta}cb}}^a\p + \frac{1}{2}e^\su{\gamma}\wedge e^\su{\delta} {R_{\su{\delta}\su{\gamma}b}}^a\p\,. \label{eq:component_fields_definition_sugra_dbp_lorentz_curvature_3}
\end{align}
If the components ${R_{\su{\delta}cb}}^a$ and ${R_{\su{\delta}\su{\gamma}b}}^a$ are written in terms of the superfields $R$, $\Rb$ and $G_a$, the corresponding bar projections are given by
\begin{align}
{R_{\delta\gamma b}}^a\p &= -\frac{4}{3}({\sig_b}^a\eps)_{\delta\gamma} \Mb\,, & \tensor{R}{^{\sd{\delta}\sd{\gamma}}_b^a}\p &= -\frac{4}{3}({\sigb_b}^a\eps)^{\sd{\delta}\sd{\gamma}} M\,, \label{eq:component_fields_definition_sugra_id_10}\\
{R_{\delta a}}^{ba}\p &= -2i\sig_{a\delta\sd{\delta}} T^{ab\sd{\delta}}\p\,, & \tensor{R}{^{\sd{\delta}}_a^{ba}}\p &= -2i\sigb_a^{\sd{\delta}\delta} {T^{ab}}_\delta\p\,, \label{eq:component_fields_definition_sugra_id_11}
\end{align}
and
\begin{align}
\tensor{R}{_\delta^{\sd{\gamma}}_b^a}\p = -\frac{2i}{3}b^c{(\sig^d\eps)_\delta}^\sd{\gamma}{\eps_{dcb}}^a\,, \label{eq:component_fields_definition_sugra_id_12}
\end{align}
from which follows
\begin{align}
\begin{split}
{R_{ba}}^{ba}\p &= \cR + 2i{e_b}^m(\psi_m\sig_a\eps)^\sd{\varphi}{T^{ab}}_\sd{\varphi}\p + 2i{e_b}^m(\psib_m\sigb_a\eps)_\varphi T^{ab\varphi}\p \\
&\quad - \frac{1}{3}\Mb(\psi_m\sig^{mn}\psi_n) - \frac{1}{3}M(\psib_m\sigb^{mn}\psib_n) - \frac{i}{3}\eps^{mnpq}b_m(\psi_n\sig_p\psib_q)\,.
\end{split} \label{eq:component_fields_definition_sugra_id_13}
\end{align}

\subsubsection{Matter and Yang-Mills multiplets}
\label{sec:component_fields_definition_matter}
The component fields of a matter multiplet are obtained by a successive application of the covariant derivative and the bar projection on the matter superfields, namely
\begin{align}
\phi^k\p &=: \varphi^k\,, & \cD_\alpha\phi^k\p &=: \sqrt{2}{\chi^k}_\alpha\,, & \cDt^\alpha\cD_\alpha\phi^k\p &=: -4F^k\,, \label{eq:component_fields_definition_matter_matter_multiplet_1}\\
\phib^\kb\p &=: \varphib^\kb\,, & \cD^\sd{\alpha}\phib^\kb\p &=: \sqrt{2}\chib^{\kb\sd{\alpha}}\,, & \cDt_\sd{\alpha}\cD^\sd{\alpha}\phib^\kb\p &=: -4\Fb^\kb\,, \label{eq:component_fields_definition_matter_matter_multiplet_2}
\end{align}
where $\varphi^k,\varphib^\kb$ are complex scalars, $\chi^k,\chib^\kb$ are Weyl spinor fermions and $F^k,\Fb^\kb$ are complex scalar auxiliary fields. Note that for the auxiliary fields the same labels as for the corresponding superfields (cf.\ Eq.~\eqref{eq:kahler_superspace_1/2_fterm_1} and \eqref{eq:kahler_superspace_1/2_fterm_2}) are used. This holds true for the K\"ahler potential and K\"ahler transformations
\begin{gather}
K(\phi,\phib)\p =: K(\varphi,\varphib)\,, \label{eq:component_fields_definition_matter_kahler_potential}\\
F(\phi)\p =: F(\varphi)\,,\hspace{4cm} \Fb(\phib)\p =: \Fb(\varphib)\label{eq:component_fields_definition_matter_kahler_transformation}\,,
\end{gather}
and derived quantities like the K\"ahler metric $g_{\kb k}$. The component fields have the following chiral weights:
\begin{align}
w(\varphi^k) &= 0\,, & w({\chi^k}_\alpha) &= -1\,, & w(F^k) &= -2\,, \label{eq:component_fields_definition_matter_matter_multiplet_weights_1}\\
w(\varphib^\kb) &= 0\,, & w(\chib^{\kb\sd{\alpha}}) &= +1\,, & w(\Fb^\kb) &= +2\,. \label{eq:component_fields_definition_matter_matter_multiplet_weights_2}
\end{align}
The component fields of a Yang-Mills multiplet, namely the vector gauge boson ${\ca_m}^\g{r}$ the Weyl spinor gaugino ${\cl_\alpha}^\g{r}$ and its conjugate $\clb^{\sd{\alpha}\g{r}}$, and the real scalar auxiliary field $\Db^\g{r}$ are defined as
\begin{gather}
\cA^\g{r}\pp =: i\exd z^m {\ca_m}^\g{r} =: i\ca^\g{r}\,, \label{eq:component_fields_definition_matter_vector_boson}\\
{\cW_\alpha}^\g{r}\p =: -i{\cl_\alpha}^\g{r}\,,\hspace{4cm} \cW^{\sd{\alpha}\g{r}}\p =: +i\clb^{\sd{\alpha}\g{r}}\,, \label{eq:component_fields_definition_matter_gaugino}\\
\cD^\alpha{\cW_\alpha}^\g{r}\p = \cD_\sd{\alpha}\cW^{\sd{\alpha}\g{r}}\p =: -2\Db^\g{r}\,, \label{eq:component_fields_definition_matter_ym_auxiliary_field}
\end{gather}
with the chiral weights
\begin{gather}
w({\ca_m}^\g{r}) = 0\,, \\
w({\cl_\alpha}^\g{r}) = +1\,,\hspace{4cm} w(\clb^{\sd{\alpha}\g{r}}) = -1\,, \\
w(\Db^\g{r}) = 0\,.
\end{gather}
Moreover, the component fields take values in the Lie algebra of the Yang-Mills gauge group, namely
\begin{gather}
\ca_m := {\ca_m}^\g{r}\gen{r}\,, \\
\cl_\alpha := {\cl_\alpha}^\g{r}\gen{r}\,,\hspace{4cm} \clb^\sd{\alpha} := \clb^{\sd{\alpha}\g{r}}\gen{r}\,, \\
\Db := \Db^\g{r}\gen{r}\,,
\end{gather}
where $\gen{r}$ are the corresponding Hermitian generators. According to the definition of the double bar projection for a $1$-superform, the double bar projection of the K\"ahler connection reads
\begin{align}
A\pp &= \exd z^m A_m\p =: \exd z^m A_m\,.
\end{align}
Since in K\"ahler superspace $A$ is a function of the matter superfields, namely
\begin{align}
A &= \frac{1}{4}(K_k\cD\phi^k - K_\kb\cD\phib^\kb) + \frac{i}{8} E^a\big(12G_a + \sigb_a^{\sd{\alpha}\alpha} g_{k\kb} \cD_\alpha\phi^k \cD_\sd{\alpha}\phib^\kb\big)\,,
\end{align}
the component $A_m$ has the form
\begin{align}
A_m +\frac{i}{2}{e_m}^a b_a &= \frac{1}{4}(K_k\cD_m \varphi^k - K_\kb\cD_m\varphib^\kb) + \frac{i}{4}g_{k\kb}(\chi^k\sig_m\chib^\kb)\,. \label{eq:component_fields_definition_matter_kahler_connection_component_1}
\end{align}
Furthermore, the bar projection of the spinor components $A_\alpha$ and $A^\sd{\alpha}$ are given by
\begin{align}
A_\alpha\p &= +\frac{1}{4}K_k\cD_\alpha\phi^k\p = +\frac{1}{2\sqrt{2}}K_k {\chi^k}_\alpha\,, \label{eq:component_fields_definition_matter_kahler_connection_component_2}\\
A^\sd{\alpha}\p &= -\frac{1}{4}K_\kb\cD^\sd{\alpha}\phib^\kb\p = -\frac{1}{2\sqrt{2}}K_\kb\chib^{\kb\sd{\alpha}}\,. \label{eq:component_fields_definition_matter_kahler_connection_component_3}
\end{align}
In the remainder of this section, some identities are derived which are used in subsequent calculations. The double bar projection of the covariant derivative of the matter superfields is given by
\begin{align}
\cD\phi^k\pp &= \exd\phi^k\pp - \cA^\g{r}(\gen{r}\phi)^k\pp = \exd z^m \cD_m \varphi^k\,, \label{eq:component_fields_definition_matter_cd_phi_1}\\
\cD\phib^\kb\pp &= \exd\phib^\kb\pp + \cA^\g{r}(\phib\gen{r})^\kb\pp = \exd z^m \cD_m \varphib^\kb\,, \label{eq:component_fields_definition_matter_cd_phib_1}
\end{align}
with the covariant derivative of the scalars
\begin{align}
\cD_m \varphi^k &:= \del_m \varphi^k - i{\ca_m}^\g{r}(\gen{r}\varphi)^k\,, \label{eq:component_fields_definition_matter_cd_scalar_field_1}\\
\cD_m \varphib^\kb &:= \del_m \varphib^\kb + i{\ca_m}^\g{r}(\varphib\gen{r})^\kb\,. \label{eq:component_fields_definition_matter_cd_scalar_field_2}
\end{align}
On the other hand Eq.~\eqref{eq:component_fields_definition_matter_cd_phi_1} and \eqref{eq:component_fields_definition_matter_cd_phib_1} can also be evaluated as
\begin{align}
\begin{split}
\cD\phi^k\pp &= e^a\cD_a\phi^k\p + e^\alpha\cD_\alpha\phi^k\p \\
&= e^a\cD_a\phi^k\p + \frac{1}{\sqrt{2}}e^a{e_a}^m{\psi_m}^\alpha{\chi^k}_\alpha\,,
\end{split} \label{eq:component_fields_definition_matter_cd_phi_2}\\
\nonumber\\
\begin{split}
\cD\phib^\kb\pp &= e^a\cD_a\phib^\kb\p + e_\sd{\alpha}\cD^\sd{\alpha}\phib^\kb\p \\
&= e^a\cD_a\phib^\kb\p + \frac{1}{\sqrt{2}}e^a{e_a}^m\psib_{m\sd{\alpha}}\chib^{\kb\sd{\alpha}}\,,
\end{split} \label{eq:component_fields_definition_matter_cd_phib_2}
\end{align}
thus
\begin{align}
\cD_a\phi^k\p &= {e_a}^m\Big(\cD_m\varphi^k - \frac{1}{\sqrt{2}}\psi_m\chi^k\Big)\,, \label{eq:component_fields_definition_matter_id_1}\\
\cD_a\phib^\kb\p &= {e_a}^m\Big(\cD_m\varphib^\kb - \frac{1}{\sqrt{2}}\psib_m\chib^\kb\Big)\,. \label{eq:component_fields_definition_matter_id_2}
\end{align}
With a similar strategy the double bar projection of the terms $\cDt\cD_\alpha\phi^k$ and $\cDt\cD^\sd{\alpha}\phib^\kb$ are evaluated. On the one hand it is given by
\begin{align}
\begin{split}
\cDt\cD_\alpha\phi^k\pp &= \exd\cD_\alpha\phi^k\pp - {\Omega_{\alpha}}^\beta\cD_\beta\phi^k\pp - \cA^\g{r}(\gen{r}\cD_\alpha\phi)^k\pp - A\cD_\alpha\phi^k\pp + {\Gamma^k}_{ij}\cD\phi^i\cD_\alpha\phi^j\pp\,, \\
&= \sqrt{2}\,\exd z^m \cDt_m{\chi^k}_\alpha\,,
\end{split} \label{eq:component_fields_definition_matter_id_3}\\
\nonumber\\
\begin{split}
\cDt\cD^\sd{\alpha}\phib^\kb\pp &= \exd\cD^\sd{\alpha}\phib^\kb\pp - {\Omega^\sd{\alpha}}_\sd{\beta}\cD^\sd{\beta}\phib^\kb\pp + \cA^\g{r}(\cD^\sd{\alpha}\phib\gen{r})^\kb\pp + A\cD^\sd{\alpha}\phib^\kb\pp + {\Gamma^\kb}_{\ib\jb}\cD\phib^\ib\cD^\sd{\alpha}\phib^\jb\pp\,, \\
&= \sqrt{2}\,\exd z^m \cDt_m\chib^{\kb\sd{\alpha}}\,,
\end{split} \label{eq:component_fields_definition_matter_id_4}
\end{align}
with the covariant derivative of the Weyl fermions
\begin{align}
\cDt_m{\chi^k}_\alpha &:= \del_m{\chi^k}_\alpha - {\omega_{m\alpha}}^\beta{\chi^k}_\beta - i{\ca_m}^\g{r}(\gen{r}\chi_\alpha)^k - A_m{\chi^k}_\alpha + {\Gamma^k}_{ij}{\chi^i}_\alpha\cD_m \varphi^j\,, \label{eq:component_fields_definition_matter_cd_fermion_1}\\
\cDt_m\chib^{\kb\sd{\alpha}} &:= \del_m\chib^{\kb\sd{\alpha}} - \tensor{\omega}{_m^{\sd{\alpha}}_{\sd{\beta}}}\chib^{\kb\sd{\beta}} + i{\ca_m}^\g{r}(\chib^\sd{\alpha}\gen{r})^\kb + A_m\chib^{\kb\sd{\alpha}} + {\Gamma^\kb}_{\ib\jb}\chib^{\ib\sd{\alpha}}\cD_m \varphib^\jb\,. \label{eq:component_fields_definition_matter_cd_fermion_2}
\end{align}
On the other hand it has the form
\begin{align}
\begin{split}
\cDt\cD_\alpha\phi^k\pp &= e^b\cDt_b\cD_\alpha\phi^k\p + e^\beta\cDt_\beta\cD_\alpha\phi^k\p + e_\sd{\beta}\cDt^\sd{\beta}\cD_\alpha\phi^k\p \\
&= e^b\cDt_b\cD_\alpha\phi^k\p + e^b{e_b}^m\psi_{m\alpha} F^k + ie^b{e_b}^m(\psib_m\sigb^n\eps)_\alpha\Big(\cD_n\varphi^k - \frac{1}{\sqrt{2}}\psi_n\chi^k\Big)\,,
\end{split} \label{eq:component_fields_definition_matter_id_5}\\
\nonumber\\
\begin{split}
\cDt\cD^\sd{\alpha}\phib^\kb\pp &= e^b\cDt_b\cD^\sd{\alpha}\phib^\kb\p + e_\sd{\beta}\cDt^\sd{\beta}\cD^\sd{\alpha}\phib^\kb\p + e^\beta\cDt_\beta\cD^\sd{\alpha}\phib^\kb\p \\
&= e^b\cDt_b\cD^\sd{\alpha}\phib^\kb\p + e^b{e_b}^m{\psib_m}^\sd{\alpha} \Fb^\kb + ie^b{e_b}^m(\psi_m\sig^n\eps)^\sd{\alpha}\Big(\cD_n\varphib^\kb - \frac{1}{\sqrt{2}}\psib_n\chib^\kb\Big)\,.
\end{split} \label{eq:component_fields_definition_matter_id_6}
\end{align}
The bar projection of the terms $\cDt_b\cD_\alpha\phi^k$ and $\cDt_b\cD^\sd{\alpha}\phi^k$ is thus given by
\begin{align}
\cDt_b\cD_\alpha\phi^k\p &= {e_b}^m\Big(\sqrt{2}\cDt_m{\chi^k}_\alpha - \psi_{m\alpha} F^k - i(\psib_m\sigb^n\eps)_\alpha\big(\cD_n\varphi^k - \frac{1}{\sqrt{2}}\psi_n\chi^k\big)\Big)\,, \label{eq:component_fields_definition_matter_id_7}\\
\cDt_b\cD^\sd{\alpha}\phi^k\p &= {e_b}^m\Big(\sqrt{2}\cDt_m\chib^{\kb\sd{\alpha}} - {\psib_m}^\sd{\alpha} \Fb^\kb - i(\psi_m\sig^n\eps)^\sd{\alpha}\big(\cD_n\varphib^\kb - \frac{1}{\sqrt{2}}\psib_n\chib^\kb\big)\Big)\,. \label{eq:component_fields_definition_matter_id_8}
\end{align}
The following two different evaluations of the double bar projection of the Yang-Mills field strength,
\begin{align}
\cF\pp &= \frac{1}{2}\exd z^m\wedge\exd z^n \cF_{nm}\p := \frac{i}{2}\exd z^m\wedge\exd z^n \cf_{nm}\,, \label{eq:component_fields_definition_matter_id_9}
\end{align}
and
\begin{align}
\begin{split}
\cF\pp &= \exd\cA\pp + \cA\wedge\cA\pp \\
&= \frac{i}{2}\exd z^m\wedge\exd z^n\big(\del_n\ca_m - \del_m\ca_n - i\com{\ca_n}{\ca_m}\big)\,,
\end{split} \label{eq:component_fields_definition_matter_id_10}
\end{align}
imply, that the field strength tensor $\cf_{nm}$ has the form
\begin{align}
\cf_{nm} &= \del_n\ca_m - \del_m\ca_n - i\com{\ca_n}{\ca_m}\,. \label{eq:component_fields_definition_matter_gauge_field_strength}
\end{align}
In addition, if $\cF_{\beta a}$ and ${\cF^\sd{\beta}}_a$ are written in terms of the superfields $\cW^\sd{\beta}$ and $\cW_\beta$, respectively, the corresponding bar projections are given by
\begin{align}
\cF_{\beta a}\p &= -\sig_{a\beta\sd{\beta}}\clb^\sd{\beta} = -(\sig_a\clb)_\beta\,, \label{eq:component_fields_definition_matter_id_11}\\
{\cF^\sd{\beta}}_a\p &= -\sigb_a^{\sd{\beta}\beta}\cl_\beta = -(\sigb_a\cl)^\sd{\beta}\,. \label{eq:component_fields_definition_matter_id_12}
\end{align}
These identities lead to the following expression for $\cF\pp$:
\begin{align}
\begin{split}
\cF\pp &= \frac{1}{2}e^a\wedge e^b \cF_{ba}\p + e^a\wedge e^\beta \cF_{\beta a}\p + e^a\wedge e_\sd{\beta} {\cF^\sd{\beta}}_a\p \\
&=  \frac{1}{2}e^a\wedge e^b \cF_{ba}\p - \frac{1}{4}e^a\wedge e^b({e_b}^m\psi_m\sig_a\clb - {e_a}^m\psi_m\sig_b\clb) - \frac{1}{4}e^a\wedge e^b({e_b}^m\psib_m\sigb_a\cl - {e_a}^m\psib_m\sigb_b\cl)\,, \label{eq:component_fields_definition_matter_id_13}
\end{split}
\end{align}
which implies
\begin{align}
\begin{split}
\cF_{ba}\p &= +i{e_b}^n{e_a}^m \cf_{nm} \\
&\quad + \frac{1}{2}{e_b}^m(\psi_m\sig_a\clb) - \frac{1}{2}{e_a}^m(\psi_m\sig_b\clb) \\
&\quad + \frac{1}{2}{e_b}^m(\psib_m\sigb_a\cl) - \frac{1}{2}{e_a}^m(\psib_m\sigb_b\cl)\,.
\end{split} \label{eq:component_fields_definition_matter_id_14}
\end{align}
The term $\cF_{ba}\p$ is used in the following to express $\cD_b\cW_\alpha\p$ and $\cD_b\cW^\sd{\alpha}\p$ in terms of component fields. The double bar projection of the covariant derivative of the superfields $\cW_\alpha$ and $\cW^\sd{\alpha}$ is given by
\begin{align}
\cD\cW_\alpha\pp &= \exd\cW_\alpha\pp - {\Omega_\alpha}^\beta\cW_\beta\pp + \com{\cW_\alpha}{\cA}\pp + A\cW_\alpha\pp = -i\exd z^m \cD_m\cl_\alpha\,, \label{eq:component_fields_definition_matter_id_15}\\
\cD\cW^\sd{\alpha}\pp &= \exd\cW^\sd{\alpha}\pp - {\Omega^\sd{\alpha}}_\sd{\beta}\cW^\sd{\beta}\pp + \com{\cW^\sd{\alpha}}{\cA}\pp - A\cW^\sd{\alpha}\pp = +i\exd z^m \cD_m\clb^\sd{\alpha}\,, \label{eq:component_fields_definition_matter_id_16}
\end{align}
with the covariant derivative of the gaugino
\begin{align}
\cD_m\cl_\alpha &:= \del_m\cl_\alpha - {\omega_{m\alpha}}^\beta \cl_\beta - i\com{\ca_m}{\cl_\alpha} + A_m\cl_\alpha\,, \label{eq:component_fields_definition_matter_cd_gaugino_1}\\
\cD_m\clb^\sd{\alpha} &:= \del_m\clb^\sd{\alpha} - \tensor{\omega}{_m^{\sd{\alpha}}_{\sd{\beta}}}\clb^\sd{\beta} - i\com{\ca_m}{\clb^\sd{\alpha}} - A_m\clb^\sd{\alpha}\,. \label{eq:component_fields_definition_matter_cd_gaugino_2}
\end{align}
Furthermore, Eq.~\eqref{eq:component_fields_definition_matter_id_15} and \eqref{eq:component_fields_definition_matter_id_16} can also be written as
\begin{align}
\begin{split}
\cD\cW_\alpha\pp &= e^b\cD_b\cW_\alpha\p + e^\beta\cD_\beta\cW_\alpha\p \\
&= e^b\cD_b\cW_\alpha\p + \frac{1}{2}e^\beta\big(-2(\sig^{ba}\eps)_{\beta\alpha}\cF_{ba} + \eps_{\beta\alpha}\cD^\varphi\cW_\varphi\big)\p\,,
\end{split} \label{eq:component_fields_definition_matter_id_17}\\
\nonumber\\
\begin{split}
\cD\cW^\sd{\alpha}\pp &= e^b\cD_b\cW^\sd{\alpha}\p + e_\sd{\beta}\cD^\sd{\beta}\cW^\sd{\alpha}\p \\
&= e^b\cD_b\cW^\sd{\alpha}\p + \frac{1}{2}e_\sd{\beta}\big(+2(\sigb^{ba}\eps)^{\sd{\beta}\sd{\alpha}}\cF_{ba} + \eps^{\sd{\beta}\sd{\alpha}}\cD_\sd{\varphi}\cW^\sd{\varphi}\big)\p\,,
\end{split} \label{eq:component_fields_definition_matter_id_18}
\end{align}
which leads to the expressions
\begin{align}
\cD_b\cW_\alpha\p &= {e_b}^m\Big(-i\cD_m\cl_\alpha + \frac{1}{2}(i\cf_{pq} + \psi_p\sig_q\clb + \psib_p\sigb_q\cl)(\psi_m\sig^{pq}\eps)_\alpha - \frac{1}{2}\Db\psi_{m\alpha}\Big)\,, \label{eq:component_fields_definition_matter_id_19}\\
\cD_b\cW^\sd{\alpha}\p &= {e_b}^m\Big(+i\cD_m\clb^\sd{\alpha} - \frac{1}{2}(i\cf_{pq} + \psi_p\sig_q\clb + \psib_p\sigb_q\cl)(\psib_m\sigb^{pq}\eps)^\sd{\alpha} - \frac{1}{2}\Db{\psib_m}^\sd{\alpha}\Big)\,. \label{eq:component_fields_definition_matter_id_20}
\end{align}

\subsection{Matter $D$-term}
In Section~\eqref{sec:component_fields_actions_sugra_matter} it will turn out that the component field Lagrangian of the supergravity+matter sector contains a matter (Fayet-Iliopoulos) $D$-term $D_\text{matter}$, given by
\begin{align}
D_\text{matter} &= -\frac{1}{2}\cD^\alpha X_\alpha\p + \frac{i}{2}(\psi_m\sig^m)_\sd{\alpha}X^\sd{\alpha}\p + \frac{i}{2}(\psib_m\sigb^m)^\alpha X_\alpha\p\,. \label{eq:component_fields_matter_dterm_definition}
\end{align}
The goal of this section is to write $D_\text{matter}$ in terms of component fields. According to the expressions of $X_\alpha$ and $X^\sd{\alpha}$ in Eq.~\eqref{eq:kahler_superspace_1/2_x_5} and \eqref{eq:kahler_superspace_1/2_x_6}, respectively, the corresponding bar projections read
\begin{align}
X_\alpha\p &= -\frac{i}{\sqrt{2}} g_{k\kb}(\sig^m\chib^\kb)_\alpha\Big(\cD_m\varphi^k - \frac{1}{\sqrt{2}}\psi_m\chi^k\Big) + \frac{1}{\sqrt{2}}g_{k\kb}{\chi^k}_\alpha\Fb^\kb -i{\cl_\alpha}^\g{r}G_\g{r}\,, \label{eq:component_fields_matter_dterm_x_components_1}\\
X^\sd{\alpha}\p &= -\frac{i}{\sqrt{2}} g_{k\kb} (\sigb^m\chi^k)^\sd{\alpha}\Big(\cD_m\varphib^\kb - \frac{1}{\sqrt{2}}\psib_m\chib^\kb\Big) + \frac{1}{\sqrt{2}}g_{k\kb}\chib^{\kb\sd{\alpha}}F^k + i\clb^{\sd{\alpha}\g{r}}G_\g{r}\,. \label{eq:component_fields_matter_dterm_x_components_2}
\end{align}
Thus, the last two terms in Eq.~\eqref{eq:component_fields_matter_dterm_definition} have the following form:
\begin{align}
\begin{split}
\frac{i}{2}(\psi_m\sig^m)_\sd{\alpha}X^\sd{\alpha}\p &= +\frac{1}{2\sqrt{2}}g_{k\kb}(\psi_m\sig^m\sigb^n\chi^k)(\cD_n\varphib^\kb - \frac{1}{\sqrt{2}}\psib_n\chib^\kb) \\
&\quad + \frac{i}{2\sqrt{2}}g_{k\kb}(\psi_m\sig^m\chib^\kb)F^k - \frac{1}{2}(\psi_m\sig^m\clb^\g{r})G_\g{r}\,,
\end{split} \label{eq:component_fields_matter_dterm_id_1}\\
\nonumber\\
\begin{split}
\frac{i}{2}(\psib_m\sigb^m)^\alpha X_\alpha\p &= +\frac{1}{2\sqrt{2}}g_{k\kb}(\psib_m\sigb^m\sig^n\chib^\kb)(\cD_n\varphi^k - \frac{1}{\sqrt{2}}\psi_n\chi^k) \\
&\quad + \frac{i}{2\sqrt{2}}g_{k\kb}(\psib_m\sigb^m\chi^k)\Fb^\kb + \frac{1}{2}(\psib_m\sigb^m\cl^\g{r})G_\g{r}\,.
\end{split} \label{eq:component_fields_matter_dterm_id_2}
\end{align}
Furthermore, the following identities
\begin{align}
\begin{split}
-g_{k\kb}\eta^{ab}\cD_a\phi^k\cD_b\phib^\kb\p &= -g_{k\kb}\,g^{mn}\cD_m\varphi^k\cD_n\varphib^\kb \\
&\quad + \frac{1}{\sqrt{2}}g_{k\kb}\,g^{mn}\cD_m\varphi^k(\psib_n\chib^\kb) + \frac{1}{\sqrt{2}}g_{k\kb}\,g^{mn}\cD_m\varphib^\kb(\psi_n\chi^k) \\
&\quad - \frac{1}{2}g_{k\kb}\,g^{mn}(\psi_m\chi^k)(\psib_n\chib^\kb)\,,
\end{split} \label{eq:component_fields_matter_dterm_id_3}
\end{align}
\begin{align}
\begin{split}
& - \frac{i}{4}g_{k\kb}\sig^a_{\alpha\sd{\alpha}}\cD^\alpha\phi^k\cDt_a\cD^\sd{\alpha}\phib^\kb\p - \frac{i}{4}g_{k\kb}\sig^a_{\alpha\sd{\alpha}}\cD^\sd{\alpha}\phib^\kb\cDt_a\cD^\alpha\phi^k\p =\\
&\quad= - \frac{i}{2}g_{k\kb}\chi^{k\alpha}\sig^m_{\alpha\sd{\alpha}}\cDt_m\chib^{\kb\sd{\alpha}} + \frac{i}{2}g_{k\kb}(\cDt_m\chi^{k\alpha})\sig^m_{\alpha\sd{\alpha}}\chib^{\kb\sd{\alpha}} \\
&\quad\quad - \frac{i}{2\sqrt{2}}g_{k\kb}(\psib_m\sigb^m\chi^k)\Fb^\kb - \frac{i}{2\sqrt{2}}g_{k\kb}(\psi_m\sig^m\chib^\kb)F^k \\
&\quad\quad - \frac{1}{2\sqrt{2}}g_{k\kb}(\psi_m\sig^n\sigb^m\chi^k)\Big(\cD_n\varphib^\kb - \frac{1}{\sqrt{2}}\psib_n\chib^\kb\Big) \\
&\quad\quad - \frac{1}{2\sqrt{2}}g_{k\kb}(\psib_m\sigb^n\sig^m\chib^\kb)\Big(\cD_n\varphi^k - \frac{1}{\sqrt{2}}\psi_n\chi^k\Big)\,,
\end{split} \label{eq:component_fields_matter_dterm_id_4}\\
\nonumber\\
\begin{split}
& g_{k\kb}(\phib\gen{r})^\kb\cW^{\alpha\g{r}}\cD_\alpha\phi^k\p + g_{k\kb}(\gen{r}\phi)^k{\cW_\sd{\alpha}}^\g{r}\cD^\sd{\alpha}\phib^\kb\p = \\
&\quad= -i\sqrt{2}g_{k\kb}(\chi^k\cl^\g{r})(\varphib\gen{r})^\kb + i\sqrt{2}g_{k\kb}(\chib^\kb\clb^\g{r})(\gen{r}\varphi)^k\,,
\end{split}
\end{align} \label{eq:component_fields_matter_dterm_id_5}
\begin{align}
\frac{1}{16} R_{k\kb j\jb}\cD^\alpha\phi^k\cD_\alpha\phi^j \cD_\sd{\alpha}\phib^\kb\cD^\sd{\alpha}\phib^\jb\p &= \frac{1}{4} R_{k\kb j\jb} (\chi^k\chi^j)(\chib^\kb\chib^\jb)\,, \label{eq:component_fields_matter_dterm_id_6}
\end{align}
imply that the first term on the right-hand side in Eq.~\eqref{eq:component_fields_matter_dterm_definition} has the form
\begin{align}
\begin{split}
-\frac{1}{2}\cD^\alpha X_\alpha\p &= -g_{k\kb}\,g^{mn}\cD_m\varphi^k\cD_n\varphib^\kb - \frac{i}{2}g_{k\kb}(\chi^k\sig^m\cDt_m\chib^\kb) + \frac{i}{2}g_{k\kb}(\cDt_m\chi^k\sig^m\chib^\kb) \\
&\quad + g_{k\kb}F^k\Fb^\kb + \frac{1}{2}g_{k\kb}\,g^{mn}(\psi_m\chi^k)(\psib_n\chib^\kb) \\
&\quad + \frac{1}{4} R_{k\kb j\jb} (\chi^k\chi^j)(\chib^\kb\chib^\jb) - i\sqrt{2}g_{k\kb}(\chi^k\cl^\g{r})(\varphib\gen{r})^\kb + i\sqrt{2}g_{k\kb}(\chib^\kb\clb^\g{r})(\gen{r}\varphi)^k \\
&\quad + \Db^\g{r}G_\g{r} - \frac{i}{2\sqrt{2}}g_{k\kb}(\psib_m\sigb^m\chi^k)\Fb^\kb - \frac{i}{2\sqrt{2}}g_{k\kb}(\psi_m\sig^m\chib^\kb)F^k \\
&\quad - \frac{1}{2\sqrt{2}}g_{k\kb}(\psib_m\sigb^n\sig^m\chib^\kb - 2g^{mn}\psib_m\chib^\kb)\Big(\cD_n\varphi^k - \frac{1}{\sqrt{2}}\psi_n\chi^k\Big) \\
&\quad - \frac{1}{2\sqrt{2}}g_{k\kb}(\psi_m\sig^n\sigb^m\chi^k - 2g^{mn}\psi_m\chi^k)\Big(\cD_n\varphib^\kb - \frac{1}{\sqrt{2}}\psib_n\chib^\kb\Big)\,.
\end{split} \label{eq:component_fields_matter_dterm_cd_x}
\end{align}
Taking the sum of Eq.~\eqref{eq:component_fields_matter_dterm_id_1}, \eqref{eq:component_fields_matter_dterm_id_2} and \eqref{eq:component_fields_matter_dterm_cd_x}, the matter $D$-term reads
\begin{align}
\begin{split}
D_\text{matter} &= -g_{k\kb}\,g^{mn}\cD_m\varphi^k\cD_n\varphib^\kb - \frac{i}{2}g_{k\kb}(\chi^k\sig^m\nabla_m\chib^\kb) + \frac{i}{2}g_{k\kb}(\nabla_m\chi^k\sig^m\chib^\kb) \\
&\quad + g_{k\kb}F^k\Fb^\kb + \frac{1}{4} R_{k\kb j\jb} (\chi^k\chi^j)(\chib^\kb\chib^\jb) - \frac{1}{2}g_{k\kb}(\chi^k\sig^a\chib^\kb)b_a \\
&\quad - \frac{1}{\sqrt{2}}g_{k\kb}(\psib_m\sigb^n\sig^m\chib^\kb)\cD_n\varphi^k - \frac{1}{\sqrt{2}}g_{k\kb}(\psi_m\sig^n\sigb^m\chi^k)\cD_n\varphib^\kb \\
&\quad - \frac{i}{2}g_{k\kb}\eps^{mnpq}(\chi^k\sig_m\chib^\kb)(\psi_n\sig_p\psib_q) - \frac{1}{2}g_{k\kb}\,g^{mn}(\psi_m\chi^k)(\psib_n\chib^\kb) \\
&\quad - i\sqrt{2}g_{k\kb}(\chi^k\cl^\g{r})(\varphib\gen{r})^\kb + i\sqrt{2}g_{k\kb}(\chib^\kb\clb^\g{r})(\gen{r}\varphi)^k \\
&\quad - \frac{1}{2}\Big(\Db^\g{r} + \frac{1}{2}(\psib_m\sigb^m\cl^\g{r} - \psi_m\sig^m\clb^\g{r})\Big)\big(K_k(\gen{r}\varphi)^k + K_\kb(\varphib\gen{r})^\kb\big)\,,
\end{split} \label{eq:component_fields_matter_dterm}
\end{align}
with the definitions
\begin{align}
\begin{split}
\nabla_m{\chi^k}_\alpha &:= \cDt_m{\chi^k}_\alpha - \frac{i}{2}{e_m}^a b_a {\chi^k}_\alpha\\
&\;= \del_m{\chi^k}_\alpha - {\omega_{m\alpha}}^\beta{\chi^k}_\beta - i{\ca_m}^\g{r}(\gen{r}\chi_\alpha)^k + {\Gamma^k}_{ij}{\chi^i}_\alpha\cD_m \varphi^j - \frac{1}{4}(K_j\cD_m\varphi^j - K_\jb\cD_m\varphib^\jb){\chi^k}_\alpha \\
&\quad - \frac{i}{4}g_{j\jb}(\chi^j\sig_m\chib^\jb){\chi^k}_\alpha\,,
\end{split} \label{eq:component_fields_matter_dterm_nabla_chi}\\
\begin{split}
\nabla_m\chib^{\kb\sd{\alpha}} &:= \cDt_m\chib^{\kb\sd{\alpha}} + \frac{i}{2}{e_m}^a b_a \chib^{\kb\sd{\alpha}}\\
&\;= \del_m\chib^{\kb\sd{\alpha}} - \tensor{\omega}{_m^{\sd{\alpha}}_{\sd{\beta}}}\chib^{\kb\sd{\beta}} + i{\ca_m}^\g{r}(\chib^\sd{\alpha}\gen{r})^\kb + {\Gamma^\kb}_{\ib\jb}\chib^{\ib\sd{\alpha}}\cD_m \varphib^\jb + \frac{1}{4}(K_j\cD_m\varphi^j - K_\jb\cD_m\varphib^\jb)\chib^{\kb\sd{\alpha}} \\
&\quad + \frac{i}{4}g_{j\jb}(\chi^j\sig_m\chib^\jb)\chib^{\kb\sd{\alpha}}\,,
\end{split} \label{eq:component_fields_matter_dterm_nabla_chib}
\end{align}
where the covariant derivatives $\cDt_m{\chi^k}_\alpha$ and $\cDt_m\chib^{\kb\sd{\alpha}}$ are stated in Eq.~\eqref{eq:component_fields_definition_matter_cd_fermion_1} and \eqref{eq:component_fields_definition_matter_cd_fermion_2}, respectively. The derivatives $\nabla_m{\chi^k}_\alpha$ and $\nabla_m\chib^{\kb\sd{\alpha}}$ are introduced to easier keep track of the auxiliary field $b_a$ in the component field Lagrangian.

\subsection{Supergravity transformations}
\label{sec:component_fields_sugra_transformations}
In this section the supergravity transformations, which correspond to local supersymmetry transformations adapted to the Wess-Zumino gauge, of the component fields in the minimal supergravity, the matter and the Yang-Mills multiplet are presented. They are derived from the supergravity transformations at the superfield level in Section~\ref{sec:sugra_transformations_superfield_level} and the identification of the component fields as lowest superspace components of superfields using the bar projection (cf.\ Section~\ref{sec:component_fields_definition}). Superspace diffeomorphisms which appear in supergravity transformations $\dsg_\xi$ are represented by a real and even supervector field $\xi=\xi^A E_A$ with the components $\xi^A=(\xi^a,\xi^\alpha,\xib_\sd{\alpha})$, where $(\xi^\alpha)^\cc=\xib^\sd{\alpha}$ and $\xi^a\p=0$. Thus, the bar projection of the components is written as
\begin{align}
(\xi^a,\xi^\alpha,\xib_\sd{\alpha})\p \equiv (0,\xi^\alpha,\xib_\sd{\alpha})\,.
\end{align}
In the explicit calculation of the supergravity transformations of the component fields the identities at the superfield level from Section~\ref{sec:sugra_transformations_superfield_level_identities} are combined with the identities at the component field level from Section~\ref{sec:component_fields_definition_sugra} and \ref{sec:component_fields_definition_matter}.
\begin{itemize}
\item \textbf{Supergravity multiplet:}\\
The supergravity transformation of the graviton ${e_m}^a$ can be read off from Eq.~\eqref{eq:sugra_transf_delta_vielbein_explicit} by applying the bar projection:
\begin{align}
\dsg_\xi{e_m}^a &= +i(\xi\sig^a\psib_m) + i(\xib\sigb^a\psi_m)\,. \label{eq:component_fields_sugra_transf_graviton}
\end{align}
From the same equation follows that the supergravity transformation of the gravitino ${\psi_m}^\alpha$ and its conjugate $\psib_{m\sd{\alpha}}$ has the form
\begin{align}
\begin{split}
\dsg_\xi{\psi_m}^\alpha &= +2\nabla_m\xi^\alpha - i{e_m}^a b_a \xi^\alpha - \frac{i}{3}(\xi\sig^a\sigb_m)^\alpha b_a \\
&\quad + \frac{i}{3}(\xib\sigb_m)^\alpha M - \frac{1}{2\sqrt{2}}{\psi_m}^\alpha(K_k\xi\chi^k - K_\kb\xib\chib^\kb)\,,
\end{split} \label{eq:component_fields_sugra_transf_gravitino_1}\\
\nonumber\\
\begin{split}
\dsg_\xi\psib_{m\sd{\alpha}} &= +2\nabla_m\xib_\sd{\alpha} + i{e_m}^a b_a \xib_\sd{\alpha} + \frac{i}{3}(\xib\sigb^a\sig_m)_\sd{\alpha}b_a \\
&\quad + \frac{i}{3}(\xi\sig_m)_\sd{\alpha}\Mb + \frac{1}{2\sqrt{2}}\psib_{m\sd{\alpha}}(K_k\xi\chi^k - K_\kb\xib\chib^\kb)\,,
\end{split} \label{eq:component_fields_sugra_transf_gravitino_2}
\end{align}
with the derivatives
\begin{align}
\begin{split}
\nabla_m\xi^\alpha &= \del_m\xi^\alpha + \xi^\beta{\omega_{m\beta}}^\alpha \\
&\quad + \frac{1}{4}(K_k\cD_n\varphi^k - K_\kb\cD_n\varphib^\kb)\xi^\alpha + \frac{i}{4}g_{k\kb}(\chi^k\sig_n\chib^\kb)\xi^\alpha\,,
\end{split} \\
\nonumber\\
\begin{split}
\nabla_m\xib_\sd{\alpha} &= \del_m\xib_\sd{\alpha} + \xib_\sd{\beta}\tensor{\omega}{_n^{\sd{\beta}}_{\sd{\alpha}}} \\
&\quad - \frac{1}{4}(K_k\cD_n\varphi^k - K_\kb\cD_n\varphib^\kb)\xib_\sd{\alpha} - \frac{i}{4}g_{k\kb}(\chi^k\sig_n\chib^\kb)\xib_\sd{\alpha}\,.
\end{split}
\end{align}
where the expression in Eq.~\eqref{eq:component_fields_definition_matter_kahler_connection_component_1} is used. Furthermore, from Eq.~\eqref{eq:sugra_transf_delta_r_explicit} and Eq.~\eqref{eq:sugra_transf_delta_rb_explicit} the supergravity transformation of the auxiliary field $M$ and its conjugate $\Mb$ can be read off by using the bar projection, namely
\begin{align}
\begin{split}
\dsg_\xi M &= - i\sqrt{2}g_{k\kb}(\xi\sig^m\chib^\kb)\Big(\cD_m\varphi^k - \frac{1}{\sqrt{2}}\psi_m\chi^k\Big) + \sqrt{2}g_{k\kb}(\xi\chi^k)\Fb^\kb \\
&\quad + i(\xi\cl^\g{r})\big(K_k(\gen{r}\varphi)^k + K_\kb(\varphib\gen{r})^\kb\big) - \frac{1}{\sqrt{2}}M(K_k\xi\chi^k - K_\kb\xib\chib^\kb) \\
&\quad + 4(\xi\sig^{nm}\cD_n\psi_m) - i(\xi\sig^m\sigb^a\psi_m)b_a - i(\xi\sig^m\psib_m)M\,,
\end{split} \label{eq:component_fields_sugra_transf_aux_m_1}\\
\nonumber\\
\begin{split}
\dsg_\xi \Mb &= - i\sqrt{2}g_{k\kb}(\xib\sigb^m\chi^k)\Big(\cD_m\varphib^\kb - \frac{1}{\sqrt{2}}\psib_m\chib^\kb\Big) + \sqrt{2}g_{k\kb}(\xib\chib^\kb)F^k \\
&\quad - i(\xib\clb^\g{r})\big(K_k(\gen{r}\varphi)^k + K_\kb(\varphib\gen{r})^\kb\big) + \frac{1}{\sqrt{2}}\Mb(K_k\xi\chi^k - K_\kb\xib\chib^\kb) \\
&\quad + 4(\xib\sigb^{nm}\cD_n\psib_m) + i(\xib\sigb^m\sig^a\psib_m)b_a - i(\xib\sigb^m\psi_m)\Mb\,,
\end{split} \label{eq:component_fields_sugra_transf_aux_m_2}
\end{align}
where the derivatives $\cD_m\varphi^k$ and $\cD_m\varphib^\kb$ are defined in Eq.~\eqref{eq:component_fields_definition_matter_cd_scalar_field_1} and \eqref{eq:component_fields_definition_matter_cd_scalar_field_2}, and $\cD_n\psi_m$ and $\cD_n\psib_m$ are defined in Eq.~\eqref{eq:component_fields_definition_sugra_cd_gravitino_1} and \eqref{eq:component_fields_definition_sugra_cd_gravitino_2}. Finally, Eq.~\eqref{eq:sugra_transf_delta_ga_explicit} implies that the supergravity transformation of the auxiliary field $b_a$ is given by
\begin{align}
\begin{split}
\dsg_\xi b_a &= + \frac{1}{2}(\xi\sig_a\sigb^{nm} - 3\xi\sig^{nm}\sig_a)\cD_n\psib_m - \frac{1}{2}(\xib\sigb_a\sig^{nm} - 3\xib\sigb^{nm}\sigb_a)\cD_n\psi_m \\
&\quad - \frac{i}{2}{e_a}^m(\xi\sig^d\psib_m + \xib\sigb^d\psi_m)b_d - \frac{i}{2}{e_a}^m(\xib\psib_m)M + \frac{i}{2}{e_a}^m(\xi\psi_m)\Mb \\
&\quad - \frac{i}{\sqrt{2}}g_{k\kb}(\xi\sig_a\sigb^m\chi^k)\Big(\cD_m\varphib^\kb - \frac{1}{\sqrt{2}}\psib_m\chib^\kb\Big) + \frac{1}{\sqrt{2}}g_{k\kb}(\xi\sig_a\chib^\kb)F^k \\
&\quad + \frac{i}{\sqrt{2}}g_{k\kb}(\xib\sigb_a\sig^m\chib^\kb)\Big(\cD_m\varphi^k - \frac{1}{\sqrt{2}}\psi_m\chi^k\Big) - \frac{1}{\sqrt{2}}g_{k\kb}(\xib\sigb_a\chi^k)\Fb^\kb \\
&\quad - \frac{i}{2}(\xi\sig_a\clb^\g{r} + \xib\sigb_a\cl^\g{r})\big(K_k(\gen{r}\varphi)^k + K_\kb(\varphib\gen{r})^\kb\big)\,.
\end{split} \label{eq:component_fields_sugra_transf_aux_m_3}
\end{align}
\item \textbf{Matter multiplet:}\\
The supergravity transformation of the matter scalar field $\varphi^k$ and its conjugate $\varphib^\kb$ can be read off from Eq.~\eqref{eq:sugra_transf_delta_phi_explicit} and \eqref{eq:sugra_transf_delta_phib_explicit} by applying the bar projection:
\begin{align}
\dsg_\xi\varphi^k &= \sqrt{2}\xi\chi^k\,, \\
\nonumber\\
\dsg_\xi\varphib^\kb &= \sqrt{2}\xib\chib^\kb\,.
\end{align}
In addition, Eq.~\eqref{eq:sugra_transf_delta_dphi_explicit} and \eqref{eq:sugra_transf_delta_dphib_explicit} imply that the corresponding Weyl fermions $\chi^k$ and $\chib^\kb$ transform under supergravity transformations as
\begin{align}
\begin{split}
\dsg_\xi{\chi^k}_\alpha &=  + i\sqrt{2}(\xib\sigb^m\eps)_\alpha\Big(\cD_m\varphi^k - \frac{1}{\sqrt{2}}\psi_m\chi^k\Big) + \sqrt{2}\xi_\alpha F^k \\
&\quad + \frac{1}{\sqrt{2}}\xi_\alpha{\Gamma^k}_{ij}(\chi^i\chi^j) + \frac{1}{2\sqrt{2}}{\chi^k}_\alpha(K_k\xi\chi^k - K_\kb\xib\chib^\kb)\,,
\end{split} \\
\nonumber\\
\begin{split}
\dsg_\xi\chib^{\kb\sd{\alpha}} &=  + i\sqrt{2}(\xi\sig^m\eps)^\sd{\alpha}\Big(\cD_m\varphib^\kb - \frac{1}{\sqrt{2}}\psib_m\chib^\kb\Big) + \sqrt{2}\xib^\sd{\alpha} \Fb^\kb \\
&\quad + \frac{1}{\sqrt{2}}\xi^\sd{\alpha}{\Gamma^\kb}_{\ib\jb}(\chib^\ib\chib^\jb) - \frac{1}{2\sqrt{2}}\chib^{k\sd{\alpha}}(K_k\xi\chi^k - K_\kb\xib\chib^\kb)\,.
\end{split}
\end{align}
Finally, according to Eq.~\eqref{eq:sugra_transf_delta_dphi_explicit} and \eqref{eq:sugra_transf_delta_dphib_explicit} the supergravity transformations of the auxiliary scalar fields $F^k$ and $\Fb^\kb$ have the form
\begin{align}
\begin{split}
\dsg_\xi F^k &= +i\sqrt{2}(\xib\sigb^m\nabla_m\chi^k) - i(\xib\sigb^m\psi_m)F^k + (\xib\sigb^m\sig^n\psib_m)\Big(\cD_n\varphi^k - \frac{1}{\sqrt{2}}\psi_n\chi^k\Big) \\
&\quad + \frac{\sqrt{2}}{3}\Mb(\xi\chi^k) - \frac{\sqrt{2}}{6}(\xib\sigb^a\chi^k)b_a - 2i(\xib\clb^\g{r})(\gen{r}\varphi)^k - \sqrt{2}{\Gamma^k}_{ij}(\xi\chi^i)F^j \\
&\quad - \frac{1}{\sqrt{2}}g^{\kb k}R_{i\kb j\lb}(\chi^i\chi^j)(\xib\chib^\lb) + \frac{1}{\sqrt{2}}F^k(K_j\xi\chi^j - K_\jb\xib\chib^\jb)\,,
\end{split} \\
\nonumber\\
\begin{split}
\dsg_\xi \Fb^\kb &= +i\sqrt{2}(\xi\sig^m\nabla_m\chib^\kb) - i(\xi\sig^m\psib_m)\Fb^\kb + (\xi\sig^m\sigb^n\psi_m)\Big(\cD_n\varphib^\kb - \frac{1}{\sqrt{2}}\psib_n\chib^\kb\Big) \\
&\quad + \frac{\sqrt{2}}{3}M(\xib\chib^\kb) + \frac{\sqrt{2}}{6}(\xi\sig^a\chib^\kb)b_a + 2i(\xi\cl^\g{r})(\varphib\gen{r})^\kb - \sqrt{2}{\Gamma^\kb}_{\ib\jb}(\xib\chib^\ib)\Fb^\jb \\
&\quad - \frac{1}{\sqrt{2}}g^{\kb k}R_{k\ib l\jb}(\chib^\ib\chib^\jb)(\xi\chi^l) - \frac{1}{\sqrt{2}}\Fb^\kb(K_j\xi\chi^j - K_\jb\xib\chib^\jb)\,,
\end{split}
\end{align}
where the derivatives $\nabla_m\chi^k$ and $\nabla_m\chib^\kb$ are defined in Eq.~\eqref{eq:component_fields_matter_dterm_nabla_chi} and \eqref{eq:component_fields_matter_dterm_nabla_chib}, respectively.
\item \textbf{Yang-Mills multiplet:}\\
The supergravity transformation of the vector gauge boson ${\ca_m}^\g{r}$ follows directly from Eq.~\eqref{eq:sugra_transf_delta_ymconnection_explicit} by using the bar projection, namely
\begin{align}
\dsg_\xi{\ca_m}^\g{r} &= i(\xi\sig_m\clb^\g{r}) + i(\xib\sigb_m\cl^\g{r})\,.
\end{align}
According to Eq.~\eqref{eq:sugra_transf_delta_w_explicit} and \eqref{eq:sugra_transf_delta_wb_explicit}, under supergravity transformations the corresponding gaugino ${\cl_\alpha}^\g{r}$ and its conjugate $\clb^{\sd{\alpha}\g{r}}$ transform as
\begin{align}
\begin{split}
\dsg_\xi{\cl_\alpha}^\g{r} &= +(\xi\sig^{mn}\eps)_\alpha ({\cf_{mn}}^\g{r} - i\psi_m\sig_n\clb^\g{r} - i\psib_m\sigb_n\cl^\g{r}) \\
&\quad + i\xi_\alpha\Db^\g{r} - \frac{1}{2\sqrt{2}}\cl^{\alpha\g{r}}(K_k\xi\chi^k - K_\kb\xib\chib^\kb)\,,
\end{split} \\
\nonumber\\
\begin{split}
\dsg_\xi\clb^{\sd{\alpha}\g{r}} &= +(\xib\sigb^{mn}\eps)^\sd{\alpha} ({\cf_{mn}}^\g{r} - i\psi_m\sig_n\clb^\g{r} - i\psib_m\sigb_n\cl^\g{r}) \\
&\quad - i\xib^\sd{\alpha}\Db^\g{r} + \frac{1}{2\sqrt{2}}\clb^{\sd{\alpha}\g{r}}(K_k\xi\chi^k - K_\kb\xib\chib^\kb)\,,
\end{split}
\end{align}
with $\cf_{mn}$ defined Eq.~\eqref{eq:component_fields_definition_matter_gauge_field_strength}. Finally, Eq.~\eqref{eq:sugra_transf_delta_dw_explicit} implies that the supergravity transformation of the auxiliary scalar field $\Db^\g{r}$ has the form
\begin{align}
\begin{split}
\dsg_\xi\Db^\g{r} &= - (\xi\sig^m\cD_m\clb^\g{r}) + (\xib\sigb^m\cD_m\cl^\g{r}) + \frac{i}{2}(\psib_m\sigb^m\xi + \psi_m\sig^m\xib)\Db^\g{r} \\
&\quad + \frac{1}{2}(\psib_m\sigb^{pq}\sigb^m\xi - \psi_m\sig^{pq}\sig^m\xib)({\cf_{pq}}^\g{r} - i\psi_p\sig_q\clb^\g{r} - i\psib_p\sigb_q\cl^\g{r})\,,
\end{split}
\end{align}
where the derivatives $\cD_m\cl^\g{r}$ and $\cD_m\clb^\g{r}$ are defined in Eq.~\eqref{eq:component_fields_definition_matter_cd_gaugino_1} and Eq.~\eqref{eq:component_fields_definition_matter_cd_gaugino_2}, respectively.
\end{itemize}

\subsection{$R$-symmetry}
\label{sec:component_fields_r_symmetry}
As discussed in Section~\ref{sec:r_symmetry_superfield_level} at the superfield level, an $R$-symmetry $\UR$ corresponds to an ordinary $\U(1)$ Yang-Mills gauge transformation accompanied with a K\"ahler transformation. In particular, $\UR$ transformations are parametrized by K\"ahler transformations, namely $\lacR=-\frac{1}{2}\im{F}$ where $-i\lacR\in\uu(1)$. At the component field level, the K\"ahler transformation $F(\varphi^k)$ is a holomorphic function of the matter scalar fields $\varphi^k$. The weight of a component field concerning the $\UR$ symmetry, also called $R$-charge, is written as $\wR$, and is given by an overall $R$-charge of the supermultiplet plus the particular chiral weight of that field. In the following, the $R$-charges of the component fields in the supergravity, the matter and the Yang-Mills multiplet are listed.
\begin{itemize}
\item \textbf{Supergravity multiplet:}\\
The overall $R$-charge of the supergravity multiplet is zero. Thus, the weights $\wR$ of the component fields take the values
\begin{gather}
\wR({e_m}^a) = 0\,, \\
\wR({\psi_m}^\alpha) = +1\,, \hspace{4cm} \wR(\psib_{m\sd{\alpha}}) = -1\,, \\
\hspace{0.33cm}\wR(M) = +2\,, \hspace{4.33cm} \wR(\Mb) = -2\,, \\
\wR(G_a) = 0\,.
\end{gather}
\item \textbf{Matter multiplet:}\\
For a matter multiplet the overall $R$-charge is not fixed and is written as $\wR(\phi^k)$. The weights $\wR$ of the component fields are then given by
\begin{align}
\wR(\varphi^k) &= \wR(\phi^k)\,, & \wR({\chi^k}_\alpha) &= \wR(\phi^k)-1\,, & \wR(F^k) &= \wR(\phi^k)-2\,, \\
\wR(\varphib^\kb) &= \wR(\phib^\kb)\,, & \wR(\chib^{\kb\sd{\alpha}}) &= \wR(\phib^\kb)+1\,, & \wR(\Fb^\kb) &= \wR(\phib^\kb)+2\,,
\end{align}
where $\wR(\phib^\kb)=-\wR(\phi^k)$. Furthermore, in order that $\UR$ transformations leave the action invariant, the K\"ahler potential $K(\varphi^k,\varphib^\kb)$, the superpotential $W(\varphi^k)$, and the gauge kinetic function $\gkf{r}{s}(\varphi^k)$, as functions of the matter scalar fields, need to have the definite weights
\begin{gather}
\wR(K) = 0\,, \\
\wR(W) = +2\,, \hspace{4cm} \wR(\Wb) = -2\,, \\
\wR(\gkf{r}{s}) = 0\,, \hspace{3.77cm} \wR(\gkfb{r}{s}) = 0\,. \hspace{0.85cm}
\end{gather}
\item \textbf{Yang-Mills multiplet:}\\
The overall $R$-charge of a Yang-Mills multiplet is always equal to zero and the corresponding component fields have the following weights $\wR$:
\begin{gather}
\wR(\ca_m) = 0\,, \\
\wR(\cl_\alpha) = +1\,,\hspace{4cm} \wR(\clb^\sd{\alpha}) = -1\,, \\
\wR(\Db) = 0\,.
\end{gather}
Note that gauge boson which belongs to the $\UR$ symmetry transforms as $\caR{}_m\mapsto\caR{}_m-\del_m\lacR$.
\end{itemize}

\subsection{Invariant actions}
\label{sec:component_fields_actions}
It is convenient to derive the action at the component field level from an action at the superfield level which is written in terms of the chiral density $\frac{E}{R}$ and a chiral superfield $\Xi$ with chiral weight $w(\Xi)=+2$ (cf.\ Section~\ref{sec:action_superfield_level}), i.e.\
\begin{align}
\cS &= \frac{1}{2}\int_*\frac{E}{R}\Xi + \frac{1}{2}\int_*\frac{E}{\Rb}\Xib\,, \label{eq:component_fields_action_general_action_superfield}
\end{align}
where $\Xib=\Xi^\cc$ is antichiral and $w(\Xib)=-2$. In Section~\ref{sec:component_fields_actions_general} the action for a general chiral superfield $\Xi$ at the component field level is computed. This result is then used in the subsequent sections to determine the action of the supergravity/matter/Yang-Mills system.

\subsubsection{General action}
\label{sec:component_fields_actions_general}
In order to derive the general action at the component field level, it is assumed that the terms in Eq.~\eqref{eq:component_fields_action_general_action_superfield} have the form \footnote{A derivation of the density formula using purely superspace methods can be found in \cite{Grisaru:1997ub}.}
\begin{align}
\frac{1}{2}\int_*\frac{E}{R}\Xi &= \int\exd^4 x\,\big(e\,f + \lambda_2^\alpha s_\alpha + \lambda_3 r\big)\,, & \frac{1}{2}\int_*\frac{E}{\Rb}\Xib &= \int\exd^4 x\,\big(e\,\fb + \lambdab_{2\sd{\alpha}}\sbb^\sd{\alpha} + \lambdab_3\rb\big)\,, \label{eq:component_fields_action_general_action_superfield_chiral}
\end{align}
with the determinant $e=\det({e_m}^a)$ and the component fields
\begin{align}
r &:= \Xi\p\,, & s_\alpha &:= \frac{1}{\sqrt{2}}\cD_\alpha\Xi\p\,, & f &:= -\frac{1}{4}\cD^\alpha\cD_\alpha\Xi\p\,, \\
\rb &:= \Xib\p\,, & \sbb^\sd{\alpha} &:= \frac{1}{\sqrt{2}}\cD^\sd{\alpha}\Xib\p\,, & \fb &:= -\frac{1}{4}\cD_\sd{\alpha}\cD^\sd{\alpha}\Xib\p\,.
\end{align}
Since at the superfield level the action is invariant under supergravity transformations, the coefficients $\lambda_2^\alpha$, $\lambdab_{2\sd{\alpha}}$, $\lambda_3$ and $\lambdab_3$, which are functions of the component fields of the supergravity multiplet, have to be chosen such that this holds true at the component field level. The coefficients are determined in the following.
\\\\
First, the supergravity transformations of the component fields of $\Xi$ and $\Xib$ are calculated. The double bar projections of the covariant derivatives $\cD\Xi$ and $\cD\Xib$ read
\begin{align}
\cD\Xi\pp &= \exd x^m\cD_m\Xi\p = \exd x^m\cD_m r\,, & \cD\Xib\pp &= \exd x^m\cD_m\Xib\p = \exd x^m\cD_m\rb\,, \label{eq:component_fields_action_general_dxibb_1}
\end{align}
with
\begin{align}
\cD_m r &= \del_m r + w(\Xi)r\,A_m\,, & \cD_m\rb &= \del_m\rb + w(\Xib)\rb\,A_m\,,
\end{align}
where $A_m$ is given in Eq.~\eqref{eq:component_fields_definition_matter_kahler_connection_component_1}. On the other hand, Eq.~\eqref{eq:component_fields_action_general_dxibb_1} can also be written as
\begin{align}
\cD\Xi\pp &= \exd x^m\Big({e_m}^a\cD_a\Xi\p + \frac{1}{\sqrt{2}}\psi_m s\Big)\,, & \cD\Xib\pp &= \exd x^m\Big({e_m}^a\cD_a\Xib\p + \frac{1}{\sqrt{2}}\psib_m\sbb\Big)\,, \label{eq:component_fields_action_general_dxibb_2}
\end{align}
thus
\begin{align}
\cD_a\Xi\p &= {e_a}^m\Big(\cD_m r - \frac{1}{\sqrt{2}}\psi_m s\Big)\,, & \cD_a\Xib\p &= {e_a}^m\Big(\cD_m\rb - \frac{1}{\sqrt{2}}\psib_m\sbb\Big)\,. \label{eq:component_fields_action_general_dxibb_3}
\end{align}
In addition, the double bar projections of the terms $\cD\cD_\alpha\Xi$ and $\cD\cD^\sd{\alpha}\Xib$ are given by
\begin{align}
\begin{split}
\cD\cD_\alpha\Xi\pp &= \exd x^m\cD_m\cD_\alpha\Xi\p = \sqrt{2}\exd x^m\cD_m s_\alpha\,, \\
\cD\cD^\sd{\alpha}\Xib\pp &= \exd x^m\cD_m\cD^\sd{\alpha}\Xib\p = \sqrt{2}\exd x^m\cD_m\sbb^\sd{\alpha}\,,
\end{split} \label{eq:component_fields_action_general_ddxibb_1}
\end{align}
where
\begin{align}
\begin{split}
\cD_m s_\alpha &= \del_m s_\alpha - {\omega_{m\alpha}}^\beta s_\beta + \big(w(\Xi)-1\big)s_\alpha A_m\,, \\
\cD_m\sbb^\sd{\alpha} &= \del_m\sbb^\sd{\alpha} - \tensor{\omega}{_m^{\sd{\alpha}}_{\sd{\beta}}}\sbb^\sd{\beta} + \big(w(\Xib)+1\big)\sbb^\sd{\alpha}A_m\,.
\end{split}
\end{align}
A comparison of Eq.~\eqref{eq:component_fields_action_general_ddxibb_1} with the identities
\begin{align}
\begin{split}
\cD\cD_\alpha\Xi\pp &= \exd x^m\Big({e_m}^a\cD_a\cD_\alpha\Xi\p + \psi_{m\alpha}f + i(\sig^a\psib_m)_\alpha\big(\cD_a\Xi\p + \frac{iw(\Xi)}{2}b_a r\big)\Big)\,, \\
\cD\cD^\sd{\alpha}\Xib\pp &= \exd x^m\Big({e_m}^a\cD_a\cD^\sd{\alpha}\Xib\p + {\psib_m}^\sd{\alpha}\fb + i(\sigb^a\psi_m)^\sd{\alpha}\big(\cD_a\Xib\p + \frac{iw(\Xib)}{2}b_a\rb\big)\Big)\,,
\end{split} \label{eq:component_fields_action_general_ddxibb_2}
\end{align}
implies that
\begin{align}
\begin{split}
\cD_a\cD_\alpha\Xi\p &= {e_a}^m\Big(\sqrt{2}\cD_m s_\alpha - \psi_{m\alpha}f - i(\sig^n\psib_m)_\alpha\big(\cD_n r - \frac{1}{\sqrt{2}}\psi_n s + \frac{iw(\Xi)}{2}{e_n}^a b_a r\big)\Big)\,, \\
\cD_a\cD^\sd{\alpha}\Xib\p &= {e_a}^m\Big(\sqrt{2}\cD_m\sbb^\sd{\alpha} - {\psi_m}^\sd{\alpha}\fb - i(\sigb^n\psi_m)^\sd{\alpha}\big(\cD_n\rb - \frac{1}{\sqrt{2}}\psib_n\sbb + \frac{iw(\Xib)}{2}{e_n}^a b_a\rb\big)\Big)\,.
\end{split} \label{eq:component_fields_action_general_ddxibb_3}
\end{align}
Using the identities in Eq.~\eqref{eq:component_fields_action_general_dxibb_3} and \eqref{eq:component_fields_action_general_ddxibb_3}, the supergravity transformations of the component fields have the following form:  Eq.~\eqref{eq:sugra_transf_delta_general_phi_explicit} and \eqref{eq:sugra_transf_delta_general_phib_explicit} imply
\begin{align}
\dsg_\xi r &= \sqrt{2}\xi s - \frac{w(\Xi)}{2\sqrt{2}}(K_k\xi\chi^k - K_\kb\xib\chib^\kb)r\,, \label{eq:component_fields_action_general_dwz_r}\\
\dsg_\xi \rb &= \sqrt{2}\xib\sbb - \frac{w(\Xib)}{2\sqrt{2}}(K_k\xi\chi^k - K_\kb\xib\chib^\kb)\rb\,, \label{eq:component_fields_action_general_dwz_rb}
\end{align}
furthermore, Eq.~\eqref{eq:sugra_transf_delta_general_dphi_explicit} and \eqref{eq:sugra_transf_delta_general_dphib_explicit} lead to
\begin{align}
\begin{split}
\dsg_\xi s_\alpha &= \sqrt{2}\xi_\alpha f + i\sqrt{2}(\sig^m\xib)_\alpha\Big(\cD_m r - \frac{1}{\sqrt{2}}\psi_m s + \frac{iw(\Xi)}{2}{e_m}^a b_a r\Big) \\
&\quad - \frac{w(\Xi)-1}{2\sqrt{2}}(K_k\xi\chi^k - K_\kb\xib\chib^\kb)s_\alpha\,,
\end{split} \label{eq:component_fields_action_general_dwz_s}\\
\nonumber\\
\begin{split}
\dsg_\xi \sbb^\sd{\alpha} &= \sqrt{2}\xib^\sd{\alpha}\fb + i\sqrt{2}(\sigb^m\xi)^\sd{\alpha}\Big(\cD_m\rb - \frac{1}{\sqrt{2}}\psib_m\sbb + \frac{iw(\Xib)}{2}{e_m}^a b_a\rb\Big) \\
&\quad - \frac{w(\Xib)+1}{2\sqrt{2}}(K_k\xi\chi^k - K_\kb\xib\chib^\kb)\sbb^\sd{\alpha}\,,
\end{split} \label{eq:component_fields_action_general_dwz_sb}
\end{align}
and from Eq.~\eqref{eq:sugra_transf_delta_general_ddphi_explicit} and \eqref{eq:sugra_transf_delta_general_ddphib_explicit} follows
\begin{align}
\begin{split}
\dsg_\xi f &= i\sqrt{2}(\xib\sigb^m\cD_m s) - i(\xib\sigb^m\psi_m)f + (\xib\sigb^m\sig^n\psib_m)\Big(\cD_n r - \frac{1}{\sqrt{2}}\psi_n s + \frac{iw(\Xi)}{2}{e_n}^a b_a r\Big) \\
&\quad + \frac{\sqrt{2}}{3}\Mb\,\xi s - \frac{\sqrt{2}}{6}\big(3w(\Xi)-2\big)(\xib\sigb^a s)b_a + w(\Xi)r\,\xib_\sd{\beta}X^\sd{\beta}\p \\
&\quad - \frac{w(\Xi)-2}{2\sqrt{2}}(K_k\xi\chi^k - K_\kb\xib\chib^\kb)f\,,
\end{split} \label{eq:component_fields_action_general_dwz_f}\\
\nonumber\\
\begin{split}
\dsg_\xi\fb &= i\sqrt{2}(\xi\sig^m\cD_m\sbb) - i(\xi\sig^m\psib_m)\fb + (\xi\sig^m\sigb^n\psi_m)\Big(\cD_n\rb - \frac{1}{\sqrt{2}}\psib_n\sbb + \frac{iw(\Xib)}{2}{e_n}^a b_a\rb\Big) \\
&\quad + \frac{\sqrt{2}}{3}M\,\xib\sbb - \frac{\sqrt{2}}{6}\big(3w(\Xib)+2\big)(\xi\sig^a\sbb)b_a + w(\Xib)\rb\,\xi^\beta X_\beta\p \\
&\quad - \frac{w(\Xib)+2}{2\sqrt{2}}(K_k\xi\chi^k - K_\kb\xib\chib^\kb)\fb\,.
\end{split} \label{eq:component_fields_action_general_dwz_fb}
\end{align}
Having determined the expressions for $\dsg_\xi r$, $\dsg_\xi s_\alpha$ and $\dsg_\xi f$, the coefficients $\lambda_2^\alpha$ and $\lambda_3$ are determined by the requirement that the supergravity transformation of the sum of the terms
\begin{align}
l_1 &:= e\,f\,, & l_2 &:= \lambda_2^\alpha\,s_\alpha\,, & l_3 := \lambda_3\,r\,,
\end{align}
vanishes up to a total derivative with respect to spacetime indices. The transformation of the term $l_1$ has the form
\begin{align}
\begin{split}
\frac{1}{e}\dsg_\xi l_1 &= + i(\xi\sig^m\psib_m)f - \frac{2\sqrt{2}}{3}(\xib\sigb^a s)b_a + i\sqrt{2}(\xib\sigb^m\cD_m s) \\
&\quad + (\xib\sigb^m\sig^n\psib_m)(\cD_n r - \frac{1}{\sqrt{2}}\psi_n s + i{e_n}^a b_a r) + \frac{\sqrt{2}}{3}\Mb\,\xi s + 2r\,\xib_\sd{\beta}X^\sd{\beta}\p\,,
\end{split} \label{eq:component_fields_action_general_dwz_l1}
\end{align}
where $\dsg_\xi e$ is computed by using the expression for $\dsg_\xi{e_m}^a$ from Eq.~\eqref{eq:component_fields_sugra_transf_graviton}, namely
\begin{align}
\dsg_\xi e &= e\,{e_a}^m\dsg_\xi{e_m}^a = ie(\xi\sig^m\psib_m + \xib\sigb^m\psi_m)\,. \label{eq:component_fields_action_general_dwz_e}
\end{align}
The transformation $\dsg_\xi s_\alpha$ in Eq.~\eqref{eq:component_fields_action_general_dwz_s} indicates that the first term in Eq.~\eqref{eq:component_fields_action_general_dwz_l1} can be cancelled by choosing
\begin{align}
\lambda_2^\alpha &= \frac{ie}{\sqrt{2}}(\psib_m\sigb^m)^\alpha\,.
\end{align}
With the expression for $\dsg_\xi\psib_{m\sd{\alpha}}$ from Eq.~\eqref{eq:component_fields_sugra_transf_gravitino_2}, the supergravity transformation of the term $l_2$ is then given by
\begin{align}
\begin{split}
\frac{1}{e}\dsg_\xi l_2 &= + i\sqrt{2}(\cD_m\xib\sigb^m s) + \frac{2\sqrt{2}}{3}(\xib\sigb^a s)b_a + \frac{2\sqrt{2}}{3}\Mb\,\xi s + i(\psib_m\sigb^m\xi)f \\
&\quad + \frac{1}{\sqrt{2}}(\psib_m\sigb^n s)(\xi\sig^m\psib_n + \xib\sigb^m\psi_n) - \frac{1}{\sqrt{2}}(\psib_m\sigb^m s)(\xi\sig^n\psib_n + \xib\sigb^n\psi_n) \\
&\quad - (\xib\sigb^n\sig^m\psib_m)\Big(\cD_n r - \frac{1}{\sqrt{2}}\psi_n s + i{e_a}^n b_a r\Big)\,,
\end{split} \label{eq:component_fields_action_general_dwz_l2}
\end{align}
thus
\begin{align}
\begin{split}
\frac{1}{e}\dsg_\xi(l_1 + l_2) &= + \sqrt{2}\,\Mb\,\xi s + 2r\,\xib_\sd{\beta}X^\sd{\beta}\p + i\sqrt{2}(\xib\sigb^m\cD_m s + \cD_m\xib\sigb^m s) \\
&\quad + \frac{1}{\sqrt{2}}(\psib_m\sigb^n s)(\xi\sig^m\psib_n + \xib\sigb^m\psi_n) - \frac{1}{\sqrt{2}}(\psib_m\sigb^m s)(\xi\sig^n\psib_n + \xib\sigb^n\psi_n) \\
&\quad + 4(\xib\sigb^{mn}\psib_m)\Big(\cD_n r - \frac{1}{\sqrt{2}}\psi_n s + i{e_a}^n b_a r\Big)\,.
\end{split} \label{eq:component_fields_action_general_dwz_l1_l2}
\end{align}
Again, the transformation $\dsg_\xi r$ in Eq.~\eqref{eq:component_fields_action_general_dwz_r} shows that the first term in Eq.~\eqref{eq:component_fields_action_general_dwz_l1_l2} is eliminated by setting
\begin{align}
\lambda_3 &= -e\,\Mb\,.
\end{align}
Using the expression for $\dsg_\xi\Mb$ from Eq.~\eqref{eq:component_fields_sugra_transf_aux_m_2}, the supergravity transformation of the term $l_3$ reads
\begin{align}
\begin{split}
\frac{1}{e}\dsg_\xi l_3 &= - \sqrt{2}\,\Mb\,\xi s - i(\xi\sig^m\psib_m)\Mb\,r - i(\xib\sigb^m\sig^a\psi_m)b_a r - 4(\xib\sigb^{nm}\cD_n\psib_m)r\\
&\quad + i\sqrt{2}g_{k\kb}(\xib\sigb^m\chi^k)\Big(\cD_m\varphib^\kb - \frac{1}{\sqrt{2}}\psib_m\chib^\kb\Big)r - \sqrt{2}g_{k\kb}F^k(\xib\chib^\kb)r \\
&\quad + i(\xib\clb^\g{r})\big(K_k(\gen{r}\varphi)^k + K_\kb(\varphib\gen{r})^\kb\big)r\,,
\end{split} \label{eq:component_fields_action_general_dwz_l3}
\end{align}
hence, with the expression for $X^\sd{\beta}\p$ from Eq.~\eqref{eq:component_fields_matter_dterm_x_components_2} follows
\begin{align}
\begin{split}
\frac{1}{e}\dsg_\xi(l_1 + l_2 + l_3) &= - i(\xi\sig^m\psib_m)\Mb\,r - i(\xib\sigb^a\sig^m\psi_m)b_a r + i\sqrt{2}\,\cD_m(\xib\sigb^m s) - i\sqrt{2}(\xib\cD_m\sigb^m s)\\
&\quad - 4\cD_n(\xib\sigb^{nm}\psib_m)r + 4(\cD_n\xib\sigb^{nm}\psib_m)r + 4(\xib\cD_n\sigb^{nm}\psib_m)r + \frac{4}{\sqrt{2}}(\xib\sigb^{nm}\psib_m)(\psi_n s) \\
&\quad + \frac{1}{\sqrt{2}}(\psib_m\sigb^n s)(\xi\sig^m\psib_n + \xib\sigb^m\psi_n) - \frac{1}{\sqrt{2}}(\psib_m\sigb^m s)(\xi\sig^n\psib_n + \xib\sigb^n\psi_n)\,.
\end{split} \label{eq:component_fields_action_general_dwz_l1_l2_l3}
\end{align}
The expression for $\dsg_\xi\psib_{m\sd{\alpha}}$ implies that the first term in Eq.~\eqref{eq:component_fields_action_general_dwz_l1_l2_l3} can be cancelled by another term $l_3^\prime:=\lambda_3^\prime\,r$, where
\begin{align}
\lambda_3^\prime &= -e\,(\psib_m\sigb^{mn}\psib_n)\,.
\end{align}
Applying the identity in Eq.~\eqref{eq:notations_cyclic_identity_pauli_matrices}, the transformation of $l_3^\prime$ is calculated as
\begin{align}
\begin{split}
\frac{1}{e}\dsg_\xi l_3^\prime &= + i(\xi\sig^m\psib_m)\Mb\,r + i(\xib\sigb^a\sig^m\psi_m)b_a r - 4(\cD_n\xib\sigb^{nm}\psib_m)r - \sqrt{2}(\psib_m\sigb^{mn}\psib_n)(\xi s) \\
&\quad + 2ir(\psib_m\sigb^{np}\psib_p)(\xib\sigb^m\psi_n) - ir(\psib_m\sigb^{mp}\psib_p)(\xib\sigb^n\psi_n)\,,
\end{split} \label{eq:component_fields_action_general_dwz_l3p}
\end{align}
thus
\begin{align}
\begin{split}
\frac{1}{e}\dsg_\xi(l_1 + l_2 + l_3 + l_3^\prime) &= + i\sqrt{2}\,\cD_m(\xib\sigb^m s) - i\sqrt{2}(\xib\cD_m\sigb^m s) \\
&\quad - 4\cD_n(\xib\sigb^{nm}\psib_m)r + 4(\xib\cD_n\sigb^{nm}\psib_m)r \\
&\quad + \frac{4}{\sqrt{2}}(\xib\sigb^{nm}\psib_m)(\psi_n s) + \frac{1}{\sqrt{2}}(\psib_m\sigb^n s)(\xib\sigb^m\psi_n) - \frac{1}{\sqrt{2}}(\psib_m\sigb^m s)(\xib\sigb^n\psi_n) \\
&\quad + 2ir(\psib_m\sigb^{np}\psib_p)(\xib\sigb^m\psi_n) - ir(\psib_m\sigb^{mp}\psib_p)(\xib\sigb^n\psi_n)\,.
\end{split} \label{eq:component_fields_action_general_dwz_l1_l2_l3_l3p_1}
\end{align}
In order to show that the expression in Eq.~\eqref{eq:component_fields_action_general_dwz_l1_l2_l3_l3p_1} is equal to a total derivative with respect to spacetime coordinates, the identities
\begin{align}
\cD_m\sigb^m &= - \frac{1}{e}(\del_m e)\sigb^m - \frac{i}{2}(\psi_n\sig^n\psib_m - \psi_m\sig^n\psib_n)\sigb^m\,, \\
\cD_n\sigb^{nm} &= - \frac{1}{e}(\del_n e)\sigb^{nm} + \frac{i}{2}(\psi_n\sig^m\psib_p)\sigb^{pn} + \frac{i}{2}(\psi_n\sig^p\psib_p - \psi_p\sig^p\psib_n)\sigb^{nm}\,,
\end{align}
are used. These identities and the fact that $\cD_m v^n=\del_m v^n$ for any field $v^n$ with a spacetime index imply 
\begin{align}
\begin{split}
+ i\sqrt{2}\,\cD_m(\xib\sigb^m s) - i\sqrt{2}(\xib\cD_m\sigb^m s) &= + \frac{i\sqrt{2}}{e}\del_m(e\,\xib\sigb^m s) - \frac{4}{\sqrt{2}}(\xib\sigb^{nm}\psib_m)(\psi_n s) \\
&\quad - \frac{1}{\sqrt{2}}(\psib_m\sigb^n s)(\xib\sigb^m\psi_n) + \frac{1}{\sqrt{2}}(\psib_n\sigb^n s)(\xib\sigb^m\psi_m)\,,
\end{split} \\
\nonumber\\
\begin{split}
- 4\cD_n(\xib\sigb^{nm}\psib_m)r + 4(\xib\cD_n\sigb^{nm}\psib_m)r &= - \frac{4}{e}\del_n(e\,r\,\xib\sigb^{nm}\psib_m) - 2ir(\psib_m\sigb^{np}\psib_p)(\xib\sigb^m\psi_n) \\
&\quad + ir(\psib_m\sigb^{mp}\psib_p)(\xib\sigb^n\psi_n)\,.
\end{split}
\end{align}
Consequently, Eq.~\eqref{eq:component_fields_action_general_dwz_l1_l2_l3_l3p_1} can be written as
\begin{align}
\dsg_\xi(l_1 + l_2 + l_3 + l_3^\prime) &= \del_m\big(i\sqrt{2}\,e\,(\xib\sigb^m s) - 4e\,(\xib\sigb^{nm}\psib_m)r\big)\,. \label{eq:component_fields_action_general_dwz_l1_l2_l3_l3p_2}
\end{align}
Since total derivatives vanish when integrated over, the first term in Eq.~\eqref{eq:component_fields_action_general_action_superfield_chiral} is given by
\begin{align}
\frac{1}{2}\int_*\frac{E}{R}\Xi &= \int\exd^4 x\,e\,\Big(f + \frac{i}{\sqrt{2}}\psib_m\sigb^m s - r(\Mb + \psib_m\sigb^{mn}\psib_n)\Big)\,,
\end{align}
where both sides are inert under supergravity transformations. A similar calculation shows that the second term in Eq.~\eqref{eq:component_fields_action_general_action_superfield_chiral} reads
\begin{align}
\frac{1}{2}\int_*\frac{E}{\Rb}\Xib &= \int\exd^4 x\,e\,\Big(\fb + \frac{i}{\sqrt{2}}\psi_m\sig^m\sbb - \rb(M + \psi_m\sig^{mn}\psi_n)\Big)\,.
\end{align}
Thus, a generic Lagrangian $\cL(\Xi,\Xib)$ which is defined by
\begin{align}
\cS &= \frac{1}{2}\int_*\frac{E}{R}\Xi + \frac{1}{2}\int_*\frac{E}{\Rb}\Xib =: \int\exd^4 x\,\cL(\Xi,\Xib)\,,
\end{align}
is invariant under supergravity transformations (up to a total derivative) and has the following form
\begin{align}
\begin{split}
\cL(\Xi,\Xib) &= e\,(f+\fb) + \frac{i}{\sqrt{2}}e\,(\psib_m\sigb^m s + \psi_m\sig^m\sbb) \\
&\quad - e\,r(\Mb + \psib_m\sigb^{mn}\psib_n) - e\,\rb(M + \psi_m\sig^{mn}\psi_n)\,.
\end{split} \label{eq:component_fields_action_general_fterm}
\end{align}

\subsubsection{Supergravity+matter part}
\label{sec:component_fields_actions_sugra_matter}
If the supergravity+matter part of the action, stated in Eq.~\eqref{eq:superfield_actions_sugra_matter}, is written in terms of the chiral density as in Eq.~\eqref{eq:superfield_actions_general_construction_2}, the corresponding chiral superfield and its conjugate have the following form
\begin{align}
\Xi_\text{sugra+matter} &= -3R\,, & \Xib_\text{sugra+matter} &= -3\Rb\,.
\end{align}
The scalar component fields $r$ and $\rb$ of $\Xi$ and $\Xib$ are then given by
\begin{align}
r &= \frac{1}{2}M\,, & \rb &= \frac{1}{2}\Mb\,.
\end{align}
Using Eq.~\eqref{eq:bianchi_sp_derivative_relation_id_3} and \eqref{eq:bianchi_sp_derivative_relation_id_4}, the Weyl spinor component fields read
\begin{align}
s_\alpha &= \frac{1}{\sqrt{2}}X_\alpha\p + \sqrt{2}(\sig^{cb}\eps)_{\alpha\varphi} {T_{cb}}^\varphi\p\,, & \sbb^\sd{\alpha} &= \frac{1}{\sqrt{2}}X^\sd{\alpha}\p + \sqrt{2}(\sig^{cb}\eps)^{\sd{\alpha}\sd{\varphi}} T_{cb\sd{\varphi}}\p\,,
\end{align}
which implies
\begin{align}
\begin{split}
\frac{i}{\sqrt{2}}(\psib_m\sigb^m s) &= + \frac{i}{2}(\psib_m\sigb^m)^\alpha X_\alpha\p + i{e_b}^m(\psib_m\sigb_a\eps)_\varphi T^{ab\varphi}\p + \frac{1}{2}\eps^{mnpq}(\psib_m\sigb_n\cD_p\psi_q) \\
&\quad + \frac{i}{6}\eps^{mnpq}(\psib_m\sigb_n\psi_q)b_p + \frac{1}{6}(\psi_n\sig^m\psib_m - \psi_m\sig^m\psib_n)b^n + \frac{1}{3}(\psib_m\sigb^{mn}\psib_n)M\,,
\end{split} \\
\nonumber\\
\begin{split}
\frac{i}{\sqrt{2}}(\psi_m\sig^m\sbb) &= + \frac{i}{2}(\psi_m\sig^m)_\sd{\alpha} X^\sd{\alpha}\p + i{e_b}^m(\psi_m\sig_a\eps)^\sd{\varphi} {T^{ab}}_\sd{\varphi}\p - \frac{1}{2}\eps^{mnpq}(\psi_m\sig_n\cD_p\psib_q) \\
&\quad + \frac{i}{6}\eps^{mnpq}(\psib_m\sigb_n\psi_q)b_p - \frac{1}{6}(\psi_n\sig^m\psib_m - \psi_m\sig^m\psib_n)b^n + \frac{1}{3}(\psi_m\sig^{mn}\psi_n)\Mb\,.
\end{split}
\end{align}
Furthermore, the two auxiliary fields $f$ and $\fb$ are given by
\begin{align}
f &= \frac{3}{4}\cD^2 R\p\,, & \fb &= \frac{3}{4}\cDb^2 \Rb\p\,,
\end{align}
thus, using Eq.~\eqref{eq:bianchi_sp_derivative_relation_id_1}, their sum is calculated as
\begin{align}
\begin{split}
f + \fb &= - \frac{1}{2}{R_{ba}}^{ba}\p - \frac{1}{2}\cD^\alpha X_\alpha\p + 3G^aG_a\p + 24R\Rb\p \\
&= - \frac{1}{2}\cR + \frac{2}{3}M\Mb + \frac{1}{3}b^a b_a - \frac{1}{2}\cD^\alpha X_\alpha\p \\
&\quad - i{e_b}^m(\psib_m\sigb_a\eps)_\varphi T^{ab\varphi}\p - i{e_b}^m(\psi_m\sig_a\eps)^\sd{\varphi}{T^{ab}}_\sd{\varphi}\p \\
&\quad + \frac{i}{6}\eps^{mnpq}(\psib_m\sigb_n\psi_q)b_p + \frac{1}{6}(\psib_m\sigb^{mn}\psib_n)M + \frac{1}{6}(\psi_m\sig^{mn}\psi_n)\Mb\,.
\end{split}
\end{align}
According to Eq.~\eqref{eq:component_fields_action_general_fterm}, the supergravity+matter part of the Lagrangian at the component field level then has the following form
\begin{align}
\begin{split}
e^{-1}\cL_\text{supergravity+matter} &= -\frac{1}{2}\cR + \frac{1}{2}\eps^{mnpq}(\psib_m\sigb_n\nabla_p\psi_q - \psi_m\sig_n\nabla_p\psib_q) \\
&\quad - \frac{1}{3}M\Mb + \frac{1}{3}b^a b_a + D_\text{matter}\,,
\end{split} \label{eq:component_fields_action_sugra_matter}
\end{align}
with the derivatives
\begin{align}
\begin{split}
\nabla_n{\psi_m}^\alpha &:= \cD_n{\psi_m}^\alpha + \frac{i}{2}{e_n}^a b_a{\psi_m}^\alpha \\
&\;= \del_n{\psi_m}^\alpha + {\psi_m}^\beta{\omega_{n\beta}}^\alpha + \frac{1}{4}(K_k\cD_n\varphi^k - K_\kb\cD_n\varphib^\kb){\psi_m}^\alpha + \frac{i}{4}g_{k\kb}(\chi^k\sig_n\chib^\kb){\psi_m}^\alpha\,,
\end{split} \\
\nonumber\\
\begin{split}
\nabla_n\psib_{m\sd{\alpha}} &:= \cD_n\psib_{m\sd{\alpha}} - \frac{i}{2}{e_n}^a b_a\psib_{m\sd{\alpha}} \\
&\;= \del_n\psib_{m\sd{\alpha}} + \psib_{m\sd{\beta}}\tensor{\omega}{_n^{\sd{\beta}}_{\sd{\alpha}}} - \frac{1}{4}(K_k\cD_n\varphi^k - K_\kb\cD_n\varphib^\kb)\psib_{m\sd{\alpha}} - \frac{i}{4}g_{k\kb}(\chi^k\sig_n\chib^\kb)\psib_{m\sd{\alpha}}\,,
\end{split}
\end{align}
where $\cD_n{\psi_m}^\alpha$ and $\cD_n\psib_{m\sd{\alpha}}$ are defined in Eq.~\eqref{eq:component_fields_definition_sugra_cd_gravitino_1} and \eqref{eq:component_fields_definition_sugra_cd_gravitino_2}. The term $D_\text{matter}$ is given in Eq.~\eqref{eq:component_fields_matter_dterm}.

\subsubsection{Superpotential part}
\label{sec:component_fields_actions_superpotential}
According to Eq.~\eqref{eq:superfield_actions_superpotential_chiral_field}, the superpotential part of the action is constructed by using the chiral and antichiral superfield
\begin{align}
\Xi_\text{superpotential} &= e^{K/2}W\,, & \Xib_\text{superpotential} &= e^{K/2}\Wb\,.
\end{align}
The corresponding scalar component fields are then given by
\begin{align}
r &= e^{K/2}W\,, & \rb &= e^{K/2}\Wb\,.
\end{align}
In order to evaluate the covariant derivatives $\cD_\alpha\Xi$ and $\cD^\sd{\alpha}\Xib$, they are explicitly written in terms of the K\"ahler connection, because the individual objects $K$, $W$ and $\Wb$ do not have well defined chiral weights:
\begin{align}
\cD_\alpha\Xi &= D_\alpha\Xi + 2A_\alpha\Xi = e^{K/2}(W_k + W K_k)\cD_\alpha\phi^k\,, \\
\cD^\sd{\alpha}\Xib &= D^\sd{\alpha}\Xi - 2A^\sd{\alpha}\Xib = e^{K/2}(\Wb_\kb + \Wb K_\kb)\cD^\sd{\alpha}\phib^\kb\,,
\end{align}
where the expressions $A_\alpha=+\frac{1}{4}K_k\cD_\alpha\phi^k$ and $A^\sd{\alpha}=-\frac{1}{4}K_\kb\cD^\sd{\alpha}\phib^\kb$, and the identities from Eq.~\eqref{eq:superfield_actions_superpotential_id_2}, \eqref{eq:kahler_superspace_1/2_derivative_k_1} and \eqref{eq:kahler_superspace_1/2_derivative_k_2} are used. Since under K\"ahler transformations the terms $(W_k + W K_k)$ and $(\Wb_\kb + \Wb K_\kb)$ transform in the same way as $W$ and $\Wb$, respectively, it is convenient to define the covariant derivatives (cf.\ the discussion in Section~\ref{sec:matter_and_yang_mills})
\begin{align}
D_k W &:= W_k + W K_k\,, & D_\kb\Wb &:= \Wb_\kb + \Wb K_\kb\,.
\end{align}
With these expressions the Weyl spinor component fields read
\begin{align}
s_\alpha &= e^{K/2}{\chi^k}_\alpha D_k W\,, & \sbb^\sd{\alpha} &= e^{K/2}\chib^{\kb\sd{\alpha}}D_\kb\Wb\,.
\end{align}
Using the identities
\begin{align}
\begin{split}
\cD^\alpha\cD_\alpha\Xi &= D^\alpha\cD_\alpha\Xi - \tensor{\Omega}{^\alpha_\alpha^\beta}\cD_\beta\Xi + A^\alpha\cD_\alpha\Xi \\
&= e^{K/2}D_k W \cDt^\alpha\cD_\alpha\phi^k + e^{K/2}\big((\del_j+K_j)D_k W - {\Gamma^i}_{jk} D_i W\big)\cD^\alpha\phi^j\cD_\alpha\phi^k\,,
\end{split} \\
\nonumber\\
\begin{split}
\cD_\sd{\alpha}\cD^\sd{\alpha}\Xib &= D_\sd{\alpha}\cD^\sd{\alpha}\Xib - \tensor{\Omega}{_{\sd{\alpha}}^{\sd{\alpha}}_{\sd{\beta}}}\cD^\sd{\beta}\Xib - A_\sd{\alpha}\cD^\sd{\alpha}\Xib \\
&= e^{K/2}D_\kb\Wb \cDt_\sd{\alpha}\cD^\sd{\alpha}\phib^\kb\p + e^{K/2}\big((\del_\jb+K_\jb)D_\kb\Wb - {\Gamma^\ib}_{\jb\kb} D_\ib\Wb\big)\cD_\sd{\alpha}\phib^\jb\cD^\sd{\alpha}\phib^\kb\,,
\end{split}
\end{align}
the auxiliary component fields have the form
\begin{align}
f &= e^{K/2}\Big(F^k D_k W - \frac{1}{2}\chi^j\chi^k D_j D_k W\Big)\,, & \fb &= e^{K/2}\Big(\Fb^\kb D_\kb \Wb - \frac{1}{2}\chib^\jb\chib^\kb D_\jb D_\kb\Wb\Big)\,,
\end{align}
with the covariant derivatives
\begin{align}
\Dt_j D_k W &:= (\del_j+K_j)D_k W - {\Gamma^i}_{jk} D_i W\,, & \Dt_\jb D_\kb\Wb &:= (\del_\jb+K_\jb)D_\kb\Wb - {\Gamma^\ib}_{\jb\kb} D_\ib\Wb\,,
\end{align}
where $\Dt_j D_k W$ and $\Dt_\jb D_\kb\Wb$ have the same transformation property under K\"ahler transformations as $W$ and $\Wb$, respectively. The tilde indicates that the derivative is also covariant with respect to Levi-Civita connection of the metric $g_{k\kb}$. The superpotential part of the Lagrangian is then given by
\begin{align}
\begin{split}
e^{-1}\cL_\text{superpotential} &= + e^{K/2}\big(F^k D_k W + \Fb^\kb D_\kb\Wb - M\Wb - \Mb W\big) \\
&\quad - \frac{1}{2}e^{K/2}\big((\chi^j\chi^k) \Dt_j D_k W + (\chib^\jb\chib^\kb) \Dt_\jb D_\kb\Wb\big) \\
&\quad + \frac{i}{\sqrt{2}}e^{K/2}\big((\psib_m\sigb^m\chi^k)D_k W + (\psi_m\sig^m\chib^\kb)D_\kb\Wb\big) \\
&\quad - e^{K/2}\big((\psib_m\sigb^{mn}\psib_n)W + (\psi_m\sig^{mn}\psi_n)\Wb\big)\,.
\end{split} \label{eq:component_fields_action_superpotential}
\end{align}

\subsubsection{Yang-Mills part}
\label{sec:component_fields_actions_yang_mills}
As stated in Eq.~\eqref{eq:superfield_actions_yang_mills_chiral_field}, the chiral and the antichiral superfield
\begin{align}
\Xi_\text{Yang-Mills} &= \frac{1}{4}\gkf{r}{s} \cW^{\beta\g{r}}{\cW_\beta}^\g{s}\,, & \Xib_\text{Yang-Mills} &= \frac{1}{4}\gkfb{r}{s} {\cW_\sd{\beta}}^\g{r}\cW^{\sd{\beta}\g{s}}\,,
\end{align}
are used to construct the Yang-Mills part of the Action. The corresponding scalar component fields thus read
\begin{align}
r &= -\frac{1}{4}\gkf{r}{s}\cl^\g{r}\cl^\g{s}\,, & \rb &= -\frac{1}{4}\gkfb{r}{s}\clb^\g{r}\clb^\g{s}\,.
\end{align}
The covariant derivatives
\begin{align}
\cD_\alpha\Xi &= \frac{1}{4}\frac{\del\gkf{r}{s}}{\del\phi^k}\cD_\alpha\phi^k\cW^{\beta\g{r}}{\cW_\beta}^\g{s} - \frac{1}{4}\gkf{r}{s}{\cW_\alpha}^\g{r}\cD^\beta{\cW_\beta}^\g{s} + \frac{1}{2}\gkf{r}{s}{(\sig^{ba})_\alpha}^\beta{\cW_\beta}^\g{r}{\cF_{ba}}^\g{s}\,, \\
\cD^\sd{\alpha}\Xib &= \frac{1}{4}\frac{\del\gkfb{r}{s}}{\del\phib^\kb}\cD^\sd{\alpha}\phib^\kb{\cW_\sd{\beta}}^\g{r}\cW^{\sd{\beta}\g{s}} - \frac{1}{4}\gkfb{r}{s}\cW^{\sd{\alpha}\g{r}}\cD_\sd{\beta}\cW^{\sd{\beta}\g{s}} - \frac{1}{2}\gkfb{r}{s}{(\sigb^{ba})^\sd{\alpha}}_\sd{\beta}\cW^{\sd{\beta}\g{r}}{\cF_{ba}}^\g{s}\,,
\end{align}
are calculated by applying Eq.~\eqref{eq:yang_mills_derivative_relations_2}--\eqref{eq:yang_mills_derivative_relations_5}. With these two identities and the expression for $\cF_{ba}\p$ in Eq.~\eqref{eq:component_fields_definition_matter_id_14} follows, that the Weyl spinor component fields are given by
\begin{align}
\begin{split}
s_\alpha &= - \frac{i}{2\sqrt{2}}\gkf{r}{s}\big({\cl_\alpha}^\g{r}\Db^\g{s} + (\sig^{mn}\cl^\g{r})_\alpha (i{\cf_{mn}}^\g{s} + \psi_m\sig_n\clb^\g{s} + \psib_m\sigb_n\cl^\g{s})\big) \\
&\quad - \frac{1}{4}\del_k\gkf{r}{s}{\chi^k}_\alpha(\cl^\g{r}\cl^\g{s})\,,
\end{split} \\
\nonumber\\
\begin{split}
\sbb^\sd{\alpha} &= + \frac{i}{2\sqrt{2}}\gkfb{r}{s}\big(\clb^{\sd{\alpha}\g{r}}\Db^\g{s} - (\sigb^{mn}\clb^\g{r})^\sd{\alpha} (i{\cf_{mn}}^\g{s} + \psi_m\sig_n\clb^\g{s} + \psib_m\sigb_n\cl^\g{s})\big) \\
&\quad - \frac{1}{4}\del_\kb\gkfb{r}{s}\chib^{\kb\sd{\alpha}}(\clb^\g{r}\clb^\g{s})\,.
\end{split}
\end{align}
Furthermore, the double covariant derivatives
\begingroup
\allowdisplaybreaks
\begin{align}
\begin{split}
\cD^\alpha\cD_\alpha\Xi &= - \frac{1}{2}\gkf{r}{s}\Big(\frac{1}{2}(\cD^\alpha{\cW_\alpha}^\g{r})(\cD^\beta{\cW_\beta}^\g{s}) + \cF^{ba\g{r}}{\cF_{ba}}^\g{s} + \frac{i}{2}\eps^{dcba}{\cF_{dc}}^\g{r}{\cF_{ba}}^\g{s}\Big) \\
&\quad + \frac{1}{2}\gkf{r}{s}\cW^{\alpha\g{r}}\big(12\Rb\,{\cW_\alpha}^\g{s} + 4i\sig^a_{\alpha\sd{\alpha}}\cD_a\cW^{\sd{\alpha}\g{s}}\big) \\
&\quad -\frac{\del\gkf{r}{s}}{\del\phi^k}\Big(\frac{1}{2}\cD^\alpha\phi^k{\cW_\alpha}^\g{r}\cD^\beta{\cW_\beta}^\g{s} - \cD^\alpha\phi^k{(\sig^{ba})_\alpha}^\beta{\cW_\beta}^\g{r}{\cF_{ba}}^\g{s}\Big) \\
&\quad + \frac{1}{4}\frac{\del\gkf{r}{s}}{\del\phi^k}\cDt^\alpha\cD_\alpha\phi^k\cW^{\beta\g{r}}{\cW_\beta}^\g{s} \\
&\quad + \frac{1}{4}\Big(\frac{\del^2\gkf{r}{s}}{\del\phi^j\del\phi^k} - {\Gamma^i}_{jk}\frac{\del\gkf{r}{s}}{\del\phi^i}\Big)\cD^\alpha\phi^j\cD_\alpha\phi^k\cW^{\beta\g{r}}{\cW_\beta}^\g{s}\,,
\end{split} \\
\nonumber\\
\begin{split}
\cD_\sd{\alpha}\cD^\sd{\alpha}\Xib &= - \frac{1}{2}\gkfb{r}{s}\Big(\frac{1}{2}(\cD_\sd{\alpha}\cW^{\sd{\alpha}\g{r}})(\cD_\sd{\beta}\cW^{\sd{\beta}\g{s}}) + \cF^{ba\g{r}}{\cF_{ba}}^\g{s} - \frac{i}{2}\eps^{dcba}{\cF_{dc}}^\g{r}{\cF_{ba}}^\g{s}\Big) \\
&\quad + \frac{1}{2}\gkfb{r}{s}{\cW_\sd{\alpha}}^\g{r}\big(12R\,\cW^{\sd{\alpha}\g{s}} + 4i\sigb^{a\sd{\alpha}\alpha}\cD_a{\cW_\alpha}^\g{s}\big) \\
&\quad - \frac{\del\gkfb{r}{s}}{\del\phib^\kb}\Big(\frac{1}{2}\cD_\sd{\alpha}\phib^\kb\cW^{\sd{\alpha}\g{r}}\cD_\sd{\beta}\cW^{\sd{\beta}\g{s}} + \cD_\sd{\alpha}\phib^\kb{(\sigb^{ba})^\sd{\alpha}}_\sd{\beta}\cW^{\sd{\beta}\g{r}}{\cF_{ba}}^\g{s}\Big) \\
&\quad + \frac{1}{4}\frac{\del\gkfb{r}{s}}{\del\phib^\kb}\cDt_\sd{\alpha}\cD^\sd{\alpha}\phib^\kb{\cW_\sd{\beta}}^\g{r}\cW^{\sd{\beta}\g{s}} \\
&\quad + \frac{1}{4}\Big(\frac{\del^2\gkfb{r}{s}}{\del\phib^\jb\del\phib^\kb} - {\Gamma^\ib}_{\jb\kb}\frac{\del\gkfb{r}{s}}{\del\phib^\ib}\Big)\cD_\sd{\alpha}\phib^\jb\cD^\sd{\alpha}\phib^\kb{\cW_\sd{\beta}}^\g{r}\cW^{\sd{\beta}\g{s}}\,,
\end{split}
\end{align}
\endgroup
in combination with the explicit form of $\cF_{ba}\p$, lead to the following expressions for the auxiliary component fields:
\begingroup
\allowdisplaybreaks
\begin{align}
\begin{split}
f &= - \frac{1}{4}\gkf{r}{s}\Big(-\Db^\g{r}\Db^\g{s} + \frac{1}{2}\cf^{mn\g{r}}{\cf_{mn}}^\g{s} + \frac{i}{4}\eps^{mnpq}{\cf_{mn}}^\g{r}{\cf_{pq}}^\g{s} \\
&\quad\quad - i\cf^{mn\g{r}}(\psi_m\sig_n\clb^\g{s} + \psib_m\sigb_n\cl^\g{s}) + \frac{1}{2}\eps^{mnpq}{\cf_{mn}}^\g{r}(\psi_p\sig_q\clb^\g{s} + \psib_p\sigb_q\cl^\g{s}) \\
&\quad\quad - \frac{1}{4}(g^{mp}g^{nq} - g^{mq}g^{np} + i\eps^{mnpq})(\psi_m\sig_n\clb^\g{r} + \psib_m\sigb_n\cl^\g{r})(\psi_p\sig_q\clb^\g{s} + \psib_p\sigb_q\cl^\g{s}) \\
&\quad\quad + \Mb(\cl^\g{r}\cl^\g{s}) + 2i(\cl^\g{r}\sig^m\cD_m\clb^\g{s}) - \Db^\g{r}(\cl^\g{s}\sig^m\psib_m) \\
&\quad\quad - (\psib_p\sigb^{mn}\sigb^p\cl^\g{r})(i{\cf_{mn}}^\g{r} + \psi_m\sig_n\clb^\g{s} + \psib_m\sigb_n\cl^\g{s})\Big) \\
&\quad - \frac{1}{4}\del_k\gkf{r}{s}\Big(F^k(\cl^\g{r}\cl^\g{s}) - i\sqrt{2}(\chi^k\cl^\g{r})\Db^\g{s} \\
&\quad\quad - i\sqrt{2} (\chi^k\sig^{mn}\cl^\g{r})(i{\cf_{mn}}^\g{s} + \psi_m\sig_n\clb^\g{s} + \psib_m\sigb_n\cl^\g{s})\Big) \\
&\quad + \frac{1}{8}\delt_j\del_k\gkf{r}{s}(\chi^j\chi^k)(\cl^\g{r}\cl^\g{s})\,,
\end{split} \\
\nonumber\\
\begin{split}
\fb &= - \frac{1}{4}\gkfb{r}{s}\Big(-\Db^\g{r}\Db^\g{s} + \frac{1}{2}\cf^{mn\g{r}}{\cf_{mn}}^\g{s} - \frac{i}{4}\eps^{mnpq}{\cf_{mn}}^\g{r}{\cf_{pq}}^\g{s} \\
&\quad\quad - i\cf^{mn\g{r}}(\psi_m\sig_n\clb^\g{s} + \psib_m\sigb_n\cl^\g{s}) - \frac{1}{2}\eps^{mnpq}{\cf_{mn}}^\g{r}(\psi_p\sig_q\clb^\g{s} + \psib_p\sigb_q\cl^\g{s}) \\
&\quad\quad - \frac{1}{4}(g^{mp}g^{nq} - g^{mq}g^{np} - i\eps^{mnpq})(\psi_m\sig_n\clb^\g{r} + \psib_m\sigb_n\cl^\g{r})(\psi_p\sig_q\clb^\g{s} + \psib_p\sigb_q\cl^\g{s}) \\
&\quad\quad + M(\clb^\g{r}\clb^\g{s}) + 2i(\clb^\g{r}\sigb^m\cD_m\cl^\g{s}) + \Db^\g{r}(\clb^\g{s}\sigb^m\psi_m) \\
&\quad\quad - (\psi_p\sig^{mn}\sig^p\clb^\g{r})(i{\cf_{mn}}^\g{r} + \psi_m\sig_n\clb^\g{s} + \psib_m\sigb_n\cl^\g{s})\Big) \\
&\quad - \frac{1}{4}\del_\kb\gkfb{r}{s}\Big(\Fb^\kb(\clb^\g{r}\clb^\g{s}) + i\sqrt{2}(\chib^\kb\clb^\g{r})\Db^\g{s} \\
&\quad\quad - i\sqrt{2}(\chib^k\sigb^{mn}\clb^\g{r})(i{\cf_{mn}}^\g{s} + \psi_m\sig_n\clb^\g{s} + \psib_m\sigb_n\cl^\g{s})\Big) \\
&\quad + \frac{1}{8}\delt_\jb\del_\kb\gkfb{r}{s}(\chib^\jb\chib^\kb)(\clb^\g{r}\clb^\g{s})\,,
\end{split}
\end{align}
\endgroup
where the covariant derivatives $\delt_j$ and $\delt_\jb$ with respect to the Levi-Civita connection of the metric $g_{k\kb}$ are defined as
\begin{align}
\delt_j\del_k &:= \del_j\del_k - {\Gamma^i}_{jk}\del_i\,, & \delt_\jb\del_\kb &:= \del_\jb\del_\kb - {\Gamma^\ib}_{\jb\kb}\del_\ib\,.
\end{align}
Using the above expressions for the component fields, the Yang-Mills part of the Lagrangian is given by
\begin{align}
\begin{split}
e^{-1}\cL_\text{YM} &= - \frac{1}{4}\re{\gkf{r}{s}}\cf^{mn\g{r}}{\cf_{mn}}^\g{s} + \frac{1}{8}\im{\gkf{r}{s}}\eps_{mnpq}\cf^{mn\g{r}}\cf^{pq\g{s}} \\
&\quad - \frac{i}{2}\big(\gkf{r}{s}\cl^\g{r}\sig^m\nabla_m\clb^\g{s} + \gkfb{r}{s}\clb^\g{r}\sigb^m\nabla_m\cl^\g{s}\big) + \frac{1}{2}\re{\gkf{r}{s}}(\cl^\g{r}\sig^a\clb^\g{s})b_a \\
&\quad + \frac{1}{2}\re{\gkf{r}{s}}\Db^\g{r}\Db^\g{s} + \frac{i}{2}\re{\gkf{r}{s}}\cf^{mn\g{r}}(\psi_m\sig_n\clb^\g{s} + \psib_m\sigb_n\cl^\g{s}) \\
&\quad - \frac{1}{4}\re{\gkf{r}{s}}\eps_{mnpq}\cf^{mn\g{r}}(\psi^p\sig^q\clb^\g{s} - \psib^p\sigb^q\cl^\g{s}) \\
&\quad - \frac{1}{16}\re{\gkf{r}{s}}\Big((\cl^\g{r}\cl^\g{s})(3g^{mn}+2\sigb^{mn})(\psib_m\psib_n) + (\clb^\g{r}\clb^\g{s})(3g^{mn}+2\sig^{mn})(\psi_m\psi_n)\Big) \\
&\quad + \frac{1}{4}\Big(\re{\gkf{r}{s}}(g^{mp}g^{nq} - g^{mq}g^{np}) + \im{\gkf{r}{s}}\eps^{mnpq}\Big)(\psib_m\sigb_n\cl^\g{r})(\psi_p\sig_q\clb^\g{s}) \\
&\quad - \frac{1}{2\sqrt{2}}\Big(\del_k\gkf{r}{s}(\chi^k\sig^{mn}\cl^\g{r}) + \del_\kb\gkfb{r}{s}(\chib^\kb\sigb^{mn}\clb^\g{r})\Big){\cf_{mn}}^\g{s} \\
&\quad + \frac{i}{2\sqrt{2}}\Big(\del_k\gkf{r}{s}(\chi^k\cl^\g{r}) - \del_\kb\gkfb{r}{s}(\chib^\kb\clb^\g{r})\Big)\Db^\g{s} \\
&\quad - \frac{1}{4}\Big(\del_k\gkf{r}{s}(\cl^\g{r}\cl^\g{s})\big(F^k - \frac{i}{2\sqrt{2}}(\psib_m\sigb^m\chi^k)\big) + \del_\kb\gkfb{r}{s}(\clb^\g{r}\clb^\g{s})\big(\Fb^\kb - \frac{i}{2\sqrt{2}}(\psi_m\sig^m\chib^\kb)\big)\Big) \\
&\quad + \frac{i}{2\sqrt{2}}\Big(\del_k\gkf{r}{s}(\psi_m\sig_n\clb^\g{r})(\chi^k\sig^{mn}\cl^\g{s}) + \del_\kb\gkfb{r}{s}(\psib_m\sigb_n\cl^\g{r})(\chib^\kb\sigb^{mn}\clb^\g{s})\Big) \\
&\quad + \frac{1}{8}\Big(\delt_j\del_k\gkf{r}{s}(\chi^j\chi^k)(\cl^\g{r}\cl^\g{s}) + \delt_\jb\del_\kb\gkfb{r}{s}(\chib^\jb\chib^\kb)(\clb^\g{r}\clb^\g{s})\Big)\,,
\end{split} \label{eq:component_fields_action_yang_mills}
\end{align}
with the derivatives
\begin{align}
\begin{split}
\nabla_m{\cl_\alpha}^\g{r} &:= \cD_m{\cl_\alpha}^\g{r} + \frac{i}{2}{e_m}^a b_a {\cl_\alpha}^\g{r}\\
&\;= \del_m{\cl_\alpha}^\g{r} - {\omega_{m\alpha}}^\beta{\cl_\beta}^\g{r} + {\ca_m}^\g{p}\sco{p}{q}{r}{\cl_\alpha}^\g{q} + \frac{1}{4}(K_k\cD_m\varphi^k - K_\kb\cD_m\varphib^\kb){\cl_\alpha}^\g{r} \\
&\quad + \frac{i}{4}g_{k\kb}(\chi^k\sig_m\chib^\kb){\cl_\alpha}^\g{r}\,,
\end{split} \\
\nonumber\\
\begin{split}
\nabla_m\clb^{\sd{\alpha}\g{r}} &:= \cD_m\clb^{\sd{\alpha}\g{r}} - \frac{i}{2}{e_m}^a b_a \clb^{\sd{\alpha}\g{r}}\\
&\;= \del_m\clb^{\sd{\alpha}\g{r}} - \tensor{\omega}{_m^{\sd{\alpha}}_{\sd{\beta}}}\clb^{\sd{\beta}\g{r}} + {\ca_m}^\g{p}\sco{p}{q}{r}\clb^{\sd{\alpha}\g{q}} - \frac{1}{4}(K_k\cD_m\varphi^k - K_\kb\cD_m\varphib^\kb)\clb^{\sd{\alpha}\g{r}} \\
&\quad - \frac{i}{4}g_{k\kb}(\chi^k\sig_m\chib^\kb)\clb^{\sd{\alpha}\g{r}}\,,
\end{split}
\end{align}
where $\cD_m{\cl_\alpha}^\g{r}$ and $\cD_m\clb^{\sd{\alpha}\g{r}}$ are defined in Eq.~\eqref{eq:component_fields_definition_matter_cd_gaugino_1} and \eqref{eq:component_fields_definition_matter_cd_gaugino_2}.

\subsubsection{Summary}
\label{sec:component_fields_actions_summary}
The Lagrangian of the supergravity/matter/Yang-Mills system reads
\begin{align}
\cL &= \cL_\text{supergravity+matter} + \cL_\text{superpotential} + \cL_\text{Yang-Mills}\,,
\end{align}
where the tree contributions are stated in Eq.~\eqref{eq:component_fields_action_sugra_matter}, \eqref{eq:component_fields_action_superpotential} and \eqref{eq:component_fields_action_yang_mills}, respectively. Thus, $\cL$ has the following form:
\begin{align}
\begin{split}
e^{-1}\cL &= -\frac{1}{2}\cR + \frac{1}{2}\eps^{mnpq}(\psib_m\sigb_n\nabla_p\psi_q - \psi_m\sig_n\nabla_p\psib_q) \\
&\quad - g_{k\kb}\,g^{mn}\cD_m\varphi^k\cD_n\varphib^\kb - \frac{i}{2}g_{k\kb}(\chi^k\sig^m\nabla_m\chib^\kb + \chib^\kb\sigb^m\nabla_m\chi^k) \\
&\quad - \frac{1}{4}\re{\gkf{r}{s}}\cf^{mn\g{r}}{\cf_{mn}}^\g{s} + \frac{1}{8}\im{\gkf{r}{s}}\eps_{mnpq}\cf^{mn\g{r}}\cf^{pq\g{s}} \\
&\quad - \frac{i}{2}\Big(\gkf{r}{s}\cl^\g{r}\sig^m\nabla_m\clb^\g{s} + \gkfb{r}{s}\clb^\g{r}\sigb^m\nabla_m\cl^\g{s}\Big) \\
&\quad + e^K\Big(3W\Wb - g^{\kb k}D_k W\,D_\kb\Wb\Big) - \frac{1}{2}e^{K/2}\Big(\Dt_j D_k W(\chi^j\chi^k) + \Dt_\jb D_\kb\Wb(\chib^\jb\chib^\kb)\Big) \\
&\quad + \frac{1}{4} \Big(R_{k\kb j\jb} + \frac{3}{2}g_{k\kb}g_{j\jb}\Big) (\chi^k\chi^j)(\chib^\kb\chib^\jb) - \frac{3}{4}g_{k\kb}\re{\gkf{r}{s}}(\chi^k\cl^\g{r})(\chib^\kb\clb^\g{s}) \\
&\quad - i\sqrt{2}g_{k\kb}(\chi^k\cl^\g{r})(\varphib\gen{r})^\kb + i\sqrt{2}g_{k\kb}(\chib^\kb\clb^\g{r})(\gen{r}\varphi)^k \\
&\quad - \frac{1}{2\sqrt{2}}\Big(\del_k\gkf{r}{s}(\chi^k\sig^{mn}\cl^\g{r}) + \del_\kb\gkfb{r}{s}(\chib^\kb\sigb^{mn}\clb^\g{r})\Big){\cf_{mn}}^\g{s} \\
&\quad + \frac{1}{8}\Big(\delt_j\del_k\gkf{r}{s}(\chi^j\chi^k) + 2g^{\kb k}e^{K/2}\del_k\gkf{r}{s}D_\kb\Wb\Big) (\cl^\g{r}\cl^\g{s}) \\
&\quad + \frac{1}{8}\Big(\delt_\jb\del_\kb\gkfb{r}{s}(\chib^\jb\chib^\kb) + 2g^{\kb k}e^{K/2}\del_\kb\gkfb{r}{s}D_k W\Big) (\clb^\g{r}\clb^\g{s}) \\
&\quad + \frac{1}{16}\Big(6\re{\gkf{r}{p}}\re{\gkf{s}{q}} - g^{\kb k}\del_k\gkf{r}{s}\del_\kb\gkfb{p}{q}\Big)(\cl^\g{r}\cl^\g{s})(\clb^\g{p}\clb^\g{q}) \\
&\quad - \frac{1}{2}\big(\re{\gkf{r}{s}}\big)^{-1}\Big(K_k(\gen{r}\varphi)^k - \frac{i}{2\sqrt{2}}\del_k\gkf{r}{p}(\chi^k\cl^\g{p}) + \frac{i}{2\sqrt{2}}\del_\kb\gkfb{r}{p}(\chib^\kb\clb^\g{p})\Big) \\
&\quad\hspace{2.56cm}\times\Big(K_\jb(\varphib\gen{s})^\jb - \frac{i}{2\sqrt{2}}\del_j\gkf{s}{q}(\chi^j\cl^\g{q}) + \frac{i}{2\sqrt{2}}\del_\jb\gkfb{s}{q}(\chib^\jb\clb^\g{q})\Big) \\
&\quad - \frac{1}{\sqrt{2}}\Big(g_{k\kb}(\psib_m\sigb^n\sig^m\chib^\kb)\cD_n\varphi^k + g_{k\kb}(\psi_m\sig^n\sigb^m\chi^k)\cD_n\varphib^\kb\Big) \\
&\quad - \frac{1}{4}\Big(\psib_m\sigb^m\cl^\g{r} - \psi_m\sig^m\clb^\g{r}\Big)\Big(K_k(\gen{r}\varphi)^k + K_\kb(\varphib\gen{r})^\kb\Big) \\
&\quad + \frac{i}{8\sqrt{2}}\Big(\del_k\gkf{r}{s}(\cl^\g{r}\cl^\g{s})(\psib_m\sigb^m\chi^k) + \del_\kb\gkfb{r}{s}(\clb^\g{r}\clb^\g{s})(\psi_m\sig^m\chib^\kb)\Big) \\
&\quad  + \frac{i}{2\sqrt{2}}\Big(\del_k\gkf{r}{s}(\psi_m\sig_n\clb^\g{r})(\chi^k\sig^{mn}\cl^\g{s}) + \del_\kb\gkfb{r}{s}(\psib_m\sigb_n\cl^\g{r})(\chib^\kb\sigb^{mn}\clb^\g{s})\Big) \\
&\quad + \frac{i}{2}\re{\gkf{r}{s}}\cf^{mn\g{r}}\Big((\psi_m\sig_n\clb^\g{s} + \psib_m\sigb_n\cl^\g{s}) + \frac{i}{2}\eps_{mnpq}(\psi^p\sig^q\clb^\g{s} - \psib^p\sigb^q\cl^\g{s})\Big) \\
&\quad + e^{K/2}\Big(\frac{i}{\sqrt{2}}(\psib_m\sigb^m\chi^k)D_k W + \frac{i}{\sqrt{2}}(\psi_m\sig^m\chib^\kb)D_\kb\Wb - (\psib_m\sigb^{mn}\psib_n)W - (\psi_m\sig^{mn}\psi_n)\Wb\Big) \\
&\quad - \frac{i}{2}g_{k\kb}\eps^{mnpq}(\chi^k\sig_m\chib^\kb)(\psi_n\sig_p\psib_q) - \frac{1}{2}g_{k\kb}\,g^{mn}(\psi_m\chi^k)(\psib_n\chib^\kb) \\
&\quad - \frac{1}{16}\re{\gkf{r}{s}}\Big((\cl^\g{r}\cl^\g{s})(3g^{mn}+2\sigb^{mn})(\psib_m\psib_n) + (\clb^\g{r}\clb^\g{s})(3g^{mn}+2\sig^{mn})(\psi_m\psi_n)\Big) \\
&\quad + \frac{1}{4}\Big(\re{\gkf{r}{s}}(g^{mp}g^{nq} - g^{mq}g^{np}) + \im{\gkf{r}{s}}\eps^{mnpq}\Big)(\psib_m\sigb_n\cl^\g{r})(\psi_p\sig_q\clb^\g{s}) \\
&\quad - \frac{1}{3}\Md\Mbd + \frac{1}{3}\bd^a\bd_a + g_{k\kb}\Fd^k\Fbd^\kb + \frac{1}{2}\re{\gkf{r}{s}}\Dd^\g{r}\Dd^\g{s}\,,
\end{split} \label{eq:component_fields_action_all}
\end{align}
where the diagonalized auxiliary fields are used, which are defined as
\begin{align}
\Md &:= M + 3e^{K/2}W\,, \label{eq:component_fields_action_aux_1}\\
\Mbd &:= \Mb + 3e^{K/2}\Wb\,, \label{eq:component_fields_action_aux_2}\\
\bd_a &:= b_a - \frac{3}{4}g_{k\kb}(\chi^k\sig_a\chib^\kb) + \frac{3}{4}\re{\gkf{r}{s}}(\cl^\g{r}\sig_a\clb^\g{s})\,, \label{eq:component_fields_action_aux_3}\\
\Fd^k &:= F^k + g^{\kb k}e^{K/2}D_\kb\Wb - \frac{1}{4}g^{\kb k}\del_\kb\gkfb{r}{s}(\clb^\g{r}\clb^\g{s})\,, \label{eq:component_fields_action_aux_4}\\
\Fbd^\kb &:= \Fb^\kb + g^{\kb k}e^{K/2}D_k W - \frac{1}{4}g^{\kb k}\del_k\gkf{r}{s}(\cl^\g{r}\cl^\g{s})\,, \label{eq:component_fields_action_aux_5}\\
\begin{split}
\Dd^\g{r} &:= \Db^\g{r} - \frac{1}{2}\big(\re{\gkf{r}{s}}\big)^{-1}\Big(K_k(\gen{s}\varphi)^k + K_\kb(\varphib\gen{s})^\kb \\
&\quad\hspace{4.3cm} - \frac{i}{\sqrt{2}}\del_k\gkf{s}{p}(\chi^k\cl^\g{p}) + \frac{i}{\sqrt{2}}\del_\kb\gkfb{s}{p}(\chib^\kb\clb^\g{p})\Big)\,.
\end{split} \label{eq:component_fields_action_aux_6}
\end{align}
It is convenient to write the Lagrangian in terms of the diagonalized auxiliary fields, since they have trivial equations of motion, which follows from Eqs.~\eqref{eq:eom_var_action_eom_1}--\eqref{eq:eom_var_action_eom_6} by projection to lowest components. Hence, Eq.~\eqref{eq:component_fields_action_all} without the last line represents the Lagrangian after integrating out the auxiliary fields.\footnote{Once the auxiliary fields are integrated out, the action specified by the Lagrangian in Eq.~\eqref{eq:component_fields_action_all} is invariant under supergravity transformations only on-shell.}. If the auxiliary fields are integrated out, the scalar part of the Lagrangian, i.e.\ the scalar potential (up to a minus sign), is given by
\begin{align}
e^{-1}\cL_\text{scalar} &= -e^K\Big(g^{\kb k}D_k W\,D_\kb\Wb - 3W\Wb\Big) - \frac{1}{2}\big(\re{\gkf{r}{s}}\big)^{-1}\Big(K_k(\gen{r}\varphi)^k K_\kb(\varphib\gen{s})^\kb\Big)\,. \label{eq:component_fields_action_scalar}
\end{align}
In the following, the quantities and expressions which are used in Eqs.~\eqref{eq:component_fields_action_all}--\eqref{eq:component_fields_action_aux_6} are summarized.
\begin{itemize}
\item \textbf{Indices:}\\
Spacetime indices: $m,n,p,q$\\
\makebox[3cm][l]{Lorentz indices:}vector: $a,b,c,d$\\
\makebox[3cm]{}Weyl spinor: $\alpha,\sd{\alpha},\beta,\sd{\beta},\gamma,\sd{\gamma},\delta,\sd{\delta}$\\
Matter field indices: $i,\ib,j,\jb,k,\kb,l,\lb$\\
Yang-Mills group (adjoint representation) indices: $\g{p},\g{q},\g{r},\g{s}$
\item \textbf{Tensors:}\\
Minkowski metric: $\eta_{ba}=\diag(-1,+1,+1,+1)$\\
\makebox[3.7cm][l]{Levi-Civita tensors:}$\eps_{abcd}$ with $\eps_{0123}=+1$, $\eps^{0123}=-1$\\
\makebox[3.7cm][l]{}$\eps^{\alpha\beta}$ ($=\eps^{\sd{\alpha}\sd{\beta}}$) with $\eps^{12}=+1$, $\eps_{12}=-1$\\
\makebox[2.9cm][l]{Pauli matrices:}$(\sigb^0,\sigb^1,\sigb^2,\sigb^3)=(+\sig^0,-\sig^1,-\sig^2,-\sig^3)$\\
\makebox[2.9cm][l]{}$\sig^{ab}=\frac{1}{4}(\sig^a\sigb^b-\sig^b\sigb^a)$, $\sigb^{ab}=\frac{1}{4}(\sigb^a\sig^b-\sigb^b\sig^a)$

\item \textbf{Component fields:}\\
Supergravity multiplet: graviton ${e_m}^a$, gravitino ${\psi_m}^\alpha\,(\psib_{m\sd{\alpha}})$, auxiliary fields $M\,(\Mb)$, $b_a$\\
Matter multiplet: complex scalar $\varphi^k\,(\varphib^\kb)$, Weyl fermion ${\chi^k}_\alpha\,(\chib^{\kb\sd{\alpha}})$, auxiliary field $F^k\,(\Fb^\kb)$\\
Yang-Mills multiplet: gauge boson ${\ca_m}^\g{r}$, gaugino ${\cl_\alpha}^\g{r}\,(\clb^{\sd{\alpha}\g{r}})$, auxiliary field $\Db^\g{r}$
\item \textbf{Scalar functions:}\\
K\"ahler potential: $K(\varphi,\varphib)$\\
Superpotential: $W(\varphi)\,\big(\Wb(\varphib)\big)$\\
Gauge kinetic function: $\gkf{r}{s}(\varphi)\,\big(\gkfb{r}{s}(\varphib)\big)$
\item \textbf{Supergravity sector:}\\
The spacetime metric is written as
\begin{align}
g_{mn} &= {e_m}^b{e_n}^a\eta_{ba}\,,
\end{align}
and the corresponding curvature scalar is given by
\begin{align}
\cR = {e_a}^n e^{bm} \big(\del_n{\omega_{mb}}^a - \del_m{\omega_{nb}}^a + {\com{\omega_m}{\omega_n}_b}^a\big)\,,
\end{align}
where the spin connection has the form
\begin{align}
\begin{split}
{e_n}^b e_{pa}{\omega_{mb}}^a = \omega_{mnp} &= +\frac{1}{2}({e_m}^a\del_n e_{pa} - {e_p}^a\del_m e_{na} - {e_n}^a\del_p e_{ma}) \\
&\quad -\frac{1}{2}({e_m}^a\del_p e_{na} - {e_n}^a\del_m e_{pa} - {e_p}^a\del_n e_{ma}) \\
&\quad +\frac{i}{4}(\psi_p\sig_m\psib_n - \psi_m\sig_n\psib_p - \psi_n\sig_p\psib_m) \\
&\quad -\frac{i}{4}(\psi_n\sig_m\psib_p - \psi_m\sig_p\psib_n - \psi_p\sig_n\psib_m)\,.
\end{split}
\end{align}
Written in terms of spinor indices, the spin connection reads
\begin{align}
\begin{gathered}
\omega_{m\sym{2}{\beta\alpha}} = +\frac{1}{2}(\sig^{ba}\eps)_{\beta\alpha} \omega_{mba}\,,\hspace{2cm} \omega_{m\sym{2}{\sd{\beta}\sd{\alpha}}} = +\frac{1}{2}(\eps\sigb^{ba})_{\sd{\beta}\sd{\alpha}} \omega_{mba}\,, \\
\omega_{mba} = -(\eps\sig_{ba})^{\beta\alpha} \omega_{m\sym{2}{\beta\alpha}} + (\sigb_{ba}\eps)^{\sd{\beta}\sd{\alpha}} \omega_{m\sym{2}{\sd{\beta}\sd{\alpha}}}\,.
\end{gathered}
\end{align}
Furthermore, the Pauli matrices with a spacetime index are defined as
\begin{align}
\sig^m &= \sig^a{e_a}^m\,, & \sigb^m &= \sigb^a{e_a}^m\,,
\end{align}
where ${e_a}^m$ is the inverse of ${e_m}^a$, and the Levi-Civita symbol with spacetime indices is specified by
\begin{align}
\eps_{mnpq} &= {e_m}^a{e_n}^b{e_p}^c{e_q}^d\eps_{abcd}\,, & \eps^{mnpq} &= {e_a}^m{e_b}^n{e_c}^p{e_d}^q\eps^{abcd}\,.
\end{align}
The canonical density of spacetime is given by
\begin{align}
e = \det({e_m}^a)\,.
\end{align}
\item \textbf{Matter sector:}\\
In terms of the K\"ahler potential, the K\"ahler metric is written as
\begin{align}
g_{k\kb} = K_{k\kb}\,,
\end{align}
and the inverse matrix is labelled as $g^{\kb k}$. The (non-vanishing) Christoffel symbols of the corresponding Levi-Civita connection are then given by
\begin{align}
{\Gamma^k}_{ij} &= g^{\kb k}g_{i\kb,j}\,, & {\Gamma^\kb}_{\ib\jb} &= g^{\kb k}g_{k\ib,\jb}\,,
\end{align}
and the Riemann tensor has the form
\begin{align}
R_{k\kb j\jb} = g_{k\kb,j\jb} - g^{\lb l}\,g_{k\lb,j}\,g_{l\kb,\jb}\,.
\end{align}
Using the notation
\begin{align}
\del_k &\equiv \frac{\del}{\del\varphi^k}\,, & \del_\kb &\equiv \frac{\del}{\del\varphib^\kb}\,,
\end{align}
the covariant derivative with respect to the Levi-Civita connection reads
\begin{align}
\delt_j\del_k &= \del_j\del_k - {\Gamma^i}_{jk}\del_i\,, & \delt_\jb\del_\kb &= \del_\jb\del_\kb - {\Gamma^\ib}_{\jb\kb}\del_\ib\,.
\end{align}
Furthermore, for the superpotential the derivatives
\begin{align}
D_k W &= W_k + W K_k\,, & D_\kb\Wb &= \Wb_\kb + \Wb K_\kb\,,
\end{align}
and
\begin{align}
\Dt_j D_k W &= (\del_j+K_j)D_k W - {\Gamma^i}_{jk} D_i W\,, & \Dt_\jb D_\kb\Wb &= (\del_\jb+K_\jb)D_\kb\Wb - {\Gamma^\ib}_{\jb\kb} D_\ib\Wb\,,
\end{align}
are defined.
\item \textbf{Yang-Mills sector:}\\
The Hermitian generators of the Yang-Mills gauge group are written as $\gen{r}$, and the (real) structure constants are defined by 
\begin{align}
\com{\gen{p}}{\gen{q}} = i\sco{p}{q}{r}\gen{r}\,.
\end{align}
Furthermore, the field strength tensor is given by
\begin{align}
{\cf_{mn}}^\g{r} &= \del_m{\ca_n}^\g{r} - \del_n{\ca_m}^\g{r} + \sco{p}{q}{r}{\ca_m}^\g{p}{\ca_n}^\g{q}\,.
\end{align}
\item \textbf{Covariant derivatives:}\\
The covariant derivatives of the matter scalar fields have the form
\begin{align}
\cD_m \varphi^k &= \del_m \varphi^k - i{\ca_m}^\g{r}(\gen{r}\varphi)^k\,, \\
\cD_m \varphib^\kb &= \del_m \varphib^\kb + i{\ca_m}^\g{r}(\varphib\gen{r})^\kb\,.
\end{align}
The following derivatives correspond to the covariant derivatives of the respective fields, extended by two terms which contain matter fields:\footnote{Note, the covariant derivatives of the gravitino and its conjugate are not covariant with respect to the Levi-Civita connection of spacetime, since the terms containing the corresponding Christoffel symbols drop out in the antisymmetric combination of the derivatives.}
\begingroup
\allowdisplaybreaks
\begin{align}
\begin{split}
\nabla_m{\chi^k}_\alpha &= \del_m{\chi^k}_\alpha - {\omega_{m\alpha}}^\beta{\chi^k}_\beta - i{\ca_m}^\g{r}(\gen{r}\chi_\alpha)^k + {\Gamma^k}_{ij}{\chi^i}_\alpha\cD_m \varphi^j \\
&\quad - \frac{1}{4}(K_j\cD_m\varphi^j - K_\jb\cD_m\varphib^\jb){\chi^k}_\alpha - \frac{i}{4}g_{j\jb}(\chi^j\sig_m\chib^\jb){\chi^k}_\alpha\,,
\end{split} \\
\nonumber\\
\begin{split}
\nabla_m\chib^{\kb\sd{\alpha}} &= \del_m\chib^{\kb\sd{\alpha}} - \tensor{\omega}{_m^{\sd{\alpha}}_{\sd{\beta}}}\chib^{\kb\sd{\beta}} + i{\ca_m}^\g{r}(\chib^\sd{\alpha}\gen{r})^\kb + {\Gamma^\kb}_{\ib\jb}\chib^{\ib\sd{\alpha}}\cD_m \varphib^\jb \\
&\quad + \frac{1}{4}(K_j\cD_m\varphi^j - K_\jb\cD_m\varphib^\jb)\chib^{\kb\sd{\alpha}} + \frac{i}{4}g_{j\jb}(\chi^j\sig_m\chib^\jb)\chib^{\kb\sd{\alpha}}\,,
\end{split} \\
\nonumber\\
\begin{split}
\nabla_m{\cl_\alpha}^\g{r} &= \del_m{\cl_\alpha}^\g{r} - {\omega_{m\alpha}}^\beta{\cl_\beta}^\g{r} + {\ca_m}^\g{p}\sco{p}{q}{r}{\cl_\alpha}^\g{q} \\
&\quad + \frac{1}{4}(K_k\cD_m\varphi^k - K_\kb\cD_m\varphib^\kb){\cl_\alpha}^\g{r} + \frac{i}{4}g_{k\kb}(\chi^k\sig_m\chib^\kb){\cl_\alpha}^\g{r}\,,
\end{split} \\
\nonumber\\
\begin{split}
\nabla_m\clb^{\sd{\alpha}\g{r}} &= \del_m\clb^{\sd{\alpha}\g{r}} - \tensor{\omega}{_m^{\sd{\alpha}}_{\sd{\beta}}}\clb^{\sd{\beta}\g{r}} + {\ca_m}^\g{p}\sco{p}{q}{r}\clb^{\sd{\alpha}\g{q}} \\
&\quad - \frac{1}{4}(K_k\cD_m\varphi^k - K_\kb\cD_m\varphib^\kb)\clb^{\sd{\alpha}\g{r}} - \frac{i}{4}g_{k\kb}(\chi^k\sig_m\chib^\kb)\clb^{\sd{\alpha}\g{r}}\,,
\end{split} \\
\nonumber\\
\begin{split}
\nabla_n{\psi_m}^\alpha &= \del_n{\psi_m}^\alpha + {\psi_m}^\beta{\omega_{n\beta}}^\alpha \\
&\quad + \frac{1}{4}(K_k\cD_n\varphi^k - K_\kb\cD_n\varphib^\kb){\psi_m}^\alpha + \frac{i}{4}g_{k\kb}(\chi^k\sig_n\chib^\kb){\psi_m}^\alpha\,,
\end{split} \\
\nonumber\\
\begin{split}
\nabla_n\psib_{m\sd{\alpha}} &= \del_n\psib_{m\sd{\alpha}} + \psib_{m\sd{\beta}}\tensor{\omega}{_n^{\sd{\beta}}_{\sd{\alpha}}} \\
&\quad - \frac{1}{4}(K_k\cD_n\varphi^k - K_\kb\cD_n\varphib^\kb)\psib_{m\sd{\alpha}} - \frac{i}{4}g_{k\kb}(\chi^k\sig_n\chib^\kb)\psib_{m\sd{\alpha}}\,.
\end{split}
\end{align}
\endgroup
\end{itemize}

\newpage

\begin{thebibliography}{00}


\bibitem{DeWitt:1992cy}
  B.~S.~DeWitt,
  ``Supermanifolds,''



\bibitem{Moura:2008ra}
  F.~Moura,
  ``Higher-order string effective actions and off-shell $d=$4 supergravity,''
  Springer Proc.\ Phys.\  {\bf 134} (2010) 317
  [arXiv:0801.4058 [hep-th]].



\bibitem{Wess:1992cp}
  J.~Wess and J.~Bagger,
  ``Supersymmetry and supergravity,''



\bibitem{Buchbinder:1998qv}
  I.~L.~Buchbinder and S.~M.~Kuzenko,
  ``Ideas and methods of supersymmetry and supergravity: Or a walk through superspace,''
  Bristol, UK: IOP (1998) 656 p



\bibitem{Binetruy:2000zx}
  P.~Binetruy, G.~Girardi and R.~Grimm,
  ``Supergravity couplings: A Geometric formulation,''
  Phys.\ Rept.\  {\bf 343} (2001) 255
  [hep-th/0005225].



\bibitem{Muller:1985vga}
  M.~Muller,
  ``Supergravity in U(1) Superspace With a Two Form Gauge Potential,''
  Nucl.\ Phys.\ B {\bf 264} (1986) 292.



\bibitem{Groeger:2010}
  J.~Groeger,
  ``Differential Geometry of Supermanifolds,''
  Lecture series (2010).
  
  
  
\bibitem{Fioresi:2010}
  R.~Fioresi,
  ``Supergeometry Lectures,''
  Lecture series (2010).



\bibitem{Helein:2008}
  F.~Hélein,
  ``An introduction to supermanifolds and supersymmetry,''
  (2008).



\bibitem{Goertsches:2006pna}
  O.~Goertsches,
  ``Riemannian Supergeometry,''
  Math.\ Z.\  {\bf 260} (2008) 557
  [math/0604143 [math.DG]].



\bibitem{Freedman:2012zz}
  D.~Z.~Freedman and A.~Van Proeyen,
  ``Supergravity,''



\bibitem{Dragon:1978nf}
  N.~Dragon,
  ``Torsion and Curvature in Extended Supergravity,''
  Z.\ Phys.\ C {\bf 2} (1979) 29.



\bibitem{Percacci:1998ag}
  R.~Percacci and E.~Sezgin,
  ``Properties of gauged sigma models,''
  hep-th/9810183.



\bibitem{Abdalla:1985nm}
  E.~Abdalla and M.~Forger,
  ``Integrable Nonlinear $\sigma$ Models With Fermions,''
  Commun.\ Math.\ Phys.\  {\bf 104} (1986) 123.



\bibitem{Bagger:1985pw}
  J.~Bagger, D.~Nemeschansky and S.~Yankielowicz,
  ``Anomaly Constraints On Nonlinear Sigma Models,''
  Nucl.\ Phys.\ B {\bf 262} (1985) 478.



\bibitem{deWit:1995tf}
  B.~de Wit and A.~Van Proeyen,
  ``Isometries of special manifolds,''
  hep-th/9505097.



\bibitem{Lindstrom:2012ci}
  U.~Lindstrom,
  ``Supersymmetric Sigma Model geometry,''
  arXiv:1207.1241 [hep-th].



\bibitem{Moroianu:2007}
  A.~Moroianu,
  ``Lectures on K\"ahler Geometry (London Mathematical Society Student Texts),''
  Cambridge University Press (2007)



\bibitem{Fre:1995dw}
  P.~Fre,
  ``Lectures on special Kahler geometry and electric - magnetic duality rotations,''
  Nucl.\ Phys.\ Proc.\ Suppl.\  {\bf 45BC} (1996) 59
  [hep-th/9512043].



\bibitem{Grisaru:1997ub}
  M.~T.~Grisaru, M.~E.~Knutt-Wehlau and W.~Siegel,
  ``A Superspace normal coordinate derivation of the density formula,''
  Nucl.\ Phys.\ B {\bf 523} (1998) 663
  [hep-th/9711120].


\end{thebibliography}

\end{document}